\documentclass[fontsize=12pt,a4paper,headings=normal,
twoside=false,leqno,parskip=half-,abstract=true]{scrartcl}

\usepackage{amssymb,amsfonts,amsthm,amsmath,mathrsfs,bm}

\usepackage{graphicx,float}
\usepackage[mathscr]{eucal}
\usepackage[all,cmtip]{xy}

\usepackage{dashbox}

\usepackage{tabularray}

\usepackage{hyperref}
\hypersetup{
 pdftitle={Bianchi VI-19},
 pdfauthor={PL,CU},
 colorlinks=true,
 linkcolor=blue,
 citecolor=blue,
 filecolor=blue,
 urlcolor=blue
 }

\usepackage{tikz}
\usepackage[format=plain,labelfont=bf,font=small]{caption}
\usepackage{subcaption}
\usepackage{xcolor,colortbl}
\definecolor{Gray}{gray}{0.9}
\usepackage{float}
\usetikzlibrary{arrows, arrows.meta, calc, intersections, patterns, decorations.pathreplacing, decorations.markings}
\tikzset{
decoration={markings,mark=at position 0.67 with {\arrow[thick,color=gray]{latex}}}
}
\tikzset{>={latex[width=1mm,length=1mm,color=gray]}}

\usepackage{cancel}
\usepackage{oubraces}
\usepackage{wasysym} 

\usepackage{comment}

\renewcommand{\u}{\mathsf{u}}

\newcommand{\textfrac}[2]{{\textstyle \frac{#1}{#2}}}

\theoremstyle{plain}
\newtheorem{theorem}{Theorem}[section]

\newtheorem{lemma}[theorem]{Lemma}
\newtheorem{proposition}[theorem]{Proposition}
\newtheorem{conjecture}[theorem]{Conjecture}

\theoremstyle{remark}

\def\bn{\pmb{n}}
\def\bxi{\pmb{\xi}}

\def\la{\langle}
\def\ra{\rangle}

\newcommand{\sfrac}[2]{{\textstyle{\frac{#1}{#2}}}}

\renewcommand{\vector}[1]{\bm{#1}}

\newcommand{\ue}{\check{u}}

\title{\LARGE{Oscillatory spacelike singularities:\\
The Bianchi type $\mathrm{VI}_{-1/9}$ vacuum models}}

\begin{document}

\author{Phillipo Lappicy* and Claes Uggla**\\}

\date{}
\maketitle
\thispagestyle{empty}

\vfill

$\ast$\\
Departamento de Análisis Matemático y Matemática Aplicada\\ Universidad Complutense de Madrid\\
28040, Madrid, Spain\\
\texttt{philemos@ucm.es}\\
(corresponding author)\\
\\
$\ast \ast$\\
Department of Physics\\
Karlstad University\\
S-65188 Karlstad, Sweden\\
\texttt{claes.uggla@kau.se}\\


\newpage
\pagestyle{plain}
\pagenumbering{arabic}
\setcounter{page}{1}

\begin{abstract}

The Bianchi type $\mathrm{VI}_{-1/9}$, $\mathrm{VIII}$ and $\mathrm{IX}$
vacuum models all have 4-dimensional Hubble-normalized state spaces and
are expected to have a generic initial oscillatory singularity, but the
invariant boundary sets responsible for the oscillations are much more
complicated for type $\mathrm{VI}_{-1/9}$ than those of type $\mathrm{VIII}$
and $\mathrm{IX}$. For the first time, we explicitly solve the equations
on these type $\mathrm{VI}_{-1/9}$ boundary sets and also introduce a new
graph representation of the associated network of heteroclinic chains
(i.e., sequences of solutions describing the oscillations). In particular,
we give examples of networks of cyclic heteroclinic chains and
show that only some of these cyclic heteroclinic chains are
asymptotically relevant.

\end{abstract}

\tableofcontents

\pagebreak

\section{Introduction\label{sec:intro}}

An important feature of Einstein's field equations is that they result in space-time
singularities under quite general conditions, as shown by the Penrose-Hawking singularity theorems and
variations thereof, see~\cite{Penrose65,HawkingPenrose70,Geroch66,HawkingEllis73,Senovilla15}.
These theorems, however, say little about the nature of generic space-time singularities.

To make mathematically rigorous progress about this issue, one has to make more
stringent conditions on the matter content. Notably scalar fields (with certain
conditions on the scalar field potential) and `stiff' perfect fluids, with an
equation of state such that the speed of sound is equal to that of light, result
in generic `quiescent' spacelike singularities. A major reason for progress in 
this area, as illustrated by the following
works~\cite{andren01,RodSpeck,FourRodSpeck,Nutzi,RingstromQuiet} and references 
therein, is that each spatial point evolves independently of its neighbours 
toward the singularity where dimensionless variables reach specific temporal 
limits (i.e., models with a quiescent singularity exhibit temporally asymptotic 
spatially local self-similarity). However, for a broad range of other types 
of matter, the situation is much more complicated.

As regards generic spacelike singularities\footnote{We focus on spacelike singularities;
for null singularities see~\cite{OriFlanagan,Luk,Dafermos}. For the interplay
between spacelike and null singularities, i.e., a singularity with both null and
spacelike pieces, see~\cite{HodPiran,BradySmith,limetal06,Chesleretal,VandeMoortel}.
For surveys on space-time singularities and their types, see~\cite{RendallSurvey,Luk2}
and references therein.}, the heuristic analysis of Belinski\v{\i}, Khalatnikov and
Lifshitz (BKL), see~\cite{lk63,bkl70,bkl82,mis69a,ugg13,ugg13c,BelHenneaux17,rin25},
suggests the following conjectures for `soft' perfect fluids (SPFs), i.e., perfect
fluids with an equation of state such that the speed of sound is less than that of
light:
\begin{itemize}
\item [(i)] \textit{Asymptotic vacuum domination}: For SPFs, generic spacelike singularities in general relativity are vacuum dominated, i.e., generically SPFs asymptotically become test fields and the dynamical effect of gravity becomes dominant over that of matter, a phenomenon that is often referred to as `matter does not matter'.
%
%
\item [(ii)] \textit{Asymptotic spatial locality}:
For SPFs and vacuum models, spatial points evolve independently of their
neighbours asymptotically toward a generic spacelike
singularity.\footnote{In the conformal Hubble-normalized orthonormal 
frame approach~\cite{rohugg05}, asymptotic spatial locality corresponds to that
the Hubble-normalized spatial frame derivatives for individual spatial
points tend to zero asymptotically. Thus, the temporal
evolution of these spatial points is asymptotically described
by the invariant `local boundary subset', where equations coincide with Bianchi cosmologies~\cite{ugg13,ugg13c,limetal06,heietal09}.}
\item [(iii)] \textit{Asymptotic temporal oscillations}: For SPFs and vacuum models, 
the spatial pointwise asymptotic generic dynamics is governed by oscillating
sequences of Bianchi type I vacuum (Kasner) states mediated by Bianchi type II vacuum
solutions (bounces), which induce a chaotic discrete \emph{BKL Kasner map},
see~\cite{heiugg09a,BegDut22} and references therein.\footnote{In addition to the
oscillatory local BKL behavior, there are also joint spike oscillations, as described
in~\cite{heietal12} where the associated inhomogeneous explicit solutions were 
found, see~\cite{lim22} and references therein for further details on spikes.
For other recent work on the existence of asymptotic inhomogeneous Kasner-like solutions,
see~\cite{FourLuk}, while~\cite{Li} proves the existence of one BKL `bounce' in
an inhomogeneous setting.}
\end{itemize}

Asymptotic spatial locality, as regards generic spacelike singularities,
suggests that an understanding of spatially homogeneous (SH) cosmologies
is essential for \emph{inhomogeneous} cosmologies. The most general SH
vacuum models are the Bianchi type $\mathrm{VIII}$, $\mathrm{IX}$ and
$\mathrm{VI}_{-1/9}$ models.
In the Hubble- (i.e expansion-) normalized orthonormal frame formalism, 
see~\cite{waiell97}, the Bianchi type $\mathrm{VIII}$, $\mathrm{IX}$ 
and $\mathrm{VI}_{-1/9}$ vacuum models possess 4-dimensional state 
spaces and are expected to generically have initial oscillatory singularities. 

All Bianchi cosmologies, except Bianchi types $\mathrm{VIII}$ and $\mathrm{IX}$, have
a $G_2$ symmetry subgroup with two commuting spacelike Killing vector fields.
This Abelian $G_2$ symmetry group acts orthogonally transitively in all cases, except
for the general type $\mathrm{VI}_{-1/9}$ models. As a consequence, the general
vacuum type $\mathrm{VI}_{-1/9}$ models are the only SH models with a SH limit
of the general \emph{inhomogeneous} vacuum models with an Abelian $G_2$ symmetry
group.\footnote{The general inhomogeneous Abelian $G_2$ models occur as a natural
special case of inhomogeneous models without any symmetries when they are expressed
in a spatial Iwasawa frame. The associated space-time symmetry hierarchies were described
in terms of invariant sets in the conformally Hubble-normalized Iwasawa frame approach
given in~\cite{heietal09}; see also~\cite{dametal03} for the configuration space
representation and cosmological billiard description.}
Since the general Abelian $G_2$ vacuum models are the simplest spatially 
inhomogeneous models that are expected to possess a spacelike initial 
singularity generically obeying the previous BKL conjectures (i), (ii), (iii), 
obtaining an understanding of the vacuum Bianchi type $\mathrm{VI}_{-1/9}$ models
(rather than those of type $\mathrm{VIII}$ and $\mathrm{IX}$) is arguably 
the most natural next step toward a deeper understanding of generic spacelike 
oscillatory singularities. However, all proofs about SH oscillatory
singularities have been obtained for Bianchi type $\mathrm{VIII}$ and
$\mathrm{IX}$ and no analogous rigorous results are known for the much
more complicated general Bianchi type $\mathrm{VI}_{-1/9}$ vacuum models.

The presently existing results about generic singularities for Bianchi type
$\mathrm{VIII}$ and $\mathrm{IX}$ illustrate how mathematics as a cumulative
science builds and relies on earlier successive contributions, involving
development of heuristics, formalisms, conjectures, and, finally, rigorous 
mathematical proofs that depend on these developments, but also on inherent 
mathematical structures of the problem at hand. The first heuristic results 
about Bianchi type $\mathrm{IX}$ were obtained by Belinski\v{\i} et 
al.~\cite{bkl70} and Misner in the late 1960s, where the latter used 
Hamiltonian methods~\cite{mis69a}. By considering the potential in the 
Hamiltonian, Misner showed that generically solutions will oscillate
between different Bianchi type $\mathrm{I}$ (i.e., Kasner) solutions by
means of Bianchi type II solutions toward the initial singularity, a
phenomenon he referred to as Mixmaster oscillations. Notably, Misner's
Hamiltonian approach was later followed up by, e.g., Jantzen and
coworkers, see~\cite{jan01,waiell97} and references therein, and much later
in the work on the Hamiltonian billiard description of singularities by
Damour et al.~\cite{dametal03}.

Instead of using a Hamiltonian approach, Belinski\v{\i} et al.~\cite{lk63,bkl70}
obtained heuristic results similar to Misner's by using a metric approach
and second order ordinary differential equations. Notably, their results
included the introduction of the gauge-invariant Kasner parameter $u$, which uniquely
describes the different Kasner solutions, and the gauge-invariant BKL Kasner map,
induced by the Bianchi type $\mathrm{II}$ solutions. This map was subsequently
shown to lead to discrete chaotic dynamics~\cite{khaetal85}, a phenomenon
that has attracted considerable attention; see also the discussion and
references in~\cite{BegDut22} for how chaos pertains to the singularity
attractor in Bianchi types $\mathrm{VIII}$ and $\mathrm{IX}$ cosmology.

An important next step to understand the Bianchi type $\mathrm{VIII}$ and
$\mathrm{IX}$ models was the introduction of dimensionless 
Hubble-normalized variables and an associated dynamical 
systems formalism for these models by Wainwright in the late 1980s, 
published in~\cite{waihsu89}, building on earlier work by, e.g., 
Collins~\cite{collins71} and Bogoyavlenskii and Novikov~\cite{BN73}
in the early 1970s. Much later, Wainwright's Hubble-normalized 
dynamical systems approach to SH cosmology was generalized by the 
geometric Hubble-normalized conformal orthonormal frame approach 
in order to deal with inhomogeneous cosmological models~\cite{rohugg05}.

Wainwright's approach to SH cosmology was eventually merged with
Hamiltonian methods in the book \emph{Dynamical Systems in
Cosmology} from the 1990s in~\cite{waiell97}. This made it possible to obtain new
monotonic functions that enabled proofs about global and asymptotic
behaviour for many models, although not for the generic singularity
of the vacuum and SPF models of Bianchi type $\mathrm{VIII}$, $\mathrm{IX}$
and $\mathrm{VI}_{-1/9}$. However, the book contained singularity attractor
conjectures for Bianchi types $\mathrm{VIII}$ and $\mathrm{IX}$. These
were based on monotonic functions and results about the lower strata of
the hierarchical Bianchi types $\mathrm{VIII}$ and $\mathrm{IX}$ invariant
boundary subset stratification and the network of heteroclinic chains\footnote{A heteroclinic
orbit is a solution trajectory that connects two different fixed points;
a heteroclinic chain consists of a concatenation of heteroclinic orbits,
where the $\omega$-limit of one heteroclinic orbit is the $\alpha$-limit
of the next one in the chain; a heteroclinic network consists of a union
of heteroclinic chains.} that was constructed on the union of the vacuum
Bianchi type $\mathrm{I}$ and $\mathrm{II}$ subsets, where the former
yielded a circle of Kasner fixed points, denoted by $\mathrm{K}^\ocircle$.
In addition, numerical investigations suggested that solutions generically
asymptotically approximate these heteroclinic chains, which described oscillations
between different Kasner fixed points on $\mathrm{K}^\ocircle$ by means of
concatenated sequences of Bianchi type $\mathrm{II}$ heteroclinic orbits.

Although no theorems about generic SH singularities were produced in
the book~\cite{waiell97}, it nevertheless provided the stage for the 
first proofs about the global asymptotics toward the generic 
singularity in Bianchi type $\mathrm{IX}$ and, to a less extent,
in type $\mathrm{VIII}$ by Ringstr\"om at the beginning of this
century~\cite{rin00,rin01}, see also~\cite{heiugg09a,heiugg09}. 
Although an impressive mathematical achievement, these results do not 
say anything about the specific asymptotics of generic solutions 
such as the attraction to any of the heteroclinic chains that
are responsible for Kasner oscillations. Lately, a series of new
types of proofs, based on these earlier works, have demonstrated the stability of
various heteroclinic chains on the union of $\mathrm{K}^\ocircle$ and the
vacuum Bianchi type $\mathrm{II}$ subsets of these models,
see~\cite{Beguin10,Lieb11,Lieb13,ReitererTrubowitz,bre16,BegDut22}.
Notably, all these theorems rely on mathematical structures that
are inherited from the properties of the space-time symmetry groups of
Bianchi types $\mathrm{VIII}$ and $\mathrm{IX}$.

Unfortunately, due to complicated, e.g. twisting, Killing vector fields (KVFs),
these mathematical structures do not exist for the general Bianchi type 
$\mathrm{VI}_{-1/9}$ vacuum model. This explains why there are no oscillatory
singularity theorems for this model, in spite of that it was first
analyzed using a Hubble-normalized dynamical systems formulation
in 2003, see~\cite{hewetal03}, following an analysis of special cases
in~\cite{hew91} and other related Bianchi models
in~\cite{hewetal93}.\footnote{The work in~\cite{hewetal93} was followed up
rigorously in~\cite{Radermacher}. Bianchi type $\mathrm{VI}_{-1/9}$ tilted
perfect fluid models were studied in~\cite{heretal07}, with a focus
on evolution at late times away from the initial singularity. Finally,
rigorous results on quiescent singularities for stiff perfect fluids
in type $\mathrm{VI}_{-1/9}$ were obtained in~\cite{Groeniger}.}
To explain the difficulties in proving oscillatory singularity
theorems for the Bianchi type $\mathrm{VI}_{-1/9}$ vacuum model, let us
list some of the main differences in innate mathematical structures between the
Bianchi type $\mathrm{VIII}$, $\mathrm{IX}$ and $\mathrm{VI}_{-1/9}$ vacuum
models in the context of the Hubble-normalized dynamical systems formulation:
\begin{itemize}
\item[(1)] Due to different KVF structures,
the type $\mathrm{VIII}$ and $\mathrm{IX}$ vacuum models
admit diagonal extrinsic and intrinsic spatial curvatures and
thereby a diagonal Einstein tensor, while this is not the case for
type $\mathrm{VI}_{-1/9}$. This leads to that the Codazzi
constraints are identically zero for the former models, while
type $\mathrm{VI}_{-1/9}$ results in a non-zero Codazzi constraint,
which complicates the dynamics.
\item[(2)] Due to different KVF structures, the Bianchi type
$\mathrm{VIII}$ and $\mathrm{IX}$ models admit closed spatial topologies,
whereas this is not the case for type $\mathrm{VI}_{-1/9}$.
As a consequence, the former admit a Hamiltonian formulation, while
this is not the case for the latter.

The combination of a Hamiltonian with the
space-time symmetry group properties of Bianchi type $\mathrm{VIII}$
and $\mathrm{IX}$ greatly simplifies the dynamics of these models.
For example, it results in a simple hierarchical stratification
of invariant sets with monotonic functions that relates the different
strata, where each strata consists of subsets that are related by
discrete symmetries. For the underlying group theoretical reasons why
the simplifying mathematical innate structures exist, which is why
singularity theorems are possible in Bianchi type $\mathrm{VIII}$
and $\mathrm{IX}$, see~\cite{heiugg10}. Unfortunately, none of 
these innate mathematical structures exist for the vacuum type 
$\mathrm{VI}_{-1/9}$ model.
\item[(3)] Due to their KVF strucure, the Bianchi type $\mathrm{VIII}$
and $\mathrm{IX}$ vacuum models have invariant boundaries that contain a 
permutation invariant union of subsets consisting of the Kasner circle
$\mathrm{K}^\ocircle$ and three identical Bianchi type $\mathrm{II}$ subsets. 
This results in that linearization at $\mathrm{K}^\ocircle$ only yields 
a single unstable eigenvalue connected to each Bianchi type $\mathrm{II}$ 
subset. These features play a crucial role for the new generation of 
type $\mathrm{VIII}$ and $\mathrm{IX}$ singularity
theorems~\cite{Beguin10,Lieb11,Lieb13,ReitererTrubowitz,bre16,BegDut22}.

None of these simplifying mathematical structures exist in Bianchi type
$\mathrm{VI}_{-1/9}$. The complicated type $\mathrm{VI}_{-1/9}$ KVF
structure results in more general representations of the Bianchi type
$\mathrm{I}$ and $\mathrm{II}$ vacuum models, for which the permutation
symmetry is broken; as a consequence, linearization at the
$\mathrm{K}^\ocircle$ subset of type $\mathrm{VI}_{-1/9}$ results in
several unstable eigenvalues instead of single ones as in type
$\mathrm{VIII}$ and $\mathrm{IX}$. These features lead to complicated networks of heteroclinic chains and phenomena that do not
occur in Bianchi types $\mathrm{VIII}$ and $\mathrm{IX}$.
\end{itemize}

The goal of this paper is to present new rigorous results about the Bianchi
type $\mathrm{VI}_{-1/9}$ vacuum models and to further highlight the differences
with type $\mathrm{VIII}$ and $\mathrm{IX}$. In particular, we will identify
new phenomena that do not exist for the latter models.
The outline and results of the paper are as follows:

\emph{Section}~\ref{sec:strat} contains a derivation of the Hubble-normalized
dynamical system for the general Bianchi type $\mathrm{VI}_{-1/9}$ vacuum model,
and a discussion about the mathematical and physical meaning of the state vector
variables. It also includes a list of the fixed points and their local stability
properties together with a description of the model's hierarchical invariant
subset stratification structure, given in Figure~\ref{FIG:hierarchy}.

Then the invariant subsets are described in more detail, which involves 
new results in Appendix~\ref{app:inv}. In particular, for the first time, we obtain  
explicit solutions for the Bianchi type $\mathrm{VI}_{-1/9}$ KVF induced 
spatially frame-rotating Bianchi type $\mathrm{I}$ and $\mathrm{II}$ subsets,
which are necessary in order to construct the heteroclinic network that 
is believed to describe the general oscillatory Bianchi type 
$\mathrm{VI}_{-1/9}$ vacuum singularity. In addition, we derive monotonic 
functions that yield new rigorous dynamical results for some of the
invariant subsets. This is akin to some of the results
in~\cite{waiell97} that formed the foundation for subsequent
singularity theorems for Bianchi types $\mathrm{VIII}$ and $\mathrm{IX}$.

\emph{Section}~\ref{sec:discdyn} begins with Conjectures~\ref{conj1} and
\ref{conj2} (motivated by earlier heuristic considerations and numerical
investigations in~\cite{hewetal03,lim04}), which can be informally
summarised as follows:

\textbf{Singularity attractor conjecture for the Bianchi type $\mathrm{VI}_{-1/9}$ vacuum model:}\\
The asymptotic limit toward the singularity for generic solutions of
the general Bianchi type $\mathrm{VI}_{-1/9}$ vacuum model
resides in the union of the invariant Bianchi type $\mathrm{I}$ and
$\mathrm{II}$ boundary subsets, or a subset thereof, leading to
an oscillatory singularity with alternating Kasner states.

This is followed by a discussion about the discrete chaotic dynamics
generated by the gauge-invariant BKL Kasner map acting on the
Kasner parameter $u$, which yield Kasner sequences $\{u_0,u_1,u_2,\dots\}$.
It is then shown how the gauge-invariant Kasner parameters $u$ and the BKL
map is translated into heteroclinic orbits on the Bianchi type
$\mathrm{I}$ and $\mathrm{II}$ subsets, thereby yielding new maps
of fixed points on $\mathrm{K}^\ocircle$. Then a new graph
representation is introduced, which translates gauge-invariant
Kasner sequences $\{u_0, u_1, u_2,\dots\}$ into directed networks of heteroclinic chains on the union of the Bianchi type
$\mathrm{I}$ and $\mathrm{II}$ subsets. 
The graph representation then leads to an algorithmic procedure showing that some heteroclinic chains in a network associated with a Kasner sequence $\{u_0, u_1, u_2,\dots\}$ cannot be reached from the rest of the network, as described in Proposition~\ref{prop:removal}. Removing these unreachable objects yields what we call a stable subnetwork of heteroclinic cycles.
This is a new phenomenon that does not exist in Bianchi type
$\mathrm{VIII}$ and $\mathrm{IX}$.


\emph{Section}~\ref{sec:cyclic} illustrates how the previously mentioned
algorithmic graph procedure can be used to produce the network of heteroclinic cycles that correspond to all possible periodic
sequences of $u$, which occur when $u$ is a quadratic irrational
number, generated by the BKL Kasner map. Moreover, we prove and discuss
the following informal theorem, stated rigorously in Theorem~\ref{thm:cyclic}:

\textbf{Theorem:} For all periodic Kasner sequences $\{u_0, u_1,\dots, u_{p-1}\}$
with a period $p\geq 2$, there are heteroclinic chains in
the network of heteroclinic cycles constructed
from this sequence that cannot be asymptotically approximated by some solutions, leaving a
stable subnetwork of heteroclinic cycles that may asymptotically
attract type $\mathrm{VI}_{-1/9}$ solutions.


\emph{Section}~\ref{sec:comparison} further discusses significant
mathematical differences between the Bianchi type $\mathrm{VI}_{-1/9}$
and the type $\mathrm{VIII}$, $\mathrm{IX}$ vacuum models. These
highlight that the type $\mathrm{VI}_{-1/9}$ vacuum model involves
features that result in completely new challenges when it comes to
attempting to prove singularity theorems for them.

\emph{Appendix~\ref{app:toy}} contains a Hamiltonian toy model that has
the same Bianchi type $\mathrm{I}$ and $\mathrm{II}$ boundary subsets 
as the Bianchi type $\mathrm{VI}_{-1/9}$ vacuum model, and thereby the 
same type $\mathrm{I}$ and $\mathrm{II}$ heteroclinic network. In contrast 
to type $\mathrm{VI}_{-1/9}$, the toy model does not have a constraint 
corresponding to the Codazzi equations. Moreover, the Hamiltonian makes 
it possible to derive a monotonic function that shows that the singularity 
attractor of the toy model must reside on a union of boundary subsets. 
This, however, is, unfortunately, not sufficiently restrictive to rule out 
other behaviour than that induced by the, with type $\mathrm{VI}_{-1/9}$ 
common, type $\mathrm{I}$ and $\mathrm{II}$ boundary subsets.

Nevertheless, the toy model is mathematically closer
to the vacuum Bianchi types $\mathrm{VIII}$ and $\mathrm{IX}$ models
than those of type $\mathrm{VI}_{-1/9}$ --- if one cannot prove
oscillatory singularity theorems for the toy model, then one is unlikely
to be able to do so for the type $\mathrm{VI}_{-1/9}$ vacuum model,
which makes it even more unlikely to obtain proofs about
the oscillatory singularity for inhomogeneous cosmologies. Thus,
the present paper provides a foundation and first step toward
solving a hierarchy of increasingly challenging future mathematical
problems.

\section{The type $\mathrm{VI}_{-1/9}$ dynamical system and its stratification\label{sec:strat}}

\subsection{Derivation of the dynamical system\label{dynsysder}}

To obtain the dynamical system of ordinary differential equations for the
Bianchi type~$\mathrm{VI}_{-1/9}$ vacuum models, we first situate these models
in the general context of the SH vacuum Bianchi models. To do so, we use
the orthonormal frame approach originally introduced by Ellis and MacCallum
in~\cite{ellmac69} and used in the dynamical
systems in cosmology book~\cite{waiell97}.

In SH cosmology the space-time manifold $M$ is regarded as a parametrized
set of copies of a 3-dimensional (3D) real Lie group $G_3$ that acts as a
simply transitive symmetry group on $M$ with 3D spacelike orbits, which form a
geodesically parallel family of SH time slices, see, e.g.~\cite{waiell97,jan01}
and also~\cite{mac73} for a detailed pedagogical group theoretical review
of SH cosmology. To choose an appropriate orthonormal frame for these so-called
Bianchi models, let $\bn$ be the unit vector field normal to the orbits of the
$G_3$ group of isometries $\bxi_\alpha$, $\alpha = 1,2,3$, where $\bn$ is invariant
under the group: $[\bxi_\alpha,\bn] = 0$. It follows that $\bn$ is tangent
to a geodesic congruence that is orthogonal to the spatial symmetry hypersurfaces,
which yields a natural choice for a synchronous clock time coordinate $t$,
where the spacelike group orbits are given by
$t = \mathrm{constant}$.\footnote{This corresponds to setting the shift vector to zero
and the lapse function to one. There is, however, an alternative by using a so-called
automorphism adapted shift vector in order to explicitly reveal the true degrees of
freedom for the various Bianchi models, see~\cite{jan01,janugg99} and references therein.}
We now choose a triad of spacelike orthonormal vectors $\vector{e}_\alpha$
that are tangent to the group orbits and commuting with the Killing vector fields,
$[\vector{e}_\alpha,\bxi_\beta] = 0$. In addition, we choose $\vector{e}_0 = \bn$,
thereby resulting in a complete group-invariant orthonormal frame.

The commutators of the orthonormal frame vectors subsequently yield the variables
in the orthonormal frame approach. Since the unit normal $\vector{e}_0$ to the
spatial symmetry hypersurfaces by definition is hypersurface forming and tangent to
a geodesic congruence due to spatial homogeneity, the commutators can be written as
\begin{subequations}\label{comm}
\begin{align}
\label{dcomts0}
[\vector{e}_{0}, \vector{e}_{\alpha}] &=
 -(H\delta_{\alpha}\!^{\beta} + \sigma_{\alpha}\!^{\beta} +
\epsilon_{\alpha}\!^{\beta}\!_{\gamma}\Omega^{\gamma})
\vector{e}_{\beta}, \\
\label{dcomtsa} [\vector{e}_{\alpha}, \vector{e}_{\beta}] &= {c^\gamma\!_{\alpha\beta}}\vector{e}_{\gamma} =
\left(\epsilon_{\alpha\beta\delta}n^{\delta\gamma} + 2a_{[\alpha}\,\delta_{\beta]}\!^{\gamma}\right)\vector{e}_{\gamma},
\end{align}
\end{subequations}
where $H(t)$ is the Hubble variable (which is $-1/3$ of the trace of the extrinsic curvature) and
$\sigma_{\alpha\beta}(t)$ is the shear (which is the negative trace-free part of the extrinsic
curvature) associated with $\bn$; $\Omega^\alpha(t)$ is the Fermi
rotation which describes how the spatial frame rotates with respect to a
gyroscopically fixed so-called Fermi frame;\footnote{The sign here in the
definition of $\Omega^\alpha$ is the same as in~\cite{heietal09,heietal12},
but opposite to that in~\cite{waiell97}.} finally, the decomposition of the
spatial commutator functions $c^\gamma\!_{\alpha\beta}$ into
$n^{\alpha\beta}(t) = n^{(\alpha\beta)}(t)$ and $a_\alpha(t) = \frac12 c^\gamma\!_{\alpha\gamma}$
is due to Sch{\"u}cking and Behr and is partly motivated by that the Jacobi identities for the spatial
triad take the simple form $n^{\alpha\beta}a_\beta = 0$.\footnote{For a historical
background of this classifications, see~\cite{Kraetal03}.} Together whether $a_\alpha$ is
zero or not, which determines the two main classes A and B introduced by Ellis and
MacCallum in~\cite{ellmac69}, the characteristic equation for $n^{\alpha\beta}$, i.e.,
$-\lambda^3 + n\lambda^2 + N\lambda + \mathrm{det}\,n^{\alpha\beta} = 0$, where
$n := n^\alpha\!_\alpha$, $N := \frac12(n^{\alpha\beta}n_{\alpha\beta} - n^2)$,
yields the modern spatially frame-invariant Bianchi classification as described
in~\cite{kingell73}, given in the following Table:

\begin{table}[H]
    \centering
    \begin{tabular}{l l l l}
    \hline
Group class & & Group type &
\\ \hline \\
Class A $\Leftrightarrow\, a_\alpha = 0\qquad$ & $\text{det}\,n^{\alpha\beta} \neq 0\qquad$ &
IX: $\text{det}\,n^{\alpha\beta} > 0\qquad$ & VIII: $\text{det}\,n^{\alpha\beta} < 0$\\
 & $\text{det}\,n^{\alpha\beta} = 0\qquad$ & VI$_0$: $N > 0\qquad $ &VII$_0$: $N < 0$\\
 &  & II: $N = 0$, $n >0\qquad$ & I: $N = n = 0$\\ \\
Class B $\Leftrightarrow\, a_\alpha \neq 0\qquad$ & $\text{det}\,n^{\alpha\beta} = 0\qquad$ &
VI$_h$: $N > 0\qquad$ & VII$_h$: $N < 0$\\
 &  & IV: $N = 0$, $n >0\qquad$ & V: $N = n = 0$
\\ \\ \hline
    \end{tabular}
    \caption{Classification of groups into classes A and B, and group types I to IX. The constant
    group invariant parameter $h$ is defined by $h := -a^2/N$, where $a^2 = a^\alpha a_\alpha$. Note that
    Bianchi type~III is type VI$_h$ with $h=-1$.}
    \label{BianchClass}
\end{table}

According to~\cite[p. 39 and 40 ]{waiell97}, modulo the sign difference for
$\Omega^\alpha$, Einstein's vacuum field equations and the Jacobi identities yield
\begin{subequations}\label{devoleq}
\begin{align}
\dot{H} &= -H^2 -\textfrac{2}{3}\sigma^2,\label{Ray}\\
\dot{\sigma}_{\alpha\beta} &= -3H\sigma_{\alpha\beta} -
2\epsilon^{\gamma\delta}{}_{\la \alpha}\,\sigma_{\beta\ra\gamma}\,\Omega_\delta
- {}^{3}\!R_{\la\alpha\beta\ra},\label{sigevol}\\
\dot{a}_{\alpha} &=  -[\,H\,\delta_{\alpha}{}^{\beta} + \sigma_{\alpha}{}^{\beta} +
\epsilon_{\alpha}{}^{\beta}{}_{\gamma}\,\Omega^{\gamma}\,] a_\beta,\label{aevol}\\
\dot{n}^{\alpha\beta} &= [-H\,\delta_{\gamma}{}^{(\alpha} + 2\sigma_{\gamma}{}^{(\alpha}
+ 2\epsilon_{\gamma}{}^{(\alpha}{}_{\delta}\,\Omega^{\delta}] n^{\beta )\gamma},\label{nevol}\\
0 &= 3H^2 - \sigma^2 + \textfrac{1}{2}{}^{3}\!R,\label{Gauss}\\
0 &=  (3\delta_\alpha{}^\gamma\,a_\beta + \epsilon_\alpha{}^{\delta\gamma}
\,n_{\delta\beta})\,\sigma^\beta{}_\gamma,\label{Codazzi}\\
0 &= n^{\alpha\beta}a_\beta,\label{Jacobi}
\end{align}
\end{subequations}
where an overdot denotes a time derivative with respect to the clock time $t$;
$\sigma^2=\frac{1}{2}\sigma_{\alpha\beta}\sigma^{\alpha\beta}$;
${}^{3}\!R_{\la\alpha\beta\ra}$ and ${}^{3}\!R$ describe the
trace-free and scalar parts of the spatial three-curvature,
respectively, according to:
\begin{equation}\label{threecurv} {}^{3}\!R_{\la\alpha\beta\ra} = B_{\la
\alpha\beta \ra} - 2\epsilon^{\gamma\delta}{}_{\la
\alpha}\,n_{\beta\ra\gamma}\,a_\delta ,\quad {}^{3}\!R =
-\textfrac{1}{2}B^\alpha{}_\alpha - 6a^2 ;\qquad
B_{\alpha\beta} := 2 n_{\alpha\gamma}\,n^\gamma{}_\beta -
n^\gamma{}_\gamma\,n_{\alpha\beta} , \end{equation}
where $a^2=a_\alpha a^\alpha$. Eq.~\eqref{Ray} is the Raychaudhuri equation
while~\eqref{sigevol} is the spatial and trace-free Einstein equation;
eqs.~\eqref{aevol} and~\eqref{nevol} are evolution equations
obtained from the Jacobi identities; eqs.~\eqref{Gauss}
and~\eqref{Codazzi} are the Gauss and Codazzi constraints,
respectively, while the constraint~\eqref{Jacobi} stems from
the spatial Jacobi identities.

There is another important classification of these models, i.e., if the 3D
symmetry group admits a 2D subgroup consisting of two spacelike commuting KVFs
or not. All Bianchi models, except for Bianchi types $\mathrm{VIII}$ and
$\mathrm{IX}$, admit such an Abelian $G_2$ subgroup and are specializations 
of the inhomogeneous Abelian $G_2$ models. For all Abelian $G_2$ models, it 
is natural to choose an orthonormal spatial Iwasawa frame: first choose one 
of the spatial frame vectors along one of the commuting spatial KVFs, then the
second orthogonally to this KVF but tangentially to the Abelian
$G_2$ orbits, and finally the third spatial frame vector orthogonally to the first two in a right hand manner. For details see~\cite{heietal12}, which we follow from now
on and we use $\vector{e}_1$ and $\vector{e}_2$ tangential to
the $G_2$ orbits. In combinations with the orthonormal spatial Iwasawa frame, equations~\eqref{devoleq} yield the following conditions:
\begin{subequations}
\begin{align}
a_1 &= a_ 2 = n^{13} = n^{23} = n^{33} = n^{22} = 0,\\
\sigma_{23} &= -\Omega^1,\qquad \sigma_{31} = \Omega^2 = 0,\qquad \sigma_{12} = -\Omega^3,
\end{align}
where the consistency of these conditions is easily seen by inserting them
into~\eqref{devoleq}.

There is nothing particular from a group theoretical point of view with the value $h = -1/9$.
Instead, its exceptional character comes from a degeneracy of the Codazzi constraints~\eqref{Codazzi}
as pointed out in~\cite[p. 123, case Bbii]{ellmac69}. In the present context this degeneracy
results in
\begin{equation}
n^{12} = 3a_3.
\end{equation}
\end{subequations}
Due to the constraint degeneracy, the Bianchi type $\mathrm{VI}_{-1/9}$ models
have an extra shear degree of freedom. This sets these models apart, since they are the
only Bianchi vacuum models with a $G_2$ symmetry subgroup for which the 2-spaces
to the $G_2$ orbits are not surface-forming (setting the extra shear degree to zero,
however, results in an invariant subset for which the $G_2$ symmetry group does act
orthogonally transitively, i.e., the 2-spaces orthogonal to the $G_2$ orbits are
surface-forming).

Next we introduce the following
\emph{dimensionless Hubble-normalized variables}\footnote{To simplify the Gauss
constraint, given below, we have rescaled some of the variables
in~\cite{heietal09,heietal12}, namely $A_3\mapsto A/2$,
$R_1\mapsto \sqrt{3} R_1$ and $R_3\mapsto \sqrt{3} R_3$ while
$N_- = N_1/2\sqrt{3} = N_{11}/2\sqrt{3}$.}
\begin{subequations}\label{Hdef}
\begin{alignat}{3}
\Sigma_1 &= \frac{\sigma_{11}}{H},&\qquad
\Sigma_2 &= \frac{\sigma_{22}}{H},&\qquad
\Sigma_3 &= \frac{\sigma_{33}}{H},\\
R_1 &= \frac{\Omega^1}{\sqrt{3}H} = -\frac{\sigma_{23}}{\sqrt{3}H},&\qquad
R_3 &= \frac{\Omega^3}{\sqrt{3}H} = -\frac{\sigma_{12}}{\sqrt{3}H},& \qquad &\\
N_- &= \frac{n^{11}}{2\sqrt{3}\,H},&\qquad
A &= \frac{2a_3}{H}. &
\end{alignat}
\end{subequations}
Thus $\Sigma_1,\Sigma_2,\Sigma_3$ correspond to the diagonal Hubble-normalized shear
while $R_1$ and $R_3$ describe frame rotations and the off-diagonal Hubble-normalized shear;
finally, $N_-$ and $A$ describe the Hubble-normalized commutator functions and thus the Hubble-normalized
spatial 3-curvature.

Thanks to the scale-invariance of the equations~\eqref{devoleq}, the introduction
of a new dimensionless \emph{time variable} $\tau$ \emph{directed toward the past} singularity,
defined by\footnote{Hence $\tau=-N$, where $N$ represents the forward directed $e$-fold time
variable used in~\cite{hewetal03}, where $N$ was called $\tau$ and thereby the negative
of the present $\tau$.}
\begin{equation}
\frac{d\tau}{dt} = -H,
\end{equation}
implies that the equation for the dimensional variable $H$ decouples from
the equations of the dimensionless Hubble-normalized variables, leading to
\begin{equation}
H^\prime = (1 + q)H = (1 + 2\Sigma^2)H.
\end{equation}
Here and henceforth a $(.)'$ denotes derivation with respect to $\tau$, while
\begin{equation}\label{deceleration}
q = 2\Sigma^2
\end{equation}
is the cosmological \emph{deceleration parameter} $q$ and
\begin{equation}
\Sigma^2 := \sfrac16\left(\Sigma_1^2 + \Sigma_2^2 + \Sigma_3^2 \right) + R_1^2 + R_3^2.
\end{equation}

It then follows from the equations~\eqref{devoleq} and the definitions~\eqref{Hdef}
that we obtain the following \emph{evolution equations}:
\begin{subequations}\label{BilliardeqsolevolHnormVI}
\begin{align}
\Sigma_1^\prime &= 2(1 - \Sigma^2)\Sigma_1 - 6R_3^2 + 8N_-^2,\label{Sig1eq}\\
\Sigma_2^\prime &= 2(1 - \Sigma^2)\Sigma_2 + 6R_3^2 - 6R_1^2 - 4N_-^2 + 3A^2,\\
\Sigma_3^\prime &= 2(1 - \Sigma^2)\Sigma_3 + 6R_1^2 - 4N_-^2 - 3A^2,\label{Sig3eq}\\
R_1^\prime &= [2(1 - \Sigma^2) + \Sigma_2 - \Sigma_3]R_1,\label{R1eq}\\
R_3^\prime &= [2(1 - \Sigma^2) + \Sigma_1 - \Sigma_2]R_3 - 4N_-A,\\
N_-^\prime &= -2(\Sigma^2 + \Sigma_1)N_- + 3R_3A,\\
A^\prime &= -(2\Sigma^2 - \Sigma_3)A,\label{Aeq}
\end{align}
\end{subequations}
where the state vector $(\Sigma_1, \Sigma_2, \Sigma_3, R_1, R_3, N_-, A)\in\mathbb{R}^7$
is subject to the following \emph{constraint equations}
\begin{subequations}\label{billiardeqsolconsHnormvacVI}
\begin{align}
1 - \Sigma^2 - N_-^2 - A^2 &= 0,\label{Gauss1}\\
2R_3N_- + \Sigma_1A &= 0,\label{Codazzi1}\\
\Sigma_1 + \Sigma_2 + \Sigma_3 &= 0.\label{Sigconstr}
\end{align}
\end{subequations}
Due to these constraints, the associated dimensionless state space is 4D 
and all Hubble-normalized variables are bounded.
Thus, these models are as general as the Bianchi type $\mathrm{VIII}$,
$\mathrm{IX}$ vacuum models, and these three models are the only SH vacuum 
models that are expected to have an oscillatory
singularity.

As shown in~\cite{hewetal03},\footnote{Apart from the sign changes in $\tau$
and $\Omega^\alpha$, our spatial frame choice is related to the one used by
Hewitt, Horwood, Wainwright (HHW) in~\cite{hewetal03} by a permutation of
the spatial axes such that the present convention with $(123)$ yields
$(231)$ in~\cite{hewetal03}, i.e., $(123) \leftrightarrow (231)_{HHW}$.
Taking these differences and appropriate scalings into account, the 
system~\eqref{BilliardeqsolevolHnormVI}, \eqref{billiardeqsolconsHnormvacVI}
can also be obtained from eqs. (2.8), (2.10), (2.2) in HHW by noticing 
the following relationship between the present variables and those in HHW:
$(\Sigma_1,\Sigma_2,\Sigma_3,R_1,R_3,N_-,A) =
(\Sigma_+ + \sqrt{3}\Sigma_-,\Sigma_+ - \sqrt{3}\Sigma_-,-2\Sigma_+,-\Sigma_2,-\Sigma_\times,N_-,2A)_{HHW}$.
\label{footHewetal}} the discrete symmetries given by
\begin{equation}
R_1 \mapsto -R_1,\qquad (R_3,A) \mapsto (-R_3,-A),\qquad (R_3,N_-) \mapsto (-R_3,-N_-)
\end{equation}
allow us to assume that
\begin{equation}
R_1 \geq 0,\qquad A \geq 0,
\end{equation}
since equations~\eqref{R1eq} and~\eqref{Aeq} do not allow either $R_1$
or $A$ to change sign along an orbit.
However, note that since $\Sigma_1$ can change sign, so can $R_3N_-$,
and hence $R_3N_-=0$ is not, in general, an invariant set when $A>0$.
This leads to multiple equivalent representations of
some of the invariant subsets.

Next we give the fixed points of the constrained dynamical system~\eqref{BilliardeqsolevolHnormVI},
\eqref{billiardeqsolconsHnormvacVI}, while we in Subsection~\ref{sec:stratification}
outline the Bianchi type $\mathrm{VI}_{-1/9}$ hierarchical stratification of invariant sets,
which are subsequently discussed individually in more detail in the following subsections.

\newpage

\subsection{Sets of fixed points}\label{sec:fixedpoints}

The system~\eqref{BilliardeqsolevolHnormVI}, \eqref{billiardeqsolconsHnormvacVI}
admits several sets of fixed points:
\begin{itemize}
\item[$\boxed{\mathrm{RT}}$] The Robinson-Trautman Bianchi type $\mathrm{VI}_{-1/9}$ fixed
point:\footnote{The Robinson-Trautman solution, belonging to Petrov type III, is also
associated with the name Collinson-French, see~\cite{hew91,heretal07}; for
a discussion about its properties see~\cite[Ch. 28, p. 200]{steetal03}.}
\begin{equation}\label{RT}
\mathrm{RT}:= \left\{ (\Sigma_1, \Sigma_2, \Sigma_3, R_1, R_3, N_-, A)
= \left(0,-\frac{2}{3},\frac{2}{3},\frac{\sqrt{5}}{3\sqrt{3}},0,0,\sqrt{\frac{2}{3}}\right) \right\},
\end{equation}
%
%
%
where the deceleration parameter $q$ takes the value $q = 2/3$,
according to~\eqref{deceleration}. Linear analysis shows that this fixed point
is a local source (with respect to the present past time direction $\tau$).
\item[$\boxed{\mathrm{PW}^\pm}$] The arc of type $\mathrm{VI}_{-1/9}$ plane
wave fixed points is parametrized by a constant value of $\Sigma_1 \in (-1,0]$
and is given by
\begin{equation}\label{PW}
\mathrm{PW}^\pm := \left\{
\begin{array}{c}
(\Sigma_2, \Sigma_3, A) = \frac12\left(-(3 + \Sigma_1), 3 - \Sigma_1, 1 + \Sigma_1\right), \\
(R_1, R_3, N_-) = \pm\frac12\sqrt{-\Sigma_1(\Sigma_1 + 1)}\,\left(0, 1, 1\right)
\end{array}
\Bigm| \Sigma_1\in (-1,0]
\right\},
\end{equation}
%
%
%
where $q = (3-\Sigma_1)/2$. 
The special $\mathrm{PW}^\pm$ fixed point with $\Sigma_1=0$, given by
\begin{equation}\label{PW0}
\mathrm{PW}^0:=\left\{(\Sigma_1,\Sigma_2, \Sigma_3,R_1, R_3, N_-,A) = \frac12(0,-3,3,0,0,0,1)\right\},
\end{equation}
divides the arc $\mathrm{PW}^\pm$ into two equivalent
branches, $\mathrm{PW}^+$ and $\mathrm{PW}^-$; one with positive
values and one with negative values of $R_3$ and $N_-$,
related by the symmetry $(R_3,N_-) \mapsto -(R_3,N_-)$.
The limit $\Sigma_1\rightarrow -1$ corresponds to that
$\mathrm{PW}^+$ and $\mathrm{PW}^-$ approach the Taub point $\mathrm{T}_3$
discussed below, which thereby leads to a closed arc of fixed points.
%

A linear analysis reveals that the fixed points $\mathrm{PW}^\pm$
is a center-source (with respect to $\tau$) in the invariant subset with $R_1=0$,
whereas $R_1$ yields a stable direction, as seen from the linearization of
the equation for $R_1$ at $\mathrm{PW}^\pm$:
\begin{equation}\label{lin:PW}
R_1^\prime =-\frac12\left(5 - \Sigma_1\right)R_1,
\end{equation}
since $-\frac12(5 - \Sigma_1) \in \left(-3,-\frac52\right]$.
\item[$\boxed{\mathrm{K}^{\ocircle}}$] Setting $R_1 = R_3 = N_- = A = 0$
yields the circle of Kasner fixed points,
\begin{equation}\label{KasnerCircdef}
\mathrm{K}^{\ocircle} := \left\{ (\Sigma_1,\Sigma_2,\Sigma_3,0,0,0,0)\in \mathbb{R}^7 \Bigm|
\begin{array}{c}
\qquad\quad\, 1-\Sigma^2 = 0, \\
\Sigma_1 + \Sigma_2 + \Sigma_3 = 0
\end{array}
\right\},
\end{equation}
for which $q=2$.
%
The Kasner circle is conveniently parametrized by the Kasner parameters
$p_\alpha$ with $\alpha=1,2,3$, which are defined by
\begin{equation}\label{def:p_alpha}
p_\alpha := \frac13\left(1 + \Sigma_\alpha\right),    
\end{equation}
where the conditions $\Sigma_1 + \Sigma_2 + \Sigma_3 = 0$ and $\Sigma^2 = 1$ respectively result in
\begin{equation}\label{KasnerExp}
p_1+p_2+p_3 = 1, \qquad p_1^2 + p_2^2 + p_3^2 = 1.
\end{equation}
\end{itemize}

The Kasner circle $\mathrm{K}^\ocircle$ is naturally divided into six equivalent
sectors, associated with permutations of the spatial axes, which corresponds
to permutations of the triple $(123)$ where each sector $(\alpha\beta\gamma)$ is
defined by
\begin{equation}
-\sfrac13 < p_\alpha < 0 < p_\beta < \sfrac23 < p_\gamma <1,
\end{equation}
see Figure~\ref{fig:sectors}. The boundaries of the sectors are six
special points that are associated with locally rotationally symmetric (LRS)
solutions (they are even plane symmetric solutions, since the spatial
curvature is flat for Bianchi type I),
\begin{subequations}
\begin{alignat}{4}
\boxed{\mathrm{T}_\alpha\!} \quad & (\Sigma_\alpha, \Sigma_\beta,
\Sigma_\gamma)\, &=&\, (2,-1,-1), &\qquad \text{ or } \qquad
(p_\alpha,p_\beta,p_\gamma)\,
&=& (1,0,0); \\
\boxed{\mathrm{Q}_\alpha\!} \quad & (\Sigma_\alpha, \Sigma_\beta,
\Sigma_\gamma)\, &=&\, (-2,1,1), &\qquad \text{ or } \qquad (p_\alpha,p_\beta,p_\gamma)\,
&=&\, (-\sfrac{1}{3},\sfrac{2}{3},\sfrac{2}{3}).
\end{alignat}
\end{subequations}
The \emph{Taub points} $\mathrm{T}_\alpha$, $\alpha = 1,2,3$,
correspond to the flat LRS solutions --- the Taub representation
of the Minkowski space-time, while $\mathrm{Q}_\alpha$, $\alpha =
1,2,3$, yield three equivalent LRS solutions with non-flat geometry. Incidentally,
note that the $(p_\alpha,p_\beta,p_\gamma)$ values for $\mathrm{Q}_\alpha$ are
those that the Schwarzschild singularity exhibit due to the spherical symmetry of
these models.

It is convenient to parametrize the Kasner parameters $p_1,p_2,p_3$
by the \emph{gauge- and spatial frame-invariant Kasner parameter} $u$, which is
spatial frame-invariantly defined by the relation\footnote{For the BKL definition
of $u$, the order of $p_\alpha$ is fixed according
to $p_1\leq p_2 \leq p_3$; we find it convenient to
use the above frame-independent definition instead, and to permute the ordering
of $p_\alpha$ according to the sector one considers when dealing
with frame-dependent matters.}
\begin{equation}\label{detSigmau}
\det(\Sigma_{\alpha\beta})= 2 + 27p_1p_2p_3 = 2 - \frac{27u^2(1+u)^2}{f^3(u)},\qquad u\in [1,\infty],
\end{equation}
where the function $f(x)$ is defined as
\begin{equation}\label{f(x)}
f(x) := 1 + x + x^2,
\end{equation}
and hence $\det(\Sigma_{\alpha\beta})$ is a
monotonically increasing function in $u$.
Due to frame invariance, the Kasner parameter $u$ naturally captures the equivalence
of the six sectors of $\mathrm{K}^{\ocircle}$. On sector $(\alpha\beta\gamma)$, where
$p_\alpha < p_\beta < p_\gamma$, we set
\begin{equation}\label{ueq}
p_\alpha = - \frac{u}{f(u)},\qquad p_\beta = \frac{1+u}{f(u)},\qquad
p_\gamma = \frac{u(1+u)}{f(u)},
\end{equation}
where $u\in(1,\infty)$. The boundary points of sector $(\alpha\beta\gamma)$,
$\mathrm{Q}_\alpha$ and $\mathrm{T}_\gamma$, are characterized
by $u=1$ and $u=\infty$, respectively.

Note that the magnetic Weyl tensor is identically zero on the Kasner subset,
but the electric part is non-trivial and leads to a non-zero Hubble-normalized
quadratic Weyl (Kretschmann, since we consider vacuum) scalar that is
related to $\det(\Sigma_{\alpha\beta})$ according to (see~\cite{heietal09})
\begin{equation}\label{Weyl}
\mathcal{W} := \frac{C_{abcd}C^{abcd}}{48H^4} =
2 - \det(\Sigma_{\alpha\beta}) = \frac{27u^2(1+u)^2}{f^3(u)} = \mathrm{const}.,
\end{equation}
which thereby explicitly yields a spatially frame-invariant description of the
various Kasner states. For large $u$, and thereby small $\mathcal{W}$, it
follows that $\mathcal{W} = 27 u^{-1}(1 + u^{-1} + O(u^{-2}))$.

In addition to the standard Kasner parameter $u$, which parametrize each
sector $(\alpha\beta\gamma)$ as in~\eqref{ueq}, it is useful in the context of
the Bianchi type $\mathrm{VI}_{-1/9}$ models to define the
\emph{extended Kasner parameter} $\ue$, which parametrizes
the whole Kasner circle, by setting
\begin{equation}\label{ueeq}
p_1 = -\frac{\ue}{f(\ue)},\qquad
p_2 = \frac{1+\ue}{f(\ue)},\qquad
p_3 = \frac{\ue(1+\ue)}{f(\ue)},
\end{equation}
where $\ue \in (-\infty,\infty)$, so that each value of $\ue$
distinguishes a unique point on the Kasner circle $\mathrm{K}^\ocircle$.
Comparing~\eqref{ueq} and~\eqref{ueeq}, we obtain a transformation 
between $u$ and $\ue$ for each sector, found in Table~\ref{ueandu}.
\begin{table}[H]
    \centering
    \begin{tabular}{|c|l|l|c|}\hline
        \rowcolor{lightgray!25}Sector & \, Parameter $\ue$ & \,\, Range of $\ue$ & $u=2$ \\ \hline
        $(213)$ & $\ue = -(1 + u)$ & $\ue \in (-\infty,-2)$ & $\ue=-3$\\ \hline
        $(231)$ & $\ue =-\sfrac{1 + u}{u}$ & $\ue \in (-2,-1)$ & $\ue=-\sfrac{3}{2}$\\ \hline
        $(321)$ & $\ue = -\sfrac{u}{1 + u}$ & $\ue \in (-1,-\sfrac{1}{2})$ & $\ue=-\sfrac{2}{3}$\\ \hline
        $(312)$ & $\ue =-\sfrac{1}{1 + u}$ & $\ue \in (-\sfrac{1}{2}, 0)$ & $\ue=-\sfrac{1}{3}$\\ \hline
        $(132)$ & $\ue = \sfrac{1}{u}$ & $\ue \in (0,1)$ & $\ue=\sfrac{1}{2}$\\ \hline
        $(123)$ & $\ue = u$ & $\ue \in (1,\infty)$ & $\ue=2$ \\ \hline
    \end{tabular}
\hspace{1cm}
        \begin{tabular}{|c|c|} \hline
        \rowcolor{lightgray!25}\multicolumn{2}{|c|}{Correspondence} \\ \hline
        \rowcolor{lightgray!25}$\ue$ & Point \\ \hline
        $-\infty$ & $\mathrm{T}_3$\\ \hline
        $-2$ & $\mathrm{Q}_2$\\ \hline
        $-1$ & $\mathrm{T}_1$\\ \hline
        $-1/2$ & $\mathrm{Q}_3$\\ \hline
        $0$ & $\mathrm{T}_2$\\ \hline
        $1$ & $\mathrm{Q}_1$ \\ \hline
        $\infty$ & $\mathrm{T}_3$ \\ \hline
    \end{tabular}
    \caption{Parametrization of each sector $(\alpha \beta \gamma)$
    of the Kasner circle $\mathrm{K}^\ocircle$ by the extended Kasner parameter $\ue$, and
    the values of $\ue$ for the different LRS points on $\mathrm{K}^\ocircle$, which
    constitute the boundaries of the different sectors $(\alpha\beta\gamma)$. We also
    describe the six representations of $u=2$, one in each sector, where $u=2$ yields the
    boundary of $1<u<2$, which signifies the last Kasner state in a Kasner era, discussed below.}
    \label{ueandu}
\end{table}
Note that even though the different sectors $(\alpha\beta\gamma)$ yield
equivalent solutions, due to the permutation symmetries inherited from
permutations of the spatial axes, we will see that they play different roles
for the global past dynamics of the present dynamical
system~\eqref{BilliardeqsolevolHnormVI}, \eqref{billiardeqsolconsHnormvacVI}.
We pursue the stability of $\mathrm{K}^{\ocircle}$ next.

Linearization of the system~\eqref{BilliardeqsolevolHnormVI} at an arbitrary point
$(p_1,p_2,p_3)\in\mathrm{K}^{\ocircle}$ yields
\begin{subequations}\label{linCircle}
\begin{alignat}{2}
R_1^\prime &= 3(p_2 - p_3)R_1 = \lambda_{R_1} R_1, \qquad\qquad &&\lambda_{R_1} := \frac{3(1 - \ue^2)}{f(\ue)},\label{linR1}\\
R_3^\prime &= 3(p_1 - p_2)R_3 = \lambda_{R_3} R_3, \qquad\qquad &&\lambda_{R_3} := -\frac{3(1 + 2\ue)}{f(\ue)},\label{linR3}\\
N_-^\prime &= -6p_1N_- = \lambda_{N_-}N_-, \qquad\qquad &&\lambda_{N_-} := \frac{6\ue}{f(\ue)},\label{linN-}\\
A^\prime &= -3(1 - p_3)A = \lambda_A A, \qquad\qquad &&\lambda_A := -\frac{3}{f(\ue)}.\label{Adecay}
\end{alignat}
\end{subequations}
The analysis of the stability of the Kasner circle $\mathrm{K}^{\ocircle}$ is summarized
in Figure~\ref{fig:sectors}, where the unstable variables in $\tau$ are given for each
sector of $\mathrm{K}^{\ocircle}$. Note that each of the variables $R_1,R_3,N_-$ trigger
instability in different parts of the Kasner circle, although there is some overlap
between these parts, which is similar to the subcritical case of the Bianchi type
$\mathrm{VIII}$ and $\mathrm{IX}$ Ho\v{r}ava-Lifshitz models discussed
in~\cite{HLU22,lapp22,Church}. On the other hand, the variable $A$ does not trigger any
instability nearby the Kasner circle, since its corresponding eigenvalue $\lambda_A<0$ for all
$\ue\in\mathbb{R}$, while $\lambda_A=0$ at $\mathrm{T}_3$ where $\ue=\pm \infty$. Since
$R_3$, $N_-$ and $A$ yield center-eigendirections at the Taub point $\mathrm{T}_3$, it
follows that understanding the dynamics in the vicinity of $\mathrm{T}_3$ poses
challenges. Note that the above remarks did not take into account the Codazzi
constraint~\eqref{Codazzi1}.\footnote{On the one hand, linearization of~\eqref{Codazzi1} at
$\mathrm{K}^{\ocircle}$ yields $(3p_1 - 1)A=0$, since $N_-R_3 = 0$ at linear order.
This implies that $A$ is zero to linear order except at $p_1 = \frac13$
(i.e., at $\Sigma_1=0$), which amounts to the Kasner fixed points
$\mathrm{K}^\ocircle_\pm$. 
On the other hand, introducing $R_3N_-$ as an auxiliary variable and linearizing at
$\mathrm{K}^{\ocircle}$ yields $(R_3N_-)^\prime = -3(1 - p_3)(R_3N_-) = \lambda_A(R_3N_-)$,
which explains why $\lambda_A = \lambda_{N_-} + \lambda_{R_3}$.  }

There are several other fixed points on $\mathrm{K}^\ocircle$ that are important
for the dynamics. The BKL Kasner map, given in~\eqref{BKLMap} below, generates
a Kasner sequence $\{u_0, u_1,\dots\}$ that is naturally decomposed into
so-called eras, where $u>2$ corresponds to a continuation of an era, while $u<2$
is the last value of $u$ in an era, which leads to a change in the BKL Kasner map
and a new era (details are given in section~\ref{subsec:frameinv}).
Since era changes represent important dynamical changes (especially for the
Bianchi type $\mathrm{VI}_{-1/9}$ models, since these
models, as we will see, have a much more complicated Bianchi type I and II
heteroclinic network than those of types $\mathrm{VIII}$ and $\mathrm{IX}$),
it is important to identify the six fixed points that correspond to $u=2$,
since they divide the six sectors into era continuing and era changing parts.
The six fixed points corresponding to $u=2$ are given in terms of $\ue$ in
Table~\ref{ueandu}, but alternatively they can also be described as follows
in each of the sectors $(\alpha\beta\gamma)$:
\begin{equation}\label{u=2}
(\Sigma_\alpha, \Sigma_\beta,
\Sigma_\gamma)\, =\, \left(-\sfrac{13}{7},\sfrac{2}{7}, \sfrac{11}{7}\right), \qquad \text{ or } \qquad
(p_\alpha,p_\beta,p_\gamma)\,
= (-\sfrac{2}{7},\sfrac{3}{7},\sfrac{6}{7}).
\end{equation}
There are also two other special Kasner fixed points, denoted by $\mathrm{K}^\ocircle_\pm$,
which characterize a special solution discussed in Section~\ref{sec:Kasner}, given by
\begin{equation}\label{Kpm}
\boxed{\mathrm{K}^\ocircle_\pm\!} \quad  (\Sigma_1, \Sigma_2,
\Sigma_3)\, =\, \left(0,\pm \sqrt{3},\mp \sqrt{3}\right), \qquad \text{ or } \qquad
(p_1,p_2,p_3)\,
= \left(\sfrac{1}{3},\sfrac{1\pm \sqrt{3}}{3},\sfrac{1\mp \sqrt{3}}{3}\right).
\end{equation}
The point $\mathrm{K}^\ocircle_+$ lies in sector $(312)$ and has the extended
Kasner parameter $\ue = -2 + \sqrt{3}$, while $\mathrm{K}^\ocircle_-$
is in sector $(213)$ with $\ue = -2 - \sqrt{3}$; both these points
correspond to the spatially frame-invariant Kasner parameter $u = 1 + \sqrt{3}$, which notably
characterizes $\lambda_{R_1} = \lambda_{R_3}$.
\begin{figure}[H]
	\centering
        \begin{tikzpicture}[scale=1.1]
        \draw[color=gray,->] (0,0) -- (-0.343,0);
        \node[color=gray] at (-0.47,0) {\scriptsize $\Sigma_{1}$};

        \draw[color=gray,->,rotate=120] (0,0) -- (-0.343,0);
        \node[color=gray] at (0.1,-0.42) {\scriptsize $\Sigma_{3}$};

        \draw[color=gray,->,rotate=240] (0,0) -- (-0.343,0);
        \node[color=gray] at (0.1,0.42) {\scriptsize $\Sigma_{2}$};

	\draw [very thin] (1,0) arc (0:360:1cm and 1cm);	 	
	
        \draw (-0.95,0) -- (-1.05,0) node[anchor= east] {\scriptsize{$\mathrm{T}_1$}};
        \draw (0.95,0) -- (1.05,0) node[anchor= west] {\scriptsize{$\mathrm{Q}_1$}};

        \draw[rotate=120] (-0.95,0) -- (-1.05,0) node[anchor= north] {\scriptsize{$\mathrm{T}_3$}};
        \draw[rotate=240] (-0.95,0) -- (-1.05,0) node[anchor= south] {\scriptsize{$\mathrm{T}_2$}};

        \draw[rotate=120] (0.95,0) -- (1.05,0) node[anchor= south] {\scriptsize{$\mathrm{Q}_3$}};
        \draw[rotate=240] (0.95,0) -- (1.05,0) node[anchor= north] {\scriptsize{$\mathrm{Q}_2$}};

        \node at (1.15,0.6) {\scriptsize $(132)$};
        \node at (1.15,-0.6) {\scriptsize $(123)$};
        \node at (0,-1.2) {\scriptsize $(213)$};
        \node at (-1.15,-0.6) {\scriptsize $(231)$};
        \node at (-1.15,0.6) {\scriptsize $(321)$};
        \node at (0,1.2) {\scriptsize $(312)$};


        \filldraw [black] (0,1) circle (1pt); \node at (0,0.8) {\scriptsize{$\mathrm{K}^{\ocircle}_+$}};
        \filldraw [black] (0,-1) circle (1pt); \node at (0,-0.8)  {\scriptsize{$\mathrm{K}^{\ocircle}_-$}};

        \end{tikzpicture}
        \hspace{0.5cm}
        \begin{tikzpicture}[scale=1.1]
       \draw [color=gray,->, domain=2.61:3.66,variable=\t,smooth] plot ({0.75*sin(\t r)},{0.75*cos(\t r)}) node[anchor= south] {\scriptsize{$\ue$}};

	\draw [very thin] (1,0) arc (0:360:1cm and 1cm);	 	
	
        \draw (-0.95,0) -- (-1.05,0) node[anchor= east] {\scriptsize{$-1$}};
        \draw (0.95,0) -- (1.05,0) node[anchor= west] {\scriptsize{$1$}};

        \draw[rotate=120] (-0.95,0) -- (-1.05,0) node[anchor= north] {\scriptsize{$\pm \infty$}};
        \draw[rotate=240] (-0.95,0) -- (-1.05,0) node[anchor= south] {\scriptsize{$0$}};

        \draw[rotate=120] (0.95,0) -- (1.05,0) node[anchor= south] {\scriptsize{$-\sfrac{1}{2}$}};
        \draw[rotate=240] (0.95,0) -- (1.05,0) node[anchor= north] {\scriptsize{$-2$}};

        \filldraw [black] (0.928,0.371) circle (1pt) node[anchor= west] {\scriptsize{$\sfrac{1}{2}$}};
        \filldraw [black] (0.928,-0.371) circle (1pt) node[anchor= west] {\scriptsize{$2$}};

        \filldraw [black] (-0.142,0.989) circle (1pt) node[anchor= south] {\scriptsize{$-\sfrac{1}{3}$}};
        \filldraw [black] (-0.142,-0.989) circle (1pt) node[anchor= north] {\scriptsize{$-3$}};

        \filldraw [black] (-0.785,0.618) circle (1pt) node[anchor= east] {\scriptsize{$-\sfrac{2}{3}$}};
        \filldraw [black] (-0.785,-0.618) circle (1pt) node[anchor= east] {\scriptsize{$-\sfrac{3}{2}$}};

        \end{tikzpicture}
        \hspace{0.5cm}
        \begin{tikzpicture}[scale=1.1]
	\draw [very thin] (1,0) arc (0:360:1cm and 1cm);	 	
	
	\draw (-0.95,0) -- (-1.05,0) node[anchor= east] {\scriptsize{$\mathrm{T}_1$}};
        \draw (0.95,0) -- (1.05,0) node[anchor= west] {\scriptsize{$\mathrm{Q}_1$}};

        \draw[rotate=120] (-0.95,0) -- (-1.05,0) node[anchor= north] {\scriptsize{$\mathrm{T}_3$}};
        \draw[rotate=240] (-0.95,0) -- (-1.05,0) node[anchor= south] {\scriptsize{$\mathrm{T}_2$}};

        \draw[rotate=120] (0.95,0) -- (1.05,0) node[anchor= south] {\scriptsize{$\mathrm{Q}_3$}};
        \draw[rotate=240] (0.95,0) -- (1.05,0) node[anchor= north] {\scriptsize{$\mathrm{Q}_2$}};

       \draw [ultra thick, domain=0.52:1.570,variable=\t,smooth] plot ({sin(\t r)},{cos(\t r)});
       \draw [ultra thick, domain=-0.52:-1.570,variable=\t,smooth] plot ({sin(\t r)},{cos(\t r)});

        \node at (0.7,-0.4) {\scriptsize $N_-$};
        \node at (0.5,0.4) {\scriptsize $R_1,N_-$};
        \node at (0,0.85) {\scriptsize $R_1$};
        \node at (-0.5,0.4) {\scriptsize $R_1,R_3$};
        \node at (-0.7,-0.4) {\scriptsize $R_3$};
        \node at (0,-0.85) {\scriptsize $R_3$};

        \end{tikzpicture}
	\caption{\textbf{Left:} The Kasner circle $\mathrm{K}^{\ocircle}$ consists of
six sectors $(\alpha\beta\gamma)$ and the boundary points $\mathrm{Q}_\alpha$
and $\mathrm{T}_\gamma$, where $\alpha\beta\gamma$ is a permutation of $123$.
The point $\mathrm{K}^{\ocircle}_+$ (resp. $\mathrm{K}^{\ocircle}_-$)
is the midpoint of the sector $(312)$ (resp. $(213)$).
\textbf{Middle:} The extended Kasner parameter $\ue$ starts at $\mathrm{T}_3$
with $\ue=-\infty$ and increases clock-wise until it reaches $\mathrm{T}_3$
again when $\ue=+\infty$, see Table~\ref{ueandu}. The values of $\ue$
for $\mathrm{Q}_\alpha$, $\mathrm{T}_\gamma$ and for the six fixed points
corresponding to $u=2$ are also given.
\textbf{Right:} The variables that trigger instability toward the singularity
on $\mathrm{K}^{\ocircle}$; sectors $(132)$ and $(321)$, denoted with bold curves,
have two unstable variables, while the remaining parts of $\mathrm{K}^{\ocircle}$
(denoted with thin curves) only have a single unstable variable (apart from
the Taub point $\mathrm{T}_3$ which has none).}\label{fig:sectors}
\end{figure}
%


\subsection{Stratification of invariant sets}\label{sec:stratification}
%
The constrained dynamical system~\eqref{BilliardeqsolevolHnormVI},
\eqref{billiardeqsolconsHnormvacVI} admits several \emph{explicitly known
invariant sets} forming a hierarchical stratification of subsets with
different dimensions:
\begin{itemize}
\item[(a)] The Bianchi type I \emph{Kasner} boundary subset, for which $A=N_-=0$,
denoted by ${\cal B}_\mathrm{I}= {\cal K}$. In this case, the constraint~\eqref{Codazzi1} is
trivially satisfied. This set consists of the union of the following subsets:
\begin{itemize}
\item[(i)] The 1D Kasner circle of fixed points, $\mathrm{K}^{\ocircle}$, for which $R_1=R_3=0$.
\item[(ii)] The 2D single frame transition subsets, denoted by $\mathcal{T}_{R_1}$
and $\mathcal{T}_{R_3}$, which respectively occur when $R_1 > 0$, $R_3=0$;
$R_1= 0,R_3 >0$ and $R_1= 0,R_3 <0$, where $\mathcal{T}_{R_3}$ consists of
two disjoint components related by the symmetry $R_3 \mapsto - R_3$. 
\item[(iii)] The 3D double frame transition subset, denoted by $\mathcal{T}_{R_1R_3}$,
consisting of two components, $R_1R_3 >0$ and $R_1R_3 <0$,
related by the symmetry $R_3 \mapsto - R_3$. 
\end{itemize}
\item[(b)] The Bianchi type II boundary subset, characterized by $A=R_3=0$ and
designated by ${\cal B}_\mathrm{II}$. Again, the constraint~\eqref{Codazzi1} is
trivially satisfied. This set is the union of the following subsets:
\begin{itemize}
\item[(i)] The 2D single curvature transition subset, denoted by
$\mathcal{T}_{N_-}$, which occurs when $R_1=0$, with two disjoint components, $N_- > 0$ and $N_-<0$,
related by the symmetry $N_-\mapsto - N_-$. 
\item[(ii)] The 3D mixed curvature-frame transition subset, denoted by $\mathcal{T}_{R_1N_-}$,
consisting of two disjoint components, $R_1N_- > 0$ and $R_1N_-<0$,
related by the symmetry $N_-\mapsto - N_-$. 
\end{itemize}
\item[(c)] The 3D \emph{orthogonally transitive} subset ${\cal OT}$ is defined by $R_1=0$
and contains the 1D arc of type $\mathrm{VI}_{-1/9}$ plane wave fixed points
$\mathrm{PW}^\pm$ given in eq.~\eqref{PW}.
\item[(d)] The 2D \emph{hypersurface orthogonal} subset, ${\cal HO}$, characterized by
$n = n^\alpha\!_\alpha=0$, see e.g.~\cite{waiell97}. This subset corresponds to that
one of the Killing vector fields of the Abelian $G_2$ subgroup of space-time
symmetries is hypersurface orthogonal. The ${\cal HO}$ subset is characterized
by $\Sigma_1=R_3=N_-=0$ and contains the Robinson-Trautman fixed point $\mathrm{RT}$
given in eq.~\eqref{RT}.
The ${\cal HO}$ subset admits the 1D
\emph{shear diagonal and Fermi-propagated} subset, ${\cal D}$, in which both of the two
Killing vector fields forming the Abelian $G_2$ symmetry group are hypersurface orthogonal,
which follows from the intersection of the ${\cal HO}$ and ${\cal OT}$ subsets. The subset
${\cal D}$ is described by $\Sigma_1=R_1 = R_3=N_-=0$ and contains
the plane wave fixed point $\mathrm{PW}^0$, given in eq.~\eqref{PW0}.
\end{itemize}
The connections between the above invariant subsets is described in the stratification diagram in Figure~\ref{FIG:hierarchy}. 
Next we turn to a more detailed description of these invariant subsets.
\begin{figure}
\centering
\begin{tikzpicture}
\node (dim) at (6, 6.6) {\underline{\footnotesize{Dimension:}}};
\node (4) at (6, 6) {\footnotesize{$4$}};
\node (3) at (6, 4.5) {\footnotesize{$3$}};
\node (2) at (6, 3) {\footnotesize{$2$}};
\node (1) at (6, 1.5) {\footnotesize{$1$}};

\node (VI19) at (-1, 6) {\boxed{\text{\footnotesize{Type $\mathrm{VI}_{-1/9}$}}}};

\node (R1R3) at (-4, 4.5) {\boxed{\text{\footnotesize{$\mathcal{T}_{R_1R_3}$}}}};
\node (R1N-) at (-2, 4.5) {\boxed{\text{\footnotesize{$\mathcal{T}_{R_1N_-}$}}}};
\node (OT) at (0, 4.5) {\boxed{\text{\footnotesize{$\mathcal{OT}$}}}};

\node (R1) at (-4, 3) {\boxed{\text{\footnotesize{$\mathcal{T}_{R_1}$}}}};
\node (R3) at (-2, 3) {\boxed{\text{\footnotesize{$\mathcal{T}_{R_3}$}}}};
\node (N-) at (-0, 3) {\boxed{\text{\footnotesize{$\mathcal{T}_{N_-}$}}}};
\node (naa) at (2, 3) {\boxed{\text{\footnotesize{$\mathcal{HO}$}}}};
\node[coordinate] (midpoint) at (2, 4.5) {};

\node (D) at (2, 1.5) {\boxed{\text{\footnotesize{$\mathcal{D}$}}}};
\node[coordinate] (midpoint2) at (1, 3) {};
\node (K0) at (-2, 1.5) {\boxed{\text{\footnotesize{$\mathrm{K}^{\ocircle}$}}}};


\draw[-] (VI19) -- (R1R3);\draw[-] (VI19) -- (R1N-);\draw[-] (VI19) -- (OT);

\draw[-] (VI19) -- (midpoint);\draw[-] (midpoint) -- (naa); 

\draw[-] (R1R3) -- (R1);\draw[-] (R1R3) -- (R3);\draw[-] (R1N-) -- (N-);\draw[-] (R1N-) -- (R1);

\draw[-] (R1) -- (K0);\draw[-] (R3) -- (K0); \draw[-] (N-) -- (K0);

\draw[-] (OT) -- (N-);\draw[-] (OT) -- (R3);\draw[-] (OT) -- (midpoint2);\draw[-] (midpoint2) -- (D);

\draw[-] (naa) -- (D);

\end{tikzpicture}
\captionof{figure}{The invariant set stratification diagram for the general Bianchi type
$\mathrm{VI}_{-1/9}$ vacuum models, where each subset of smaller dimension is
obtained by setting certain variables to zero. Subsequent intersections of
the closure of these subsets yield lower dimensional subsets, e.g. intersections of
two of the $\overline{{\cal T}}_{R_1}$, $\overline{{\cal T}}_{R_3}$,
$\overline{{\cal T}}_{N_-}$ subsets yield the 1D Kasner circle
$\mathrm{K}^{\ocircle}$ of fixed points, while the intersection of the ${\cal OT}$
and ${\cal HO}$ subsets yields the 1D subset ${\cal D}$, where the intersection
of $\overline{{\cal D}}$ with $\mathrm{K}^{\ocircle}$ yields the special Kasner
fixed points $\mathrm{K}^{\ocircle}_\pm$.}\label{FIG:hierarchy}
\end{figure}
%

\subsection{The Kasner subset ${\cal K}$}\label{sec:Kasner}

The shear-diagonalized Fermi-propagated representation of
the Kasner solutions on the Kasner circle $\mathrm{K}^\ocircle$ of the
Kasner boundary subset ${\cal K}$ is not the only one, as they can also
be expressed in a rotating spatial frame. This leads to time
dependent off-diagonal Hubble-normalized shear components with non-zero $R_1$
or/and non-zero $R_3$, but $N_-=A=0$. However, the
shear frame-invariants, i.e., the trace, the trace of its square,
and the determinant of the shear, are constant, since they can also be
expressed in the shear-diagonalized Fermi-propagated frame and
thereby subsequently by the spatially frame-invariant parameter $u$.

It follows from~\eqref{billiardeqsolconsHnormvacVI}
and~\eqref{detSigmau} that the Kasner solutions, irrespective of spatial frame,
obey the relations: 
\begin{subequations}\label{Kasnerrot}
\begin{align}
0 &= \Sigma_1 + \Sigma_2 + \Sigma_3,\\
1 &= \sfrac16 \left(\Sigma_1^2 + \Sigma_2^2 + \Sigma_3^2\right) + R_1^2 + R_3^2,\label{K2}\\
\det \Sigma_{\alpha\beta} &= \Sigma_1 \Sigma_2 \Sigma_3 - 3\Sigma_1 R_1^2 - 3\Sigma_3 R_3^2
= 2 - \frac{27u^2(1+u)^2}{f^3(u)},\label{R1R3const}
\end{align}
\end{subequations}
where we recall that $u\in [1,\infty]$.\footnote{Using the variables of Hewitt et al. given in
footnote~\ref{footHewetal} and solving for $R_3^2 = \Sigma_\times^2$ in eq.~\eqref{K2} and inserting
the result into~\eqref{R1R3const} leads to an expression that is equivalent to eq. (5.5)
in~\cite{hewetal03}.}

We refer to heteroclinic orbits in ${\cal K}$ that connect two
different fixed points on $\mathrm{K}^\ocircle$ with either $R_1\neq 0$,
$R_3=0$ or $R_1=0$, $R_3 \neq 0$  as \emph{single frame transitions}.
These individual orbits and collectively as 2D invariant subsets,
are denoted by $\mathcal{T}_{R_1}$ and $\mathcal{T}_{R_3}$, respectively.
We now describe the dynamics of these subsets.

The single frame transitions `triggered' by the instability of $R_1>0$ yield
a frame rotation in the $(2-3)$-plane. This implies that
$\Sigma_1 = 3p_1 - 1 = -(1 + 4\ue + \ue^2)/f(\ue) = \mathrm{constant}$
is invariant, where $\ue = \ue_- \in (-1,1)$ denotes the fixed point on
$\mathrm{K}^\ocircle$ from where a particular heteroclinic orbit
$\mathcal{T}_{R_1}$ originates, as follows from~\eqref{Sig1eq}
and~\eqref{R1R3const}, where the latter can be rewritten as
$\det \Sigma_{\alpha\beta} = \Sigma_1(\Sigma_1^2 - 3) + 3(\Sigma_1 - \Sigma_3)R_3^2$,
in which setting $R_3=0$ yields $\Sigma_1 = \mathrm{constant}$.
Such an orbit is represented by a single straight line
in $(\Sigma_1,\Sigma_2,\Sigma_3)$-space with $\Sigma_3$ (resp. $\Sigma_2$) monotonically
increasing (resp. decreasing) in $\tau$, resulting in a total frame rotation of $\pi/2$
in the $(2-3)$-plane, thereby permuting the indices $2$ and $3$.
The frame transitions $\mathcal{T}_{R_1}$ thereby originate from the Kasner arc
$\mathbf{A}^-_{R_1}$ (i.e., the frame transitions $\mathcal{T}_{R_1}$ converge to
$\mathbf{A}^-_{R_1}$ as $\tau\to -\infty$), on which $R_1$ is unstable on
$\mathrm{K}^\ocircle$, which is parametrized by $\ue_-\in (-1,1)$ and consisting
of the sectors $(132)$, $(312)$, $(321)$ and the fixed points $\mathrm{Q}_3$,
$\mathrm{T}_2$, whereas they end at the Kasner arc $\mathbf{A}^+_{R_1}$ (i.e., the frame
transitions $\mathcal{T}_{R_1}$ converge to $\mathbf{A}^+_{R_1}$ as $\tau\to +\infty$),
which is parametrized by $\ue_+$ such that $\mid\ue_+\mid > 1$,
obtained by interchanging $2$ and $3$ in the description of $\mathbf{A}^-_{R_1}$,
see Figure~\ref{KMaps}.

Analogous statements hold for the $\mathcal{T}_{R_3}$ transitions in either of
the two equivalent disjoint subsets, $R_3>0$ and $R_3<0$. Thus
frame rotations in the $(1-2)$-plane result in straight lines in
$(\Sigma_1,\Sigma_2,\Sigma_3)$-space such that
$\Sigma_3 = 3p_3 - 1 = -(1 - 2\ue - 2\ue^2)/f(\ue) = \mathrm{constant}$,
$\ue = \ue_- \in (-\infty,-1/2)$ where $\Sigma_2$ (resp. $\Sigma_1$) is
monotonically increasing (resp. decreasing) in $\tau$. These frame transitions
hence originate from the Kasner arc $\mathbf{A}^-_{R_3}$ (i.e., the frame
transitions $\mathcal{T}_{R_3}$ converge to $\mathbf{A}^-_{R_3}$ as $\tau\to -\infty$),
consisting of the sectors $(213)$, $(231)$, $(321)$ and the fixed points
$\mathrm{Q}_2$, $\mathrm{T}_1$, and end at the Kasner arc $\mathbf{A}^+_{R_3}$
(i.e., the frame transitions $\mathcal{T}_{R_3}$ converge to $\mathbf{A}^+_{R_3}$
as $\tau\to +\infty$) which is parametrized by $\ue = \ue_+ \in (-1/2,+\infty)$,
obtained from $\mathbf{A}^-_{R_3}$ by interchanging 1 and 2, see Figure~\ref{KMaps}.
Finally, note that the $\mathcal{T}_{R_3}$ transitions become tangential to the Kasner
circle $\mathrm{K}^{\ocircle}$ at $\mathrm{T}_3$, 
which is the reason for the zero eigenvalue at $\mathrm{T}_3$ associated with $R_3$,
see Figure~\ref{KMaps}.

Heteroclinic orbits with $R_1 R_3 >0$ or $R_1 R_3 <0$ are called
\emph{double frame transitions}. These individual orbits and collectively as
two equivalent invariant 3D subsets are denoted by $\mathcal{T}_{R_1 R_3}$.
The variable $\Sigma_3$ (resp. $\Sigma_1$) is monotonically increasing
(resp. decreasing) for these transitions, which together with the linear
analysis of $\mathrm{K}^\ocircle$ establishes the following: The 2-parameter
sets of orbits $\mathcal{T}_{R_1 R_3}$ are described by two equivalent one-parameter
sets of orbits (one with $R_3<0$ and one with $R_3>0$) originating from each
of the fixed points on the Kasner arc $\mathbf{A}^-_{R_1R_3}$ (i.e., the
double frame transitions $\mathcal{T}_{R_1R_3}$ converge to
$\mathbf{A}^-_{R_1R_3}$ as $\tau\to -\infty$), given by sector $(321)$
with $\ue_- \in (-1,-1/2)$; they end at the arc
$\mathbf{A}^+_{R_1R_3}$, given by sector $(123)$ with
$\ue_+ \in (1,+\infty)$, as $\tau\to +\infty$. Note that the
$\mathcal{T}_{R_1 R_3}$ set is bounded by the single frame transition
$\mathcal{T}_{R_1}$ and $\mathcal{T}_{R_3}$ Kasner subsets. 

These results can be summarized as follows:
\begin{lemma}\label{lem:frameT}
The single frame transitions subset $\mathcal{T}_{R_1}$ (resp. $\mathcal{T}_{R_3}$ and the double frame transitions subset $\mathcal{T}_{R_1R_3}$)
consists of heteroclinic orbits such that their $\alpha$-limit set resides in the subset
$\mathbf{A}^-_{R_1}\subseteq \mathrm{K}^{\ocircle}$
(resp. $\mathbf{A}^-_{R_3}\subseteq \mathrm{K}^{\ocircle}$ and 
$\mathbf{A}^-_{R_1R_3}\subseteq \mathrm{K}^{\ocircle}$), whereas their $\omega$-limit set
resides in $\mathbf{A}^+_{R_1}\subseteq \mathrm{K}^{\ocircle}$
(resp. $\mathbf{A}^+_{R_3}\subseteq \mathrm{K}^{\ocircle}$ and
$\mathbf{A}^+_{R_1R_3}\subseteq \mathrm{K}^{\ocircle}$).
An explicit form for these solutions can be found in equation~\eqref{AppB:solTR1R3} of Appendix~\ref{app:Kasnersubset}.
\end{lemma}
\begin{figure}[H]\centering
\begin{tikzpicture}[scale=1.1]
        \draw [white]  (-0.5392,-1.4660) circle (0.001pt);
        \draw [white]  (1,1.2) circle (0.001pt);


\draw [gray, dashed, - ]  (1,1.2)--(1,0);
\draw [gray, dashed, - ]  (-1,1.2)--(-1,0);

\draw (1,0) arc (0:360:1cm and 1cm);	 	

       \draw [ultra thick, white, dashed, domain=-1.570:2.617-1.047,variable=\t,smooth] plot ({sin(\t r)},{-cos(\t r)});

        \node at (-0.7,1.15) {\scriptsize $\mathbf{A}^-_{R_1}$};
        \node at (-0.7,-1.15) {\scriptsize $\mathbf{A}^+_{R_1}$};


	\draw (-0.95,0) -- (-1.05,0);
        \draw (0.95,0) -- (1.05,0);

        \draw[rotate=120] (-0.95,0) -- (-1.05,0);
        \draw[rotate=240] (-0.95,0) -- (-1.05,0);

        \draw[rotate=120] (0.95,0) -- (1.05,0);
        \draw[rotate=240] (0.95,0) -- (1.05,0);

\draw [dotted, thick, postaction={decorate}]  (0,1)--(0,-1);

\draw [dotted, thick, postaction={decorate}]  (0.4,0.9165)--(0.4,-0.9165);
\draw [dotted, thick, postaction={decorate}]  (-0.4,0.9165)--(-0.4,-0.9165);
	
\draw [dotted, thick, postaction={decorate}]  (0.75,0.66143)--(0.75,-0.66143);
\draw [dotted, thick, postaction={decorate}]  (-0.75,0.66143)--(-0.75,-0.66143);

\filldraw [black] (0,1) circle (1pt) node[anchor= south] {\scriptsize{$\mathrm{K}^{\ocircle}_+$}};
\filldraw [black] (0,-1) circle (1pt) node[anchor= north] {\scriptsize{$\mathrm{K}^{\ocircle}_-$}};

 \end{tikzpicture}
\hspace{1cm}
\begin{tikzpicture}[scale=1.1]
        \draw [white]  (-0.5392,1.4660) circle (0.001pt);
        \draw [white]  (1,-1.2) circle (0.001pt);


        \draw [gray,dashed, - ]  (-0.5392,-1.4660)--(0.5,-0.8660);
        \draw [gray,dashed, - ]  (-1.5392,0.2660)--(-0.5,0.8660);

        \draw (1,0) arc (0:360:1cm and 1cm);	 	

       \draw [ultra thick, white, dashed, domain=-1.570+1.047:2.617,variable=\t,smooth] plot ({sin(\t r)},{cos(\t r)});

        \node at (-1.42,-0.75) {\scriptsize $\mathbf{A}^-_{R_3}$};
        \node at (1.3,0.6) {\scriptsize $\mathbf{A}^+_{R_3}$};

	\draw (-0.95,0) -- (-1.05,0);
        \draw (0.95,0) -- (1.05,0);

        \draw[rotate=120] (-0.95,0) -- (-1.05,0);
        \draw[rotate=240] (-0.95,0) -- (-1.05,0);

        \draw[rotate=120] (0.95,0) -- (1.05,0);
        \draw[rotate=240] (0.95,0) -- (1.05,0);

\draw [dotted, thick, postaction={decorate}]  (-0.8660,-0.5000)--(0.8660,0.5000);

\draw [dotted, thick, postaction={decorate}]  (-0.5937,-0.8047)--(0.9937,0.1118);
\draw [dotted, thick, postaction={decorate}]  (-0.9937,-0.1118)--(0.5937,0.8047);

\draw [dotted, thick, postaction={decorate}]  (-0.1978,-0.9802)--(0.9478,-0.3188);
\draw [dotted, thick, postaction={decorate}]  (-0.9478,0.3188)--(0.1978,0.9802);

	\end{tikzpicture}
\hspace{0.8cm}
\begin{tikzpicture}[scale=1.1]
        \draw [white]  (-0.5392,-1.4660) circle (0.001pt);
        \draw [white]  (1,1.2) circle (0.001pt);

        \draw [gray,dashed, - ]  (-0.5392,-1.4660)--(0.5,-0.8660);
        \draw [gray,dashed, - ]  (-1.5392,0.2660)--(-0.5,0.8660);
        \draw [gray, dashed, - ]  (1,1.2)--(1,0);
        \draw [gray, dashed, - ]  (-1,1.2)--(-1,0);

        \draw (1,0) arc (0:360:1cm and 1cm);	 	

       \draw [ultra thick, white, dashed, domain=1.570:2.617,variable=\t,smooth] plot ({sin(\t r)},{cos(\t r)});

	\draw (-0.95,0) -- (-1.05,0);
        \draw (0.95,0) -- (1.05,0);

        \draw[rotate=120] (-0.95,0) -- (-1.05,0);
        \draw[rotate=240] (-0.95,0) -- (-1.05,0);

        \draw[rotate=120] (0.95,0) -- (1.05,0);
        \draw[rotate=240] (0.95,0) -- (1.05,0);

        \draw [ultra thick, domain=-0.52:-1.570,variable=\t,smooth] plot ({sin(\t r)},{cos(\t r)});

        \node at (-1.42,0.75) {\scriptsize $\mathbf{A}^-_{R_1R_3}$};
        \node at (1.3,-0.6) {\scriptsize $\mathbf{A}^+_{R_1R_3}$};


    \draw [dotted, thick, postaction={decorate}]  (-0.866,0.5)--(-0.866,-0.5);\draw [dotted, gray]  (-0.866,0.5)--(-0.866,1.2);
    \draw [dotted, thick, postaction={decorate}]  (0.866,0.5)--(0.866,-0.5);\draw [dotted, gray]  (0.866,0.5)--(0.866,1.2);

    \draw [dotted, thick, postaction={decorate}]  (0,1)--(0,-1);\draw [dotted, gray]  (0,1)--(0,1.2);

    \draw [rotate=-240,dotted, thick, postaction={decorate}]  (-0.866,0.5)--(-0.866,-0.5);\draw [rotate=-240,dotted, gray]  (-0.866,0.5)--(-0.866,1.2);
    \draw [rotate=-240,dotted, thick, postaction={decorate}]  (0.866,0.5)--(0.866,-0.5);\draw [rotate=-240,dotted, gray]  (0.866,0.5)--(0.866,1.2);
    \draw [rotate=-240,dotted, thick, postaction={decorate}]  (0,1)--(0,-1);\draw [rotate=-240,dotted, gray]  (0,1)--(0,1.2);

    \def\u{-0.732}
    \foreach \a in {0.3, 0.7, 1.6} 
        \foreach \b in {0.7} 
        {
        \newcommand \n {(-\a*\b*(1-\u*\u)/((2+\u)*\u))}; 
        \newcommand \pone {(-\u/(1+\u+\u*\u))}; 
        \newcommand \ptwo {((1+\u)/(1+\u+\u*\u))}; 
        \newcommand \pthree {((\u*(1+\u))/(1+\u+\u*\u))}; 
        \newcommand \Sone {(-1+3*(\pone*\n*\n*exp(-6*\pone*(\t))+\ptwo*\b*\b*exp(-6*\ptwo*(\t))+\pthree*exp(-6*\pthree*(\t)))/(\n*\n*exp(-6*\pone*(\t))+\b*\b*exp(-6*\ptwo*(\t))+exp(-6*\pthree*(\t))))}; 
        \newcommand \Sthree {(-1+3*(\pone*exp(6*\pone*(\t))+\ptwo*\a*\a*exp(6*\ptwo*(\t))+\pthree*(\a*\b-\n)*(\a*\b-\n)*exp(6*\pthree*(\t)))/(exp(6*\pone*(\t))+\a*\a*exp(6*\ptwo*(\t))+(\a*\b-\n)*(\a*\b-\n)*exp(6*\pthree*(\t))))}; 
        \draw[dotted, ultra thick, postaction={decorate},variable=\t,domain=-1.75:0.995] plot ({-0.5*\Sone},{-(0.5*\Sone+\Sthree)/1.732});
        }

    \filldraw [black] (-0.866,0.5) circle (1pt);

    \filldraw [black] (0,1) circle (1pt);
    \filldraw [black] (0.866,0.5) circle (1pt);

    \filldraw [black] (-0.866,-0.5) circle (1pt);
    \filldraw [black] (0,-1) circle (1pt);

    \filldraw [black] (0.866,-0.5) circle (1pt);

	\end{tikzpicture}
\caption{Projection of heteroclinic frame transition orbits onto the plane
$\Sigma_1 + \Sigma_2 + \Sigma_3 = 0$ in $(\Sigma_1,\Sigma_2,\Sigma_3)$-space.
\textbf{Left and Middle:} The single frame transitions occur when either
$R_1 > 0$ (left) or $R_3 \neq0$ (middle).
The $\alpha$-limit set when $R_1>0$ (resp. $R_3\neq 0$) is a fixed point
in the arc $\mathbf{A}^-_{R_1}$ (resp. $\mathbf{A}^-_{R_3}$), while the
$\omega$-limit set is a fixed point in the (dashed) 
set denoted by $\mathbf{A}^+_{R_1}$ (resp. $\mathbf{A}^+_{R_3}$).
The special heteroclinic orbit
$\mathrm{K}^{\ocircle}_+\to \mathrm{K}^{\ocircle}_-$, which occurs when
$\Sigma_1=0$, plays a special role, since it forms one of the boundaries of
the ${\cal HO}$ subset, see Figure~\ref{fig:naa=0}. \textbf{Right:} A double frame transition,
which occurs when $R_1R_3 > 0$ or $R_1R_3 < 0$ (right), yields two equivalent one-parameter
sets of orbits with an $\alpha$-limit set given by a fixed point in the arc
$\mathbf{A}^-_{R_1R_3}:=(321)$, whereas the $\omega$-limit of these orbits is
a fixed point in the Kasner set denoted by $\mathbf{A}^+_{R_1R_3}:=(123)$. Thin (resp. bold)
dotted lines denote single (resp. double) frame transitions. Moreover, single transitions
are straight lines, whereas multiple transitions are curved (except when $R_1=R_3$).}\label{KMaps}
\end{figure}
As shown in Appendix~\ref{app:Kasnersubset}, there exists a
discrete symmetry, which corresponds to interchanging the indices $1$ and $3$
together with $(\Sigma_1,\Sigma_2,\Sigma_3) = - (\Sigma_1,\Sigma_2,\Sigma_3)$,
or letting $\tau \rightarrow - \tau$. Moreover, there exists a special 
orbit, for which $R_1 = R_3$, $\Sigma_2=0$, $\Sigma_3=-\Sigma_1$,
that originates from $\Sigma_1=\sqrt{3}$ and ends at $\Sigma_1 = - \sqrt{3}$ (where,
as for all double frame transitions, $\Sigma_1$ is monotonically decreasing), which
correspond to $\ue_- = 1 - \sqrt{3}$ and $\ue_+ = 1 + \sqrt{3}$, respectively, where
$\ue_+$ and $\ue_-$ correspond to $u = 1 + \sqrt{3}$, since $\ue_+$ is in sector $(123)$.
The projection of this orbit in $(\Sigma_1,\Sigma_2,\Sigma_3)$-space
is a straight line, see Figure~\ref{KMaps} (right).
Although the double frame transitions $\mathcal{T}_{R_1 R_3}$ satisfy~\eqref{Kasnerrot},
one more constant of the motion is needed to describe them completely. It is possible to
obtain all of the double frame transition orbits explicitly by means of a coordinate
transformation of the Kasner solutions in a shear diagonalized Fermi-propagated frame with
$R_1=R_3=0$, as shown by Lim~\cite{lim15}, see also~\cite{limmou22}. We derive the
solutions in a simple explicit form in the present variables in
Appendix~\ref{app:Kasnersubset}.

\subsection{The Bianchi type II subset ${\cal B}_\mathrm{II}$}\label{sec:BII}

The Bianchi type II subset ${\cal B}_\mathrm{II}$ is obtained by setting $A=0=R_3$.
Fermi-propagated solutions ($R_1=R_3=0$) such that $N_-> 0$ or $N_-<0$ yield
heteroclinic orbits triggered by the instability of the variable $N_-$. We refer
to these individual orbits and collectively as an invariant subset
with two disjoint components, as \emph{single curvature transitions} that
are denoted by $\mathcal{T}_{N_-}$. The heteroclinic orbits originate from
the Kasner arc $\mathbf{A}^-_{N_-}$ (i.e., the curvature transitions $\mathcal{T}_{N_-}$
converge to $\mathbf{A}^-_{N_-}$ as $\tau\to -\infty$) parametrized by $\ue_- \in (0,+\infty)$,
which consists of the sectors $(123)$, $(132)$ and the fixed point $\mathrm{Q}_1$,
and then end at the Kasner arc $\mathbf{A}^+_{N_-}$ (i.e., the curvature transitions
$\mathcal{T}_{N_-}$ converge to $\mathbf{A}^+_{N_-}$ as $\tau\to +\infty$), which
is the open complement of the $\mathbf{A}^-_{N_-}$ arc on $\mathrm{K}^\ocircle$ parametrized by
$ \ue_+ \in (-\infty,0)$. They are straight lines in $(\Sigma_1, \Sigma_2, \Sigma_3)$-space,
see Figure~\ref{fig:CurvatureTrans}, conveniently parametrized as follows
(see~\cite{heietal09}):
\begin{equation}\label{typeIIlines}
\Sigma_1 = -4 + (1+\ue_-^2)\zeta,\qquad
\Sigma_2 = 2- \ue_-^2\zeta,\qquad \Sigma_3 = 2 - \zeta,
\end{equation}
where $\ue_- \in (0,+\infty)$ parametrizes the initial
Kasner states, and the parameter $\zeta$ evolves according to
\begin{equation}\label{typeIIetaeq}
\zeta^\prime =  2(1-\Sigma^2) =
6\left(1 - \frac{\zeta}{\zeta_+}\right)\left(\frac{\zeta}{\zeta_-} - 1\right),
\end{equation}
where $\zeta_\pm := 3/f(\mp \ue_-)$. Thus, $\zeta$ is monotonically 
increasing from $\zeta_-$ to $\zeta_+$.
The variable $N_-$ is determined by the constraint
\begin{equation}
\sfrac16 \left(\Sigma_1^2 + \Sigma_2^2 + \Sigma_3^2\right) + N_-^2 = 1.
\end{equation}
Finally note that the single curvature transitions $\mathcal{T}_{N_-}$
become tangential to $\mathrm{T}_3$, 
which explains the zero eigenvalue associated with $N_-$ at $\mathrm{T}_3$.

The Bianchi type~II vacuum models assume a less transparent form in an
Iwasawa frame that is rotating with respect to a Fermi-propagated frame with $R_1 > 0$.
In this case the solutions are called \emph{mixed curvature-frame transitions},
which we denote by $\mathcal{T}_{R_1 N_-}$ since $N_-\neq 0$ and $R_1 \neq 0$
simultaneously. Combining the linearization results at $\mathrm{K}^\ocircle$ with
the fact that $\Sigma_1$ is monotonically increasing\footnote{The
equation for $\Sigma_1$ is the same for single curvature and mixed
curvature-frame transitions, due to that $\Sigma_1$ is unaffected by
rotations induced by $R_1$ in the $(2-3)$-plane. For details, see
Appendix D in~\cite{heietal09}.} leads to the conclusion that mixed curvature-frame
transitions originate from the Kasner arc $\mathbf{A}^-_{R_1N_-}$ (i.e., the mixed
curvature-frame transitions $\mathcal{T}_{R_1N_-}$ converge to
$\mathbf{A}^-_{R_1N_-}$ as $\tau\to -\infty$), which is the Kasner sector
$(132)$, parametrized by $\ue_- \in (0,1)$, and end at the Kasner arc
$\mathbf{A}^+_{R_1N_-}$ (i.e., the mixed curvature-frame transitions $\mathcal{T}_{R_1N_-}$
converge to $\mathbf{A}^+_{R_1N_-}$ as $\tau\to +\infty$), given by the Kasner sectors $(213)$,
$(231)$ and $\mathrm{Q}_2$, parametrized by $\ue_+ \in (-\infty,-1)$.
Since the boundary of the Bianchi type II set with $R_1N_-\neq 0$
is given by the union of the subsets $R_1=0$ and $N_-=0$, the final state on
$\mathrm{K}^\ocircle$  of a $\mathcal{T}_{R_1 N_-}$ transition
coincides with the Kasner circle point obtained by successively
applying the single curvature and frame transitions of the boundary
of $\mathcal{T}_{R_1 N_-}$. As in the case of double frame
transitions, single transitions act as building blocks that determine
the final result of multiple transitions, see Figure~\ref{fig:CurvatureTrans}.
These results are summarized as follows:
\begin{lemma}\label{lem:curvT}
The single curvature transitions subset $\mathcal{T}_{N_-}$ (resp. the mixed curvature-frame transitions $\mathcal{T}_{R_1N_-}$) consists of heteroclinic orbits such that their $\alpha$-limit set lies in
$\mathbf{A}^-_{N_-}\subseteq \mathrm{K}^{\ocircle}$ (resp. $\mathbf{A}^-_{R_1N_-}\subseteq \mathrm{K}^{\ocircle}$),
whereas their $\omega$-limit set resides in
$\mathbf{A}^+_{N_-}\subseteq \mathrm{K}^{\ocircle}$ (resp. $\mathbf{A}^+_{R_1N_-}\subseteq \mathrm{K}^{\ocircle}$).
An explicit form for these solutions is in equation~\eqref{solII} of Appendix~\ref{app:BII}.
\end{lemma}
\begin{figure}[H]
	\centering
\begin{tikzpicture}[scale=1.1]
        \draw [white]  (-0.5392,-1.4660) circle (0.001pt);
        \draw [white]  (1,1.2) circle (0.001pt);

	\draw[gray,dashed, -] (2,0) -- (0.5,-0.866);
	\draw[gray,dashed, -] (2,0) -- (0.5, 0.866);
	
	\draw [very thin] (1,0) arc (0:360:1cm and 1cm);	 	

       \draw [ultra thick, white, dashed, domain=2.617:4.713+2.094,variable=\t,smooth] plot ({sin(\t r)},{cos(\t r)});

    \draw[  postaction={decorate}] (1,0) -- (-1,0);\draw[color=gray,dotted] (2,0) -- (1,0);

    \draw[  postaction={decorate}] (0.99,0.124) -- (-0.93,0.35);\draw[color=gray,dotted] (2,0) -- (0.99,0.124);

    \draw[  postaction={decorate}] (0.95,0.32) -- (-0.6,0.8); \draw[color=gray,dotted] (2,0) -- (0.95,0.32);

    \draw[  postaction={decorate}] (0.86,0.51) -- (-0.2,0.98); \draw[color=gray,dotted] (2,0) -- (0.86,0.51);

    \draw[  postaction={decorate}] (0.99,-0.124) -- (-0.93,-0.35);\draw[color=gray,dotted] (2,0) -- (0.99,-0.124);

    \draw[  postaction={decorate}] (0.95,-0.32) -- (-0.6,-0.8); \draw[color=gray,dotted] (2,0) -- (0.95,-0.32);

    \draw[  postaction={decorate}] (0.86,-0.51) -- (-0.2,-0.98); \draw[color=gray,dotted] (2,0) -- (0.86,-0.51);

    \filldraw [black] (2,0) circle (0.7pt);
    \draw [black] (2,0) circle (0.1pt) node[anchor=west] {\scriptsize{$2\mathrm{Q}_1$}};

        \node at (1.3,0.35) {\scriptsize $\mathbf{A}^-_{N_-}$};
        \node at (-1.15,0.75) {\scriptsize $\mathbf{A}^+_{N_-}$};

   	\draw (-0.95,0) -- (-1.05,0) node[anchor= east] {\scriptsize{$\mathrm{T}_1$}};
        \draw (0.95,0) -- (1.05,0) node[anchor= west] {\scriptsize{$\mathrm{Q}_1$}};

        \draw[rotate=120] (-0.95,0) -- (-1.05,0) node[anchor= north] {\scriptsize{$\mathrm{T}_3$}};
        \draw[rotate=240] (-0.95,0) -- (-1.05,0) node[anchor= south] {\scriptsize{$\mathrm{T}_2$}};

        \draw[rotate=120] (0.95,0) -- (1.05,0) node[anchor= south] {\scriptsize{$\mathrm{Q}_3$}};
        \draw[rotate=240] (0.95,0) -- (1.05,0) node[anchor= north] {\scriptsize{$\mathrm{Q}_2$}};

        \end{tikzpicture}
\hspace{1cm}
\begin{tikzpicture}[scale=1.1]
        \draw [white]  (-0.5392,-1.4660) circle (0.001pt);
        \draw [white]  (1,1.2) circle (0.001pt);

    \draw (1,0) arc (0:360:1cm and 1cm);	 	

       \draw [ultra thick, white, dashed, domain=2.617:4.713,variable=\t,smooth] plot ({sin(\t r)},{cos(\t r)});

        \draw [gray, dashed, - ]  (1,1.2)--(1,0);
        \draw [gray, dashed, - ]  (-1,1.2)--(-1,0);
	\draw[gray,dashed, -] (2,0) -- (0.5,-0.866);
	\draw[gray,dashed, -] (2,0) -- (0.5, 0.866);

	\draw (-0.95,0) -- (-1.05,0);
        \draw (0.95,0) -- (1.05,0);

        \draw[rotate=120] (-0.95,0) -- (-1.05,0);
        \draw[rotate=240] (-0.95,0) -- (-1.05,0);

        \draw[rotate=120] (0.95,0) -- (1.05,0);
        \draw[rotate=240] (0.95,0) -- (1.05,0);

        \node at (1.45,0.75) {\scriptsize $\mathbf{A}^-_{R_1N_-}$};
        \draw[rotate=240] (0.95,0) -- (1.05,0) node[anchor= north] {\scriptsize{$\mathbf{A}^+_{R_1N_-}$}};

    \def\u{0.414}
    \foreach \a in {-0.9,0,0.9}
    {\draw[very thick ,postaction={decorate},variable=\t,domain=-2.4:3.2] plot ({(1+2*(2*\u/(1+\u*\u))*tanh(-\t))/(2+(2*\u/(1+\u*\u))*tanh(-\t))},{(1.732*((1-\u*\u)/(1+\u*\u))*tanh((((1-\u*\u)/(1+\u*\u))/(2*\u/(1+\u*\u)))*(-\t)-\a))/(2+(2*\u/(1+\u*\u))*tanh(-\t))});}

    \draw[  postaction={decorate}] (0.891,0.454) -- (-0.32,0.947);\draw [dotted, gray]  (0.891,0.454)--(2,0);

    \draw[  postaction={decorate}] (0.891,-0.454) -- (-0.32,-0.947);\draw [dotted, gray]  (0.891,-0.454)--(2,0);

    \draw [dotted, thick, postaction={decorate}]  (-0.32,0.947)--(-0.32,-0.947);\draw [dotted, gray]  (-0.32,0.947)--(-0.32,1.2);

    \draw [dotted, thick, postaction={decorate}]  (0.891,0.454)--(0.891,-0.454);\draw [dotted, gray]  (0.891,0.454)--(0.891,1.2);

    \draw [ultra thick, domain=0.52:1.570,variable=\t,smooth] plot ({sin(\t r)},{cos(\t r)});

    \filldraw [black] (-0.32,0.947) circle (1pt);
    \filldraw [black] (-0.32,-0.947) circle (1pt);

    \filldraw [black] (0.891,0.454) circle (1pt);
    \filldraw [black] (0.891,-0.454) circle (1pt);

\end{tikzpicture}
\hspace{1cm}
\begin{tikzpicture}[scale=1.1]
        \draw [white]  (-0.5392,-1.4660) circle (0.001pt);
        \draw [white]  (1,1.2) circle (0.001pt);

        \draw [gray, dashed, - ]  (1,1.2)--(1,0);
        \draw [gray, dashed, - ]  (-1,1.2)--(-1,0);
	\draw[gray,dashed, -] (2,0) -- (0.5,-0.866);
	\draw[gray,dashed, -] (2,0) -- (0.5, 0.866);

        \draw (1,0) arc (0:360:1cm and 1cm);	 	

       \draw [ultra thick, white, dashed, domain=2.617:4.713,variable=\t,smooth] plot ({sin(\t r)},{cos(\t r)});
	
	\draw (-0.95,0) -- (-1.05,0);
        \draw (0.95,0) -- (1.05,0);

        \draw[rotate=120] (-0.95,0) -- (-1.05,0);
        \draw[rotate=240] (-0.95,0) -- (-1.05,0);

        \draw[rotate=120] (0.95,0) -- (1.05,0);
        \draw[rotate=240] (0.95,0) -- (1.05,0);

        \node at (1.45,0.75) {\scriptsize $\mathbf{A}^-_{R_1N_-}$};
        \draw[rotate=240] (0.95,0) -- (1.05,0) node[anchor= north] {\scriptsize{$\mathbf{A}^+_{R_1N_-}$}};

    \def\u{0.55}
    \foreach \a in {-0.9,0,0.9}
    {\draw[very thick ,postaction={decorate},variable=\t,domain=-3:4.4] plot ({(1+2*(2*\u/(1+\u*\u))*tanh(-\t))/(2+(2*\u/(1+\u*\u))*tanh(-\t))},{(1.732*((1-\u*\u)/(1+\u*\u))*tanh((((1-\u*\u)/(1+\u*\u))/(2*\u/(1+\u*\u)))*(-\t)-\a))/(2+(2*\u/(1+\u*\u))*tanh(-\t))});}



    \draw[  postaction={decorate}] (0.95,0.32) -- (-0.6,0.8); \draw[color=gray,dotted] (2,0) -- (0.95,0.32);


    \draw[  postaction={decorate}] (0.95,-0.32) -- (-0.6,-0.8); \draw[color=gray,dotted] (2,0) -- (0.95,-0.32);

    \draw [dotted, thick, postaction={decorate}]  (-0.6,0.8)--(-0.6,-0.8);\draw [dotted, gray]  (-0.6,0.8)--(-0.6,1.2);

    \draw [dotted, thick, postaction={decorate}]  (0.95,0.32)--(0.95,-0.32);\draw [dotted, gray]  (0.95,0.32)--(0.95,1.2);

       \draw [ultra thick, domain=0.52:1.570,variable=\t,smooth] plot ({sin(\t r)},{cos(\t r)});

    \filldraw [black] (-0.6,0.8) circle (1pt);
    \filldraw [black] (-0.6,-0.8) circle (1pt);

    \filldraw [black] (0.95,0.32) circle (1pt);
    \filldraw [black] (0.95,-0.32) circle (1pt);

\end{tikzpicture}
\caption {Projection of Bianchi type II heteroclinic orbits onto the plane
$\Sigma_1 + \Sigma_2 + \Sigma_3 = 0$ in $(\Sigma_1,\Sigma_2,\Sigma_3)$-space.
\textbf{Left:} the single curvature transitions occur when $N_-\neq 0$: the
$\alpha$-limit set is a fixed point in $\mathbf{A}^-_{N_-}:=(132)\cup (123)\cup \mathrm{Q}_1 $,
whereas the $\omega$-limit set is a fixed point in the (dashed) 
set denoted by $\mathbf{A}^+_{N_-}$.
\textbf{Middle:} the mixed curvature-frame transitions occur when $R_1N_-\neq 0$.
A mixed curvature-frame transition yields two equivalent one-parameter set of orbits,
related by $N_- \mapsto - N_-$,
for which the $\alpha$-limit set is a fixed point in $\mathbf{A}^-_{R_1N_-}:=(132)$,
whereas the $\omega$-limit set is a fixed point in the (dashed) 
set denoted by
$\mathbf{A}^+_{R_1N_-}:=(231)\cup (213)\cup\mathrm{Q}_2$. We display the case of mixed curvature-frame
transitions with the $\alpha$-limit given by the Kasner fixed point with $\ue_-=\sqrt{2}-1$
and hence $u_- = 1 + \sqrt{2} > 2$,
which admit a special mixed curvature-frame transition orbit that is a straight line
when projected onto the plane $\Sigma_1 + \Sigma_2 + \Sigma_3 = 0$
in $(\Sigma_1,\Sigma_2,\Sigma_3)$-space.
\textbf{Right:} mixed curvature-frame transitions corresponding to $1<u_-<2$,
where the $\omega$-limit set is contained in sector $(231)$ instead of sector $(213)$, as in the
middle Figure. The thin (resp. bold) solid curves denote single (resp. mixed)
curvature transitions while the dotted lines stand for frame transitions, see Figure~\ref{KMaps}.
Note that the projections of single transitions are straight lines while this is not the case
for multiple transitions (except for one special orbit).
}\label{fig:CurvatureTrans}
\end{figure}
In Appendix~\ref{app:BII}, we derive the explicit form of solutions,
as well as the following integral,
\begin{equation}
\frac{(1 + \Sigma_1)^2 + 3N_-^2}{(4 + \Sigma_1)^2} = \left(\frac{\ue}{1 + \ue^2}\right)^2,
\end{equation}
where $\ue=\ue_-$ is the initial Kasner state of the transition and where this expression
also holds for the single curvature transitions $\mathcal{T}_{N_-}$ for which
$\ue = \ue_- \in (0,+\infty)$.\footnote{This integral corresponds to the one given in eq. (5.4)
in~\cite{hewetal03}. Note, however, that this integral does not suffice to determine the
mixed curvature-frame transitions orbits.}


We note that at $\ue_- = \sqrt{2} - 1 \in (132)$, there are two equal
unstable eigenvalues $\lambda_{R_1}=\lambda_{N_-}$. At this value of $\ue_-$,
as for all values of $\ue_- \in (132)$, there are two equivalent one-parameter sets
of mixed curvature-frame transitions, one for $N_-<0$ and one for $N_->0$.
There is a special orbit in the set for $\ue_- = \sqrt{2} - 1 \in (132)$ that
is a straight line in the projection onto the plane
$\Sigma_1 + \Sigma_2 + \Sigma_3 = 0$ in $(\Sigma_1,\Sigma_2,\Sigma_3)$-space,
described by
$\Sigma_1 = -2(1 - \Sigma_3)$, $\Sigma_2 = 2 - 3\Sigma_3$,
see Figure~\ref{fig:CurvatureTrans} (middle).

\subsection{The orthogonally transitive subset ${\cal OT}$}

The invariant ${\cal OT}$ subset is defined by $R_1=0$.
For this subset it is convenient to solve the constraint $\Sigma_1 + \Sigma_2 + \Sigma_3 = 0$
by introducing the following Misner parametrization adapted to the third spatial direction:
\begin{equation}\label{Misner3}
\Sigma_1 = \Sigma_+ + \sqrt{3}\Sigma_-,\qquad \Sigma_2 = \Sigma_+ - \sqrt{3}\Sigma_-,\qquad \Sigma_3 = -2\Sigma_+.
\end{equation}
This implies that the equations~\eqref{BilliardeqsolevolHnormVI} with $R_1=0$ lead
to the following evolution equations
\begin{subequations}\label{eq:OTspecSigma3}
\begin{align}
\Sigma_+^\prime &= 2(1 - \Sigma^2)\Sigma_+ + 2N_-^2 + \sfrac{3}{2}A^2,\\
\Sigma_-^\prime &= 2(1 - \Sigma^2)\Sigma_- + \sqrt{3} \left[ 2N_-^2 - 2R_3^2 - \sfrac{1}{2}A^2\right],\\
R_3^\prime &= 2[1 - \Sigma^2 + \sqrt{3}\Sigma_-]R_3 - 4N_-A,\\
N_-^\prime &= -2(\Sigma^2 + \Sigma_+ + \sqrt{3}\Sigma_-)N_- + 3R_3A,\\
A^\prime &= -2(\Sigma^2 +\Sigma_+)A,
\end{align}
\end{subequations}
while the remaining constraints are given by
\begin{subequations}\label{eq:OTspecSigma3constr}
\begin{align}
1 - \Sigma^2 - N_-^2 - A^2 &= 0, \label{OTGauss1}\\
2R_3N_- + (\Sigma_+ + \sqrt{3}\Sigma_-)A &= 0,\label{OTCodazzi1}
\end{align}
\end{subequations}
where $\Sigma^2 = \Sigma_+^2 + \Sigma_-^2 + R_3^2$.

The subset ${\cal OT}$ contains the plane wave fixed points $\mathrm{PW}^\pm$ in~\eqref{PW},
which in the Misner variables~\eqref{Misner3} yields
\begin{equation}\label{PWOT}
\mathrm{PW}^\pm := \left\{
\begin{array}{c}
(\Sigma_-, A) = (1 + \Sigma_+)(\sqrt{3}, 2), \\
(R_1, R_3, N_-) = \pm\sqrt{-(3 + 4\Sigma_+)(1 + \Sigma_+)}\,\left(0, 1, 1\right)
\end{array}
\Bigm| \Sigma_+\in \left(-1,-\frac{3}{4}\right]
\right\},
\end{equation}
where $\Sigma_+ = \mathrm{constant}$.
In~\cite{hewetal93}, it was shown that the ${\cal OT}$ case admits a bounded
(from above and below, since $|\Sigma_+|\leq 1$, $A\leq 1$) \emph{non-negative}\footnote{Since $Z$
is invariant under frame rotations in the $(1-2)$-plane, it is not a coincidence that the
proof that $Z\geq 0$ in~\cite{hewetal93} is also given in
terms of such invariants.} monotonic function, which in the present variables takes the form
\begin{equation}\label{lyap:OT}
Z := (1+\Sigma_+)^2-\frac{A^2}{4},
\end{equation}
for which
\begin{equation}\label{lyap:OTeq}
Z^\prime = 4 (1-\Sigma^2) Z.
\end{equation}
Hence, the quantity $Z$ is monotonically increasing, except when $Z=0$ or $\Sigma^2=1$,
where $Z=0$ implies that $A=2(1+\Sigma_+)$, which characterizes the set
$\overline{\mathrm{PW}}^\pm = \mathrm{PW}^\pm\cup \mathrm{T}_3$,
while $\Sigma^2=1$ implies that $N_-=A=0$ due to the
constraint~\eqref{OTGauss1}.
The set $\Sigma^2=1$ consists of the union of the Kasner circle, $\mathrm{K}^\ocircle$ (when $R_3=0$)
and the single transition subset, $\mathcal{T}_{R_3}$ (when $R_3\neq 0$).
It then follows from the stability properties of $\mathrm{PW}^\pm$ and $\mathrm{K}^\ocircle$ that
the $\alpha$- and $\omega$-limit sets for these orbits reside in
$\overline{\mathrm{PW}}^\pm$ and the stable sector $S_{\mathcal{OT}}:= (312)$ of the Kasner
circle $\mathrm{K}^{\ocircle}$, respectively, see Figure~\ref{OTasymp}. Therefore, due to the
monotonicity principle in~\cite{waiell97}, the following result:\footnote{One can also use the bounded monotone function:
$\bar{Z} := \frac{(1+\Sigma_+)^2 - \sfrac14A^2}{(1+\Sigma_+)^2 + 
\sfrac34A^2}\geq 0$ so that $\bar{Z}^\prime = 4(1 + \Sigma_+)\bar{Z}(1 - \bar{Z})$.
}
\begin{lemma}\label{lem:OT}
The orthogonally transitive subset $\mathcal{OT}$ consists of heteroclinic orbits such that their $\alpha$-limit set resides in
$\overline{\mathrm{PW}}^\pm$, whereas their $\omega$-limit set resides in the sector
$S_{\mathcal{OT}} := (312)$ of the Kasner circle $\mathrm{K}^{\ocircle}$.
\end{lemma}
The asymptotic subsets of the $\mathcal{OT}$ subset are depicted in
Figure~\ref{OTasymp}.
We note that the features of the monotone function $Z$ do not prove if
there are (or are not) any $\mathcal{OT}$ orbits that possess the Taub
point $\mathrm{T}_3$ as its $\alpha$-limit set.
%
%
\begin{figure}[H]
	\centering
       \begin{tikzpicture}[scale=1.1]
        \draw[color=gray,->,rotate=210] (0,0) -- (-0.343,0)  node[anchor= west] {\scriptsize{$\Sigma_{-}$}};
        \draw[color=gray,->,rotate=-60] (0,0) -- (-0.343,0)  node[anchor= south] {\scriptsize{$\Sigma_{+}$}};

	\draw[gray,dashed, -] (2,0) -- (0.41,-0.91);
	\draw[gray,dashed, -] (2,0) -- (0.41, 0.91);

        \draw [gray,dashed, - ]  (-0.5392,-1.4660)--(0.5,-0.8660);
        \draw [gray,dashed, - ]  (-1.5392,0.2660)--(-0.5,0.8660);

	\draw [very thin] (1,0) arc (0:360:1cm and 1cm);	 	
       \draw [ultra thick, white, dashed, domain=-0.523:0.523,variable=\t,smooth] plot ({sin(\t r)},{cos(\t r)});
        \node at (0, 1.2) {\scriptsize{$\mathcal{S}_\mathcal{OT}$}};

        \draw [very thick]  (0.5,-0.866)--(0,-0.8660) node[anchor= south] {\scriptsize{$\mathrm{PW}^\pm$}};

        \draw (-0.95,0) -- (-1.05,0);
        \draw (0.95,0) -- (1.05,0);

        \filldraw [rotate=120,black] (-1,0) circle (1.25pt) node[anchor=north west]{\scriptsize $\mathrm{T}_3$};

        \draw[rotate=240] (-0.95,0) -- (-1.05,0);

        \draw[rotate=120] (0.95,0) -- (1.05,0);
        \draw[rotate=240] (0.95,0) -- (1.05,0);


	\end{tikzpicture}
\caption {Projection onto the $(\Sigma_+, \Sigma_-)$-plane of the asymptotic subsets for the
$\mathcal{OT}$ subset, which are described by the minima and maxima of the monotone function $Z$. The
$\alpha$-limits are contained in the closure of the plane wave fixed points
$\overline{\mathrm{PW}}^\pm$. It is unknown if there are $\mathcal{OT}$ orbits
that have $\mathrm{T}_3$ as their $\alpha$-limit. The $\omega$-limits are contained
in the (dashed) stable set $\mathcal{S}_{OT}\subseteq \mathrm{K}^{\ocircle}$. }\label{OTasymp}
\end{figure}
%

\subsection{The hypersurface orthogonal subset ${\cal HO}$}\label{sec:naa=0}

The 2D hypersurface orthogonal invariant subset ${\cal HO}$ occurs 
when $n^\alpha\!_\alpha=0$, see e.g.~\cite{waiell97}, which in the
present variables results in $\Sigma_1=R_3=N_-= 0$. In 
Appendix~\ref{app:D}, we show that the dynamics in this invariant 
subset can be described by introducing 
$\sqrt{3}\Sigma_-:=\Sigma_2 = -\Sigma_3$, which leads to
\begin{subequations}\label{ODE:D}
\begin{align}
\Sigma_-^\prime &= (1-\Sigma^2)(2\Sigma_- + \sqrt{3}) - 2\sqrt{3} R_1^2,\\
R_1^\prime &= 2(1-\Sigma^2 + \sqrt{3}\Sigma_-)R_1,
\end{align}
\end{subequations}
where $\Sigma^2 = \Sigma_-^2 + R_1^2$, while the constraint~\eqref{Gauss1} 
implies that $A^2 = 1 - \Sigma^2$.
Setting $R_1=0$ yields the 1D diagonal shear and Fermi-propagated subset
${\cal D}$, on which
\begin{equation}\label{Devol}
\Sigma_-^\prime = (1-\Sigma_-^2)(2\Sigma_- + \sqrt{3}). 
\end{equation}
This subset contains three fixed points: $\mathrm{PW}^0$ with $\Sigma_-=-\sqrt{3}/2$ and
the two Kasner fixed points with $\Sigma_-=\pm 1$, denoted by $\mathrm{K}^\ocircle_\pm$, given by~\eqref{Kpm}.
Moreover, the invariant set ${\cal D}$ with the evolution equation~\eqref{Devol}
consists of two heteroclinic orbits $\mathrm{PW}^0 \rightarrow \mathrm{K}^\ocircle_\pm$,
see Figure \ref{fig:naa=0}.

To obtain proofs about the two-dimensional dynamical system described by~\eqref{ODE:D}
with $R_1>0$, we use the following new monotonic function that is derived in Appendix~\ref{app:D}
using Hamiltonian techniques:
\begin{equation}\label{LyapMinHO}
M := \frac{(1 - v\Sigma_-)^2}{[(1 - \Sigma^2)^{9}R_1^5]^{\sfrac{2}{23}}},
\end{equation}
where $v := -2/(3\sqrt{3})$. The variable $M$ evolves according to
\begin{equation}
M^\prime = \frac{27}{23}\frac{\left(\Sigma_- - v\right)^2}{\left(1 - v\Sigma_-\right)}M,
\end{equation}
where $1-v\Sigma_->0$, since $|\Sigma_-| \leq 1$, due to the constraint~\eqref{Gauss1}.

Hence $M$ is monotonically increasing in the interior of the ${\cal HO}$ state space, i.e.,
when $R_1>0$ and $\Sigma^2 = \Sigma_-^2 + R_1^2 < 1$. This implies that the fixed point
$\mathrm{RT}$, given by $(\Sigma_-,R_1) = \left(-2/(3\sqrt{3}), \sqrt{5}/(3\sqrt{3})\right)$,
is a global minimum of $M$ and thereby $\mathrm{RT}$ is a global source. Moreover, since
$M\rightarrow \infty$, it follows that $R_1=0$ or $\Sigma^2=1$ when $\tau\rightarrow \infty$.
The former consists of the invariant set ${\cal D}$, whereas the latter consists of a
frame transition ${\cal T}_{R_1}$ (and hence $A=0$) given by
$\mathrm{K}^\ocircle_+ \rightarrow \mathrm{K}^\ocircle_-$. Incidentally, this
heteroclinic orbit is part of a heteroclinic network on the boundary of the
full state space playing an important role.

In combination with the heteroclinic orbits on the boundary 
and the local fixed point analysis, $M$ determines the global dynamics of the
invariant ${\cal HO}$ subset. In particular, $\mathrm{PW}^0$ is a
saddle that attracts a single interior heteroclinic orbit coming from $\mathrm{RT}$,
i.e., $\mathrm{RT} \rightarrow \mathrm{PW}^0$. Lastly, $\mathrm{K}^\ocircle_-$
is a sink that attracts all interior orbits, except for the heteroclinic orbit
$\mathrm{RT} \rightarrow \mathrm{PW}^0$, see Figure \ref{fig:naa=0}.
We therefore obtain the following result:
\begin{lemma}\label{lem:HO}
Apart from the fixed points and the heteroclinic orbits given by 
$\mathrm{RT} \rightarrow \mathrm{PW}^0$, $\mathrm{PW}^0 \rightarrow \mathrm{K}^\ocircle_\pm$,
$\mathrm{K}^\ocircle_+ \rightarrow \mathrm{K}^\ocircle_-$,
the hypersurface orthogonal subset $\mathcal{HO}$ consists of heteroclinic orbits such that
their $\alpha$-limit set is $\mathrm{RT}$, whereas their
$\omega$-limit set is $\mathrm{K}^\ocircle_-$. 
\end{lemma}
Due to the constraint $R_1^2 = 1 - \Sigma_-^2 - A^2$, one can alternatively
use $\Sigma_-$ and $A$ as variables, which yields the dynamical system
\begin{subequations}\label{ODE:D2}
\begin{align}
\Sigma_-^\prime &= -2\sqrt{3}(1-\Sigma_-^2) + (2\Sigma_- + 3\sqrt{3})A^2,\\
A^\prime &= -[2(1-A^2) + \sqrt{3}\Sigma_-]A.
\end{align}
\end{subequations}
For the details of the dynamical systems features in the two representations
$(\Sigma_-,R_1)$ and $(\Sigma_-,A)$ for the invariant subset ${\cal HO}$,
see Figure~\ref{fig:naa=0}.
%
%
\begin{figure}[H]
\centering
\begin{tikzpicture}[scale=1.1]
        \draw [lightgray,-]  (0,-1.25)--(0,1.25) node[anchor= south] {\scriptsize{$\Sigma_{-}$}};
        \draw [lightgray,-]  (0,0)--(1.25,0) node[anchor= west] {\scriptsize{$R_{1}$}};

       \draw [dotted, thick, domain=0:3.14,variable=\t,smooth] plot ({sin(\t r)},{cos(\t r)});
        \draw [postaction={decorate}]  (0.976,0.18)--(0.977,0.175);

        \draw [postaction={decorate}]  (0,-0.866)--(0,1);
        \draw [postaction={decorate}]  (0,-0.98)--(0,-1);

        \draw [domain=0:-18.6,variable=\t,smooth, samples=75,shift={(0.43,-0.384)}] plot ({\t r}: {0.001*\t*\t});
        \draw [domain=2.03:3.14,variable=\t,smooth] plot ({0.86*sin(\t r)},{0.86*cos(\t r)});
        \draw [postaction={decorate}]  (0.77,-0.4)--(0.77,-0.401);


        \filldraw [black] (0,1) circle (1pt) node[anchor= east] {\scriptsize{$\mathrm{K}^\ocircle_+$}};
        \filldraw [black] (0,-1) circle (1pt) node[anchor= east] {\scriptsize{$\mathrm{K}^\ocircle_-$}};

        \filldraw [black] (0,-0.866) circle (0.5pt) node[anchor= south east] {\scriptsize{$PW^0$}};

        \filldraw [black] (0.430,-0.384) circle (0.5pt) node[anchor= south] {\scriptsize{$RT$}};

 \end{tikzpicture}
\hspace{2.5cm}
\begin{tikzpicture}[scale=1.1]
        \draw [lightgray,-]  (0,-1.25)--(0,1.25) node[anchor= south] {\scriptsize{$\Sigma_{-}$}};
        \draw [lightgray,-]  (0,0)--(1.25,0) node[anchor= west] {\scriptsize{$A$}};

       \draw [domain=0:3.14,variable=\t,smooth] plot ({sin(\t r)},{cos(\t r)});
        \draw [postaction={decorate}]  (0.977,0.175)--(0.976,0.18);
        \draw [postaction={decorate}]  (0.2,-0.99)--(0.199,-0.9904);


        \draw [domain=0:15.4,variable=\t,smooth, samples=75,shift={(0.816,-0.384)}] plot ({\t r}: {0.0005*\t*\t});
        \draw [domain=-0.9:-1.57,variable=\t,smooth] plot ({1.5+sin(\t r)},{-(0.866)+0.866*cos(\t r)});
        \draw [postaction={decorate}]  (0.62,-0.44)--(0.619,-0.441);


        \filldraw [black] (0,1) circle (1pt) node[anchor= east] {\scriptsize{$\mathrm{K}^\ocircle_+$}};
        \filldraw [black] (0,-1) circle (1pt) node[anchor= east] {\scriptsize{$\mathrm{K}^\ocircle_-$}};

        \filldraw [black] (0.5,-0.866) circle (0.5pt) node[anchor= north west] {\scriptsize{$PW^0$}};

        \filldraw [black] (0.816,-0.384) circle (0.5pt) node[anchor= south] {\scriptsize{$RT$}}; 

        \draw [dotted, thick, postaction={decorate}]  (0,1)--(0,-1);

 \end{tikzpicture}
\caption{Schematic orbit structure of the 2D invariant set ${\cal HO}$ using different variables.
Note that the (dotted) heteroclinic orbit $\mathrm{K}^\ocircle_+\to \mathrm{K}^\ocircle_-$
is also contained in the set $\mathcal{T}_{R_1}$, see Figure~\ref{KMaps}.
\textbf{Left:} the orbit structure resulting from the dynamical system~\eqref{ODE:D}
using the variables $(\Sigma_-,R_1)$.
\textbf{Right:} the orbit structure resulting from the dynamical system~\eqref{ODE:D2}
using $(\Sigma_-,A)$.}\label{fig:naa=0}
\end{figure}
%

\subsection{The $\mathcal{W}^s(\mathrm{PW}^\pm)$ subset
}\label{RTtoPWpm}

As follows from the linear analysis of $\mathrm{PW}^\pm$ (including $\mathrm{PW}^0$)
in~\eqref{lin:PW}, there exists an invariant 2D subset that is not explicitly known.
Moreover, it shows that this is a transversally hyperbolic
1D set of fixed points, where each fixed point attracts a single orbit from the
interior generic 4D set, which amounts to a foliation of a 2D invariant set
$\mathcal{W}^s(\mathrm{PW}^\pm)$. We now make the following conjecture about the
global properties of $\mathcal{W}^s(\mathrm{PW}^\pm)$ and its boundary characterization.
\begin{conjecture}\label{conj:Ws}
\hfill \break\vspace{-5mm}
\begin{itemize}
\item[(i)] The orbits in the manifold $\mathcal{W}^s(\mathrm{PW}^\pm)$
originate from the local fixed point source $\mathrm{RT}$, i.e.,
the $\alpha$-limit set of all solutions in $\mathcal{W}^s(\mathrm{PW}^\pm)$
is the fixed point $\mathrm{RT}$, where the orbit
$\mathrm{RT} \rightarrow \mathrm{PW}^0$ contained in the ${\cal HO}$
subset divides $\mathcal{W}^s(\mathrm{PW}^\pm)$ into two parts:
$\mathrm{RT} \rightarrow \mathrm{PW}^+$
and $\mathrm{RT} \rightarrow \mathrm{PW}^-$.
\item[(ii)] The boundaries of the two parts $\mathrm{RT} \rightarrow \mathrm{PW}^\pm$,
and thereby the whole subset $\mathcal{W}^s(\mathrm{PW}^\pm)$, 
are given by two heteroclinic orbits $\mathrm{RT} \rightarrow \mathrm{T}_3$,
where one orbit is approaching $\mathrm{T}_3$ from the positive $R_3$, $N_-$ direction
and the other from the negative $R_3$, $N_-$ direction, due to the discrete
symmetry $(R_3,N_-) \mapsto -(R_3,N_-)$.
\end{itemize}
\end{conjecture}
%

\section{Discrete dynamics and heteroclinic networks}\label{sec:discdyn}

It is generally believed that the attractor ${\cal A}$ for the Bianchi type
$\mathrm{VI}_{-1/9}$ vacuum models, which describes the generic behaviour in
the vicinity of the initial singularity of these models, resides in the union of
the Bianchi type I Kasner subset, ${\cal K}$, and the Bianchi type II subset,
${\cal B}_\mathrm{II}$, which leads to the following conjecture introduced
in~\cite{hewetal03}:
\begin{conjecture}\label{conj1}
%
${\cal A} \subseteq {\cal K}\cup{\cal B}_\mathrm{II} = 
\mathrm{K}^\ocircle\cup{\cal T}_{R_1}\cup{\cal T}_{R_3}\cup{\cal T}_{R_1R_3}\cup{\cal T}_{N_-}\cup{\cal T}_{R_1N_-}$.
%
\end{conjecture}
Supported by the numerical investigations in~\cite{hewetal03}, which suggests
that multiple transitions become increasingly rare toward the singularity and thus that only
single transitions play an asymptotic role. This suggests the stronger conjecture:
\begin{conjecture} \label{conj2}
%
${\cal A} \subseteq 
\mathrm{K}^\ocircle\cup{\cal T}_{R_1}\cup{\cal T}_{R_3}\cup{\cal T}_{N_-}$.
%
\end{conjecture}
However, as discussed in~\cite{lim04}, numerical experiments suggest that although multiple transitions become increasingly rare they still may asymptotically occur.  
In addition, heuristic dynamical arguments are provided in~\cite{heietal09} for the general inhomogeneous case, of which Bianchi $\mathrm{VI}_{-1/9}$ is a special case, that favor Conjecture~\ref{conj2}.

For Bianchi types $\mathrm{VIII}$ and $\mathrm{IX}$, the attractor theorems in~\cite{rin01,heiugg09,bre16} establish
that generic solutions converge to the union of the Bianchi types I and II subsets. However, it is still an open question which of the heteroclinic chains on these subsets are asymptotically approximated (or not). 
For Bianchi type~$\mathrm{VI}_{-1/9}$, we will see in Section~\ref{sec:cyclic} that there are isolated heteroclinic chains that are not asymptotically relevant for a certain class of solutions. This raises the question of whether ${\cal A}$ is a strict subset of Bianchi types I and II.
This new phenomena observed for type~$\mathrm{VI}_{-1/9}$ does not occur for types $\mathrm{VIII}$ and $\mathrm{IX}$.

In view of the new generation of proofs for Bianchi types $\mathrm{VIII}$ and $\mathrm{IX}$, which uses the properties of the Bianchi type $\mathrm{I}$ and $\mathrm{II}$ heteroclinic network~\cite{Beguin10,Lieb11,Lieb13,BegDut22}, any asymptotic
proofs about the singularity for the Bianchi type $\mathrm{VI}_{-1/9}$ vacuum model is
likely to require a thorough understanding of the heteroclinic Bianchi type $\mathrm{I}$ and $\mathrm{II}$ network
of these models. To obtain this network we will use the new frame rotating Bianchi type
$\mathrm{I}$ and $\mathrm{II}$ solutions to translate the spatially frame-invariant Kasner
parameter $u$ and the consequences of the iterated BKL Kasner map obtained by
Belinski\v{\i} et al.~\cite{bkl70}, which we turn to next.

%
%
%
%

\subsection{Frame-invariant discrete dynamics\label{subsec:frameinv}}


According to Belinski\v{\i} et al.~\cite{bkl70}, the Bianchi type II solutions
induce a gauge- and spatially frame-independent map between different Kasner states,
described by the gauge- and spatially frame-independent parameter $u$, given by
\begin{equation}\label{BKLMap}
u_+  = \left\{
\begin{array}{ll}
u_- - 1 & \qquad \text{if}\quad u_- \in [2, \infty), \\[1ex]
\frac{1}{u_- - 1} &\qquad  \text{if} \quad u_- \in [1,2],
\end{array}\right.
\end{equation}
which is known as the \emph{BKL Kasner map}.

A sequence of transitions corresponds to an iteration of~\eqref{BKLMap}.
The curvature transition from $u_i$ to $u_{i+1}$, $i=0,1,2,\ldots$,
\begin{equation}
u_i \:\,\xrightarrow{\;\text{$i^{th}$ curvature
transition}\;}\:\, u_{i+1},
\end{equation}
is thereby described by the BKL Kasner map,
\begin{equation}\label{Kasnermap}
u_{i+1} \:= \: \left\{\begin{array}{ll}
u_i - 1 & \qquad \text{if}\quad u_i \in[2,\infty), \\[1ex]
\frac{1}{u_i - 1} &\qquad  \text{if} \quad u_i \in [1,2].
\end{array}\right.
\end{equation}
%
The information contained in this map suffices to
represent the collection of all transition orbits (as a whole)
that undergo a single or mixed curvature-frame transition.
%


In a sequence $(u_i)_{i\in\mathbb{N}_0}$ that is generated by an iteration
of the BKL Kasner map~\eqref{Kasnermap}, each Kasner state $u_i$ is called
a (spatially frame-invariant) \emph{Kasner epoch}. A given sequence,
$(u_i)_{i\in\mathbb{N}_0}$ of Kasner epochs possesses a natural partition
into pieces called (spatially frame-invariant) \emph{Kasner eras},
based on the properties of~\eqref{Kasnermap}, see~\cite{bkl70}:
\begin{equation}\label{EraPartition}
\{u_0,u_1,u_2,\ldots\} = \{\underbrace{u^0_{0},\ldots , u^0_{\mathrm{f}}}_{\text{\scriptsize Initial era}}, \underbrace{u^1_{0},\ldots , u^1_f}_{\text{\scriptsize $1^{st}$ era}}, \underbrace{u^2_{0},\ldots , u^2_f}_{\text{\scriptsize $2^{nd}$ era}}, \ldots, \underbrace{u^j_{0},\ldots , u^j_f}_{\text{\scriptsize $j^{th}$ era}},\ldots\},
\end{equation}
where $u^0_0=u_0$ and where each $j^{th}$ era begins with the largest Kasner
epoch value $u^j_0$ in the era followed by a finite monotonically decreasing sequence of
Kasner parameters, obtained from $u^j_{k+1} = u^j_k - 1$, until the era reaches
the final smallest value $u^j_f \in (1, 2)$, which in turn yields the initial Kasner parameter
value $u^{j+1}_0 = 1/(u^j_f -1)$ of the next $(j+1)^{th}$ era (thus $u=2$ plays a
special role for the map~\eqref{Kasnermap} and identifying era changes).
For example,
\begin{equation}\label{phases}
\underbrace{
5.29  \rightarrow 4.29 \rightarrow 3.29 \rightarrow 2.29 \rightarrow 1.29}_{\text{\scriptsize Initial era}}\rightarrow
\underbrace{3.44 \rightarrow 2.44 \rightarrow 1.44}_{\text{\scriptsize $1^{st}$ era}}  \rightarrow
\underbrace{2.27 \rightarrow 1.27}_{\text{\scriptsize $2^{nd}$ era}}  \rightarrow \ldots
\end{equation}
Following~\cite{bkl70,khaetal85}, we decompose each $u^j_0$ into its integer,
$\lfloor u^j_0\rfloor\in\mathbb{N}$, and
fractional, $\{u^j_0\}\in (0,1)$, parts. Hence, in the
$j^{th}$ era, the Kasner parameter $u^j_{0}$ denotes the maximal value
of the parameter among the $\lfloor u^j_0\rfloor$ Kasner epochs in the $j^{th}$ era,
which decreases $\lfloor u^j_0\rfloor - 1$ times by $1$ until its minimal value,
$u^j_f = 1+\{u^j_0\}$.
%
%
This induces the following definition, called the \emph{Era map},
\begin{equation}\label{Tt}
u^j_0 \mapsto u^{j+1}_{0} = \frac{1}{\{u_0^j\}},
\end{equation}
which thereby generates a sequence $(u_0^j)_{j\in\mathbb{N}_0}$.
It is known that the transformation~\eqref{Tt}
is associated with exponential instability: the distance between two
sufficiently close points grows exponentially with
the number of iterations $j$, see~\cite{khaetal85}.

Next, we give an interpretation of the Kasner parameter $u$ and the
map~\eqref{Kasnermap} in terms of continued fraction representations. Before
doing so, however, we introduce the following notation for the set of
Kasner parameter values in the $j^{th}$ era:
\begin{equation}\label{EraPoints}
\mathbb{E}^j:=\{u^j_{0},\ldots, u^j_f\}.
\end{equation}
Now, consider a Kasner sequence $\{u_0,u_1,\ldots\}$, partitioned into different
eras according to~\eqref{EraPartition}. The initial Kasner epoch value, which
is the first epoch in the initial era, i.e., $u_0=u_0^0\in\mathbb{R}$, is
represented by the continued fraction:
%
%
\begin{equation}\label{ContFrac}
u_0 = k_0 + \cfrac{1}{k_1 + \cfrac{1}{k_2 + \dotsb}} =: [k_0; k_1,k_2,k_3,\dotsc],
\end{equation}
where $\lfloor u_0\rfloor=k_0$ and $\{u_0\}=[0;k_1,k_2,k_3,\dotsc]$
and where $k_j = \lfloor u_0^j\rfloor$ is the length of the $j^{th}$ era.
According to the BKL Kasner map~\eqref{Kasnermap}, the value $u_0$ decreases by $1$
a total of $k_0 - 1$ times until it reaches a final value $u_f^0\in (1,2)$.
The BKL Kasner map~\eqref{Kasnermap} then maps the final epoch of the initial era
to the initial epoch of the $1^{st}$ era,
$u_f^0 \in \mathbb{E}^0\mapsto u_0^1 = 1/(u_f^0-1) \in \mathbb{E}^1$,
which leads to the continued fraction representation
\begin{equation}
u_0^1 = k_1 + \cfrac{1}{k_2 + \cfrac{1}{k_3 + \dotsb}} = [k_1; k_2,k_3,k_4,\dotsc].
\end{equation}
The \emph{Era map} is thereby simply a shift to the left in the continued fraction expansion,
\begin{equation}\label{EraMap}
u_0^j = [k_j; k_{j+1}, k_{j+2}, \dotsc] \:\mapsto\:
u_0^{j+1} = [k_{j+1}; k_{j+2}, k_{j+3},\dotsc].
\end{equation}
It thus follows that the Kasner sequence $(u_i)_{i\in\mathbb{N}_0}$,
generated by $u_0 =  \big[ k_0; k_1, k_2 , \dotsc \big]$, amounts to the following
concatenation of maps in terms of continued fractions:
%
%
%
%
\begin{align}\label{mapCF}
u_0 = u_0^0 & =
\big[ k_0; k_1, k_2,  \dotsc \big] \rightarrow  \big[ k_0 -1 ; k_1, k_2 , \dotsc \big]
\rightarrow \ldots \rightarrow \big[1 ; k_1, k_2 , \dotsc \big] \nonumber\\
\rightarrow u_0^1 & = \big[ k_1; k_2 , k_3, \dotsc \big]
\rightarrow  \big[  k_1-1; k_2 , k_3, \dotsc \big]
\rightarrow \ldots \rightarrow \big[1 ; k_2, k_3 , \dotsc \big] \\
\rightarrow u_0^2 & = \big[ k_2; k_3 , k_4, \dotsc \big]
\rightarrow  \big[ k_2-1; k_3 , k_4, \dotsc \big]
\rightarrow \ldots \nonumber
\end{align}
Below we relate properties of Kasner sequences to properties of $u_0$
by inferring these from the continued fraction expansion of $u_0$.
We are particularly interested in generic dynamics of~\eqref{mapCF}, but note
that genericity is a delicate issue since there are several notions of genericity,
such as measure theoretical genericity (with respect to the Lebesgue measure), topological
genericity (with respect to the Baire measure) or dimensional genericity (with respect to the Hausdorff dimension).
\begin{itemize}
\item[(i)] The initial Kasner parameter is a \emph{rational number},
$u_0 \in \mathbb{Q}$, if, and only if, its continued fraction
representation is finite:
\begin{equation}\label{terminalki}
u_0 = [k_0; k_1, k_2, \dotsc, k_n],
\end{equation}
such that $k_n > 1$; \cite[Lemma 4B]{Schmidt}. The Kasner sequence is finite
with $n+1$ eras where the last era begins with $k_n$. At the end of the last era the
Kasner parameter reaches $u^n_f=1$, which subsequently terminates the
recursion~\eqref{Kasnermap} at a Taub state/epoch with $u = \infty$.
Even though $\mathbb{Q}$ is a dense set in $\mathbb{R}$, it is a set of Lebesgue measure zero. 

On the other hand, the irrational numbers, $\mathbb{R}\backslash\mathbb{Q}$, consist
of a dense set of full Lebesgue measure in $\mathbb{R}$. We will now describe several
interesting subcases. 
%
\item[(ii)] The initial Kasner parameter $u_0$ is a \emph{badly approximable irrational
number}\footnote{A number $u_0\in\mathbb{R}\backslash\mathbb{Q}$ is called a
\emph{badly approximable irrational number} if there is a constant $c>0$ such that
for all rational numbers, $p/q$, we have that $\left| u_0-p/q\right|>c/q^2$.} 
if, and only if, its continued fraction representation is bounded, i.e.,
\begin{equation}\label{boundedki}
u_0 = \big[k_0; k_1, k_2, k_3, \dotsc \big] \qquad \text{with}\quad k_i \leq K,\:\quad \forall i=0,1,2,\ldots
\end{equation}
for some constant $K>0$, see \cite[Theorem 5F]{Schmidt}.
Consequently, the 
Kasner sequence $(u_i)_{i\in\mathbb{N}_0}$ is bounded,
i.e., $u_i \leq (K+1), \forall\, i\in\mathbb{N}_0$.
The set of badly approximable irrational numbers is a set of Lebesgue measure zero
(see~\cite[Page 60]{Schmidt}), which thereby is a non-generic set in a measure
theoretic sense, but it has full Hausdorff dimension that equals 1, see \cite{Schmidt69}.
%

Of particular interest is the following subcase. The initial Kasner parameter $u_0$ is
a \emph{quadratic irrational number}\footnote{A \emph{quadratic irrational number} (quadratic surd) is an
algebraic number of degree $2$, i.e., an irrational solution of a quadratic equation with integer
coefficients. In particular, such a solution can be written as $q_1 + \sqrt{q_2}$,
where $q_1 \in \mathbb{Q}$ and where $q_2 \in \mathbb{Q}$ is not a perfect square,
i.e., $\sqrt{q_2} \not\in \mathbb{Q}$.}
if, and only if, its continued fraction representation possess a periodic tail, i.e., when
\begin{equation}\label{periodicki}
u_0 = \big[k_0; k_1,\ldots,k_n, \,\overline{k_{n+1},\ldots, k_{n+p}}\,\big],
\end{equation}
where $k_1,\ldots,k_n$ represents a preperiodic piece and $\overline{k_{n+1},\ldots, k_{n+p}}$
is repeated ad infinitum, see~\cite[Theorem 6A]{Schmidt}, where $\overline{k_{n+1},\ldots, k_{n+p}}$
yields both a periodic Kasner sequence with the period $k_{n+1} + \ldots + k_{n+p}$ and a
periodic sequence of eras with the period $p$. 
%
\item[(iii)] If the initial Kasner parameter $u_0$ is a \emph{well approximable irrational
number}\footnote{A number $u_0\in\mathbb{R}\backslash\mathbb{Q}$ is called
\emph{well approximable irrational number} if it is not a badly approximable irrational number.}, then
the partial quotients $k_i$ in the continued fraction representation
\begin{equation}\label{wellki}
u_0 = \big[ k_0; k_1, k_2 , k_3, \dotsc \big]
\end{equation}
are unbounded. In particular, we can construct a diverging subsequence of
$(k_i)_{i\in\mathbb{N}_0}$ and hence such sequences come arbitrarily close to
a Taub state, with $u=\infty$. Since this is the complementary
set of badly approximable irrational numbers (which is of Lebesgue measure zero),
this set is of full Lebesgue measure. This is the generic case, in terms of
Lebesgue measure, and hence generically the Kasner sequence
$(u_i)_{i\in\mathbb{N}_0}$ is infinite and unbounded.
%
%

Of particular interest are the following subcases.
First, the works~\cite{BegDut22} and~\cite{ReitererTrubowitz} introduced
a moderate growth condition (MGC) on the coefficients of the continued fraction expansion
of $u_0$.  
%
%
The set of $u_0$ satisfying such MGC has full Lebesgue measure and it possesses a stable manifold of positive Lebesgue measure within the state space of Bianchi types $\mathrm{VIII}$ and $\mathrm{IX}$, see~\cite{BegDut22}.
%
Second, the work~\cite{ReitererTrubowitz} describes a different set of full Lebesgue
measure consisting of values of $u_0$ that satisfies a subpolynomial bound
on the coefficients of its continued fraction expansion. 
Other examples of subcases that still have no implications in Bianchi models are \emph{very well approximable irrational
numbers} or \emph{Liouville numbers}, see~\cite[Corollary 7.2 and Corollary 2.8]{Fishman}. 
\end{itemize}

Next, as a first step to translate Kasner sequences $\{u_0,u_1,u_2,\dots\}$ to
the Bianchi type $\mathrm{VI}_{-1/9}$ type $\mathrm{I}$ and $\mathrm{II}$ heteroclinic network,
we show how the new Bianchi type I and II frame transition solutions
induce Kasner circle $\mathrm{K}^\ocircle$ transition maps.

\subsection{Kasner circle transition maps}

The single frame transitions $\mathcal{T}_{R_1}$ and $\mathcal{T}_{R_3}$
induce a map of the Kasner circle $\mathrm{K}^\ocircle$ that maps the $\alpha$- limit set to
the $\omega$-limit set of each heteroclinic orbit. Using the same nomenclature to denote
such maps, they are given by the extended Kasner parameter $\ue$ according to
\begin{subequations}
\begin{alignat}{2}
\mathcal{T}_{R_1}(\ue)&:= \frac{1}{\ue}, \qquad & \text{ for } \ue &\in (-1,1),\\
\mathcal{T}_{R_3}(\ue)&:= -(1 + \ue),\qquad &\text{ for } \ue &\in \left(-\infty,-\sfrac12\right).
\end{alignat}
\end{subequations}
Note that for any point $\ue \in (-1,-\frac12)$, which corresponds to
sector $(321)$ and thereby the multivalued region $\mathbf{A}^-_{R_1R_3}$,
the frame transitions commute after three iterates, i.e.,
$\mathcal{T}_{R_1}\circ\mathcal{T}_{R_3}\circ\mathcal{T}_{R_1} (\ue)=  -\sfrac{\ue}{1 + \ue} =   \mathcal{T}_{R_3}\circ\mathcal{T}_{R_1}\circ\mathcal{T}_{R_3}(\ue)$,
see Figure~\ref{fig:Mapsymmetries} (left). The corresponding orbits form
the boundary of the double frame transitions, and hence
uniquely defines the Kasner circle double frame transition map, 
\begin{equation}\label{R1R3map}
\mathcal{T}_{R_1R_3}(\ue):= -\frac{\ue}{1 + \ue}, \qquad  \text{ for } \ue \in \left(-1,-\sfrac12\right).
\end{equation}
The effect of a double frame transition is to interchange the first and third
directions while $\Sigma_2$ is unchanged, i.e.,
$(\Sigma_2)_- = 3(p_2)_- - 1 = 3(p_2)_+ - 1 = (\Sigma_2)_+$.

Next we show how to replace the BKL Kasner map~\eqref{BKLMap} with
the map of the Kasner circle $\mathrm{K}^\ocircle$ induced by
the curvature transitions $\mathcal{T}_{N_-}$ and $\mathcal{T}_{R_1N_-}$,
which both result in the BKL Kasner map~\eqref{BKLMap} when one quotients out
the spatial frame rotations by translating the extended Kasner parameter
$\ue$ to $u$ according to Table~\ref{ueandu}. Similarly to the frame transitions,
the single curvature transitions $\mathcal{T}_{N_-}$ induce a
map of $\mathrm{K}^\ocircle$, which when expressed in $\ue$ is given by:
\begin{equation}
\mathcal{T}_{N_-}(\ue) := -\ue, \qquad \text{ for } \ue \in (0,+\infty).
\end{equation}
For any point $\ue \in (0,1)$, which corresponds to sector $(132)$ and
thereby the multivalued region $\mathbf{A}^-_{R_1N_-}$, note that a curvature
transition $\mathcal{T}_{N_-}$ and a frame transition $\mathcal{T}_{R_1}$ commute after two iterates, i.e.,
$\mathcal{T}_{R_1}\circ\mathcal{T}_{N_-}(\ue) = -\frac{1}{\ue} = \mathcal{T}_{N_-}\circ\mathcal{T}_{R_1}(\ue)$,
see Figure~\ref{fig:Mapsymmetries} (right), which uniquely defines the
mixed curvature-frame transition map,
\begin{equation}\label{NmR1map}
\mathcal{T}_{R_1N_-}(\ue) :=  -\frac{1}{\ue}, \qquad \text{ for } \ue \in (0,1).
\end{equation}
\begin{figure}[H]
\centering
\begin{tikzpicture}[scale=1.1]

        \draw [gray, dashed, - ]  (1,1.2)--(1,0);
        \draw [gray, dashed, - ]  (-1,1.2)--(-1,0);
        \draw [gray,dashed, - ]  (-0.5392,-1.4660)--(0.5,-0.8660);
        \draw [gray,dashed, - ]  (-1.5392,0.2660)--(-0.5,0.8660);
	\draw[gray,dashed, -] (2,0) -- (0.5,-0.866);
	\draw[gray,dashed, -] (2,0) -- (0.5, 0.866);

        \draw (1,0) arc (0:360:1cm and 1cm);	 	

	\draw (-0.95,0) -- (-1.05,0);
        \draw (0.95,0) -- (1.05,0);
        \draw[rotate=120] (-0.95,0) -- (-1.05,0);
        \draw[rotate=240] (-0.95,0) -- (-1.05,0);
        \draw[rotate=120] (0.95,0) -- (1.05,0);
        \draw[rotate=240] (0.95,0) -- (1.05,0);

       \draw [ultra thick, domain=0.52:1.570,variable=\t,smooth] plot ({sin(\t r)},{cos(\t r)});
       \draw [ultra thick, domain=-0.52:-1.570,variable=\t,smooth] plot ({sin(\t r)},{cos(\t r)});


    \draw [dotted, thick, postaction={decorate}]  (0.96,0.28)--(0.96,-0.28);\draw [dotted, gray]  (0.96,0.28)--(0.96,1.2);
    \draw [dotted, thick, postaction={decorate}]  (-0.71,0.71)--(-0.71,-0.71);\draw [dotted, gray]  (-0.71,0.71)--(-0.71,1.2);

    \draw [rotate=240,dotted, thick, postaction={decorate}]  (-0.71,-0.71)--(0.96,0.28);\draw [rotate=240,dotted, gray]  (-0.91,-0.825)--(-0.71,-0.71);

    \draw [dotted, thick, postaction={decorate}]  (-0.71,-0.71)--(0.96,0.28);\draw [dotted, gray]  (-0.91,-0.825)--(-0.71,-0.71);

    \draw [rotate=-240,dotted, thick, postaction={decorate}]  (0.96,0.28)--(0.96,-0.28);\draw [rotate=-240,dotted, gray]  (0.96,0.28)--(0.96,1.2);
    \draw [rotate=-240,dotted, thick, postaction={decorate}]  (-0.71,0.71)--(-0.71,-0.71);\draw [rotate=-240,dotted, gray]  (-0.71,0.71)--(-0.71,1.2);

    \filldraw [black] (-0.71,0.71) circle (1pt);
    \filldraw [black] (-0.71,-0.71) circle (1pt);

    \filldraw [black] (0.96,0.28) circle (1pt);
    \filldraw [black] (0.96,-0.28) circle (1pt);

    \filldraw [black] (-0.28,0.96) circle (1pt);
    \filldraw [black] (-0.28,-0.96) circle (1pt);

 \end{tikzpicture}
\hspace{2cm}
\begin{tikzpicture}[scale=1.1]

        \draw [gray, dashed, - ]  (1,1.2)--(1,0);
        \draw [gray, dashed, - ]  (-1,1.2)--(-1,0);
        \draw [gray,dashed, - ]  (-0.5392,-1.4660)--(0.5,-0.8660);
        \draw [gray,dashed, - ]  (-1.5392,0.2660)--(-0.5,0.8660);
	\draw[gray,dashed, -] (2,0) -- (0.5,-0.866);
	\draw[gray,dashed, -] (2,0) -- (0.5, 0.866);

        \draw (1,0) arc (0:360:1cm and 1cm);	 	

       \draw [ultra thick, domain=0.52:1.570,variable=\t,smooth] plot ({sin(\t r)},{cos(\t r)});
       \draw [ultra thick, domain=-0.52:-1.570,variable=\t,smooth] plot ({sin(\t r)},{cos(\t r)});

	\draw (-0.95,0) -- (-1.05,0);
        \draw (0.95,0) -- (1.05,0);
        \draw[rotate=120] (-0.95,0) -- (-1.05,0);
        \draw[rotate=240] (-0.95,0) -- (-1.05,0);
        \draw[rotate=120] (0.95,0) -- (1.05,0);
        \draw[rotate=240] (0.95,0) -- (1.05,0);

    \draw[  postaction={decorate}] (0.96,0.28) -- (-0.71,0.71);\draw [dotted, gray]  (0.96,0.28)--(2,0);
    \draw[  postaction={decorate}] (0.96,-0.28) -- (-0.71,-0.71);\draw [dotted, gray]  (0.96,-0.28)--(2,0);

    \draw [dotted, thick, postaction={decorate}]  (0.96,0.28)--(0.96,-0.28);\draw [dotted, gray]  (0.96,0.28)--(0.96,1.2);
    \draw [dotted, thick, postaction={decorate}]  (-0.71,0.71)--(-0.71,-0.71);\draw [dotted, gray]  (-0.71,0.71)--(-0.71,1.2);


    \filldraw [black] (-0.71,0.71) circle (1pt);
    \filldraw [black] (-0.71,-0.71) circle (1pt);

    \filldraw [black] (0.96,0.28) circle (1pt);
    \filldraw [black] (0.96,-0.28) circle (1pt);

 \end{tikzpicture}
	\caption{\textbf{Left:} Commutation of frame transitions, $\mathcal{T}_{R_1}\circ\mathcal{T}_{R_3}\circ\mathcal{T}_{R_1}
= \mathcal{T}_{R_3}\circ\mathcal{T}_{R_1}\circ\mathcal{T}_{R_3}$,
which defines a unique double frame transition map, $\mathcal{T}_{R_1R_3}$.
\textbf{Right:} Commutation of curvature and frame transitions,
$\mathcal{T}_{R_1}\circ\mathcal{T}_{N_-}=\mathcal{T}_{N_-}\circ\mathcal{T}_{R_1}$,
which defines a unique mixed curvature-frame transition map $\mathcal{T}_{R_1N_-}$.}\label{fig:Mapsymmetries}
\end{figure}
%

Recall that the unstable modes trigger transitions from each corresponding
sector and note that the stable modes are complementary to those modes, see
Figure~\ref{fig:sectors}. This serves as the foundation for the Bianchi
type I and II heteroclinic network.
There are two possible single transitions in the sectors $(321)$ and $(132)$.
To obtain unique maps, we view the one-parameter sets of heteroclinic orbits at
each point on $\mathrm{K}^\ocircle$ associated with $\mathcal{T}_{R_1R_3}$ and
$\mathcal{T}_{R_1N_-}$, including their boundaries, as equivalence classes, and
use the maps~\eqref{R1R3map} and~\eqref{NmR1map}. This is in contrast to the
iterated function system analysis developed in~\cite{lapp22} and the description
of Kasner epochs using symbolic dynamics in~\cite{HLU22}. Note that the column
for the Domain in Table~\ref{tab:maps} then covers all the sectors of
$\mathrm{K}^\ocircle$ with a unique map for each sector. 
\begin{table}[H]
    \centering
    \begin{tabular}{|c|c|l|l|}\hline
        \rowcolor{lightgray!25} Map & Image of $\ue$ & \hspace{1.25cm} Domain & \hspace{1.4cm} Range \\ \hline
        $\mathcal{T}_{R_3}$ & $-(1+\ue)$ & $(-\infty,-1)\in (213)\cup (231)$ & \hspace{0.5cm} $(0,\infty)\in (123)\cup (132) $\\ \hline
        $\mathcal{T}_{R_1R_3}$ & $-\frac{\ue}{1+\ue}$ & \hspace{0.05cm} $(-1,-\sfrac12)\in (321)$ & \hspace{0.5cm}  $(1,\infty)\in (123)$\\ \hline
        $\mathcal{T}_{R_1}$ & $\frac{1}{\ue}$ & \hspace{0.4cm} $(-\sfrac12,0)\in (312) $ & $(-\infty,-2)\in (213) $\\ \hline
        $\mathcal{T}_{R_1N_-}$ & $-\frac{1}{\ue}$ & \hspace{0.73cm} $(0,1)\in (132)$ & $(-\infty,-1)\in (213)\cup (231)$ \\ \hline
        $\mathcal{T}_{N_-}$ & $-\ue$ & \hspace{0.55cm} $(1,\infty) \in (123)$ & $ (-\infty,-1)\in (213)\cup (231)$\\ \hline
    \end{tabular}
    \caption{Unique maps for each sector, e.g., the single curvature transitions are given as follows: any $\ue\in (1,\infty)$,
    which parametrizes sector $(123)$, yields the map $\mathcal{T}_{N_-}(\ue)=-\ue \in (-\infty,-1)$,
    which parametrizes sectors $(213)$, on which $\ue \in (-\infty,-2)$, and $(231)$,
    for which $\ue \in (-2,-1)$.}\label{tab:maps}
\end{table}
%

\subsection{From Kasner sequences $\{u_0,u_1,u_2,\dots\}$ to networks of heteroclinic chains}\label{subsec:relevant}


It follows straightforwardly from the dynamical system~\eqref{BilliardeqsolevolHnormVI},
\eqref{billiardeqsolconsHnormvacVI} that each spatially frame-invariant Kasner epoch,
described by a Kasner parameter $u\neq 1, \infty$, corresponds to the invariant
set given by the closure of the double
frame transitions, $\overline{\mathcal{T}}_{R_1R_3}(\ue)$, originating from the
fixed point with $\ue = -u/(1+u)$ in sector $(321)$, see Table~\ref{ueandu}
and Figure~\ref{KMaps}. The invariant set
$\overline{\mathcal{T}}_{R_1R_3}(\ue)$ consists of:
\begin{itemize}
\item[(i)] Six fixed points on $\mathrm{K}^\ocircle$, one in each sector,
given by $\ue=\ue(u)$ according to Table~\ref{ueandu}.
\item[(ii)] Two equivalence classes, related by $R_3 \rightarrow - R_3$,
of single frame transition heteroclinic chains, given by
$\mathcal{T}_{R_1}\circ\mathcal{T}_{R_3}\circ\mathcal{T}_{R_1}(\ue)$
and $\mathcal{T}_{R_3}\circ\mathcal{T}_{R_1}\circ\mathcal{T}_{R_3}(\ue)$,
where each chain connects four fixed points on
$\mathrm{K}^\ocircle$ according to (see the figure to the right
in Figure~\ref{KMaps})
\begin{equation}\nonumber
\begin{split} &(321)\cdots\hspace{-0.1cm}\color{gray}{\small{\RHD}} \color{black}{(312)}
\cdots\hspace{-0.1cm}\color{gray}{\small{\RHD}} \color{black}{(213)}
\cdots\hspace{-0.1cm}\color{gray}{\small{\RHD}} \color{black}{(123)},\\
&(321)\cdots\hspace{-0.1cm}\color{gray}{\small{\RHD}} \color{black}{(231)}
\cdots\hspace{-0.1cm}\color{gray}{\small{\RHD}} \color{black}{(132)}
\cdots\hspace{-0.1cm}\color{gray}{\small{\RHD}} \color{black}{(123)},
\end{split}
\end{equation}
respectively, where $\ue(u)$ for the different sectors are given in
Table~\ref{ueandu}. Note that both types of heteroclinic chains connect
the common first and final fixed points in sector $(321)$, where
$\ue = -u/(1+u)$, and sector $(123)$, where $\ue = u$, respectively,
which reflects the commuting property
$\mathcal{T}_{R_1}\circ\mathcal{T}_{R_3}\circ\mathcal{T}_{R_1}(\ue)
= \mathcal{T}_{R_3}\circ\mathcal{T}_{R_1}\circ\mathcal{T}_{R_3}(\ue)$,
$\ue = -u/(1+u)$, illustrated to the left in
Figure~\ref{fig:Mapsymmetries}.
\item[(iii)] An equivalence class, related by $R_3 \rightarrow - R_3$,
of one-parameter sets of heteroclinic double frame transition orbits
$\mathcal{T}_{R_1R_3}(\ue)$, connecting the fixed point $\ue = -u/(1+u)$
in sector $(321)$ with that in sector $(123)$ given by $\ue = u$, where the
heteroclinic chains in (ii) form the boundary of $\mathcal{T}_{R_1R_3}(\ue)$.
\end{itemize}
It becomes quite complicated to describe the network of heteroclinic chains corresponding to a Kasner sequence
$\{u_0,u_1,u_2,\ldots\}$ by projecting orbits onto the plane $\Sigma_1+\Sigma_2+\Sigma_3=0$
in $(\Sigma_1,\Sigma_2,\Sigma_3)$-space after a few Kasner epochs. For this reason, we introduce a new simplifying representation,
which emphasizes that such a heteroclinic network is a directed graph, where the vertices
are points on $\mathrm{K}^\ocircle$, while the edges are the heteroclinic
orbits that connect them. 
This representation is obtained by arranging the six fixed points 
$\ue=\ue(u)$ by means of the heteroclinic single transition chains in (ii) above
so that they form a hexagon, illustrating the commutating feature
$\mathcal{T}_{R_1}\circ\mathcal{T}_{R_3}\circ\mathcal{T}_{R_1}(\ue \in (321)) =
\mathcal{T}_{R_3}\circ\mathcal{T}_{R_1}\circ\mathcal{T}_{R_3}(\ue \in (321))$,
while the double frame transition orbits $\mathcal{T}_{R_1R_3}(\ue \in (321))$ in
(iii) are represented by an arrow inside the hexagon, symbolizing that
the hexagon chains in (ii) form the boundary of $\mathcal{T}_{R_1R_3}(\ue \in (321))$,
see Figure~\ref{FIG:hex} and Figure~\ref{KMaps} (right).
\begin{figure}[H]\centering
\scriptsize
\begin{tikzpicture}[scale=0.65]
\node (1) at (-2, 0) {\boxed{\text{\tiny{$(321)$}}}};
\node (2a) at (-1, 1.73) {\boxed{\text{\tiny{$(312)$}}}};
\node (3a) at (1, 1.73) {\boxed{\text{\tiny{$(213)$}}}};
\node (2b) at (-1, -1.73) {\boxed{\text{\tiny{$(231)$}}}};
\node (3b) at (1, -1.73) {\boxed{\text{\tiny{$(132)$}}}};
\node (4) at (2, 0) {\boxed{\text{\tiny{$(123)$}}}};

\draw[dotted, ultra thick, ->] (1) -- (4);

\draw[dotted, thick, ->] (1) -- (2a);
\draw[dotted, thick, ->] (2a) -- (3a);
\draw[dotted, thick, ->] (3a) -- (4);

\draw[dotted, thick, ->] (1) -- (2b);
\draw[dotted, thick, ->] (2b) -- (3b);
\draw[dotted, thick, ->] (3b) -- (4);

\end{tikzpicture}
\hspace{2.5cm}
\begin{tikzpicture}[scale=0.65]
\node (1) at (-2, 0) {\textbf{\textbullet}};
\node (2a) at (-1, 1.73) {\textbf{\textbullet}};
\node (3a) at (1, 1.73) {\textbf{\textbullet}};
\node (2b) at (-1, -1.73) {\textbf{\textbullet}};
\node (3b) at (1, -1.73) {\textbf{\textbullet}};
\node (4) at (2, 0) {\textbf{\textbullet}};

\draw[dotted, ultra thick, ->] (1) -- (4);

\draw[dotted, thick, ->] (1) -- (2a);
\draw[dotted, thick, ->] (2a) -- (3a);
\draw[dotted, thick, ->] (3a) -- (4);

\draw[dotted, thick, ->] (1) -- (2b);
\draw[dotted, thick, ->] (2b) -- (3b);
\draw[dotted, thick, ->] (3b) -- (4);

\end{tikzpicture}
\captionof{figure}{Graph representation of a heteroclinic network
corresponding to a 
frame-invariant Kasner parameter
$u\neq 1, \infty$, obtained from $\overline{\mathcal{T}}_{R_1R_3}(\ue)$,
where $\ue = \ue(u) = -u/(1+u) \in (321)$.
\textbf{Left:} The six fixed points $\ue$ in the different sectors,
given by $\ue$ in Table~\ref{ueandu}, are arranged as the vertices of a
hexagon-shaped graph, where the edges are given by the heteroclinic
chains of single frame transition,
$\mathcal{T}_{R_3}\circ\mathcal{T}_{R_1}\circ\mathcal{T}_{R_3}(\ue \in (321))$ and
$\mathcal{T}_{R_1}\circ\mathcal{T}_{R_3}\circ\mathcal{T}_{R_1}(\ue \in (321))$,
represented by thin dotted arrows.
The one-parameter family of double frame transitions,
${\cal T}_{R_1R_3}(\ue \in (321))$, are represented by a bold dotted arrow
inside the hexagon, 
which is consistent with the representation in Figure~\ref{KMaps}.
\textbf{Right:} To avoid clutter when representing Kasner sequences
as sequences of hexagons, the sector notation for the fixed points is
omitted and replaced with dots. 
}\label{FIG:hex}
\end{figure}
%

The BKL Kasner map~\eqref{BKLMap} result in two different cases for a
Kasner parameter $u_-$: when $u_- \in (1, 2)$
yields an era change according to $u_+ = 1/(u_- - 1)$, while $u_- \in (2, \infty)$
an era continues according to $u_+ = u_- - 1$. The two different intervals
for $u_-$ partition each sector on $\mathrm{K}^{\ocircle}$ into two
disjoint parts (for a description of the six representations in $\ue$ for
$u=2$, recall Table~\ref{ueandu} and Figure~\ref{fig:sectors}). For example,
sector $(132)$ is divided into $(132)_{\mathrm{f}}$ with
$\ue_- \in (\frac12,1)$, corresponding to $u_-\in(1,2)$,
and its open complement $(132)_c$ with $\ue_- \in (0,\frac12)$ for which
$u_-\in (2,\infty)$. In the hexagonal representation the two Kasner epochs
$u_-$ and $u_+$ correspond to two hexagons, where the two different BKL Kasner
map~\eqref{BKLMap} rules for $u_- \in (1,2)$ and $u_- \in (2,\infty)$ amount
to connecting the two hexagons in two different ways:


\begin{itemize}
\item \textbf{Era change:} A change of a Kasner era takes place when $u_- \in (1,2)$,
where the BKL Kasner map~\eqref{BKLMap} yields $u_+ = 1/(u_- - 1)$.
This amounts to connecting the two hexagon graphs corresponding to $u_-$ and $u_+ $,
according to:
\begin{itemize}
\item[(i)] The single curvature transition $\mathcal{T}_{N_-}(\ue_-= u_-)$,
$\ue_-= u_- \in (1,2) = (123)_{\mathrm{f}}$, from sector $(123)_{\mathrm{f}}$ to sector $(231)$.
\item[(ii)] The single curvature transition $\mathcal{T}_{N_-}(\ue_-)$,
$\ue_- = 1/u_- \in (\frac12,1) = (132)_{\mathrm{f}}$, from sector $(132)_{\mathrm{f}}$ to sector $(321)$.
\item[(iii)] The mixed curvature-frame transitions
$\mathcal{T}_{R_1N_-}(\ue_-)$, where $\ue_- = 1/u_- \in (\frac12,1) = (132)_{\mathrm{f}}$,
from sector $(132)_{\mathrm{f}}$ to sector $(231)$,
\end{itemize}
as illustrated in Figure~\ref{FIG:BKLmapshex} (left), see also
Figure~\ref{fig:CurvatureTrans} (middle).
\item \textbf{No era change:} A change of Kasner epochs without a change of Kasner era
takes place when $u_- \in (2,\infty)$, for which the BKL Kasner map~\eqref{BKLMap}
yields $u_+ = u_- -1$. This corresponds to connecting the two hexagons with
$u_-$ and $u_+$ according to:
\begin{itemize}
\item[(i)] The single curvature transition $\mathcal{T}_{N_-}(\ue_-)$,
$\ue_- = u_- \in (2,\infty) = (123)_c$, from sector $(123)_c$ to sector $(213)$.
\item[(ii)] The single curvature transition $\mathcal{T}_{N_-}(\ue_-)$,
where $\ue_- = 1/u_- \in (0,\frac12) = (132)_c$, from sector $(132)_c$ to sector $(312)$.
\item[(iii)] The mixed curvature-frame transitions
$\mathcal{T}_{R_1N_-}(\ue_-)$, where $\ue_- = 1/u_- \in (0,\frac12) = (132)_c$,
from sector $(132)_c$ to sector $(213)$,
\end{itemize}
as in Figure~\ref{FIG:BKLmapshex} (right), see also
Figure~\ref{fig:CurvatureTrans} (right).
The heteroclinic chain given by
$(321)\cdots\hspace{-0.1cm}\color{gray}{\small{\RHD}} \color{black}{(231)}
\cdots\hspace{-0.1cm}\color{gray}{\small{\RHD}} \color{black}{(132)}
\cdots\hspace{-0.1cm}\color{gray}{\small{\RHD}} \color{black}{(123)}$,
the heteroclinic orbit
$(321)\cdots\hspace{-0.1cm}\color{gray}{\small{\RHD}} \color{black}{(312)}$,
and the heteroclinic double frame transition orbits
$(321)\boldsymbol{\cdots}\hspace{-0.05cm}\color{gray}{\small{\RHD}} \color{black}{(123)}$
in the hexagon associated with $u_+$ are not part of any
heteroclinic chain that originates from the hexagon corresponding to $u_-$.
Therefore we say that the fixed points $\ue_+$ in sectors
$(321)$, $(231)$, $(132)$ and the heteroclinic orbits that originate from them
are \emph{isolated}, see Figure~\ref{FIG:BKLmapshex} (right).
%
%
%
\end{itemize}
\begin{figure}[H]\centering
\scriptsize
\begin{tikzpicture}[scale=0.4]
\node (1) at (-2, 0) {\textbf{\textbullet}};
\node (2a) at (-1, 1.73) {\textbf{\textbullet}};
\node (3a) at (1, 1.73) {\textbf{\textbullet}};
\node (2b) at (-1, -1.73) {\textbf{\textbullet}};
\node (3b) at (1, -1.73) {\textbf{\textbullet}};
\node (4) at (2, 0) {\textbf{\textbullet}};

\node (R1) at (4, 0) {\textbf{\textbullet}};
\node (R2a) at (5, 1.73) {\textbf{\textbullet}};
\node (R3a) at (7, 1.73) {\textbf{\textbullet}};
\node (R2b) at (5, -1.73) {\textbf{\textbullet}};
\node (R3b) at (7, -1.73) {\textbf{\textbullet}};
\node (R4) at (8, 0) {\textbf{\textbullet}};

\node at (-2.75,0) {\scriptsize $\ue_-$};
\node at (8.75,0) {\scriptsize $\ue_+$};

\draw[dotted, ultra thick, ->] (1) -- (4);

\draw[dotted, thick, ->] (1) -- (2a);
\draw[dotted, thick, ->] (2a) -- (3a);
\draw[dotted, thick, ->] (3a) -- (4);

\draw[dotted, thick, ->] (1) -- (2b);
\draw[dotted, thick, ->] (2b) -- (3b);
\draw[dotted, thick, ->] (3b) -- (4);

\draw[dotted, ultra thick, ->] (R1) -- (R4);

\draw[dotted, thick, ->] (R1) -- (R2a);
\draw[dotted, thick, ->] (R2a) -- (R3a);
\draw[dotted, thick, ->] (R3a) -- (R4);

\draw[dotted, thick, ->] (R1) -- (R2b);
\draw[dotted, thick, ->] (R2b) -- (R3b);
\draw[dotted, thick, ->] (R3b) -- (R4);

\draw[bend right=15,->] (4) to (R2b);
\draw [very thick, ->] (3b) to (R2b);
\draw [bend right=15,->] (3b) to (R1);

\draw [thick,decorate,decoration={brace,amplitude=10pt,mirror},xshift=0.4pt,yshift=-0.4pt](1,-3) -- (5,-3) node[black,midway,yshift=-0.6cm] {\tiny Era change.};

\end{tikzpicture}
\hspace{2cm}
\begin{tikzpicture}[scale=0.4]
\node (1) at (-2, 0) {\textbf{\textbullet}};
\node (2a) at (-1, 1.73) {\textbf{\textbullet}};
\node (3a) at (1, 1.73) {\textbf{\textbullet}};
\node (2b) at (-1, -1.73) {\textbf{\textbullet}};
\node (3b) at (1, -1.73) {\textbf{\textbullet}};
\node (4) at (2, 0) {\textbf{\textbullet}};

\node (R1) at (4, 0) {$\circ$};
\node (R2a) at (5, 1.73) {\textbf{\textbullet}};
\node (R3a) at (7, 1.73) {\textbf{\textbullet}};
\node (R2b) at (5, -1.73) {$\circ$};
\node (R3b) at (7, -1.73) {$\circ$};
\node (R4) at (8, 0) {\textbf{\textbullet}};

\node at (-2.75,0) {\scriptsize $\ue_-$};
\node at (8.75,0) {\scriptsize $\ue_+$};

\draw[dotted, ultra thick, ->] (1) -- (4);

\draw[dotted, thick, ->] (1) -- (2a);
\draw[dotted, thick, ->] (2a) -- (3a);
\draw[dotted, thick, ->] (3a) -- (4);

\draw[dotted, thick, ->] (1) -- (2b);
\draw[dotted, thick, ->] (2b) -- (3b);
\draw[dotted, thick, ->] (3b) -- (4);

\draw[dotted, ultra thick, ->] (R1) -- (R4);

\draw[dotted, thick, ->] (R1) -- (R2a);
\draw[dotted, thick, ->] (R2a) -- (R3a);
\draw[dotted, thick, ->] (R3a) -- (R4);

\draw[dotted, thick, ->] (R1) -- (R2b);
\draw[dotted, thick, ->] (R2b) -- (R3b);
\draw[dotted, thick, ->] (R3b) -- (R4);

\draw[bend left=50,->] (4) to (R3a);
\draw[->,very thick] (3b) to[out=0,in=-75]++ (1.75,2) to[out=100,in=-205] (R3a);
\path[->] (3b) edge [->,out=-5,in=-155,looseness=1] (R2a);

\draw [thick,decorate,decoration={brace,amplitude=10pt,mirror},xshift=0.4pt,yshift=-0.4pt](1,-3) -- (5,-3) node[black,midway,yshift=-0.6cm] {\tiny No era change.};
\end{tikzpicture}
\captionof{figure}{The Kasner parameters $u_-$ and $u_+$ yield two hexagons 
using the graph representations in Figure~\ref{FIG:hex}.
The BKL Kasner map~\eqref{BKLMap}, $u_- \mapsto u_+$, yields two rules for how to connect
such hexagons with heteroclinic orbits. We represent single curvature
(mixed curvature-frame) transitions with thin (thick) solid arrows. 
\textbf{Left:} Representation of the heteroclinic orbit connections between the two hexagons when there is an era change, i.e., when $u_- \in (1,2)$ and $u_+ = 1/(u_- - 1)$.
\textbf{Right:} Representation of the heteroclinic connections between the two hexagons when there is no era change, i.e., when $u_- \in (2,\infty)$ and $u_+ = u_- - 1$.
Note that in the hexagon corresponding to $u_+$, there are fixed points
(and thereby heteroclinic orbits originating from them) that are not connected to any
point in the hexagon associated with $u_-$. This yields an \emph{isolated} structure,
and for this reason, we depict such fixed points with hollow circles.
}\label{FIG:BKLmapshex}
\end{figure}
Note that an era change does not yield any isolated structure, except when $u_-$ is the last value in a Kasner era with length two or longer, then this gives rise to additional isolated heteroclinic structures, as we will see. 

Next we consider Kasner sequences
$(u_i)_{i\in\mathbb{N}_0}$ partitioned into (frame-invariant) Kasner eras
according to~\eqref{EraPartition}, which we for convenience repeat:
\begin{equation}\label{EraPartition2}
\{u_0,u_1,u_2,\ldots\} = \{\underbrace{u^0_{0},\ldots , u^0_{\mathrm{f}}}_{\text{\scriptsize Initial era}}, \underbrace{u^1_{0},\ldots , u^1_{\mathrm{f}}}_{\text{\scriptsize $1^{st}$ era}}, \underbrace{u^2_{0},\ldots , u^2_{\mathrm{f}}}_{\text{\scriptsize $2^{nd}$ era}}, \ldots, \underbrace{u^j_{0},\ldots , u^j_{\mathrm{f}}}_{\text{\scriptsize $j^{th}$ era}},\ldots\},
\end{equation}
where $u^0_0 = u_0$. Due to the definition of the BKL Kasner map~\eqref{Kasnermap},
points within $(132)_{\mathrm{f}}$ correspond to the final values
$u_{\mathrm{f}}^j$ for all $j=0,1,2,\ldots$ in all eras, whereas points within
$(132)_c$ correspond to all other Kasner parameter values, not being a final era value,
i.e., $\mathbb{E}^j \backslash \{ u^j_{\mathrm{f}}\}$ for all $j=0,1,2,\ldots$
(recall~\eqref{EraPoints}, which introduced the notation $\mathbb{E}^j$ for the number of
Kasner epochs, given by $\lfloor u^j_0\rfloor$, for the $j^{th}$ era).
For each $u^j_i\in \mathbb{E}^j$, $i=0\dots {\mathrm{f}}$, there is one Kasner fixed point
in each sector given by $\ue^j_i(u^j_i)$ in Table~\ref{ueandu}, which thereby leads to
$6\lfloor u^j_0\rfloor$ fixed points on $\mathrm{K}^\ocircle$ for the $j^{th}$ Kasner era,
denoted by $\check{\mathbb{E}}^j$.

As follows from the above, a change of Kasner epochs without a change of Kasner eras
with $u^j_i \in (2,\infty)$ yields a piece of a heteroclinic network corresponding to
$\overline{\mathcal{T}}_{R_1N_-}(\ue^j_i = 1/u^j_i)$,
where $\ue^j_i\in (132)_c$, see Figure~\ref{FIG:BKLmapshex} (right),
while a change of the $j^{th}$ era to the $(j+1)^{th}$ era corresponds to
$\overline{\mathcal{T}}_{R_1N_-}(\ue^j_f = 1/u^j_f)$
where $\ue^j_f \in (\frac12,1) = (132)_f$,
see Figure~\ref{FIG:BKLmapshex} (left).
Hence, the full heteroclinic network associated with
the $j^{th}$ era of length $\lfloor u^j_0\rfloor$ is obtained by taking the following union:
\begin{equation}\label{eranetwork}
\mathbb{H}^j := \left(\bigcup_{\ue \in \check{\mathbb{E}}^j \cap (321)}\overline{\mathcal{T}}_{R_1R_3}(\ue)\right)\bigcup\left(\bigcup_{\ue \in \check{\mathbb{E}}^j \cap (132)}\overline{\mathcal{T}}_{R_1N_-}(\ue)\right),
\end{equation}
where $\overline{\mathcal{T}}_{R_1R_3}(\ue),\overline{\mathcal{T}}_{R_1N_-}(\ue)$
denote the closure of the respective multiple transition sets. 

Since all heteroclinic orbits arising from points
$\ue \in \check{\mathbb{E}}^j \backslash[(321) \cup (132)]$ are contained in one of the
closures above, the \emph{single transition network} of the $j^{th}$ era 
is given by the heteroclinic chains that form the boundaries of the era
network~\eqref{eranetwork}, i.e.,
\begin{equation}\label{eranetworksingle}
\partial \mathbb{H}^j := \left(\bigcup_{\ue \in \check{\mathbb{E}}^j \cap (321)}\partial
\overline{\mathcal{T}}_{R_1R_3}(\ue)\right)\bigcup\left(\bigcup_{\ue \in \check{\mathbb{E}}^j \cap (132)}
\partial \overline{\mathcal{T}}_{R_1N_-}(\ue)\right).
\end{equation}
Finally, the entire heteroclinic network associated with a
Kasner sequence $\{u_0,u_1,u_2,\ldots\}$ is obtained by taking the union
of the heteroclinic networks, i.e., $\cup_j \mathbb{H}^j$,
while the heteroclinic network of single transitions is given by
$\cup_j \partial\mathbb{H}^j$.

Thus, in the hexagonal representation there are two main steps to construct
heteroclinic chains corresponding to a Kasner sequence $\{u_0,u_1,u_2,\ldots\}$, which in turn is
partitioned into a sequence of eras according to~\eqref{EraPartition2}. First, we
identify the hexagons corresponding to each value of $u$ in the Kasner sequence and then
divide them according to the different eras they belong to. Second, we connect the different
hexagons with the BKL Kasner map rules depicted in Figure~\ref{FIG:BKLmapshex},
where the rule for era continuation (Figure~\ref{FIG:BKLmapshex} (right)) yields a
representation for the era network in~\eqref{eranetwork}, while the rule for era change
(Figure~\ref{FIG:BKLmapshex} (left)) enables us to take the union of two different
era networks. This allows us to obtain a representation for the entire
heteroclinic network $\cup_j \mathbb{H}^j$. 

The golden ratio, $u_0=u_\mathrm{f} = (1 + \sqrt{5})/2$, discussed
in the next section, is the only Kasner sequence with $u^{j+1}_0 = u^j_\mathrm{f}$
for all $j=0,1,2,\ldots$; all other Kasner sequences involve eras
where $u^{j+1}_0 \neq u^j_\mathrm{f}$ for some $j$. Such Kasner sequences
yield \emph{isolated fixed points} from which
\emph{isolated heteroclinic orbits} originate, which gives rise to
\emph{isolated heteroclinic chains}, where the notation isolated stands
for that these structures are not connected to preceding non-isolated
heteroclinic chains in the heteroclinic network corresponding to such Kasner sequences.
To illustrate these features by means of the
hexagonal representation, we consider an example corresponding to a Kasner sequence characterized by
a continued fraction given by $u_0 = [2;4,2,1,2,1,2,2,2,\dots]$, which can correspond to finite, periodic, or aperiodic Kasner sequences, depending on the meaning of $\dots$ in the expression $u_0 = [2;4,2,1,2,1,2,2,2,\dots]$; 
see Figure~\ref{FIG:hexEX}.
This example illustrates all the possibilities for how isolated
heteroclinic chains occur. It also provides the ingredients to obtain the
network of heteroclinic cycles and their properties for the periodic Kasner sequences
in the next section.
\begin{figure}[H]\centering
\scriptsize
\begin{tikzpicture}[scale=0.4]
\node (1) at (-2, 0) {\textbf{\textbullet}};
\node (2a) at (-1, 1.73) {\textbf{\textbullet}};
\node (3a) at (1, 1.73) {\textbf{\textbullet}};
\node (2b) at (-1, -1.73) {\textbf{\textbullet}};
\node (3b) at (1, -1.73) {\textbf{\textbullet}};
\node (4) at (2, 0) {\textbf{\textbullet}};

\node (R1) at (4, 0) {$\circ$};
\node (R2a) at (5, 1.73) {\textbf{\textbullet}};
\node (R3a) at (7, 1.73) {\textbf{\textbullet}};
\node (R2b) at (5, -1.73) {$\circ$};
\node (R3b) at (7, -1.73) {$\circ$};
\node (R4) at (8, 0) {\textbf{\textbullet}};

\node (RR1) at (10, 0) {$\circ$};
\node (RR2a) at (11, 1.73) {$\circ$};
\node (RR3a) at (13, 1.73) {$\circ$};
\node (RR2b) at (11, -1.73) {\textbf{\textbullet}};
\node (RR3b) at (13, -1.73) {\textbf{\textbullet}};
\node (RR4) at (14, 0) {\textbf{\textbullet}};

\node (RRR1) at (16, 0) {$\circ$};
\node (RRR2a) at (17, 1.73) {\textbf{\textbullet}};
\node (RRR3a) at (19, 1.73) {\textbf{\textbullet}};
\node (RRR2b) at (17, -1.73) {$\circ$};
\node (RRR3b) at (19, -1.73) {$\circ$};
\node (RRR4) at (20, 0) {\textbf{\textbullet}};

\node (RRRR1) at (22, 0) {$\circ$};
\node (RRRR2a) at (23, 1.73) {$\circ$};
\node (RRRR3a) at (25, 1.73) {\textbf{\textbullet}};
\node (RRRR2b) at (23, -1.73) {$\circ$};
\node (RRRR3b) at (25, -1.73) {$\circ$};
\node (RRRR4) at (26, 0) {\textbf{\textbullet}};

\node (RRRRR1) at (28, 0) {$\circ$};
\node (RRRRR2a) at (29, 1.73) {$\circ$};
\node (RRRRR3a) at (31, 1.73) {\textbf{\textbullet}};
\node (RRRRR2b) at (29, -1.73) {$\circ$};
\node (RRRRR3b) at (31, -1.73) {$\circ$};
\node (RRRRR4) at (32, 0) {\textbf{\textbullet}};

\draw[dotted, ultra thick, ->] (1) -- (4);

\draw[dotted, thick, ->] (1) -- (2a);
\draw[dotted, thick, ->] (2a) -- (3a);
\draw[dotted, thick, ->] (3a) -- (4);

\draw[dotted, thick, ->] (1) -- (2b);
\draw[dotted, thick, ->] (2b) -- (3b);
\draw[dotted, thick, ->] (3b) -- (4);

\draw[dotted, ultra thick, ->] (R1) -- (R4);

\draw[dotted, thick, ->] (R1) -- (R2a);
\draw[dotted, thick, ->] (R2a) -- (R3a);
\draw[dotted, thick, ->] (R3a) -- (R4);

\draw[dotted, thick, ->] (R1) -- (R2b);
\draw[dotted, thick, ->] (R2b) -- (R3b);
\draw[dotted, thick, ->] (R3b) -- (R4);

\draw[dotted, ultra thick, ->] (RR1) -- (RR4);

\draw[dotted, thick, ->] (RR1) -- (RR2a);
\draw[dotted, thick, ->] (RR2a) -- (RR3a);
\draw[dotted, thick, ->] (RR3a) -- (RR4);

\draw[dotted, thick, ->] (RR1) -- (RR2b);
\draw[dotted, thick, ->] (RR2b) -- (RR3b);
\draw[dotted, thick, ->] (RR3b) -- (RR4);

\draw[dotted, ultra thick, ->] (RRR1) -- (RRR4);

\draw[dotted, thick, ->] (RRR1) -- (RRR2a);
\draw[dotted, thick, ->] (RRR2a) -- (RRR3a);
\draw[dotted, thick, ->] (RRR3a) -- (RRR4);

\draw[dotted, thick, ->] (RRR1) -- (RRR2b);
\draw[dotted, thick, ->] (RRR2b) -- (RRR3b);
\draw[dotted, thick, ->] (RRR3b) -- (RRR4);

\draw[dotted, ultra thick, ->] (RRRR1) -- (RRRR4);

\draw[dotted, thick, ->] (RRRR1) -- (RRRR2a);
\draw[dotted, thick, ->] (RRRR2a) -- (RRRR3a);
\draw[dotted, thick, ->] (RRRR3a) -- (RRRR4);

\draw[dotted, thick, ->] (RRRR1) -- (RRRR2b);
\draw[dotted, thick, ->] (RRRR2b) -- (RRRR3b);
\draw[dotted, thick, ->] (RRRR3b) -- (RRRR4);

\draw[dotted, ultra thick, ->] (RRRRR1) -- (RRRRR4);

\draw[dotted, thick, ->] (RRRRR1) -- (RRRRR2a);
\draw[dotted, thick, ->] (RRRRR2a) -- (RRRRR3a);
\draw[dotted, thick, ->] (RRRRR3a) -- (RRRRR4);

\draw[dotted, thick, ->] (RRRRR1) -- (RRRRR2b);
\draw[dotted, thick, ->] (RRRRR2b) -- (RRRRR3b);
\draw[dotted, thick, ->] (RRRRR3b) -- (RRRRR4);

\draw[bend left=50,->] (4) to (R3a);
\draw[->,very thick] (3b) to[out=0,in=-75]++ (1.75,2) to[out=100,in=-205] (R3a);
\path[->] (3b) edge [->,out=-5,in=-155,looseness=1] (R2a);

\draw[bend right=15,->] (R4) to (RR2b);
\draw [very thick, ->] (R3b) to (RR2b);
\draw [bend right=15,->] (R3b) to (RR1);

\draw[bend left=50,->] (RR4) to (RRR3a);
\draw[->,very thick] (RR3b) to[out=0,in=-75]++ (1.75,2) to[out=100,in=-205] (RRR3a);
\path[->] (RR3b) edge [->,out=-5,in=-155,looseness=1] (RRR2a);

\draw[bend left=50,->] (RRR4) to (RRRR3a);
\draw[->,very thick] (RRR3b) to[out=0,in=-75]++ (1.75,2) to[out=100,in=-205] (RRRR3a);
\path[->] (RRR3b) edge [->,out=-5,in=-155,looseness=1] (RRRR2a);

\draw[bend left=50,->] (RRRR4) to (RRRRR3a);
\draw[->,very thick] (RRRR3b) to[out=0,in=-75]++ (1.75,2) to[out=100,in=-205] (RRRRR3a);
\path[->] (RRRR3b) edge [->,out=-5,in=-155,looseness=1] (RRRRR2a);

\node (1cont) at (34, 0) {...};
\node (2cont) at (35, -1.73) {...};

\draw[bend right=15,->] (RRRRR4) to (2cont);
\draw [very thick, ->] (RRRRR3b) to (2cont);
\draw [bend right=15,->] (RRRRR3b) to (1cont);

\draw [thick,decorate,decoration={brace,amplitude=10pt,mirror},xshift=0.4pt,yshift=-0.4pt](-2,-3) -- (7.5,-3) node[black,midway,yshift=-0.6cm] {\tiny Initial Era.};

\draw [thick,decorate,decoration={brace,amplitude=10pt,mirror},xshift=0.4pt,yshift=-0.4pt](10.5,-3) -- (32,-3) node[black,midway,yshift=-0.6cm] {\tiny Era 1.};

\end{tikzpicture}
\begin{tikzpicture}[scale=0.4]
\node (1) at (-2, 0) {$\circ$};
\node (2a) at (-1, 1.73) {$\circ$};
\node (3a) at (1, 1.73) {$\circ$};
\node (2b) at (-1, -1.73) {\textbf{\textbullet}};
\node (3b) at (1, -1.73) {\textbf{\textbullet}};
\node (4) at (2, 0) {\textbf{\textbullet}};

\node (R1) at (4, 0) {$\circ$};
\node (R2a) at (5, 1.73) {\textbf{\textbullet}};
\node (R3a) at (7, 1.73) {\textbf{\textbullet}};
\node (R2b) at (5, -1.73) {$\circ$};
\node (R3b) at (7, -1.73) {$\circ$};
\node (R4) at (8, 0) {\textbf{\textbullet}};

\node (RR1) at (10, 0) {$\circ$};
\node (RR2a) at (11, 1.73) {$\circ$};
\node (RR3a) at (13, 1.73) {$\circ$};
\node (RR2b) at (11, -1.73) {\textbf{\textbullet}};
\node (RR3b) at (13, -1.73) {\textbf{\textbullet}};
\node (RR4) at (14, 0) {\textbf{\textbullet}};

\node (RRR1) at (16, 0) {\textbf{\textbullet}};
\node (RRR2a) at (17, 1.73) {\textbf{\textbullet}};
\node (RRR3a) at (19, 1.73) {\textbf{\textbullet}};
\node (RRR2b) at (17, -1.73) {\textbf{\textbullet}};
\node (RRR3b) at (19, -1.73) {\textbf{\textbullet}};
\node (RRR4) at (20, 0) {\textbf{\textbullet}};

\node (RRRR1) at (22, 0) {$\circ$};
\node (RRRR2a) at (23, 1.73) {\textbf{\textbullet}};
\node (RRRR3a) at (25, 1.73) {\textbf{\textbullet}};
\node (RRRR2b) at (23, -1.73) {$\circ$};
\node (RRRR3b) at (25, -1.73) {$\circ$};
\node (RRRR4) at (26, 0) {\textbf{\textbullet}};

\node (RRRRR1) at (28, 0) {$\circ$};
\node (RRRRR2a) at (29, 1.73) {$\circ$};
\node (RRRRR3a) at (31, 1.73) {$\circ$};
\node (RRRRR2b) at (29, -1.73) {\textbf{\textbullet}};
\node (RRRRR3b) at (31, -1.73) {\textbf{\textbullet}};
\node (RRRRR4) at (32, 0) {\textbf{\textbullet}};

\draw[dotted, ultra thick, ->] (1) -- (4);

\draw[dotted, thick, ->] (1) -- (2a);
\draw[dotted, thick, ->] (2a) -- (3a);
\draw[dotted, thick, ->] (3a) -- (4);

\draw[dotted, thick, ->] (1) -- (2b);
\draw[dotted, thick, ->] (2b) -- (3b);
\draw[dotted, thick, ->] (3b) -- (4);

\draw[dotted, ultra thick, ->] (R1) -- (R4);

\draw[dotted, thick, ->] (R1) -- (R2a);
\draw[dotted, thick, ->] (R2a) -- (R3a);
\draw[dotted, thick, ->] (R3a) -- (R4);

\draw[dotted, thick, ->] (R1) -- (R2b);
\draw[dotted, thick, ->] (R2b) -- (R3b);
\draw[dotted, thick, ->] (R3b) -- (R4);

\draw[dotted, ultra thick, ->] (RR1) -- (RR4);

\draw[dotted, thick, ->] (RR1) -- (RR2a);
\draw[dotted, thick, ->] (RR2a) -- (RR3a);
\draw[dotted, thick, ->] (RR3a) -- (RR4);

\draw[dotted, thick, ->] (RR1) -- (RR2b);
\draw[dotted, thick, ->] (RR2b) -- (RR3b);
\draw[dotted, thick, ->] (RR3b) -- (RR4);

\draw[dotted, ultra thick, ->] (RRR1) -- (RRR4);

\draw[dotted, thick, ->] (RRR1) -- (RRR2a);
\draw[dotted, thick, ->] (RRR2a) -- (RRR3a);
\draw[dotted, thick, ->] (RRR3a) -- (RRR4);

\draw[dotted, thick, ->] (RRR1) -- (RRR2b);
\draw[dotted, thick, ->] (RRR2b) -- (RRR3b);
\draw[dotted, thick, ->] (RRR3b) -- (RRR4);

\draw[dotted, ultra thick, ->] (RRRR1) -- (RRRR4);

\draw[dotted, thick, ->] (RRRR1) -- (RRRR2a);
\draw[dotted, thick, ->] (RRRR2a) -- (RRRR3a);
\draw[dotted, thick, ->] (RRRR3a) -- (RRRR4);

\draw[dotted, thick, ->] (RRRR1) -- (RRRR2b);
\draw[dotted, thick, ->] (RRRR2b) -- (RRRR3b);
\draw[dotted, thick, ->] (RRRR3b) -- (RRRR4);

\draw[dotted, ultra thick, ->] (RRRRR1) -- (RRRRR4);

\draw[dotted, thick, ->] (RRRRR1) -- (RRRRR2a);
\draw[dotted, thick, ->] (RRRRR2a) -- (RRRRR3a);
\draw[dotted, thick, ->] (RRRRR3a) -- (RRRRR4);

\draw[dotted, thick, ->] (RRRRR1) -- (RRRRR2b);
\draw[dotted, thick, ->] (RRRRR2b) -- (RRRRR3b);
\draw[dotted, thick, ->] (RRRRR3b) -- (RRRRR4);

\draw[bend left=50,->] (4) to (R3a);
\draw[->,very thick] (3b) to[out=0,in=-75]++ (1.75,2) to[out=100,in=-205] (R3a);
\path[->] (3b) edge [->,out=-5,in=-155,looseness=1] (R2a);

\draw[bend right=15,->] (R4) to (RR2b);
\draw [very thick, ->] (R3b) to (RR2b);
\draw [bend right=15,->] (R3b) to (RR1);

\draw[bend right=15,->] (RR4) to (RRR2b);
\draw [very thick, ->] (RR3b) to (RRR2b);
\draw [bend right=15,->] (RR3b) to (RRR1);

\draw[bend left=50,->] (RRR4) to (RRRR3a);
\draw[->,very thick] (RRR3b) to[out=0,in=-75]++ (1.75,2) to[out=100,in=-205] (RRRR3a);
\path[->] (RRR3b) edge [->,out=-5,in=-155,looseness=1] (RRRR2a);

\draw[bend right=15,->] (RRRR4) to (RRRRR2b);
\draw [very thick, ->] (RRRR3b) to (RRRRR2b);
\draw [bend right=15,->] (RRRR3b) to (RRRRR1);

\node (1past) at (-5, -1.73) {...};
\node (2past) at (-4, 0) {...};

\draw[bend right=15,->] (2past) to (2b);
\draw [very thick, ->] (1past) to (2b);
\draw [bend right=15,->] (1past) to (1);

\node (1cont) at (34, 0) {...};
\node (2cont) at (35, -1.73) {...};

\draw[bend right=15,->] (RRRRR4) to (2cont);
\draw [very thick, ->] (RRRRR3b) to (2cont);
\draw [bend right=15,->] (RRRRR3b) to (1cont);

\draw [thick,decorate,decoration={brace,amplitude=10pt,mirror},xshift=0.4pt,yshift=-0.4pt](-2,-3) -- (7.5,-3) node[black,midway,yshift=-0.6cm] {\tiny Era 2.};

\draw [thick,decorate,decoration={brace,amplitude=10pt,mirror},xshift=0.4pt,yshift=-0.4pt](10.5,-3) -- (13.5,-3) node[black,midway,yshift=-0.6cm] {\tiny Era 3.};

\draw [thick,decorate,decoration={brace,amplitude=10pt,mirror},xshift=0.4pt,yshift=-0.4pt](16.5,-3) -- (25.5,-3) node[black,midway,yshift=-0.6cm] {\tiny Era 4.};

\draw [thick,decorate,decoration={brace,amplitude=10pt,mirror},xshift=0.4pt,yshift=-0.4pt](28.5,-3) -- (31.5,-3) node[black,midway,yshift=-0.6cm] {\tiny Era 5.};

\end{tikzpicture}
\begin{tikzpicture}[scale=0.4]
\node (1) at (-2, 0) {\textbf{\textbullet}};
\node (2a) at (-1, 1.73) {\textbf{\textbullet}};
\node (3a) at (1, 1.73) {\textbf{\textbullet}};
\node (2b) at (-1, -1.73) {\textbf{\textbullet}};
\node (3b) at (1, -1.73) {\textbf{\textbullet}};
\node (4) at (2, 0) {\textbf{\textbullet}};

\node (R1) at (4, 0) {$\circ$};
\node (R2a) at (5, 1.73) {\textbf{\textbullet}};
\node (R3a) at (7, 1.73) {\textbf{\textbullet}};
\node (R2b) at (5, -1.73) {$\circ$};
\node (R3b) at (7, -1.73) {$\circ$};
\node (R4) at (8, 0) {\textbf{\textbullet}};

\node (RR1) at (10, 0) {$\circ$};
\node (RR2a) at (11, 1.73) {$\circ$};
\node (RR3a) at (13, 1.73) {$\circ$};
\node (RR2b) at (11, -1.73) {\textbf{\textbullet}};
\node (RR3b) at (13, -1.73) {\textbf{\textbullet}};
\node (RR4) at (14, 0) {\textbf{\textbullet}};

\node (RRR1) at (16, 0) {$\circ$};
\node (RRR2a) at (17, 1.73) {\textbf{\textbullet}};
\node (RRR3a) at (19, 1.73) {\textbf{\textbullet}};
\node (RRR2b) at (17, -1.73) {$\circ$};
\node (RRR3b) at (19, -1.73) {$\circ$};
\node (RRR4) at (20, 0) {\textbf{\textbullet}};

\node (RRRR1) at (22, 0) {$\circ$};
\node (RRRR2a) at (23, 1.73) {$\circ$};
\node (RRRR3a) at (25, 1.73) {$\circ$};
\node (RRRR2b) at (23, -1.73) {\textbf{\textbullet}};
\node (RRRR3b) at (25, -1.73) {\textbf{\textbullet}};
\node (RRRR4) at (26, 0) {\textbf{\textbullet}};

\node (RRRRR1) at (28, 0) {$\circ$};
\node (RRRRR2a) at (29, 1.73) {\textbf{\textbullet}};
\node (RRRRR3a) at (31, 1.73) {\textbf{\textbullet}};
\node (RRRRR2b) at (29, -1.73) {$\circ$};
\node (RRRRR3b) at (31, -1.73) {$\circ$};
\node (RRRRR4) at (32, 0) {\textbf{\textbullet}};

\draw[dotted, ultra thick, ->] (1) -- (4);

\draw[dotted, thick, ->] (1) -- (2a);
\draw[dotted, thick, ->] (2a) -- (3a);
\draw[dotted, thick, ->] (3a) -- (4);

\draw[dotted, thick, ->] (1) -- (2b);
\draw[dotted, thick, ->] (2b) -- (3b);
\draw[dotted, thick, ->] (3b) -- (4);

\draw[dotted, ultra thick, ->] (R1) -- (R4);

\draw[dotted, thick, ->] (R1) -- (R2a);
\draw[dotted, thick, ->] (R2a) -- (R3a);
\draw[dotted, thick, ->] (R3a) -- (R4);

\draw[dotted, thick, ->] (R1) -- (R2b);
\draw[dotted, thick, ->] (R2b) -- (R3b);
\draw[dotted, thick, ->] (R3b) -- (R4);

\draw[dotted, ultra thick, ->] (RR1) -- (RR4);

\draw[dotted, thick, ->] (RR1) -- (RR2a);
\draw[dotted, thick, ->] (RR2a) -- (RR3a);
\draw[dotted, thick, ->] (RR3a) -- (RR4);

\draw[dotted, thick, ->] (RR1) -- (RR2b);
\draw[dotted, thick, ->] (RR2b) -- (RR3b);
\draw[dotted, thick, ->] (RR3b) -- (RR4);

\draw[dotted, ultra thick, ->] (RRR1) -- (RRR4);

\draw[dotted, thick, ->] (RRR1) -- (RRR2a);
\draw[dotted, thick, ->] (RRR2a) -- (RRR3a);
\draw[dotted, thick, ->] (RRR3a) -- (RRR4);

\draw[dotted, thick, ->] (RRR1) -- (RRR2b);
\draw[dotted, thick, ->] (RRR2b) -- (RRR3b);
\draw[dotted, thick, ->] (RRR3b) -- (RRR4);

\draw[dotted, ultra thick, ->] (RRRR1) -- (RRRR4);

\draw[dotted, thick, ->] (RRRR1) -- (RRRR2a);
\draw[dotted, thick, ->] (RRRR2a) -- (RRRR3a);
\draw[dotted, thick, ->] (RRRR3a) -- (RRRR4);

\draw[dotted, thick, ->] (RRRR1) -- (RRRR2b);
\draw[dotted, thick, ->] (RRRR2b) -- (RRRR3b);
\draw[dotted, thick, ->] (RRRR3b) -- (RRRR4);

\draw[dotted, ultra thick, ->] (RRRRR1) -- (RRRRR4);

\draw[dotted, thick, ->] (RRRRR1) -- (RRRRR2a);
\draw[dotted, thick, ->] (RRRRR2a) -- (RRRRR3a);
\draw[dotted, thick, ->] (RRRRR3a) -- (RRRRR4);

\draw[dotted, thick, ->] (RRRRR1) -- (RRRRR2b);
\draw[dotted, thick, ->] (RRRRR2b) -- (RRRRR3b);
\draw[dotted, thick, ->] (RRRRR3b) -- (RRRRR4);

\draw[bend left=50,->] (4) to (R3a);
\draw[->,very thick] (3b) to[out=0,in=-75]++ (1.75,2) to[out=100,in=-205] (R3a);
\path[->] (3b) edge [->,out=-5,in=-155,looseness=1] (R2a);

\draw[bend right=15,->] (R4) to (RR2b);
\draw [very thick, ->] (R3b) to (RR2b);
\draw [bend right=15,->] (R3b) to (RR1);

\draw[bend left=50,->] (RR4) to (RRR3a);
\draw[->,very thick] (RR3b) to[out=0,in=-75]++ (1.75,2) to[out=100,in=-205] (RRR3a);
\path[->] (RR3b) edge [->,out=-5,in=-155,looseness=1] (RRR2a);

\draw[bend right=15,->] (RRR4) to (RRRR2b);
\draw [very thick, ->] (RRR3b) to (RRRR2b);
\draw [bend right=15,->] (RRR3b) to (RRRR1);

\draw[bend left=50,->] (RRRR4) to (RRRRR3a);
\draw[->,very thick] (RRRR3b) to[out=0,in=-75]++ (1.75,2) to[out=100,in=-205] (RRRRR3a);
\path[->] (RRRR3b) edge [->,out=-5,in=-155,looseness=1] (RRRRR2a);

\node (1past) at (-5, -1.73) {...};
\node (2past) at (-4, 0) {...};

\draw[bend right=15,->] (2past) to (2b);
\draw [very thick, ->] (1past) to (2b);
\draw [bend right=15,->] (1past) to (1);

\draw [thick,decorate,decoration={brace,amplitude=10pt,mirror},xshift=0.4pt,yshift=-0.4pt](-2,-3) -- (7.5,-3) node[black,midway,yshift=-0.6cm] {\tiny Era 6.};

\draw [thick,decorate,decoration={brace,amplitude=10pt,mirror},xshift=0.4pt,yshift=-0.4pt](10.5,-3) -- (19.5,-3) node[black,midway,yshift=-0.6cm] {\tiny Era 7.};

\draw [thick,decorate,decoration={brace,amplitude=10pt,mirror},xshift=0.4pt,yshift=-0.4pt](22.5,-3) -- (31.5,-3) node[black,midway,yshift=-0.6cm] {\tiny Era 8.};

\end{tikzpicture}
\captionof{figure}{An example corresponding to a partial Kasner sequence characterized by
a continued fraction given by $u_0 = [2;4,2,1,2,1,2,2,2,\dots]$. When an era has a length
longer than one, this gives rise to isolated fixed points, denoted by hollow circles,
from which isolated heteroclinic orbits and chains originate. These isolated structures
are not connected to earlier non-isolated heteroclinic chains. Since the isolated structures
cannot be approximated by Bianchi type $\mathrm{VI}_{-1/9}$ solutions that already
asymptotically approximates preceding non-isolated heteroclinic chains (especially for
cyclic chains), they are not expected to play an asymptotic role for such solutions.
Hence, they should be removed to yield the relevant heteroclinic network. 
}\label{FIG:hexEX}
\end{figure}

As illustrated by the example in Figure~\ref{FIG:hexEX}, the main ingredient for how one constructs general heteroclinic chains is how to connect hexagons, according to an era change or not of the BKL Kasner map.
In particular, concatenations of such connections between hexagons describe the isolated pieces
that should be removed.
The identification of isolated pieces depends on the length of an era and if the preceding era has a length longer
than one or not. This yields the following general rules for identifying isolated structures:
\begin{itemize}
\item[($\mathrm{r}_1$)] If the $j^{th}$ era has length $\lfloor u^j_0\rfloor = k_j = 1$, then the first
hexagon in the subsequent $(j+1)^{th}$ era has no isolated fixed points, heteroclinic
orbits and chains, as exemplified in Figure~\ref{FIG:hexEX} in the transition from
era 3 to 4 and era 5 to 6.
\item[($\mathrm{r}_2$)] If the $j^{th}$ era has length $\lfloor u^j_0\rfloor= k_j \geq 2$,
then all hexagons within the $j^{th}$ era, except the first (which corresponds
to the largest value of the Kasner parameter in that era, i.e., $u^j_0$),
contain an \emph{isolated heteroclinic chain}
$(321)\cdots\hspace{-0.1cm}\color{gray}{\small{\RHD}} \color{black}{(231)}
\cdots\hspace{-0.1cm}\color{gray}{\small{\RHD}} \color{black}{(132)}
\cdots\hspace{-0.1cm}\color{gray}{\small{\RHD}} \color{black}{(123)}$,
an isolated heteroclinic orbit
$(321)\cdots\hspace{-0.1cm}\color{gray}{\small{\RHD}} \color{black}{(312)}$,
and isolated double frame transition orbits
$(321)\boldsymbol{\cdots}\hspace{-0.05cm}\color{gray}{\small{\RHD}} \color{black}{(123)}$,
with no previous non-isolated heteroclinic orbit connections in
the Kasner sequence heteroclinic network.
\item[($\mathrm{r}_3$)] If the $j^{th}$ era has length $\lfloor u^j_0\rfloor= k_j \geq 2$,
then the mixed curvature-frame transition orbits
$(132)\boldsymbol{-}\hspace{-0.2cm}\color{black}{\small{\RHD}} \color{black}{(231)}$
from the $j^{th}$ to the $(j+1)^{th}$ era are isolated, as is the
single curvature transition orbit
$(132){-}\hspace{-0.2cm}\color{gray}{\small{\RHD}} \color{black}{(123)}$
which is followed by the thereby isolated heteroclinic chain
$(321)\cdots\hspace{-0.1cm}\color{gray}{\small{\RHD}} \color{black}{(312)}
\cdots\hspace{-0.1cm}\color{gray}{\small{\RHD}} \color{black}{(213)}
\cdots\hspace{-0.1cm}\color{gray}{\small{\RHD}} \color{black}{(123)}$,
the heteroclinic orbit
$(321)\cdots\hspace{-0.1cm}\color{gray}{\small{\RHD}} \color{black}{(231)}$,
and the double frame transition orbits
$(321)\boldsymbol{\cdots}\hspace{-0.05cm}\color{gray}{\small{\RHD}} \color{black}{(123)}$
in the first hexagon with $u^{j+1}_0$ in the $(j+1)^{th}$ era,
see era 0 to 1, era 1 to 2, era 2 to 3, era 4 to 5, era 6 to 7, and era 7 to 8
in Figure~\ref{FIG:hexEX}.
\item[($\mathrm{r}_4$)] When the $j^{th}$ era has length
$\lfloor u^j_0\rfloor = k_j \geq 3$, in addition to the isolated
structures given in ($r_2$) and ($r_3$), all
hexagons within the $j^{th}$ era, except the first (with $u^j_0$) and second
(with $u^j_1$), prolong the isolated heteroclinic orbit
$(321)\cdots\hspace{-0.1cm}\color{gray}{\small{\RHD}} \color{black}{(312)}$
in ($r_2$) to the isolated heteroclinic chain
$(321)\cdots\hspace{-0.1cm}\color{gray}{\small{\RHD}} \color{black}{(312)}
\cdots\hspace{-0.1cm}\color{gray}{\small{\RHD}} \color{black}{(213)}$. Moreover,
this also leads to that the single heteroclinic curvature transition
$(132)\sim\hspace{-0.26cm}\color{gray}{\small{\RHD}} \color{black}{(312)}$
and the mixed curvature-frame transition orbits
$(132)\boldsymbol{\sim}\hspace{-0.26cm}\color{black}{\small{\RHD}} \color{black}{(213)}$
that connect the hexagons after the first two Kasner epochs within the era
become isolated, see era 1 in Figure~\ref{FIG:hexEX}.
\end{itemize}
The above discussion takes into account general infinite
heteroclinic chains, regardless if they are aperiodic or periodic. The
heteroclinic network corresponding to an aperiodic Kasner sequence is constructed
according to the above guidelines while the network corresponding to a
periodic Kasner sequence is obtained by considering an infinite
sequence of hexagons with repeated copies of heteroclinic subnetworks, which
amounts to appropriately identifying repeated objects, once a period is realized,
in the quotient topology.

We summarize the above
in the following Proposition, which describes the construction of
heteroclinic chains on the union of the Bianchi type $\mathrm{I}$ and
$\mathrm{II}$ subsets of Bianchi type $\mathrm{VI}_{-1/9}$:
\begin{proposition}\label{prop:removal}
Consider an initial Kasner parameter $u_0 \in [1,\infty)$
with orbit under the BKL Kasner map~\eqref{Kasnermap} given by
$\{u_i\}_{i\in\mathbb{N}_0}=\{u_0,u_1,...\}$, partitioned into eras according to~\eqref{EraPartition}.
The following algorithm yields a construction of general heteroclinic chains:
\begin{itemize}
\item[(i)] Each $u_i$ yields six values for the generalized Kasner
parameter $\ue_{i_n}\in\mathbb{R},n=1,..., 6,$ connected by heteroclinic orbits into the
hexagon graph  in Figure~\ref{FIG:hex}.
\item[(ii)] Hexagons are connected by two possible configurations
depending on if the BKL Kasner map~\eqref{Kasnermap} changes era or not,
shown in Figure~\ref{FIG:BKLmapshex}.
\item[(iii)] Apart from the golden ratio\footnote{This is the only heteroclinic network without isolated structures and it is dealt with in the next section.}, $u_0=(1+\sqrt{5})/2$, only a subset of
the heteroclinic network generated by 
Kasner sequence $\{u_i\}_{i\in\mathbb{N}_0}$, called the stable heteroclinic
subnetwork, is non-isolated, 
where this subnetwork thereby is the complement to the isolated structures
described in $(\mathrm{r}_1)$, $(\mathrm{r}_2)$, $(\mathrm{r}_3)$
and $(\mathrm{r}_4)$.
\end{itemize}
\end{proposition}
%


As a consequence of the construction of the hexagon
representation, Proposition~\ref{prop:removal} and the removal rules
$(\mathrm{r}_1)-(\mathrm{r}_4)$, it follows that there always exists a heteroclinic
network of directed heteroclinic chains within and between hexagons, even after
the removal of isolated heteroclinic chains. Moreover, the remaining stable
subnetwork always contains one-parameter sets of multiple transition orbits, as
illustrated in Figure~\ref{FIG:hexEX}.

The work~\cite{heietal09} introduced the concept of
large and small curvature phases and generalized the stochastic analysis
in~\cite{khaetal85}, which shows that a Kasner
epoch in a small curvature phase is more likely, which entails a large value of $u$. Due
to Proposition~\ref{prop:removal} and the removal rules
$(\mathrm{r}_1)-(\mathrm{r}_4)$, it follows that most multiple transitions are
removed. In the present state space, a small curvature phase corresponds to being
mostly in a small neighborhood of the Taub point $\mathrm{T}_3$, which leads
to a sequence of alternating single frame transitions $\mathcal{T}_{R_3}$ and single curvature transitions
$\mathcal{T}_{N_-}$ between the sectors $(213)$ and $(123)$, as illustrated
by Era 1 in Figure~\ref{FIG:hexEX}. Thus most of the multiple transitions
are statistically expected not to occur. This, however, is not the case for
networks of heteroclinic cycles discussed below.

Finally, using the requirement given in~\cite[Def. 1.4]{BegDut22} for
solutions to approximate a heteroclinic chain, no solutions
can approximate the isolated heteroclinic chains in a heteroclinic network that
is constructed from a Kasner sequence $\{u_0,u_1,\dots u_i,u_{i+1}, \dots\}$.
Note that if the network corresponding to a Kasner epoch $u_i$
contains isolated heteroclinic chains, then in a Kasner sequence $\{u_i,u_{i+1}, \dots\}$,
where $u_i$ now is regarded as having no isolated chains, this results in
a subsequent change of isolated chains for $u_{i+1}$ (consider, e.g.,
the example in Figure~\ref{FIG:hexEX}; if $u_0$ in that sequence is
preceded by an era with a length larger than one, then there would have
been isolated chains in the network corresponding to $u_0$, although
subsequently the isolated chains would quickly coincide for the two
Kasner sequences). This small ambiguity does not exist for
periodic Kasner sequences $\{u_0, u_1,\dots, u_p\}$, for which the
existence of isolated heteroclinic chains have particularly far
ranging consequences.

\section{Networks of cyclic heteroclinic chains}\label{sec:cyclic}


First, note that a periodic Kasner sequence $\{u_0,u_1,\dots,u_{n-1}\}$ with length $n$
results in $n$ fixed points in each of the six sectors of $\mathrm{K}^\ocircle$
and thereby $6n$ fixed points on $\mathrm{K}^\ocircle$, described by the
extended Kasner parameter $\ue(u)$. Four of the six sectors only yield single
transitions and thereby $n$ single transitions each, while the two remaining
sectors with multiple transitions give rise to two single transitions for each fixed point
in those sectors. This yields $8n$ single transitions, while the $n$ fixed points
in each of the two multiple transition sectors yield $2n$ families of heteroclinic
multiple transition orbits. However, from an asymptotic point of view, it is
only the stable subnetwork that is of interest for periodic Kasner sequences,
as described by the following theorem:


%
\begin{theorem}\label{thm:cyclic}
Consider the heteroclinic network that is obtained from a periodic Kasner sequence
with period two or more, generated by $u_0 = \big[\,\overline{k_{1},k_{2},\dots}\,\big]$.
Suppose that a Bianchi type $\mathrm{VI}_{-1/9}$ solution has an $\omega$-limit
set that resides in this network. Then the $\omega$-limit set only resides in the
stable heteroclinic subnetwork (i.e., the $\omega$-limit set cannot include the
isolated part of the network removed by the rules $(\mathrm{r}_1)-(\mathrm{r}_4)$).
\end{theorem}
For a given $u_0 = \big[\,\overline{k_{1},k_{2},\dots}\,\big]$, note that all heteroclinic cycles are contained in the stable heteroclinic subnetwork, since the isolated part of the network consists of directed heteroclinic paths that do not admit recurrence, by definition.

\begin{proof}
To obtain a contradiction, suppose that the $\omega$-limit set contains the removed isolated part of the heteroclinic network. This implies that the type $\mathrm{VI}_{-1/9}$ solution repeatedly must come back to a neighborhood of the isolated part of the network, but it is impossible to do so through a path in this network, by construction (i.e., it is impossible to go back to the removed part because of its isolated nature), which thereby yields a contradiction.
As a consequence, a type $\mathrm{VI}_{-1/9}$ solution cannot come arbitrarily close to the heteroclinic orbits that forms the isolated heteroclinic subnetwork, which thereby cannot be part of the $\omega$-limit set of such a type $\mathrm{VI}_{-1/9}$ solution.
\end{proof}
%

%
%

We expect that there are Bianchi type $\mathrm{VI}_{-1/9}$ solutions that
have an $\omega$-limit set residing within the networks generated by a
periodic Kasner sequence, i.e., that an analogous result to that
in~\cite{Lieb11} is valid for type $\mathrm{VI}_{-1/9}$ solutions.
This justifies the hypothesis in Theorem~\ref{thm:cyclic}. However,
note that a similar phenomenon as described in Theorem~\ref{thm:cyclic}
does not exist for Bianchi types $\mathrm{VIII}$ and $\mathrm{IX}$, since there
are no isolated heteroclinic structures for these models. This is due to the
permutation symmetries of the equations, which, when quotient out, yields a
one-to-one correspondence between Kasner sequences $\{u_0,u_1,u_2,\dots\}$
and heteroclinic chains.
%
This is illustrated in detail in connection with the silver ratio discussed below.

%

Finally, note that until now there exist no descriptions and results about
heteroclinic cycles for Bianchi type $\mathrm{VI}_{-1/9}$, except for
that periodic cycles of period 13 and 18 were found numerically in~\cite{buc13}.
In contrast, we here have presented an algorithmic procedure for constructing all
possible cyclic networks and their properties from the quadratic irrational
values of $u_0$. Next we will illustrate the construction of networks of
heteroclinic cycles from periodic Kasner sequences $\{u_0,u_1,u_2,\dots\}$
with three examples: the metallic ratios given by the golden and silver ratios
with an era sequence with period one, $u_0 = \sfrac{1 + \sqrt{5}}{2}$ and
$u_0 = 1 + \sqrt{2}$, respectively, and one with an era sequence with period
two, $u_0 = 1 + \sqrt{3}$. The two last examples exemplify the removal
of the isolated part of a network of heteroclinic cycles, thereby
illustrating the key element in Theorem~\ref{thm:cyclic}.

\subsection{The golden ratio $u_0 = \sfrac{1 + \sqrt{5}}{2}$}\label{sec:goldrat}

The golden ratio $u_0 = [(1)] = (1+\sqrt{5})/2$ leads to that both the
frame-invariant Kasner sequence and the era sequence have period $1$,
\begin{equation}
(u_i)_{i\in\mathbb{N}_0}:\quad
\sfrac{1 + \sqrt{5}}{2} \rightarrow
\sfrac{1 + \sqrt{5}}{2} \rightarrow
\sfrac{1 + \sqrt{5}}{2} \rightarrow
\sfrac{1 + \sqrt{5}}{2} \rightarrow
\ldots ,
\end{equation}
where the (Hubble-normalized) Kretschmann scalar in~\eqref{Weyl} for
$u_0 = (1 + \sqrt{5})/2$ is given by
$\mathcal{W}((1 + \sqrt{5})/2) = 27 (5-2\sqrt{5})/8\approx 1.781$.

As described in ($\mathrm{r}_1$), an era that follows an era of length
one has no isolated fixed points, heteroclinic orbits and chains.
Since the golden ratio consists of a sequence of eras with length one,
there are thereby no isolated structures in this case. Note again that
having a sequence of eras with length one only is the only way of
producing a heteroclinic network without any isolated chains. Thus,
the golden ratio is the only infinite Kasner sequence that does not
yield isolated structures.

It follows straightforwardly by applying the rules
given in the previous section that the heteroclinic network, when expressed
in the hexagonal representation, takes the form given in
Figure~\ref{FIG:graphper3}.
\begin{figure}[H]\centering
\scriptsize
\begin{tikzpicture}[scale=0.5]
\node (1) at (-2, 0) {\textbf{\textbullet}};
\node (2a) at (-1, 1.73) {\textbf{\textbullet}};
\node (3a) at (1, 1.73) {\textbf{\textbullet}};
\node (2b) at (-1, -1.73) {\textbf{\textbullet}};
\node (3b) at (1, -1.73) {\textbf{\textbullet}};
\node (4) at (2, 0) {\textbf{\textbullet}};

\draw[dotted, ultra thick, ->] (1) -- (4);

\draw[dotted, thick, ->] (1) -- (2a);
\draw[dotted, thick, ->] (2a) -- (3a);
\draw[dotted, thick, ->] (3a) -- (4);

\draw[dotted, thick, ->] (1) -- (2b);
\draw[dotted, thick, ->] (2b) -- (3b);
\draw[dotted, thick, ->] (3b) -- (4);

\draw [very thick, bend left=30,->] (3b) to (2b);

\path[->] (3b) edge [->,out=-120,in=-90,looseness=1.6] (1);
\path[->] (4) edge [->,out=-90,in=-60,looseness=1.5] (2b);



\end{tikzpicture}
\captionof{figure}{A graph that describes the possible itineraries of the
heteroclinic network associated with $u_0=(1+\sqrt{5})/2$.
Thin dotted arrows describe single frame transitions while the bold dotted arrow depict double
frame transitions; single curvature transitions are described by thin curved arrows while
mixed curvature-frame transitions are depicted by a thick curved arrow. The leftmost point denotes
$\ue=-(\sqrt{5}-1)/2$ in sector $(321)$ while the rightmost point corresponds to
$\ue = u=(\sqrt{5}+1)/2$ in sector $(123)$.}\label{FIG:graphper3}
\end{figure}

One of the applications of the hexagonal representation is
to easily identify different cyclic subnetworks. This is illustrated by Figure~\ref{FIG:graphper3}
from which we easily see that $u_0=\sfrac{1 + \sqrt{5}}{2}$ leads to that single
transitions form heteroclinic cycles with period 3 and 6; double frame transitions
yield heteroclinic cycles with period 4 (replacing the former period 6 chains);
mixed curvature-frame transitions leads to heteroclinic cycles with period 2.

The hexagonal representation in Figure~\ref{FIG:graphper3} for the
golden ratio $u_0 = \sfrac{1 + \sqrt{5}}{2}$ amounts to the
following full heteroclinic network for $u_0 = (1+\sqrt{5})/2$ according
to~\eqref{eranetwork}:
\begin{equation}\label{eranetworkgold}
\mathbb{H}_\mathrm{Gold} = \overline{\mathcal{T}}_{R_1R_3}(\ue=-(\sqrt{5}-1)/2\in (321))\,\bigcup\,
\overline{\mathcal{T}}_{R_1N_-}(\ue=(\sqrt{5}-1)/2\in (132)),
\end{equation}
while the network of single transitions described in~\eqref{eranetworksingle} is given by
\begin{equation}\label{singleeranetworkgold}
\partial\mathbb{H}_\mathrm{Gold} = \partial\overline{\mathcal{T}}_{R_1R_3}(\ue=-(\sqrt{5}-1)/2\in (321))
\,\bigcup\,\partial\overline{\mathcal{T}}_{R_1N_-}(\ue=(\sqrt{5}-1)/2\in (132)).
\end{equation}
Using these results in combination with the projection rules onto
the plane $\Sigma_1 + \Sigma_2 + \Sigma_3 =0$ in $(\Sigma_1,\Sigma_2,\Sigma_3)$-space
described in Sections~\ref{sec:Kasner} (see Figure~\ref{KMaps}) and~\ref{sec:BII}
(see Figure~\ref{fig:CurvatureTrans}) for the various transitions lead to
the illustrations in Figure~\ref{fig:period3network} and the heteroclinic cycles in Figure~\ref{fig:period3}.

The heteroclinic chains arising from the golden ratio
are the simplest Bianchi type $\mathrm{VI}_{-1/9}$ candidates for
a stable manifold proof, akin to
in~\cite{Beguin10,Lieb11,Lieb13,BegDut22} for Bianchi types $\mathrm{VIII}$ and
$\mathrm{IX}$. However, the methods used in those works are not applicable to
the present case. The methods in~\cite{Lieb11,Lieb13,BegDut22} require an
ordering and conditions on the relative size of the eigenvalues,
which, according to Table~\ref{tab:per3EV}, does not occur here; nor is the
Sternberg Non-Resonance Conditions fulfilled and thus
the Takens linearization used in the proof of Béguin in~\cite{Beguin10} does not apply either.
Worse of all, note that all heteroclinic cycles generated from the golden ratio possess a fixed point with two unstable variables, yielding a one-parameter set of mixed curvature-frame transition
orbits, which do not occur for Bianchi types $\mathrm{VIII}$ and $\mathrm{IX}$. This exemplify entirely new challenges for proofs for Bianchi type $\mathrm{VI}_{-1/9}$.

\begin{figure}[H]
\centering
\begin{tikzpicture}[scale=1.1]
        \draw (1,0) arc (0:360:1cm and 1cm);	 	

        \draw [gray, dashed, - ]  (1,1.2)--(1,0);
        \draw [gray, dashed, - ]  (-1,1.2)--(-1,0);
        \draw [gray,dashed, - ]  (-0.5392,-1.4660)--(0.5,-0.8660);
        \draw [gray,dashed, - ]  (-1.5392,0.2660)--(-0.5,0.8660);
	\draw[gray,dashed, -] (2,0) -- (0.5,-0.866);
	\draw[gray,dashed, -] (2,0) -- (0.5, 0.866);

       \draw [ultra thick, domain=0.52:1.570,variable=\t,smooth] plot ({sin(\t r)},{cos(\t r)});
       \draw [ultra thick, domain=-0.52:-1.570,variable=\t,smooth] plot ({sin(\t r)},{cos(\t r)});

	\draw (-0.95,0) -- (-1.05,0);
        \draw (0.95,0) -- (1.05,0);
        \draw[rotate=120] (-0.95,0) -- (-1.05,0);
        \draw[rotate=240] (-0.95,0) -- (-1.05,0);
        \draw[rotate=120] (0.95,0) -- (1.05,0);
        \draw[rotate=240] (0.95,0) -- (1.05,0);


    \def\u{-0.618}
    \foreach \a in {0.3, 1.3} 
        \foreach \b in {0.5} 
        {
        \newcommand \n {(-\a*\b*(1-\u*\u)/((2+\u)*\u))}; 
        \newcommand \pone {(-\u/(1+\u+\u*\u))}; 
        \newcommand \ptwo {((1+\u)/(1+\u+\u*\u))}; 
        \newcommand \pthree {((\u*(1+\u))/(1+\u+\u*\u))}; 
        \newcommand \Sone {(-1+3*(\pone*\n*\n*exp(-6*\pone*(\t))+\ptwo*\b*\b*exp(-6*\ptwo*(\t))+\pthree*exp(-6*\pthree*(\t)))/(\n*\n*exp(-6*\pone*(\t))+\b*\b*exp(-6*\ptwo*(\t))+exp(-6*\pthree*(\t))))}; 
        \newcommand \Sthree {(-1+3*(\pone*exp(6*\pone*(\t))+\ptwo*\a*\a*exp(6*\ptwo*(\t))+\pthree*(\a*\b-\n)*(\a*\b-\n)*exp(6*\pthree*(\t)))/(exp(6*\pone*(\t))+\a*\a*exp(6*\ptwo*(\t))+(\a*\b-\n)*(\a*\b-\n)*exp(6*\pthree*(\t))))}; 
        \draw[dotted, ultra thick, postaction={decorate},variable=\t,domain=-1.9:1.2] plot ({-0.5*\Sone},{-(0.5*\Sone+\Sthree)/1.732});
        }

    \draw [dotted, thick, postaction={decorate}]  (0.96,0.28)--(0.96,-0.28);\draw [dotted, gray]  (0.96,0.28)--(0.96,1.2);
    \draw [dotted, thick, postaction={decorate}]  (-0.71,0.71)--(-0.71,-0.71);\draw [dotted, gray]  (-0.71,0.71)--(-0.71,1.2);

    \draw [rotate=240,dotted, thick, postaction={decorate}]  (-0.71,-0.71)--(0.96,0.28);\draw [rotate=240,dotted, gray]  (-0.91,-0.825)--(-0.71,-0.71);

    \draw [dotted, thick, postaction={decorate}]  (-0.71,-0.71)--(0.96,0.28);\draw [dotted, gray]  (-0.91,-0.825)--(-0.71,-0.71);

    \draw [rotate=-240,dotted, thick, postaction={decorate}]  (0.96,0.28)--(0.96,-0.28);\draw [rotate=-240,dotted, gray]  (0.96,0.28)--(0.96,1.2);
    \draw [rotate=-240,dotted, thick, postaction={decorate}]  (-0.71,0.71)--(-0.71,-0.71);\draw [rotate=-240,dotted, gray]  (-0.71,0.71)--(-0.71,1.2);

    \filldraw [black] (0.96,-0.28) circle (1.5pt);
    \filldraw [black] (0.96,0.28) circle (1.5pt);
    \filldraw [rotate=240,black] (-0.71,-0.71) circle (1.5pt);
    \filldraw [black] (-0.71,0.71) circle (1.5pt);
    \filldraw [black] (-0.71,-0.71) circle (1.5pt);
    \filldraw [rotate=120,black] (-0.71,0.71) circle (1.5pt);
 \end{tikzpicture}
\hspace{0.5cm}
\begin{tikzpicture}[scale=1.1]
        \draw (1,0) arc (0:360:1cm and 1cm);	 	

        \draw [gray, dashed, - ]  (1,1.2)--(1,0);
        \draw [gray, dashed, - ]  (-1,1.2)--(-1,0);
        \draw [gray,dashed, - ]  (-0.5392,-1.4660)--(0.5,-0.8660);
        \draw [gray,dashed, - ]  (-1.5392,0.2660)--(-0.5,0.8660);
	\draw[gray,dashed, -] (2,0) -- (0.5,-0.866);
	\draw[gray,dashed, -] (2,0) -- (0.5, 0.866);

       \draw [ultra thick, domain=0.52:1.570,variable=\t,smooth] plot ({sin(\t r)},{cos(\t r)});
       \draw [ultra thick, domain=-0.52:-1.570,variable=\t,smooth] plot ({sin(\t r)},{cos(\t r)});

	\draw (-0.95,0) -- (-1.05,0);
        \draw (0.95,0) -- (1.05,0);
        \draw[rotate=120] (-0.95,0) -- (-1.05,0);
        \draw[rotate=240] (-0.95,0) -- (-1.05,0);
        \draw[rotate=120] (0.95,0) -- (1.05,0);
        \draw[rotate=240] (0.95,0) -- (1.05,0);

    \def\u{1.618}
    \foreach \a in {-0.5,0.2,0.8}
    {\draw[very thick , postaction={decorate},variable=\t,domain=-2.6:4.4] plot ({(1+2*(2*\u/(1+\u*\u))*tanh(-\t))/(2+(2*\u/(1+\u*\u))*tanh(-\t))},{(1.732*((1-\u*\u)/(1+\u*\u))*tanh((((1-\u*\u)/(1+\u*\u))/(2*\u/(1+\u*\u)))*(-\t)-\a))/(2+(2*\u/(1+\u*\u))*tanh(-\t))});}

    \draw[  postaction={decorate}] (0.96,0.28) -- (-0.71,0.71);\draw [dotted, gray]  (0.96,0.28)--(2,0);
    \draw[  postaction={decorate}] (0.96,-0.28) -- (-0.71,-0.71);\draw [dotted, gray]  (0.96,-0.28)--(2,0);

    \draw [dotted, thick, postaction={decorate}]  (0.96,0.28)--(0.96,-0.28);\draw [dotted, gray]  (0.96,0.28)--(0.96,1.2);
    \draw [dotted, thick, postaction={decorate}]  (-0.71,0.71)--(-0.71,-0.71);\draw [dotted, gray]  (-0.71,0.71)--(-0.71,1.2);

    \filldraw [black] (0.96,-0.28) circle (1.5pt);
    \filldraw [black] (0.96,0.28) circle (1.5pt);
    \filldraw [black] (-0.71,0.71) circle (1.5pt);
    \filldraw [black] (-0.71,-0.71) circle (1.5pt);

 \end{tikzpicture}
\hspace{0.5cm}
\begin{tikzpicture}[scale=1.1]
        \draw (1,0) arc (0:360:1cm and 1cm);	 	

        \draw [gray, dashed, - ]  (1,1.2)--(1,0);
        \draw [gray, dashed, - ]  (-1,1.2)--(-1,0);
        \draw [gray,dashed, - ]  (-0.5392,-1.4660)--(0.5,-0.8660);
        \draw [gray,dashed, - ]  (-1.5392,0.2660)--(-0.5,0.8660);
	\draw[gray,dashed, -] (2,0) -- (0.5,-0.866);
	\draw[gray,dashed, -] (2,0) -- (0.5, 0.866);

       \draw [ultra thick, domain=0.52:1.570,variable=\t,smooth] plot ({sin(\t r)},{cos(\t r)});
       \draw [ultra thick, domain=-0.52:-1.570,variable=\t,smooth] plot ({sin(\t r)},{cos(\t r)});
	
	\draw (-0.95,0) -- (-1.05,0);
        \draw (0.95,0) -- (1.05,0);
        \draw[rotate=120] (-0.95,0) -- (-1.05,0);
        \draw[rotate=240] (-0.95,0) -- (-1.05,0);
        \draw[rotate=120] (0.95,0) -- (1.05,0);
        \draw[rotate=240] (0.95,0) -- (1.05,0);

    \draw[  postaction={decorate}] (0.96,0.28) -- (-0.71,0.71);\draw [dotted, gray]  (0.96,0.28)--(2,0);
    \draw[  postaction={decorate}] (0.96,-0.28) -- (-0.71,-0.71);\draw [dotted, gray]  (0.96,-0.28)--(2,0);

    \draw [dotted, thick, postaction={decorate}]  (0.96,0.28)--(0.96,-0.28);\draw [dotted, gray]  (0.96,0.28)--(0.96,1.2);
    \draw [dotted, thick, postaction={decorate}]  (-0.71,0.71)--(-0.71,-0.71);\draw [dotted, gray]  (-0.71,0.71)--(-0.71,1.2);

    \draw [rotate=240,dotted, thick, postaction={decorate}]  (-0.71,-0.71)--(0.96,0.28);\draw [rotate=240,dotted, gray]  (-0.91,-0.825)--(-0.71,-0.71);

    \draw [dotted, thick, postaction={decorate}]  (-0.71,-0.71)--(0.96,0.28);\draw [dotted, gray]  (-0.91,-0.825)--(-0.71,-0.71);

    \draw [rotate=-240,dotted, thick, postaction={decorate}]  (0.96,0.28)--(0.96,-0.28);\draw [rotate=-240,dotted, gray]  (0.96,0.28)--(0.96,1.2);
    \draw [rotate=-240,dotted, thick, postaction={decorate}]  (-0.71,0.71)--(-0.71,-0.71);\draw [rotate=-240,dotted, gray]  (-0.71,0.71)--(-0.71,1.2);


    \filldraw [black] (0.96,-0.28) circle (1.5pt);
    \filldraw [black] (0.96,0.28) circle (1.5pt);
    \filldraw [rotate=240,black] (-0.71,-0.71) circle (1.5pt);
    \filldraw [black] (-0.71,0.71) circle (1.5pt);
    \filldraw [black] (-0.71,-0.71) circle (1.5pt);
    \filldraw [rotate=120,black] (-0.71,0.71) circle (1.5pt);



 \end{tikzpicture}
\caption {Projection onto the $\Sigma_1 + \Sigma_2 + \Sigma_3 = 0$ plane in $(\Sigma_1,\Sigma_2,\Sigma_3)$-space
of the heteroclinic network induced by the golden ratio Kasner parameter $u_0=\sfrac{1 + \sqrt{5}}{2}$.
\textbf{Left:} The frame transition network $\overline{\mathcal{T}}_{R_1R_3}(\ue)$ for $\ue=-(\sqrt{5}-1)/2 \in (321)$;
single frame transitions are denoted by thin dotted directed lines while double frame transitions
are illustrated by two representative solutions denoted by directed thick dotted lines.
\textbf{Middle:} The curvature transition network $\overline{\mathcal{T}}_{R_1N_-}(\ue)$ for $\ue=(\sqrt{5}-1)/2\in (132)$;
single curvature transitions are given by directed thin lines while mixed curvature-frame solutions are
depicted as directed thicker lines. The full heteroclinic network of the golden ratio Kasner parameter
is given by the union of the frame and curvature transition networks.
\textbf{Right:} The single transition network $\partial\overline{\mathcal{T}}_{R_1R_3}(\ue\in (321))
\,\bigcup\,\partial\overline{\mathcal{T}}_{R_1N_-}(\ue\in (132))$.}\label{fig:period3network}
\end{figure}
%
%
%
\begin{figure}[H]
\centering
\begin{tikzpicture}[scale=1.1]
        \draw (1,0) arc (0:360:1cm and 1cm);	 	

        \draw [gray, dashed, - ]  (1,1.2)--(1,0);
        \draw [gray, dashed, - ]  (-1,1.2)--(-1,0);
        \draw [gray,dashed, - ]  (-0.5392,-1.4660)--(0.5,-0.8660);
        \draw [gray,dashed, - ]  (-1.5392,0.2660)--(-0.5,0.8660);
	\draw[gray,dashed, -] (2,0) -- (0.5,-0.866);
	\draw[gray,dashed, -] (2,0) -- (0.5, 0.866);

       \draw [ultra thick, domain=0.52:1.570,variable=\t,smooth] plot ({sin(\t r)},{cos(\t r)});
       \draw [ultra thick, domain=-0.52:-1.570,variable=\t,smooth] plot ({sin(\t r)},{cos(\t r)});

	\draw (-0.95,0) -- (-1.05,0);
        \draw (0.95,0) -- (1.05,0);
        \draw[rotate=120] (-0.95,0) -- (-1.05,0);
        \draw[rotate=240] (-0.95,0) -- (-1.05,0);
        \draw[rotate=120] (0.95,0) -- (1.05,0);
        \draw[rotate=240] (0.95,0) -- (1.05,0);


    \draw[  postaction={decorate}] (0.96,-0.28) -- (-0.71,-0.71);\draw [dotted, gray]  (0.96,-0.28)--(2,0);

    \draw [dotted, thick, postaction={decorate}]  (0.96,0.28)--(0.96,-0.28);\draw [dotted, gray]  (0.96,0.28)--(0.96,1.2);

    \draw [dotted, thick, postaction={decorate}]  (-0.71,-0.71)--(0.96,0.28);\draw [dotted, gray]  (-0.91,-0.825)--(-0.71,-0.71);


    \filldraw [black] (0.96,-0.28) circle (1.5pt);
    \filldraw [black] (0.96,0.28) circle (1.5pt);
    \filldraw [black] (-0.71,-0.71) circle (1.5pt);

 \end{tikzpicture}
\hspace{0.5cm}
\begin{tikzpicture}[scale=1.1]
        \draw (1,0) arc (0:360:1cm and 1cm);	 	

        \draw [gray, dashed, - ]  (1,1.2)--(1,0);
        \draw [gray, dashed, - ]  (-1,1.2)--(-1,0);
        \draw [gray,dashed, - ]  (-0.5392,-1.4660)--(0.5,-0.8660);
        \draw [gray,dashed, - ]  (-1.5392,0.2660)--(-0.5,0.8660);
	\draw[gray,dashed, -] (2,0) -- (0.5,-0.866);
	\draw[gray,dashed, -] (2,0) -- (0.5, 0.866);

       \draw [ultra thick, domain=0.52:1.570,variable=\t,smooth] plot ({sin(\t r)},{cos(\t r)});
       \draw [ultra thick, domain=-0.52:-1.570,variable=\t,smooth] plot ({sin(\t r)},{cos(\t r)});

	\draw (-0.95,0) -- (-1.05,0);
        \draw (0.95,0) -- (1.05,0);
        \draw[rotate=120] (-0.95,0) -- (-1.05,0);
        \draw[rotate=240] (-0.95,0) -- (-1.05,0);
        \draw[rotate=120] (0.95,0) -- (1.05,0);
        \draw[rotate=240] (0.95,0) -- (1.05,0);


    \draw[  postaction={decorate}] (0.96,0.28) -- (-0.71,0.71);\draw [dotted, gray]  (0.96,0.28)--(2,0);

    \draw [dotted, thick, postaction={decorate}]  (-0.71,0.71)--(-0.71,-0.71);\draw [dotted, gray]  (-0.71,0.71)--(-0.71,1.2);

    \draw [dotted, thick, postaction={decorate}]  (-0.71,-0.71)--(0.96,0.28);\draw [dotted, gray]  (-0.91,-0.825)--(-0.71,-0.71);

    \filldraw [black] (0.96,0.28) circle (1.5pt);
    \filldraw [black] (-0.71,0.71) circle (1.5pt);
    \filldraw [black] (-0.71,-0.71) circle (1.5pt);

 \end{tikzpicture}
\hspace{0.5cm}
\begin{tikzpicture}[scale=1.1]
        \draw (1,0) arc (0:360:1cm and 1cm);	 	

        \draw [gray, dashed, - ]  (1,1.2)--(1,0);
        \draw [gray, dashed, - ]  (-1,1.2)--(-1,0);
        \draw [gray,dashed, - ]  (-0.5392,-1.4660)--(0.5,-0.8660);
        \draw [gray,dashed, - ]  (-1.5392,0.2660)--(-0.5,0.8660);
	\draw[gray,dashed, -] (2,0) -- (0.5,-0.866);
	\draw[gray,dashed, -] (2,0) -- (0.5, 0.866);

       \draw [ultra thick, domain=0.52:1.570,variable=\t,smooth] plot ({sin(\t r)},{cos(\t r)});
       \draw [ultra thick, domain=-0.52:-1.570,variable=\t,smooth] plot ({sin(\t r)},{cos(\t r)});

	\draw (-0.95,0) -- (-1.05,0);
        \draw (0.95,0) -- (1.05,0);
        \draw[rotate=120] (-0.95,0) -- (-1.05,0);
        \draw[rotate=240] (-0.95,0) -- (-1.05,0);
        \draw[rotate=120] (0.95,0) -- (1.05,0);
        \draw[rotate=240] (0.95,0) -- (1.05,0);

    \draw[  postaction={decorate}] (0.96,-0.28) -- (-0.71,-0.71);\draw [dotted, gray]  (0.96,-0.28)--(2,0);
    \draw[  postaction={decorate}] (0.96,0.28) -- (-0.71,0.71);\draw [dotted, gray]  (0.96,0.28)--(2,0);

    \draw [rotate=240,dotted, thick, postaction={decorate}]  (-0.71,-0.71)--(0.96,0.28);\draw [rotate=240,dotted, gray]  (-0.91,-0.825)--(-0.71,-0.71);

    \draw [dotted, thick, postaction={decorate}]  (-0.71,-0.71)--(0.96,0.28);\draw [dotted, gray]  (-0.91,-0.825)--(-0.71,-0.71);

    \draw [rotate=-240,dotted, thick, postaction={decorate}]  (0.96,0.28)--(0.96,-0.28);\draw [rotate=-240,dotted, gray]  (0.96,0.28)--(0.96,1.2);
    \draw [rotate=-240,dotted, thick, postaction={decorate}]  (-0.71,0.71)--(-0.71,-0.71);\draw [rotate=-240,dotted, gray]  (-0.71,0.71)--(-0.71,1.2);


    \filldraw [black] (0.96,-0.28) circle (1.5pt);
    \filldraw [black] (0.96,0.28) circle (1.5pt);
    \filldraw [rotate=240,black] (-0.71,-0.71) circle (1.5pt);
    \filldraw [black] (-0.71,0.71) circle (1.5pt);
    \filldraw [black] (-0.71,-0.71) circle (1.5pt);
    \filldraw [rotate=120,black] (-0.71,0.71) circle (1.5pt);

 \end{tikzpicture}
\hspace{0.5cm}
\begin{tikzpicture}[scale=1.1]
        \draw (1,0) arc (0:360:1cm and 1cm);	 	

        \draw [gray, dashed, - ]  (1,1.2)--(1,0);
        \draw [gray, dashed, - ]  (-1,1.2)--(-1,0);
        \draw [gray,dashed, - ]  (-0.5392,-1.4660)--(0.5,-0.8660);
        \draw [gray,dashed, - ]  (-1.5392,0.2660)--(-0.5,0.8660);
	\draw[gray,dashed, -] (2,0) -- (0.5,-0.866);
	\draw[gray,dashed, -] (2,0) -- (0.5, 0.866);

       \draw [ultra thick, domain=0.52:1.570,variable=\t,smooth] plot ({sin(\t r)},{cos(\t r)});
       \draw [ultra thick, domain=-0.52:-1.570,variable=\t,smooth] plot ({sin(\t r)},{cos(\t r)});

	\draw (-0.95,0) -- (-1.05,0);
        \draw (0.95,0) -- (1.05,0);
        \draw[rotate=120] (-0.95,0) -- (-1.05,0);
        \draw[rotate=240] (-0.95,0) -- (-1.05,0);
        \draw[rotate=120] (0.95,0) -- (1.05,0);
        \draw[rotate=240] (0.95,0) -- (1.05,0);



    \def\u{1.618}
    \foreach \a in {-1,1.8}
    {\draw[very thick, postaction={decorate},variable=\t,domain=-4.2:7] plot ({(1+2*(2*\u/(1+\u*\u))*tanh(-\t))/(2+(2*\u/(1+\u*\u))*tanh(-\t))},{(1.732*((1-\u*\u)/(1+\u*\u))*tanh((((1-\u*\u)/(1+\u*\u))/(2*\u/(1+\u*\u)))*(-\t)-\a))/(2+(2*\u/(1+\u*\u))*tanh(-\t))});}

    \draw [dotted, thick, postaction={decorate}]  (-0.71,-0.71)--(0.96,0.28);\draw [dotted, gray]  (-0.91,-0.825)--(-0.71,-0.71);

    \filldraw [black] (0.96,0.28) circle (1.5pt);
    \filldraw [black] (-0.71,-0.71) circle (1.5pt);

 \end{tikzpicture}
\hspace{0.5cm}
\begin{tikzpicture}[scale=1.1]
        \draw (1,0) arc (0:360:1cm and 1cm);	 	

        \draw [gray, dashed, - ]  (1,1.2)--(1,0);
        \draw [gray, dashed, - ]  (-1,1.2)--(-1,0);
        \draw [gray,dashed, - ]  (-0.5392,-1.4660)--(0.5,-0.8660);
        \draw [gray,dashed, - ]  (-1.5392,0.2660)--(-0.5,0.8660);
	\draw[gray,dashed, -] (2,0) -- (0.5,-0.866);
	\draw[gray,dashed, -] (2,0) -- (0.5, 0.866);

       \draw [ultra thick, domain=0.52:1.570,variable=\t,smooth] plot ({sin(\t r)},{cos(\t r)});
       \draw [ultra thick, domain=-0.52:-1.570,variable=\t,smooth] plot ({sin(\t r)},{cos(\t r)});

	\draw (-0.95,0) -- (-1.05,0);
        \draw (0.95,0) -- (1.05,0);
        \draw[rotate=120] (-0.95,0) -- (-1.05,0);
        \draw[rotate=240] (-0.95,0) -- (-1.05,0);
        \draw[rotate=120] (0.95,0) -- (1.05,0);
        \draw[rotate=240] (0.95,0) -- (1.05,0);

    \draw[  postaction={decorate}] (0.96,-0.28) -- (-0.71,-0.71);\draw [dotted, gray]  (0.96,-0.28)--(2,0);
    \draw[  postaction={decorate}] (0.96,0.28) -- (-0.71,0.71);\draw [dotted, gray]  (0.96,0.28)--(2,0);


    \draw [dotted, thick, postaction={decorate}]  (-0.71,-0.71)--(0.96,0.28);\draw [dotted, gray]  (-0.91,-0.825)--(-0.71,-0.71);



    \def\u{-0.618}
    \foreach \a in {0.3, 1.3} 
        \foreach \b in {0.5} 
        {
        \newcommand \n {(-\a*\b*(1-\u*\u)/((2+\u)*\u))}; 
        \newcommand \pone {(-\u/(1+\u+\u*\u))}; 
        \newcommand \ptwo {((1+\u)/(1+\u+\u*\u))}; 
        \newcommand \pthree {((\u*(1+\u))/(1+\u+\u*\u))}; 
        \newcommand \Sone {(-1+3*(\pone*\n*\n*exp(-6*\pone*(\t))+\ptwo*\b*\b*exp(-6*\ptwo*(\t))+\pthree*exp(-6*\pthree*(\t)))/(\n*\n*exp(-6*\pone*(\t))+\b*\b*exp(-6*\ptwo*(\t))+exp(-6*\pthree*(\t))))}; 
        \newcommand \Sthree {(-1+3*(\pone*exp(6*\pone*(\t))+\ptwo*\a*\a*exp(6*\ptwo*(\t))+\pthree*(\a*\b-\n)*(\a*\b-\n)*exp(6*\pthree*(\t)))/(exp(6*\pone*(\t))+\a*\a*exp(6*\ptwo*(\t))+(\a*\b-\n)*(\a*\b-\n)*exp(6*\pthree*(\t))))}; 
        \draw[dotted, ultra thick, postaction={decorate},variable=\t,domain=-1.9:1.2] plot ({-0.5*\Sone},{-(0.5*\Sone+\Sthree)/1.732});
        }

    \filldraw [black] (0.96,-0.28) circle (1.5pt);
    \filldraw [black] (0.96,0.28) circle (1.5pt);
    \filldraw [black] (-0.71,0.71) circle (1.5pt);
    \filldraw [black] (-0.71,-0.71) circle (1.5pt);

 \end{tikzpicture}
\caption {Subnetworks of heteroclinic cycles generated by
$u_0=\sfrac{1 + \sqrt{5}}{2}$ in Figure~\ref{fig:period3network}.
\textbf{Top: } three cycles with period 3 and 6 consisting of single transitions,
which correspond to the single transition cycles in Figure~\ref{FIG:graphper3}.
\textbf{Bottom: } two families of cycles containing a
mixed curvature-frame transition, with period 2, or a double frame transition,
with a period 4. Note that each of the bottom one-parameter
families of heteroclinic cycles possesses the two above adjacent
single heteroclinic transition subnetworks as boundaries.
}\label{fig:period3}
\end{figure}
\begin{table}[H]
\centering
\footnotesize
\begin{tabular}{|c|c|c|c|c|c|c|}\hline
        \rowcolor{lightgray!25} $\ue$ & Sector & $\lambda_{R_1}$ & $\lambda_{R_3}$ &
        $\lambda_{N_-}$ & $\lambda_{A}$ & Eigenvalues Order\\ \hline
        $-\frac{\left(3+\sqrt{5}\right)}{2}$ & (213) & $-\frac{3\sqrt{5}}{2}$ & $\frac{3(1+\sqrt{5})}{4}$ & $-3$ & $-\frac{3(3-\sqrt{5})}{4}$ & 
        $-\lambda_{A} < \lambda_{R_3}< -\lambda_{N_-} < -\lambda_{R_1}$\\ \hline
        $-\frac{\left(1+\sqrt{5}\right)}{2}$ & (231) & $-\frac{3(1+\sqrt{5})}{4}$ & $\frac{3\sqrt{5}}{2}$ & $-\frac{3(1+\sqrt{5})}{2}$ & $-\frac{3}{2}$ & 
        $-\lambda_{A} < -\lambda_{R_1} < \lambda_{R_3} < -\lambda_{N_-}$\\ \hline
        $-\frac{\left(\sqrt{5}-1\right)}{2}$ & (321) & $\frac{3(1+\sqrt{5})}{4}$ & $\frac{3(\sqrt{5}-1)}{4}$ & $-\frac{3(1+\sqrt{5})}{2}$ & $-\frac{3(3+\sqrt{5})}{4}$ & 
        $\lambda_{R_3} < \lambda_{R_1} < -\lambda_{A} < -\lambda_{N_-}$\\ \hline
        $-\frac{\left(3-\sqrt{5}\right)}{2}$ & (312) & $\frac{3\sqrt{5}}{2}$ & $-\frac{3(\sqrt{5}-1)}{4}$ & $-3$ & $-\frac{3(3+\sqrt{5})}{4}$ & 
        $-\lambda_{R_3} < -\lambda_{N_-} < \lambda_{R_1}  < -\lambda_{A}$\\ \hline  %
        $\frac{\left(\sqrt{5}-1\right)}{2}$ & (132) & $\frac{3(\sqrt{5}-1)}{4}$ & $-\frac{3\sqrt{5}}{2}$ & $\frac{3(\sqrt{5}-1)}{2}$ & $-\frac{3}{2}$ & 
        $\lambda_{R_1} < -\lambda_{A} < \lambda_{N_-}  < -\lambda_{R_3}$\\ \hline
        $\frac{\left(1+\sqrt{5}\right)}{2}$ & (123) & $-\frac{3(\sqrt{5}-1)}{4}$ & $-\frac{3(1+\sqrt{5})}{4}$ & $\frac{3(\sqrt{5}-1)}{2}$ & $-\frac{3(3-\sqrt{5})}{4}$ & 
        $-\lambda_{A} < -\lambda_{R_1} < \lambda_{N_-}  < -\lambda_{R_3}$\\ \hline
\end{tabular}
\caption{The $\ue$ parametrization of each fixed point of the heteroclinic network
corresponding to the golden ratio $u_0=\sfrac{1 + \sqrt{5}}{2}$ in the 6 different
sectors, ordered by an increasing $\ue$ parameter, 
and the corresponding eigenvalues and their ordering.}\label{tab:per3EV}
\end{table}
%


\subsection{The silver ratio $u_0 = 1 + \sqrt{2}$}\label{sec:silverrat}
%
The silver ratio $u_0 = [(2)] = 1 +\sqrt{2}$ yields an era sequence of
period $1$, but a Kasner sequence with period $2$:
\begin{equation}
(u_i)_{i\in\mathbb{N}_0}:\quad
\underbrace{(1+\sqrt{2}) \rightarrow {\color{gray!75}\sqrt{2}}}_{\text{\scriptsize{era}}} \rightarrow
\underbrace{(1+\sqrt{2}) \rightarrow {\color{gray!75}\sqrt{2}}}_{\text{\scriptsize{era}}} \rightarrow
\underbrace{(1+\sqrt{2}) \rightarrow {\color{gray!75}\sqrt{2}}}_{\text{\scriptsize{era}}} \rightarrow
\ldots ,
\end{equation}
for which the Kretschmann scalar in~\eqref{Weyl} is given by
$\mathcal{W}(1+\sqrt{2})=54(44 - 25 \sqrt{2})/343\approx 1.360$ and
$\mathcal{W}({\color{gray!75}\sqrt{2}})=162/(3 + \sqrt{2})^3\approx 1.883$.
Throughout, to simplify the identification of the six values of $\ue$ belonging to $u=1+\sqrt{2}$ and $u={\color{gray!75}\sqrt{2}}$, we respectively denote them in black and in light gray, both in the text and in the figures below (apart from the hollow dots that correspond to the isolated fixed
points discussed below).

%
%
%
Each of the two Kasner parameter values,
$u\in \{ 1 + \sqrt{2}$ and ${\color{gray!75}\sqrt{2}} \}$,
yields six values of $\ue$, one in each sector,
resulting in a pair of fixed points in each sector and a total of
twelve points on the Kasner circle $\mathrm{K}^{\ocircle}$,
described in Table~\ref{tab:per5EV}:
%

Next we will use the results in Sections~\ref{sec:Kasner}, \ref{sec:BII},
\ref{subsec:relevant} and Proposition~\ref{prop:removal} and apply
them to the silver ratio $u_0 = 1 + \sqrt{2}$ as an illustration of
how to construct a network of heteroclinic cycles and how to remove
its isolated parts. Using the last two eras of length 2 in the example
in Figure~\ref{FIG:hexEX} lead to Figure~\ref{FIG:graphper5}.
%


Note that it follows from the heteroclinic network structure, consisting of directed
heteroclinic orbits, that if a solution has been asymptotically attracted to the
heteroclinic network, then it cannot be attracted to the isolated part of it; it is
only the stable `attractor' part of the cyclic network that asymptotically can attract
solutions. For this reason we remove the isolated structure to the left in
Figure~\ref{FIG:graphper5} and thereby obtain the stable part to the
right. As is easily seen from this hexagonal representation, the stable
attractor part consists of the following heteroclinic cycles: two
heteroclinic cycles consisting of single transitions with period 5 and period
4 heteroclinic cycles involving the one-parameter set of mixed curvature-frame transition
orbits.

Using the values for $\ue$ in sectors $(321)$ and $(132)$ in Table~\ref{tab:per5EV}
for $u = 1+\sqrt{2}$ and ${\color{gray!75}\sqrt{2}}$ leads to the general
construction of the heteroclinic network~\eqref{eranetwork} that takes the form
\begin{equation}\label{eranetworksilver}
\begin{split} \mathbb{H}_\mathrm{Silver} =\, &
\overline{\mathcal{T}}_{R_1R_3}\left(\ue=-\sfrac{1}{\sqrt{2}}\right) \,\bigcup\,
\overline{\mathcal{T}}_{R_1R_3}({\color{gray!75}\ue=-(2 - \sqrt{2})})\\
& \bigcup \overline{\mathcal{T}}_{R_1N_-}(\ue=\sqrt{2} - 1)\,\bigcup\,
\overline{\mathcal{T}}_{R_1N_-}({\color{gray!75}\ue=\sfrac{1}{\sqrt{2}}})
\end{split},
\end{equation}
while the single transition network in~\eqref{eranetworksingle} result in
\begin{equation}\label{eranetworksilverbdry}
\begin{split} \partial\mathbb{H}_\mathrm{Silver} = \, &
\partial \overline{\mathcal{T}}_{R_1R_3}(\ue=-\sfrac{1}{\sqrt{2}}) \,\bigcup\,
\partial\overline{\mathcal{T}}_{R_1R_3}({\color{gray!75}\ue=-(2 - \sqrt{2})})\\
& \bigcup\partial
\overline{\mathcal{T}}_{R_1N_-}(\ue=\sqrt{2} - 1)\,\bigcup\,\partial
\overline{\mathcal{T}}_{R_1N_-}({\color{gray!75}\ue=\sfrac{1}{\sqrt{2}}}).
\end{split},
\end{equation}

Using the projection rules onto the plane
$\Sigma_1 + \Sigma_2 + \Sigma_3 =0$ in $(\Sigma_1,\Sigma_2,\Sigma_3)$-space
in sections~\ref{sec:Kasner} (see Figure~\ref{KMaps}) and~\ref{sec:BII}
(see Figure~\ref{fig:CurvatureTrans}) for the above various transitions
yield the depiction in Figure~\ref{fig:period5network}.
\begin{table}[H]
    \centering
    \footnotesize
    \begin{tabular}{|c|c|c|}\hline
        \rowcolor{lightgray!25} $u$ & $\ue$ & Sector \\ \hline
        \begin{tabular}{c}
            \textcolor{gray!75}{$\sqrt{2}$}\\
            $1+\sqrt{2}$
        \end{tabular}
         &
        \begin{tabular}{c}
            \textcolor{gray!75}{$-(\sqrt{2}+1)$} \\
            $-(\sqrt{2}+2)$
        \end{tabular}
         & (213) \\ \hline
        \begin{tabular}{c}
            $1+\sqrt{2}$ \\
            \textcolor{gray!75}{$\sqrt{2}$}
        \end{tabular}
         &
        \begin{tabular}{c}
            $-\sqrt{2}$ \\
            \textcolor{gray!75}{$-(\sfrac{1}{\sqrt{2}}+1)$}
        \end{tabular}
         & (231) \\ \hline
        \begin{tabular}{c}
            \textcolor{gray!75}{$\sqrt{2}$}\\
            $1+\sqrt{2}$
        \end{tabular}
         &
        \begin{tabular}{c}
            \textcolor{gray!75}{$-(2-\sqrt{2})$} \\
            $-\sfrac{1}{\sqrt{2}}$
        \end{tabular}
         & (321) \\ \hline
        \begin{tabular}{c}
            $1+\sqrt{2}$ \\
            \textcolor{gray!75}{$\sqrt{2}$}
        \end{tabular}
         &
        \begin{tabular}{c}
            $-(1-\sfrac{1}{\sqrt{2}})$ \\
            \textcolor{gray!75}{$-(\sqrt{2}-1)$}
        \end{tabular}
         & (312) \\ \hline
        \begin{tabular}{c}
            \textcolor{gray!75}{$\sqrt{2}$}\\
            $1+\sqrt{2}$
        \end{tabular}
         &
        \begin{tabular}{c}
            \textcolor{gray!75}{$\sfrac{1}{\sqrt{2}}$} \\
            $\sqrt{2}-1$
        \end{tabular}
         & (132)\\ \hline
        \begin{tabular}{c}
            $1+\sqrt{2}$ \\
            \textcolor{gray!75}{$\sqrt{2}$}
        \end{tabular}
         &
        \begin{tabular}{c}
            $1+\sqrt{2}$ \\
            \textcolor{gray!75}{$\sqrt{2}$}
        \end{tabular}
         & (123) \\ \hline
    \end{tabular}
    \caption{The values of $\ue$ for each fixed point
    in the six sectors on the Kasner circle $\mathrm{K}^{\ocircle}$ corresponding to
    $u = 1 + \sqrt{2}$ and $u = {\color{gray!75}\sqrt{2}}$.
    }\label{tab:per5EV}
\end{table}
\vspace{-0.5cm}
\begin{figure}[H]\centering
\tiny
\begin{tikzpicture}[scale=0.4]
\node (1) at (-2, 0) {$\circ$};
\node (2a) at (-1, 1.73) {$\circ$};
\node (3a) at (1, 1.73) {$\circ$};
\node (2b) at (-1, -1.73) {\textbf{\textbullet}};
\node (3b) at (1, -1.73) {\textbf{\textbullet}};
\node (4) at (2, 0) {\textbf{\textbullet}};

\node (R1) at (4, 0) {$\circ$};
\node[lightgray] (R2a) at (5, 1.73) {\textbf{\textbullet}};
\node[lightgray] (R3a) at (7, 1.73) {\textbf{\textbullet}};
\node (R2b) at (5, -1.73) {$\circ$};
\node (R3b) at (7, -1.73) {$\circ$};
\node[lightgray] (R4) at (8, 0) {\textbf{\textbullet}};

\draw[dotted, ultra thick, ->] (1) -- (4);

\draw[dotted, thick, ->] (1) -- (2a);
\draw[dotted, thick, ->] (2a) -- (3a);
\draw[dotted, thick, ->] (3a) -- (4);

\draw[dotted, thick, ->] (1) -- (2b);
\draw[dotted, thick, ->] (2b) -- (3b);
\draw[dotted, thick, ->] (3b) -- (4);

\draw[dotted, ultra thick, ->] (R1) -- (R4);

\draw[dotted, thick, ->] (R1) -- (R2a);
\draw[dotted, thick, ->] (R2a) -- (R3a);
\draw[dotted, thick, ->] (R3a) -- (R4);

\draw[dotted, thick, ->] (R1) -- (R2b);
\draw[dotted, thick, ->] (R2b) -- (R3b);
\draw[dotted, thick, ->] (R3b) -- (R4);

\draw[bend left=50,->] (4) to (R3a);
\draw[->,very thick] (3b) to[out=0,in=-75]++ (1.75,2) to[out=100,in=-205] (R3a);
\path[->] (3b) edge [->,out=-5,in=-155,looseness=1] (R2a);

\draw [very thick, bend left=25,->] (R3b) to (2b);

\path[->] (R3b) edge [->,out=-120,in=-90,looseness=1.2] (1);
\path[->] (R4) edge [->,out=-90,in=-60,looseness=1.2] (2b);


\end{tikzpicture}
\hspace{2.5cm}
\begin{tikzpicture}[scale=0.4]
\node (2b) at (-1, -1.73) {\textbf{\textbullet}};
\node (3b) at (1, -1.73) {\textbf{\textbullet}};
\node (4) at (2, 0) {\textbf{\textbullet}};

\node[lightgray] (R2a) at (5, 1.73) {\textbf{\textbullet}};
\node[lightgray] (R3a) at (7, 1.73) {\textbf{\textbullet}};
\node[lightgray] (R4) at (8, 0) {\textbf{\textbullet}};

\draw[dotted, thick, ->] (2b) -- (3b);
\draw[dotted, thick, ->] (3b) -- (4);

\draw[dotted, thick, ->] (R2a) -- (R3a);
\draw[dotted, thick, ->] (R3a) -- (R4);

\draw[bend left=50,->] (4) to (R3a);
\draw[->,very thick] (3b) to[out=0,in=-75]++ (1.75,2) to[out=100,in=-205] (R3a);
\path[->] (3b) edge [->,out=-5,in=-155,looseness=1] (R2a);

\path[->] (R4) edge [->,out=-90,in=-60,looseness=1.2] (2b);


\end{tikzpicture}
{\setlength{\abovecaptionskip}{-7pt}
\setlength{\belowcaptionskip}{0pt}
\captionof{figure}{\textbf{Left:} The hexagonal representation describing
the heteroclinic network of the periodic Kasner sequence induced by the silver
ratio $u_0 = 1 + \sqrt{2}$ (the hexagon to the left) and its iterate
${\color{gray!75}u=\sqrt{2}} $ (the hexagon to the right).
\textbf{Right:} The stable `attractor' network obtained by removing the isolated parts of
the heteroclinic network, which yields the heteroclinic cycles described in
Figure~\ref{fig:period5}. }\label{FIG:graphper5}
}
\end{figure}
\begin{figure}[H]
\centering
\begin{tikzpicture}[scale=1.1]
        \draw (1,0) arc (0:360:1cm and 1cm);	 	

        \draw [gray, dashed, - ]  (1,1.2)--(1,0);
        \draw [gray, dashed, - ]  (-1,1.2)--(-1,0);
        \draw [gray,dashed, - ]  (-0.5392,-1.4660)--(0.5,-0.8660);
        \draw [gray,dashed, - ]  (-1.5392,0.2660)--(-0.5,0.8660);
	\draw[gray,dashed, -] (2,0) -- (0.5,-0.866);
	\draw[gray,dashed, -] (2,0) -- (0.5, 0.866);

       \draw [ultra thick, domain=0.52:1.570,variable=\t,smooth] plot ({sin(\t r)},{cos(\t r)});
       \draw [ultra thick, domain=-0.52:-1.570,variable=\t,smooth] plot ({sin(\t r)},{cos(\t r)});

	\draw (-0.95,0) -- (-1.05,0);
        \draw (0.95,0) -- (1.05,0);
        \draw[rotate=120] (-0.95,0) -- (-1.05,0);
        \draw[rotate=240] (-0.95,0) -- (-1.05,0);
        \draw[rotate=120] (0.95,0) -- (1.05,0);
        \draw[rotate=240] (0.95,0) -- (1.05,0);

    \filldraw [black] (0.891,0.454) circle (1.5pt);
    \filldraw [lightgray] (0.98,0.198) circle (1.5pt);

    \filldraw [lightgray] (0.98,-0.198) circle (1.5pt);
    \filldraw [black] (0.891,-0.454) circle (1.5pt);

    \filldraw [lightgray] (-0.32,-0.947) circle (1.5pt);
    \filldraw [black] (-0.054,-0.998) circle (1.5pt);

    \filldraw [lightgray] (-0.660,-0.751) circle (1.5pt);
    \filldraw [black] (-0.837,-0.547) circle (1.5pt);

    \filldraw [black] (-0.837,0.547) circle (1.5pt);
    \filldraw [lightgray] (-0.660,0.751) circle (1.5pt);

    \filldraw [lightgray] (-0.32,0.947) circle (1.5pt);
    \filldraw [black] (-0.054,0.998) circle (1.5pt);

 \end{tikzpicture}
\hspace{0.5cm}
\begin{tikzpicture}[scale=1.1]
        \draw (1,0) arc (0:360:1cm and 1cm);	 	

        \draw [gray, dashed, - ]  (1,1.2)--(1,0);
        \draw [gray, dashed, - ]  (-1,1.2)--(-1,0);
        \draw [gray,dashed, - ]  (-0.5392,-1.4660)--(0.5,-0.8660);
        \draw [gray,dashed, - ]  (-1.5392,0.2660)--(-0.5,0.8660);
	\draw[gray,dashed, -] (2,0) -- (0.5,-0.866);
	\draw[gray,dashed, -] (2,0) -- (0.5, 0.866);

       \draw [ultra thick, domain=0.52:1.570,variable=\t,smooth] plot ({sin(\t r)},{cos(\t r)});
       \draw [ultra thick, domain=-0.52:-1.570,variable=\t,smooth] plot ({sin(\t r)},{cos(\t r)});

	\draw (-0.95,0) -- (-1.05,0);
        \draw (0.95,0) -- (1.05,0);
        \draw[rotate=120] (-0.95,0) -- (-1.05,0);
        \draw[rotate=240] (-0.95,0) -- (-1.05,0);
        \draw[rotate=120] (0.95,0) -- (1.05,0);
        \draw[rotate=240] (0.95,0) -- (1.05,0);

    \def\u{-0.707}
    \foreach \a in {0.3, 0.7, 1.6} 
        \foreach \b in {0.7} 
        {
        \newcommand \n {(-\a*\b*(1-\u*\u)/((2+\u)*\u))}; 
        \newcommand \pone {(-\u/(1+\u+\u*\u))}; 
        \newcommand \ptwo {((1+\u)/(1+\u+\u*\u))}; 
        \newcommand \pthree {((\u*(1+\u))/(1+\u+\u*\u))}; 
        \newcommand \Sone {(-1+3*(\pone*\n*\n*exp(-6*\pone*(\t))+\ptwo*\b*\b*exp(-6*\ptwo*(\t))+\pthree*exp(-6*\pthree*(\t)))/(\n*\n*exp(-6*\pone*(\t))+\b*\b*exp(-6*\ptwo*(\t))+exp(-6*\pthree*(\t))))}; 
        \newcommand \Sthree {(-1+3*(\pone*exp(6*\pone*(\t))+\ptwo*\a*\a*exp(6*\ptwo*(\t))+\pthree*(\a*\b-\n)*(\a*\b-\n)*exp(6*\pthree*(\t)))/(exp(6*\pone*(\t))+\a*\a*exp(6*\ptwo*(\t))+(\a*\b-\n)*(\a*\b-\n)*exp(6*\pthree*(\t))))}; 
        \draw[dotted, ultra thick, postaction={decorate},variable=\t,domain=-1.75:1] plot ({-0.5*\Sone},{-(0.5*\Sone+\Sthree)/1.732});
        }

    \draw [dotted, thick, postaction={decorate}]  (-0.837,0.547)--(-0.837,-0.547);\draw [dotted, gray]  (-0.837,0.547)--(-0.837,1.2);

    \draw [dotted, thick, postaction={decorate}]  (-0.054,0.998)--(-0.054,-0.998);\draw [dotted, gray]  (-0.054,0.998)--(-0.054,1.2);

    \draw [dotted, thick, postaction={decorate}]  (0.891,0.454)--(0.891,-0.454);\draw [dotted, gray]  (0.891,0.454)--(0.891,1.2);

    \draw [dotted, thick, postaction={decorate}]  (-0.054,-0.998)--(0.891,-0.454);\draw [dotted, gray]  (-0.625,-1.33)--(-0.054,-0.998);

    \draw [dotted, thick, postaction={decorate}]  (-0.837,-0.547)--(0.891,0.454);\draw [dotted, gray]  (-1,-0.64)--(-0.837,-0.547);

    \draw [dotted, thick, postaction={decorate}]  (-0.837,0.547)--(-0.054,0.998);\draw [dotted, gray]  (-1.49,0.17)--(-0.837,0.547);

    \filldraw [black] (0.891,0.454) circle (1.5pt);

    \filldraw [black] (0.891,-0.454) circle (1.5pt);

    \filldraw [black] (-0.054,-0.998) circle (1.5pt);

    \filldraw [black] (-0.837,-0.547) circle (1.5pt);

    \filldraw [black] (-0.837,0.547) circle (1.5pt);

    \filldraw [black] (-0.054,0.998) circle (1.5pt);

 \end{tikzpicture}
\hspace{0.5cm}
\begin{tikzpicture}[scale=1.1]
        \draw (1,0) arc (0:360:1cm and 1cm);	 	

        \draw [gray, dashed, - ]  (1,1.2)--(1,0);
        \draw [gray, dashed, - ]  (-1,1.2)--(-1,0);
        \draw [gray,dashed, - ]  (-0.5392,-1.4660)--(0.5,-0.8660);
        \draw [gray,dashed, - ]  (-1.5392,0.2660)--(-0.5,0.8660);
	\draw[gray,dashed, -] (2,0) -- (0.5,-0.866);
	\draw[gray,dashed, -] (2,0) -- (0.5, 0.866);

       \draw [ultra thick, domain=0.52:1.570,variable=\t,smooth] plot ({sin(\t r)},{cos(\t r)});
       \draw [ultra thick, domain=-0.52:-1.570,variable=\t,smooth] plot ({sin(\t r)},{cos(\t r)});

	\draw (-0.95,0) -- (-1.05,0);
        \draw (0.95,0) -- (1.05,0);
        \draw[rotate=120] (-0.95,0) -- (-1.05,0);
        \draw[rotate=240] (-0.95,0) -- (-1.05,0);
        \draw[rotate=120] (0.95,0) -- (1.05,0);
        \draw[rotate=240] (0.95,0) -- (1.05,0);

    \def\u{-0.585}
    \foreach \a in {0.7} 
        \foreach \b in {0.5} 
        {
        \newcommand \n {(-\a*\b*(1-\u*\u)/((2+\u)*\u))}; 
        \newcommand \pone {(-\u/(1+\u+\u*\u))}; 
        \newcommand \ptwo {((1+\u)/(1+\u+\u*\u))}; 
        \newcommand \pthree {((\u*(1+\u))/(1+\u+\u*\u))}; 
        \newcommand \Sone {(-1+3*(\pone*\n*\n*exp(-6*\pone*(\t))+\ptwo*\b*\b*exp(-6*\ptwo*(\t))+\pthree*exp(-6*\pthree*(\t)))/(\n*\n*exp(-6*\pone*(\t))+\b*\b*exp(-6*\ptwo*(\t))+exp(-6*\pthree*(\t))))}; 
        \newcommand \Sthree {(-1+3*(\pone*exp(6*\pone*(\t))+\ptwo*\a*\a*exp(6*\ptwo*(\t))+\pthree*(\a*\b-\n)*(\a*\b-\n)*exp(6*\pthree*(\t)))/(exp(6*\pone*(\t))+\a*\a*exp(6*\ptwo*(\t))+(\a*\b-\n)*(\a*\b-\n)*exp(6*\pthree*(\t))))}; 
        \draw[dotted, ultra thick, postaction={decorate},variable=\t,domain=-1.75:1.1] plot ({-0.5*\Sone},{-(0.5*\Sone+\Sthree)/1.732});
        }

    \draw [dotted, thick, postaction={decorate}]  (-0.660,0.751)--(-0.660,-0.751);\draw [dotted, gray]  (-0.660,0.751)--(-0.660,1.2);

    \draw [dotted, thick, postaction={decorate}]  (-0.32,0.947)--(-0.32,-0.947);\draw [dotted, gray]  (-0.32,0.947)--(-0.32,1.2);

    \draw [dotted, thick, postaction={decorate}]  (0.98,0.198)--(0.98,-0.198);\draw [dotted, gray]  (0.98,0.198)--(0.98,1.2);

    \draw [dotted, thick, postaction={decorate}]  (-0.32,-0.947)--(0.98,-0.198);\draw [dotted, gray]  (-0.7,-1.17)--(-0.32,-0.947);

    \draw [dotted, thick, postaction={decorate}]  (-0.660,-0.751)--(0.98,0.198);\draw [dotted, gray]  (-0.87,-0.88)--(-0.660,-0.751);

    \draw [dotted, thick, postaction={decorate}]  (-0.660,0.751)--(-0.32,0.947);\draw [dotted, gray]  (-1.51,0.25)--(-0.660,0.751);

    \filldraw [lightgray] (0.98,0.198) circle (1.5pt);

    \filldraw [lightgray] (0.98,-0.198) circle (1.5pt);

    \filldraw [lightgray] (-0.32,-0.947) circle (1.5pt);

    \filldraw [lightgray] (-0.660,-0.751) circle (1.5pt);

    \filldraw [lightgray] (-0.660,0.751) circle (1.5pt);

    \filldraw [lightgray] (-0.32,0.947) circle (1.5pt);

 \end{tikzpicture}
\hspace{0.5cm}
\begin{tikzpicture}[scale=1.1]
        \draw (1,0) arc (0:360:1cm and 1cm);	 	

        \draw [gray, dashed, - ]  (1,1.2)--(1,0);
        \draw [gray, dashed, - ]  (-1,1.2)--(-1,0);
        \draw [gray,dashed, - ]  (-0.5392,-1.4660)--(0.5,-0.8660);
        \draw [gray,dashed, - ]  (-1.5392,0.2660)--(-0.5,0.8660);
	\draw[gray,dashed, -] (2,0) -- (0.5,-0.866);
	\draw[gray,dashed, -] (2,0) -- (0.5, 0.866);

       \draw [ultra thick, domain=0.52:1.570,variable=\t,smooth] plot ({sin(\t r)},{cos(\t r)});
       \draw [ultra thick, domain=-0.52:-1.570,variable=\t,smooth] plot ({sin(\t r)},{cos(\t r)});

	\draw (-0.95,0) -- (-1.05,0);
        \draw (0.95,0) -- (1.05,0);
        \draw[rotate=120] (-0.95,0) -- (-1.05,0);
        \draw[rotate=240] (-0.95,0) -- (-1.05,0);
        \draw[rotate=120] (0.95,0) -- (1.05,0);
        \draw[rotate=240] (0.95,0) -- (1.05,0);

    \def\u{0.707}
    \foreach \a in {-1.1,-0.3,0.5}
    {\draw[very thick, postaction={decorate},variable=\t,domain=-2:5.3] plot ({(1+2*(2*\u/(1+\u*\u))*tanh(-\t))/(2+(2*\u/(1+\u*\u))*tanh(-\t))},{(1.732*((1-\u*\u)/(1+\u*\u))*tanh((((1-\u*\u)/(1+\u*\u))/(2*\u/(1+\u*\u)))*(-\t)-\a))/(2+(2*\u/(1+\u*\u))*tanh(-\t))});}

    \draw[  postaction={decorate}] (0.98,-0.198) -- (-0.837,-0.547);\draw [dotted, gray]  (0.98,-0.198)--(2,0);

    \draw[  postaction={decorate}] (0.98,0.198) -- (-0.837,0.547);\draw [dotted, gray]  (0.98,0.198)--(2,0);

    \draw [dotted, thick, postaction={decorate}]  (-0.837,0.547)--(-0.837,-0.547);\draw [dotted, gray]  (-0.837,0.547)--(-0.837,1.2);

    \draw [dotted, thick, postaction={decorate}]  (0.98,0.198)--(0.98,-0.198);\draw [dotted, gray]  (0.98,0.198)--(0.98,1.2);

    \filldraw [lightgray] (0.98,0.198) circle (1.5pt);

    \filldraw [lightgray] (0.98,-0.198) circle (1.5pt);


    \filldraw [black] (-0.837,-0.547) circle (1.5pt);

    \filldraw [black] (-0.837,0.547) circle (1.5pt);


 \end{tikzpicture}
\hspace{0.5cm}
\begin{tikzpicture}[scale=1.1]
        \draw (1,0) arc (0:360:1cm and 1cm);	 	

        \draw [gray, dashed, - ]  (1,1.2)--(1,0);
        \draw [gray, dashed, - ]  (-1,1.2)--(-1,0);
        \draw [gray,dashed, - ]  (-0.5392,-1.4660)--(0.5,-0.8660);
        \draw [gray,dashed, - ]  (-1.5392,0.2660)--(-0.5,0.8660);
	\draw[gray,dashed, -] (2,0) -- (0.5,-0.866);
	\draw[gray,dashed, -] (2,0) -- (0.5, 0.866);

       \draw [ultra thick, domain=0.52:1.570,variable=\t,smooth] plot ({sin(\t r)},{cos(\t r)});
       \draw [ultra thick, domain=-0.52:-1.570,variable=\t,smooth] plot ({sin(\t r)},{cos(\t r)});

	\draw (-0.95,0) -- (-1.05,0);
        \draw (0.95,0) -- (1.05,0);
        \draw[rotate=120] (-0.95,0) -- (-1.05,0);
        \draw[rotate=240] (-0.95,0) -- (-1.05,0);
        \draw[rotate=120] (0.95,0) -- (1.05,0);
        \draw[rotate=240] (0.95,0) -- (1.05,0);

    \def\u{0.414}
    \foreach \a in {-0.9,0,0.9}
    {\draw[very thick , postaction={decorate},variable=\t,domain=-2.2:3] plot ({(1+2*(2*\u/(1+\u*\u))*tanh(-\t))/(2+(2*\u/(1+\u*\u))*tanh(-\t))},{(1.732*((1-\u*\u)/(1+\u*\u))*tanh((((1-\u*\u)/(1+\u*\u))/(2*\u/(1+\u*\u)))*(-\t)-\a))/(2+(2*\u/(1+\u*\u))*tanh(-\t))});}

    \draw[  postaction={decorate}] (0.891,0.454) -- (-0.32,0.947);\draw [dotted, gray]  (0.891,0.454)--(2,0);

    \draw[  postaction={decorate}] (0.891,-0.454) -- (-0.32,-0.947);\draw [dotted, gray]  (0.891,-0.454)--(2,0);

    \draw [dotted, thick, postaction={decorate}]  (-0.32,0.947)--(-0.32,-0.947);\draw [dotted, gray]  (-0.32,0.947)--(-0.32,1.2);

    \draw [dotted, thick, postaction={decorate}]  (0.891,0.454)--(0.891,-0.454);\draw [dotted, gray]  (0.891,0.454)--(0.891,1.2);

    \filldraw [black] (0.891,0.454) circle (1.5pt);

    \filldraw [black] (0.891,-0.454) circle (1.5pt);

    \filldraw [lightgray] (-0.32,-0.947) circle (1.5pt);



    \filldraw [lightgray] (-0.32,0.947) circle (1.5pt);

 \end{tikzpicture}
\hspace{0.5cm}
\begin{tikzpicture}[scale=1.1]
        \draw (1,0) arc (0:360:1cm and 1cm);	 	

        \draw [gray, dashed, - ]  (1,1.2)--(1,0);
        \draw [gray, dashed, - ]  (-1,1.2)--(-1,0);
        \draw [gray,dashed, - ]  (-0.5392,-1.4660)--(0.5,-0.8660);
        \draw [gray,dashed, - ]  (-1.5392,0.2660)--(-0.5,0.8660);
	\draw[gray,dashed, -] (2,0) -- (0.5,-0.866);
	\draw[gray,dashed, -] (2,0) -- (0.5, 0.866);

       \draw [ultra thick, domain=0.52:1.570,variable=\t,smooth] plot ({sin(\t r)},{cos(\t r)});
       \draw [ultra thick, domain=-0.52:-1.570,variable=\t,smooth] plot ({sin(\t r)},{cos(\t r)});

	\draw (-0.95,0) -- (-1.05,0);
        \draw (0.95,0) -- (1.05,0);
        \draw[rotate=120] (-0.95,0) -- (-1.05,0);
        \draw[rotate=240] (-0.95,0) -- (-1.05,0);
        \draw[rotate=120] (0.95,0) -- (1.05,0);
        \draw[rotate=240] (0.95,0) -- (1.05,0);

    \draw[  postaction={decorate}] (0.891,-0.454) -- (-0.32,-0.947);\draw [dotted, gray]  (0.891,-0.454)--(2,0);

    \draw[  postaction={decorate}] (0.98,-0.198) -- (-0.837,-0.547);\draw [dotted, gray]  (0.98,-0.198)--(2,0);

    \draw[  postaction={decorate}] (0.98,0.198) -- (-0.837,0.547);\draw [dotted, gray]  (0.98,0.198)--(2,0);

    \draw[  postaction={decorate}] (0.891,0.454) -- (-0.32,0.947);\draw [dotted, gray]  (0.891,0.454)--(2,0);

    \draw [dotted, thick, postaction={decorate}]  (-0.837,0.547)--(-0.837,-0.547);\draw [dotted, gray]  (-0.837,0.547)--(-0.837,1.2);

    \draw [dotted, thick, postaction={decorate}]  (-0.660,0.751)--(-0.660,-0.751);\draw [dotted, gray]  (-0.660,0.751)--(-0.660,1.2);

    \draw [dotted, thick, postaction={decorate}]  (-0.32,0.947)--(-0.32,-0.947);\draw [dotted, gray]  (-0.32,0.947)--(-0.32,1.2);

    \draw [dotted, thick, postaction={decorate}]  (-0.054,0.998)--(-0.054,-0.998);\draw [dotted, gray]  (-0.054,0.998)--(-0.054,1.2);

    \draw [dotted, thick, postaction={decorate}]  (0.891,0.454)--(0.891,-0.454);\draw [dotted, gray]  (0.891,0.454)--(0.891,1.2);

    \draw [dotted, thick, postaction={decorate}]  (0.98,0.198)--(0.98,-0.198);\draw [dotted, gray]  (0.98,0.198)--(0.98,1.2);

    \draw [dotted, thick, postaction={decorate}]  (-0.054,-0.998)--(0.891,-0.454);\draw [dotted, gray]  (-0.625,-1.33)--(-0.054,-0.998);

    \draw [dotted, thick, postaction={decorate}]  (-0.32,-0.947)--(0.98,-0.198);\draw [dotted, gray]  (-0.7,-1.17)--(-0.32,-0.947);

    \draw [dotted, thick, postaction={decorate}]  (-0.660,-0.751)--(0.98,0.198);\draw [dotted, gray]  (-0.87,-0.88)--(-0.660,-0.751);

    \draw [dotted, thick, postaction={decorate}]  (-0.837,-0.547)--(0.891,0.454);\draw [dotted, gray]  (-1,-0.64)--(-0.837,-0.547);

    \draw [dotted, thick, postaction={decorate}]  (-0.837,0.547)--(-0.054,0.998);\draw [dotted, gray]  (-1.49,0.17)--(-0.837,0.547);

    \draw [dotted, thick, postaction={decorate}]  (-0.660,0.751)--(-0.32,0.947);\draw [dotted, gray]  (-1.51,0.25)--(-0.660,0.751);

    \filldraw [black] (0.891,0.454) circle (1.5pt);
    \filldraw [lightgray] (0.98,0.198) circle (1.5pt);

    \filldraw [lightgray] (0.98,-0.198) circle (1.5pt);
    \filldraw [black] (0.891,-0.454) circle (1.5pt);

    \filldraw [lightgray] (-0.32,-0.947) circle (1.5pt);
    \filldraw [black] (-0.054,-0.998) circle (1.5pt);

    \filldraw [lightgray] (-0.660,-0.751) circle (1.5pt);
    \filldraw [black] (-0.837,-0.547) circle (1.5pt);

    \filldraw [black] (-0.837,0.547) circle (1.5pt);
    \filldraw [lightgray] (-0.660,0.751) circle (1.5pt);

    \filldraw [lightgray] (-0.32,0.947) circle (1.5pt);
    \filldraw [black] (-0.054,0.998) circle (1.5pt);
 \end{tikzpicture}
\caption {The Kasner fixed points associated with $1+\sqrt{2}$ and \textcolor{lightgray}{$\sqrt{2}$} in black
and lightgray dots, respectively, and representations of the heteroclinic networks of
the closure of the various double frame and mixed curvature-frame transitions
(the two upper right and the two lower left figures) induced by the silver ratio $u_0 = 1+\sqrt{2}$.
The entire heteroclinic network is obtained by taking the union of these closures, i.e.,
by superimposing the network figures, while the single transition network (lower
figure to the right) is obtained by superimposing their single transition boundaries.
There are thereby 4 one-parameter families of multiple
transitions containing a total of 16 single transitions on their boundaries.}
\label{fig:period5network}
\end{figure}

Combining the information in Figure~\ref{FIG:graphper5}
with Figure~\ref{fig:period5network} yields Figure~\ref{fig:period5}, which
contains two heteroclinic chains consisting of single transitions
which yield period 5 cycles, whereas the mixed curvature-frame transitions
${\cal T}_{R_1N_-}$ result in heteroclinic cycles with period 4.
\begin{figure}[H]
\centering
\hspace{0.5cm}
\begin{tikzpicture}[scale=1.1]
        \draw (1,0) arc (0:360:1cm and 1cm);	 	

        \draw [gray, dashed, - ]  (1,1.2)--(1,0);
        \draw [gray, dashed, - ]  (-1,1.2)--(-1,0);
        \draw [gray,dashed, - ]  (-0.5392,-1.4660)--(0.5,-0.8660);
        \draw [gray,dashed, - ]  (-1.5392,0.2660)--(-0.5,0.8660);
	\draw[gray,dashed, -] (2,0) -- (0.5,-0.866);
	\draw[gray,dashed, -] (2,0) -- (0.5, 0.866);

       \draw [ultra thick, domain=0.52:1.570,variable=\t,smooth] plot ({sin(\t r)},{cos(\t r)});
       \draw [ultra thick, domain=-0.52:-1.570,variable=\t,smooth] plot ({sin(\t r)},{cos(\t r)});

	\draw (-0.95,0) -- (-1.05,0);
        \draw (0.95,0) -- (1.05,0);
        \draw[rotate=120] (-0.95,0) -- (-1.05,0);
        \draw[rotate=240] (-0.95,0) -- (-1.05,0);
        \draw[rotate=120] (0.95,0) -- (1.05,0);
        \draw[rotate=240] (0.95,0) -- (1.05,0);


    \draw[  postaction={decorate}] (0.98,-0.198) -- (-0.837,-0.547);\draw [dotted, gray]  (0.98,-0.198)--(2,0);

    \draw[  postaction={decorate}] (0.891,-0.454) -- (-0.32,-0.947);\draw [dotted, gray]  (0.891,-0.454)--(2,0);

    \draw [dotted, thick, postaction={decorate}]  (0.891,0.454)--(0.891,-0.454);\draw [dotted, gray]  (0.891,0.454)--(0.891,1.2);

    \draw [dotted, thick, postaction={decorate}]  (-0.32,-0.947)--(0.98,-0.198);\draw [dotted, gray]  (-0.7,-1.17)--(-0.32,-0.947);

    \draw [dotted, thick, postaction={decorate}]  (-0.837,-0.547)--(0.891,0.454);\draw [dotted, gray]  (-1,-0.64)--(-0.837,-0.547);

    \filldraw [black] (0.891,0.454) circle (1.5pt);

    \filldraw [lightgray] (0.98,-0.198) circle (1.5pt);
    \filldraw [black] (0.891,-0.454) circle (1.5pt);

    \filldraw [lightgray] (-0.32,-0.947) circle (1.5pt);

    \filldraw [black] (-0.837,-0.547) circle (1.5pt);



 \end{tikzpicture}
\hspace{1cm}
\begin{tikzpicture}[scale=1.1]
        \draw (1,0) arc (0:360:1cm and 1cm);	 	

        \draw [gray, dashed, - ]  (1,1.2)--(1,0);
        \draw [gray, dashed, - ]  (-1,1.2)--(-1,0);
        \draw [gray,dashed, - ]  (-0.5392,-1.4660)--(0.5,-0.8660);
        \draw [gray,dashed, - ]  (-1.5392,0.2660)--(-0.5,0.8660);
	\draw[gray,dashed, -] (2,0) -- (0.5,-0.866);
	\draw[gray,dashed, -] (2,0) -- (0.5, 0.866);

       \draw [ultra thick, domain=0.52:1.570,variable=\t,smooth] plot ({sin(\t r)},{cos(\t r)});
       \draw [ultra thick, domain=-0.52:-1.570,variable=\t,smooth] plot ({sin(\t r)},{cos(\t r)});

	\draw (-0.95,0) -- (-1.05,0);
        \draw (0.95,0) -- (1.05,0);
        \draw[rotate=120] (-0.95,0) -- (-1.05,0);
        \draw[rotate=240] (-0.95,0) -- (-1.05,0);
        \draw[rotate=120] (0.95,0) -- (1.05,0);
        \draw[rotate=240] (0.95,0) -- (1.05,0);


    \draw[  postaction={decorate}] (0.98,-0.198) -- (-0.837,-0.547);\draw [dotted, gray]  (0.98,-0.198)--(2,0);


    \draw [dotted, thick, postaction={decorate}]  (-0.32,-0.947)--(0.98,-0.198);\draw [dotted, gray]  (-0.7,-1.17)--(-0.32,-0.947);

    \draw [dotted, thick, postaction={decorate}]  (-0.837,-0.547)--(0.891,0.454);\draw [dotted, gray]  (-1,-0.64)--(-0.837,-0.547);

    \def\u{0.414}
    \foreach \a in {-1.4,0,1.8}
    {\draw[very thick, postaction={decorate},variable=\t,domain=-3.2:3.2] plot ({(1+2*(2*\u/(1+\u*\u))*tanh(-\t))/(2+(2*\u/(1+\u*\u))*tanh(-\t))},{(1.732*((1-\u*\u)/(1+\u*\u))*tanh((((1-\u*\u)/(1+\u*\u))/(2*\u/(1+\u*\u)))*(-\t)-\a))/(2+(2*\u/(1+\u*\u))*tanh(-\t))});}

    \filldraw [black] (0.891,0.454) circle (1.5pt);

    \filldraw [lightgray] (0.98,-0.198) circle (1.5pt);

    \filldraw [lightgray] (-0.32,-0.947) circle (1.5pt);

    \filldraw [black] (-0.837,-0.547) circle (1.5pt);



 \end{tikzpicture}
\hspace{1cm}
\begin{tikzpicture}[scale=1.1]
        \draw (1,0) arc (0:360:1cm and 1cm);	 	

        \draw [gray, dashed, - ]  (1,1.2)--(1,0);
        \draw [gray, dashed, - ]  (-1,1.2)--(-1,0);
        \draw [gray,dashed, - ]  (-0.5392,-1.4660)--(0.5,-0.8660);
        \draw [gray,dashed, - ]  (-1.5392,0.2660)--(-0.5,0.8660);
	\draw[gray,dashed, -] (2,0) -- (0.5,-0.866);
	\draw[gray,dashed, -] (2,0) -- (0.5, 0.866);

       \draw [ultra thick, domain=0.52:1.570,variable=\t,smooth] plot ({sin(\t r)},{cos(\t r)});
       \draw [ultra thick, domain=-0.52:-1.570,variable=\t,smooth] plot ({sin(\t r)},{cos(\t r)});

	\draw (-0.95,0) -- (-1.05,0);
        \draw (0.95,0) -- (1.05,0);
        \draw[rotate=120] (-0.95,0) -- (-1.05,0);
        \draw[rotate=240] (-0.95,0) -- (-1.05,0);
        \draw[rotate=120] (0.95,0) -- (1.05,0);
        \draw[rotate=240] (0.95,0) -- (1.05,0);


    \draw[  postaction={decorate}] (0.891,0.454) -- (-0.32,0.947);\draw [dotted, gray]  (0.891,0.454)--(2,0);

    \draw[  postaction={decorate}] (0.98,-0.198) -- (-0.837,-0.547);\draw [dotted, gray]  (0.98,-0.198)--(2,0);

    \draw [dotted, thick, postaction={decorate}]  (-0.32,0.947)--(-0.32,-0.947);\draw [dotted, gray]  (-0.32,0.947)--(-0.32,1.2);

    \draw [dotted, thick, postaction={decorate}]  (-0.32,-0.947)--(0.98,-0.198);\draw [dotted, gray]  (-0.7,-1.17)--(-0.32,-0.947);

    \draw [dotted, thick, postaction={decorate}]  (-0.837,-0.547)--(0.891,0.454);\draw [dotted, gray]  (-1,-0.64)--(-0.837,-0.547);

    \filldraw [black] (0.891,0.454) circle (1.5pt);

    \filldraw [lightgray] (0.98,-0.198) circle (1.5pt);

    \filldraw [lightgray] (-0.32,-0.947) circle (1.5pt);

    \filldraw [black] (-0.837,-0.547) circle (1.5pt);


    \filldraw [lightgray] (-0.32,0.947) circle (1.5pt);

 \end{tikzpicture}
\hspace{1cm}
 \caption {The stable subnetwork of heteroclinic cycles associated
 with the silver ratio, obtained from Figures~\ref{FIG:graphper5}
 and~\ref{fig:period5network}; it consists of two period 5 heteroclinic single
 transition cycles (left and right) and period 4 heteroclinic cykles
 involving mixed curvature-frame transitions (middle). 
	}\label{fig:period5}
\end{figure}

The heteroclinic network arising from $u = 1 + \sqrt{2}$ is of particular
dynamical importance. This is due to that the eigenvalues in sector $(132)$ associated with the
instability of $R_1$ and $N_-$, given by~\eqref{linR1} and~\eqref{linN-}, respectively,
are equal. This follows from that $\lambda_{R_1} = 3(1 - \ue^2)/f(\ue)$ and
$\lambda_{N_-} = 6\ue/f(\ue)$, where $\ue = \sqrt{2}-1$ in sector $(132)$
(see Table~\ref{tab:per5EV}) yields $\lambda_{R_1} = \lambda_{N_-}$,
while $\lambda_{R_1}>\lambda_{N_-}$
when $\ue \in (0,\sqrt{2}-1)$ and $\lambda_{N_-}>\lambda_{R_1}$
when $\ue \in (\sqrt{2}-1,1)$. Note that when $\lambda_{R_1} = \lambda_{N_-}$
for $\ue = \sqrt{2}-1$ there exists a special orbit that is a straight line
given by $\Sigma_1 = -2(1 - \Sigma_3)$ in $(\Sigma_1,\Sigma_2,\Sigma_3)$-space, as
shown in Appendix~\ref{app:BII}. Due to this, these chains may play a role in a possible
construction of a stable manifolds, akin to that in~\cite{Beguin10,Lieb11,Lieb13,BegDut22}.




\subsection{The alloy ratio $u_0 = 1 + \sqrt{3}$}\label{sec:otherrat}

The alloy ratio\footnote{The continued fraction expansions given by $[(n)]$
for $n\in\mathbb{N}$ are called the \emph{metallic ratios} (e.g., the golden,
silver and bronze ratios occur respectively for $n=1,2,3$). We therefore
introduce this new nomenclature for the impure ratio $u_0=[2;(1,2)]=1+\sqrt{3}$.}
$u_0 = [2;(1,2)] = 1 +\sqrt{3}$ yields an era sequence of
period $2$ and a Kasner sequence with period $3$:
\begin{equation}\label{app:sqrt3}
(u_i)_{i\in\mathbb{N}}:\quad
\underbrace{1 + \sqrt{3} \rightarrow \textcolor{gray}{\sqrt{3}}}_{\text{\scriptsize era}} \rightarrow
\underbrace{\textcolor{gray!75}{\sfrac{1+\sqrt{3}}{2}}}_{\text{\scriptsize era}} \rightarrow
\underbrace{1+\sqrt{3} \rightarrow \textcolor{gray}{\sqrt{3}}}_{\text{\scriptsize era}} \rightarrow
\underbrace{\textcolor{gray!75}{\sfrac{1+\sqrt{3}}{2}}}_{\text{\scriptsize era}} \rightarrow \ldots ,
\end{equation}
with Kretschmann scalar in~\eqref{Weyl} given by
$\mathcal{W}(1+\sqrt{3})=22 - 12 \sqrt{3}\approx 1.215$,
$\mathcal{W}({\color{gray}\sqrt{3}})=324/(4 + \sqrt{3})^3\approx 1.720$ and
$\mathcal{W}({\color{gray!75}(1+\sqrt{3})/2})=54 (223 - 84 \sqrt{3})/2197\approx 1.905$.

The case $u = 1 + \sqrt{3}$ gives rise to two periodic eras and a periodic
Kasner sequence with three values of $u\in \{1 + \sqrt{3}$, \textcolor{gray}{$\sqrt{3}$},
$\textcolor{gray!75}{\sfrac{1 + \sqrt{3}}{2}}\}$ such that each value yields six fixed point
on the Kasner circle $\mathrm{K}^{\ocircle}$, one in
each sector, resulting in the values of $\ue$ given in Table~\ref{tab:per8EV}.
The hexagonal representation of the heteroclinic network follows straightforwardly
from the example given in Figure~\ref{FIG:hexEX} by considering eras 2 to 5,
which results in Figure~\ref{FIG:graphper8}.
%

The entire cyclic heteroclinic network is obtained by taking the union
$\mathbb{H}^0\bigcup\mathbb{H}^1$ of the initial and the subsequent (first) era, where
the relevant values for $\ue$ for sectors $(321)$ and $(132)$ are to be inserted into
equation~\eqref{eranetwork}, while equation~\eqref{eranetworksingle} results in
the single transition network, where the different transition elements are
depicted in Figure~\ref{fig:period8}.

\begin{table}[H]
    \centering
    \footnotesize
    \begin{tabular}{|c|c|c|}\hline
        \rowcolor{lightgray!25} $u$ & $\ue$ & Sector \\ \hline
        \begin{tabular}{c}
            \textcolor{gray!75}{$\sfrac{1+\sqrt{3}}{2}$}\\
            \textcolor{gray}{$\sqrt{3}$}\\
            $1+\sqrt{3}$
        \end{tabular}
         &
        \begin{tabular}{c}
            \textcolor{gray!75}{$-\sfrac{3+\sqrt{3}}{2}$}\\
            \textcolor{gray}{$-(1+\sqrt{3})$}\\
            $-(2+\sqrt{3})$
        \end{tabular}
         & (213) \\ \hline
        \begin{tabular}{c}
            $1+\sqrt{3}$\\
            \textcolor{gray}{$\sqrt{3}$}\\
            \textcolor{gray!75}{$\sfrac{1+\sqrt{3}}{2}$}
        \end{tabular}
         &
        \begin{tabular}{c}
            $-\sfrac{1+\sqrt{3}}{2}$\\
            \textcolor{gray}{$-(1+\sfrac{1}{\sqrt{3}})$}\\
            \textcolor{gray!75}{$-\sqrt{3}$}
        \end{tabular}
         & (231) \\ \hline
        \begin{tabular}{c}
            \textcolor{gray!75}{$\sfrac{1+\sqrt{3}}{2}$}\\
            \textcolor{gray}{$\sqrt{3}$}\\
            $1+\sqrt{3}$
        \end{tabular}
         &
        \begin{tabular}{c}
            \textcolor{gray!75}{$-\sfrac{1}{\sqrt{3}}$}\\
            \textcolor{gray}{$-\sfrac{3-\sqrt{3}}{2}$}\\
            $-(\sqrt{3}-1)$
        \end{tabular}
        & (321) \\ \hline
        \begin{tabular}{c}
            $1+\sqrt{3}$\\
            \textcolor{gray}{$\sqrt{3}$}\\
            \textcolor{gray!75}{$\sfrac{1+\sqrt{3}}{2}$}
        \end{tabular}
         &
        \begin{tabular}{c}
            $-(2-\sqrt{3})$\\
            \textcolor{gray}{$-\sfrac{\sqrt{3}-1}{2}$}\\
            \textcolor{gray!75}{$-(1-\sfrac{1}{\sqrt{3}})$}
        \end{tabular}
         & (312) \\ \hline
        \begin{tabular}{c}
            \textcolor{gray!75}{$\sfrac{1+\sqrt{3}}{2}$}\\
            \textcolor{gray}{$\sqrt{3}$}\\
            $1+\sqrt{3}$
        \end{tabular}
         &
        \begin{tabular}{c}
            \textcolor{gray!75}{$\sqrt{3}-1$}\\
            \textcolor{gray}{$\sfrac{1}{\sqrt{3}}$}\\
            $\sfrac{\sqrt{3}-1}{2}$
        \end{tabular}
         & (132) \\ \hline
        \begin{tabular}{c}
            $1+\sqrt{3}$\\
            \textcolor{gray}{$\sqrt{3}$}\\
            \textcolor{gray!75}{$\sfrac{1+\sqrt{3}}{2}$}
        \end{tabular}
         &
        \begin{tabular}{c}
            $1+\sqrt{3}$\\
            \textcolor{gray}{$\sqrt{3}$}\\
            \textcolor{gray!75}{$\sfrac{1+\sqrt{3}}{2}$}
        \end{tabular}
         & (123) \\ \hline
    \end{tabular}
    \caption{The values of $u$ and $\ue$ for the Kasner sequence generated by $u = 1 + \sqrt{3}$.
    The table is organized by increasing $\ue$. 
    We respectively denote the points associated with $1+\sqrt{3}$,\textcolor{gray}{$\sqrt{3}$},\textcolor{gray!75}{$\sfrac{1+\sqrt{3}}{2}$} by black, gray and light gray.}\label{tab:per8EV}
\end{table}
\vspace{-0.5cm}
\begin{figure}[H]\centering
\tiny
\begin{tikzpicture}[scale=0.4]
\node (1) at (-2, 0) {\textbf{\textbullet}};
\node (2a) at (-1, 1.73) {\textbf{\textbullet}};
\node (3a) at (1, 1.73) {\textbf{\textbullet}};
\node (2b) at (-1, -1.73) {\textbf{\textbullet}};
\node (3b) at (1, -1.73) {\textbf{\textbullet}};
\node (4) at (2, 0) {\textbf{\textbullet}};

\node (R1) at (4, 0) {$\circ$};
\node[gray] (R2a) at (5, 1.73) {\textbf{\textbullet}};
\node[gray] (R3a) at (7, 1.73) {\textbf{\textbullet}};
\node (R2b) at (5, -1.73) {$\circ$};
\node (R3b) at (7, -1.73) {$\circ$};
\node[gray] (R4) at (8, 0) {\textbf{\textbullet}};

\node (RR1) at (10, 0) {$\circ$};
\node (RR2a) at (11, 1.73) {$\circ$};
\node (RR3a) at (13, 1.73) {$\circ$};
\node[lightgray] (RR2b) at (11, -1.73) {\textbf{\textbullet}};
\node[lightgray] (RR3b) at (13, -1.73) {\textbf{\textbullet}};
\node[lightgray] (RR4) at (14, 0) {\textbf{\textbullet}};

\draw[dotted, ultra thick, ->] (1) -- (4);

\draw[dotted, thick, ->] (1) -- (2a);
\draw[dotted, thick, ->] (2a) -- (3a);
\draw[dotted, thick, ->] (3a) -- (4);

\draw[dotted, thick, ->] (1) -- (2b);
\draw[dotted, thick, ->] (2b) -- (3b);
\draw[dotted, thick, ->] (3b) -- (4);

\draw[dotted, ultra thick, ->] (R1) -- (R4);

\draw[dotted, thick, ->] (R1) -- (R2a);
\draw[dotted, thick, ->] (R2a) -- (R3a);
\draw[dotted, thick, ->] (R3a) -- (R4);

\draw[dotted, thick, ->] (R1) -- (R2b);
\draw[dotted, thick, ->] (R2b) -- (R3b);
\draw[dotted, thick, ->] (R3b) -- (R4);

\draw[dotted, ultra thick, ->] (RR1) -- (RR4);

\draw[dotted, thick, ->] (RR1) -- (RR2a);
\draw[dotted, thick, ->] (RR2a) -- (RR3a);
\draw[dotted, thick, ->] (RR3a) -- (RR4);

\draw[dotted, thick, ->] (RR1) -- (RR2b);
\draw[dotted, thick, ->] (RR2b) -- (RR3b);
\draw[dotted, thick, ->] (RR3b) -- (RR4);

\draw[bend left=50,->] (4) to (R3a);
\draw[->,very thick] (3b) to[out=0,in=-75]++ (1.75,2) to[out=100,in=-205] (R3a);
\path[->] (3b) edge [->,out=-5,in=-155,looseness=1] (R2a);

\draw[bend right=15,->] (R4) to (RR2b);
\draw [very thick, ->] (R3b) to (RR2b);
\draw [bend right=15,->] (R3b) to (RR1);

\draw [very thick, bend left=25,->] (RR3b) to (2b);

\path[->] (RR3b) edge [->,out=-120,in=-90,looseness=1.2] (1);
\path[->] (RR4) edge [->,out=-80,in=-60,looseness=1.2] (2b);


\end{tikzpicture}
\hspace{1cm}
\begin{tikzpicture}[scale=0.4]
\node (1) at (-2, 0) {\textbf{\textbullet}};
\node (2a) at (-1, 1.73) {\textbf{\textbullet}};
\node (3a) at (1, 1.73) {\textbf{\textbullet}};
\node (2b) at (-1, -1.73) {\textbf{\textbullet}};
\node (3b) at (1, -1.73) {\textbf{\textbullet}};
\node (4) at (2, 0) {\textbf{\textbullet}};

\node[gray] (R2a) at (5, 1.73) {\textbf{\textbullet}};
\node[gray] (R3a) at (7, 1.73) {\textbf{\textbullet}};
\node[gray] (R4) at (8, 0) {\textbf{\textbullet}};

\node[lightgray] (RR2b) at (11, -1.73) {\textbf{\textbullet}};
\node[lightgray] (RR3b) at (13, -1.73) {\textbf{\textbullet}};
\node[lightgray] (RR4) at (14, 0) {\textbf{\textbullet}};

\draw[dotted, ultra thick, ->] (1) -- (4);

\draw[dotted, thick, ->] (1) -- (2a);
\draw[dotted, thick, ->] (2a) -- (3a);
\draw[dotted, thick, ->] (3a) -- (4);

\draw[dotted, thick, ->] (1) -- (2b);
\draw[dotted, thick, ->] (2b) -- (3b);
\draw[dotted, thick, ->] (3b) -- (4);

\draw[dotted, thick, ->] (R2a) -- (R3a);
\draw[dotted, thick, ->] (R3a) -- (R4);

\draw[dotted, thick, ->] (RR2b) -- (RR3b);
\draw[dotted, thick, ->] (RR3b) -- (RR4);

\draw[bend left=50,->] (4) to (R3a);
\draw[->,very thick] (3b) to[out=0,in=-75]++ (1.75,2) to[out=100,in=-205] (R3a);
\path[->] (3b) edge [->,out=-5,in=-155,looseness=1] (R2a);

\draw[bend right=15,->] (R4) to (RR2b);

\draw [very thick, bend left=25,->] (RR3b) to (2b);

\path[->] (RR3b) edge [->,out=-120,in=-90,looseness=1.2] (1);
\path[->] (RR4) edge [->,out=-90,in=-60,looseness=1.2] (2b);


\end{tikzpicture}
{\setlength{\abovecaptionskip}{-7pt}
\setlength{\belowcaptionskip}{0pt}
\captionof{figure}{\textbf{Left:} Hexagonal representation describing
the network of heteroclinic cycles of the periodic Kasner sequence
generated by $u = 1 + \sqrt{3}$.
\textbf{Right:} The stable subnetwork, obtained by removing the isolated
heteroclinic structure.}\label{FIG:graphper8}
}
\end{figure}
\begin{figure}[H]
\centering
\begin{tikzpicture}[scale=1.1]
        \draw (1,0) arc (0:360:1cm and 1cm);	 	

        \draw [gray, dashed, - ]  (1,1.2)--(1,0);
        \draw [gray, dashed, - ]  (-1,1.2)--(-1,0);
        \draw [gray,dashed, - ]  (-0.5392,-1.4660)--(0.5,-0.8660);
        \draw [gray,dashed, - ]  (-1.5392,0.2660)--(-0.5,0.8660);
	\draw[gray,dashed, -] (2,0) -- (0.5,-0.866);
	\draw[gray,dashed, -] (2,0) -- (0.5, 0.866);

       \draw [ultra thick, domain=0.52:1.570,variable=\t,smooth] plot ({sin(\t r)},{cos(\t r)});
       \draw [ultra thick, domain=-0.52:-1.570,variable=\t,smooth] plot ({sin(\t r)},{cos(\t r)});

	\draw (-0.95,0) -- (-1.05,0);
        \draw (0.95,0) -- (1.05,0);
        \draw[rotate=120] (-0.95,0) -- (-1.05,0);
        \draw[rotate=240] (-0.95,0) -- (-1.05,0);
        \draw[rotate=120] (0.95,0) -- (1.05,0);
        \draw[rotate=240] (0.95,0) -- (1.05,0);

    \filldraw [black] (0.866,-0.5) circle (1.5pt);
    \filldraw [gray] (0.953,0.302) circle (1.5pt);
    \filldraw [lightgray] (0.984,0.178) circle (1.5pt);

    \filldraw [lightgray] (0.984,-0.178) circle (1.5pt);
    \filldraw [gray] (0.953,-0.302) circle (1.5pt);
    \filldraw [black] (0.866,0.5) circle (1.5pt);

    \filldraw [black] (0,-1) circle (1.5pt);
    \filldraw [gray] (-0.214,-0.97) circle (1.5pt);
    \filldraw [lightgray] (-0.338,-0.941) circle (1.5pt);

    \filldraw [black] (-0.866,-0.5) circle (1.5pt);
    \filldraw [lightgray] (-0.645,-0.764) circle (1.5pt);
    \filldraw [gray] (-0.738,-0.674) circle (1.5pt);

    \filldraw [black] (-0.866,0.5) circle (1.5pt);
    \filldraw [lightgray] (-0.645,0.764) circle (1.5pt);
    \filldraw [gray] (-0.738,0.674) circle (1.5pt);

    \filldraw [black] (0,1) circle (1.5pt);
    \filldraw [gray] (-0.214,0.97) circle (1.5pt);
    \filldraw [lightgray] (-0.338,0.941) circle (1.5pt);


 \end{tikzpicture}
\hspace{-0.3cm}
\begin{tikzpicture}[scale=1.1]
        \draw (1,0) arc (0:360:1cm and 1cm);	 	

        \draw [gray, dashed, - ]  (1,1.2)--(1,0);
        \draw [gray, dashed, - ]  (-1,1.2)--(-1,0);
        \draw [gray,dashed, - ]  (-0.5392,-1.4660)--(0.5,-0.8660);
        \draw [gray,dashed, - ]  (-1.5392,0.2660)--(-0.5,0.8660);
	\draw[gray,dashed, -] (2,0) -- (0.5,-0.866);
	\draw[gray,dashed, -] (2,0) -- (0.5, 0.866);

       \draw [ultra thick, domain=0.52:1.570,variable=\t,smooth] plot ({sin(\t r)},{cos(\t r)});
       \draw [ultra thick, domain=-0.52:-1.570,variable=\t,smooth] plot ({sin(\t r)},{cos(\t r)});

	\draw (-0.95,0) -- (-1.05,0);
        \draw (0.95,0) -- (1.05,0);
        \draw[rotate=120] (-0.95,0) -- (-1.05,0);
        \draw[rotate=240] (-0.95,0) -- (-1.05,0);
        \draw[rotate=120] (0.95,0) -- (1.05,0);
        \draw[rotate=240] (0.95,0) -- (1.05,0);


    \def\u{-0.732}
    \foreach \a in {0.25, 0.5, 1} 
        \foreach \b in {0.5} 
        {
        \newcommand \n {(-\a*\b*(1-\u*\u)/((2+\u)*\u))}; 
        \newcommand \pone {(-\u/(1+\u+\u*\u))}; 
        \newcommand \ptwo {((1+\u)/(1+\u+\u*\u))}; 
        \newcommand \pthree {((\u*(1+\u))/(1+\u+\u*\u))}; 
        \newcommand \Sone {(-1+3*(\pone*\n*\n*exp(-6*\pone*(\t))+\ptwo*\b*\b*exp(-6*\ptwo*(\t))+\pthree*exp(-6*\pthree*(\t)))/(\n*\n*exp(-6*\pone*(\t))+\b*\b*exp(-6*\ptwo*(\t))+exp(-6*\pthree*(\t))))}; 
        \newcommand \Sthree {(-1+3*(\pone*exp(6*\pone*(\t))+\ptwo*\a*\a*exp(6*\ptwo*(\t))+\pthree*(\a*\b-\n)*(\a*\b-\n)*exp(6*\pthree*(\t)))/(exp(6*\pone*(\t))+\a*\a*exp(6*\ptwo*(\t))+(\a*\b-\n)*(\a*\b-\n)*exp(6*\pthree*(\t))))}; 
        \draw[dotted, ultra thick, postaction={decorate},variable=\t,domain=-1.75:1] plot ({-0.5*\Sone},{-(0.5*\Sone+\Sthree)/1.732});
        }

    \draw [dotted, thick, postaction={decorate}]  (-0.866,0.5)--(-0.866,-0.5);\draw [dotted, gray]  (-0.866,0.5)--(-0.866,1.2);
    \draw [dotted, thick, postaction={decorate}]  (0.866,0.5)--(0.866,-0.5);\draw [dotted, gray]  (0.866,0.5)--(0.866,1.2);

    \draw [dotted, thick, postaction={decorate}]  (0,1)--(0,-1);\draw [dotted, gray]  (0,1)--(0,1.2);

    \draw [rotate=-240,dotted, thick, postaction={decorate}]  (-0.866,0.5)--(-0.866,-0.5);\draw [rotate=-240,dotted, gray]  (-0.866,0.5)--(-0.866,1.2);
    \draw [rotate=-240,dotted, thick, postaction={decorate}]  (0.866,0.5)--(0.866,-0.5);\draw [rotate=-240,dotted, gray]  (0.866,0.5)--(0.866,1.2);
    \draw [rotate=-240,dotted, thick, postaction={decorate}]  (0,1)--(0,-1);\draw [rotate=-240,dotted, gray]  (0,1)--(0,1.2);

    \filldraw [black] (0.866,-0.5) circle (1.5pt);

    \filldraw [black] (0.866,0.5) circle (1.5pt);

    \filldraw [black] (0,-1) circle (1.5pt);

    \filldraw [black] (-0.866,-0.5) circle (1.5pt);

    \filldraw [black] (-0.866,0.5) circle (1.5pt);

    \filldraw [black] (0,1) circle (1.5pt);

 \end{tikzpicture}
\hspace{-0.3cm}
\begin{tikzpicture}[scale=1.1]
        \draw (1,0) arc (0:360:1cm and 1cm);	 	

        \draw [gray, dashed, - ]  (1,1.2)--(1,0);
        \draw [gray, dashed, - ]  (-1,1.2)--(-1,0);
        \draw [gray,dashed, - ]  (-0.5392,-1.4660)--(0.5,-0.8660);
        \draw [gray,dashed, - ]  (-1.5392,0.2660)--(-0.5,0.8660);
	\draw[gray,dashed, -] (2,0) -- (0.5,-0.866);
	\draw[gray,dashed, -] (2,0) -- (0.5, 0.866);

       \draw [ultra thick, domain=0.52:1.570,variable=\t,smooth] plot ({sin(\t r)},{cos(\t r)});
       \draw [ultra thick, domain=-0.52:-1.570,variable=\t,smooth] plot ({sin(\t r)},{cos(\t r)});

	\draw (-0.95,0) -- (-1.05,0);
        \draw (0.95,0) -- (1.05,0);
        \draw[rotate=120] (-0.95,0) -- (-1.05,0);
        \draw[rotate=240] (-0.95,0) -- (-1.05,0);
        \draw[rotate=120] (0.95,0) -- (1.05,0);
        \draw[rotate=240] (0.95,0) -- (1.05,0);


    \def\u{-0.633}
    \foreach \a in {0.3, 0.9}
        \foreach \b in {0.5} 
        {
        \newcommand \n {(-\a*\b*(1-\u*\u)/((2+\u)*\u))}; 
        \newcommand \pone {(-\u/(1+\u+\u*\u))}; 
        \newcommand \ptwo {((1+\u)/(1+\u+\u*\u))}; 
        \newcommand \pthree {((\u*(1+\u))/(1+\u+\u*\u))}; 
        \newcommand \Sone {(-1+3*(\pone*\n*\n*exp(-6*\pone*(\t))+\ptwo*\b*\b*exp(-6*\ptwo*(\t))+\pthree*exp(-6*\pthree*(\t)))/(\n*\n*exp(-6*\pone*(\t))+\b*\b*exp(-6*\ptwo*(\t))+exp(-6*\pthree*(\t))))}; 
        \newcommand \Sthree {(-1+3*(\pone*exp(6*\pone*(\t))+\ptwo*\a*\a*exp(6*\ptwo*(\t))+\pthree*(\a*\b-\n)*(\a*\b-\n)*exp(6*\pthree*(\t)))/(exp(6*\pone*(\t))+\a*\a*exp(6*\ptwo*(\t))+(\a*\b-\n)*(\a*\b-\n)*exp(6*\pthree*(\t))))}; 
        \draw[dotted, ultra thick, postaction={decorate},variable=\t,domain=-1.9:1] plot ({-0.5*\Sone},{-(0.5*\Sone+\Sthree)/1.732});
        }

    \draw [dotted, thick, postaction={decorate}]  (-0.738,0.674)--(-0.738,-0.674);\draw [dotted, gray]  (-0.738,0.674)--(-0.738,1.2);
    \draw [dotted, thick, postaction={decorate}]  (0.953,0.302)--(0.953,-0.302);\draw [dotted, gray]  (0.953,0.302)--(0.953,1.2);

    \draw [dotted, thick, postaction={decorate}]  (-0.214,0.97)--(-0.214,-0.97);\draw [dotted, gray]  (-0.214,0.97)--(-0.214,1.2);

    \draw [rotate=-240,dotted, thick, postaction={decorate}]  (-0.738,0.674)--(-0.738,-0.674);\draw [rotate=-240,dotted, gray]  (-0.738,0.674)--(-0.738,1.2);
    \draw [rotate=-240,dotted, thick, postaction={decorate}]  (0.953,0.302)--(0.953,-0.302);\draw [rotate=-240,dotted, gray]  (0.953,0.302)--(0.953,1.2);

    \draw [rotate=-240,dotted, thick, postaction={decorate}]  (-0.214,0.97)--(-0.214,-0.97);\draw [rotate=-240,dotted, gray]  (-0.214,0.97)--(-0.214,1.2);

    \filldraw [gray] (0.953,0.302) circle (1.5pt);

    \filldraw [gray] (0.953,-0.302) circle (1.5pt);

    \filldraw [gray] (-0.214,-0.97) circle (1.5pt);

    \filldraw [gray] (-0.738,-0.674) circle (1.5pt);

    \filldraw [gray] (-0.738,0.674) circle (1.5pt);

    \filldraw [gray] (-0.214,0.97) circle (1.5pt);


 \end{tikzpicture}
\hspace{-0.3cm}
\begin{tikzpicture}[scale=1.1]
        \draw (1,0) arc (0:360:1cm and 1cm);	 	

        \draw [gray, dashed, - ]  (1,1.2)--(1,0);
        \draw [gray, dashed, - ]  (-1,1.2)--(-1,0);
        \draw [gray,dashed, - ]  (-0.5392,-1.4660)--(0.5,-0.8660);
        \draw [gray,dashed, - ]  (-1.5392,0.2660)--(-0.5,0.8660);
	\draw[gray,dashed, -] (2,0) -- (0.5,-0.866);
	\draw[gray,dashed, -] (2,0) -- (0.5, 0.866);

       \draw [ultra thick, domain=0.52:1.570,variable=\t,smooth] plot ({sin(\t r)},{cos(\t r)});
       \draw [ultra thick, domain=-0.52:-1.570,variable=\t,smooth] plot ({sin(\t r)},{cos(\t r)});

	\draw (-0.95,0) -- (-1.05,0);
        \draw (0.95,0) -- (1.05,0);
        \draw[rotate=120] (-0.95,0) -- (-1.05,0);
        \draw[rotate=240] (-0.95,0) -- (-1.05,0);
        \draw[rotate=120] (0.95,0) -- (1.05,0);
        \draw[rotate=240] (0.95,0) -- (1.05,0);


    \def\u{-0.577}
    \foreach \a in {0.7} 
        \foreach \b in {0.5} 
        {
        \newcommand \n {(-\a*\b*(1-\u*\u)/((2+\u)*\u))}; 
        \newcommand \pone {(-\u/(1+\u+\u*\u))}; 
        \newcommand \ptwo {((1+\u)/(1+\u+\u*\u))}; 
        \newcommand \pthree {((\u*(1+\u))/(1+\u+\u*\u))}; 
        \newcommand \Sone {(-1+3*(\pone*\n*\n*exp(-6*\pone*(\t))+\ptwo*\b*\b*exp(-6*\ptwo*(\t))+\pthree*exp(-6*\pthree*(\t)))/(\n*\n*exp(-6*\pone*(\t))+\b*\b*exp(-6*\ptwo*(\t))+exp(-6*\pthree*(\t))))}; 
        \newcommand \Sthree {(-1+3*(\pone*exp(6*\pone*(\t))+\ptwo*\a*\a*exp(6*\ptwo*(\t))+\pthree*(\a*\b-\n)*(\a*\b-\n)*exp(6*\pthree*(\t)))/(exp(6*\pone*(\t))+\a*\a*exp(6*\ptwo*(\t))+(\a*\b-\n)*(\a*\b-\n)*exp(6*\pthree*(\t))))}; 
        \draw[dotted, ultra thick, postaction={decorate},variable=\t,domain=-1.75:1] plot ({-0.5*\Sone},{-(0.5*\Sone+\Sthree)/1.732});
        }

    \draw [dotted, thick, postaction={decorate}]  (0.984,0.178)--(0.984,-0.178);\draw [dotted, gray]  (0.984,0.178)--(0.984,1.2);
    \draw [dotted, thick, postaction={decorate}]  (-0.645,0.764)--(-0.645,-0.764);\draw [dotted, gray]  (-0.645,0.764)--(-0.645,1.2);

    \draw [dotted, thick, postaction={decorate}]  (-0.338,0.941)--(-0.338,-0.941);\draw [dotted, gray]  (-0.338,0.941)--(-0.338,1.2);

    \draw [rotate=-240,dotted, thick, postaction={decorate}]  (0.984,0.178)--(0.984,-0.178);\draw [rotate=-240,dotted, gray]  (0.984,0.178)--(0.984,1.2);
    \draw [rotate=-240,dotted, thick, postaction={decorate}]  (-0.645,0.764)--(-0.645,-0.764);\draw [rotate=-240,dotted, gray]  (-0.645,0.764)--(-0.645,1.2);

    \draw [rotate=-240,dotted, thick, postaction={decorate}]  (-0.338,0.941)--(-0.338,-0.941);\draw [rotate=-240,dotted, gray]  (-0.338,0.941)--(-0.338,1.2);

    \filldraw [lightgray] (0.984,0.178) circle (1.5pt);

    \filldraw [lightgray] (0.984,-0.178) circle (1.5pt);

    \filldraw [lightgray] (-0.338,-0.941) circle (1.5pt);

    \filldraw [lightgray] (-0.645,-0.764) circle (1.5pt);

    \filldraw [lightgray] (-0.645,0.764) circle (1.5pt);

    \filldraw [lightgray] (-0.338,0.941) circle (1.5pt);

 \end{tikzpicture}
\hspace{-0.3cm}
\begin{tikzpicture}[scale=1.1]
        \draw (1,0) arc (0:360:1cm and 1cm);	 	

        \draw [gray, dashed, - ]  (1,1.2)--(1,0);
        \draw [gray, dashed, - ]  (-1,1.2)--(-1,0);
        \draw [gray,dashed, - ]  (-0.5392,-1.4660)--(0.5,-0.8660);
        \draw [gray,dashed, - ]  (-1.5392,0.2660)--(-0.5,0.8660);
	\draw[gray,dashed, -] (2,0) -- (0.5,-0.866);
	\draw[gray,dashed, -] (2,0) -- (0.5, 0.866);

       \draw [ultra thick, domain=0.52:1.570,variable=\t,smooth] plot ({sin(\t r)},{cos(\t r)});
       \draw [ultra thick, domain=-0.52:-1.570,variable=\t,smooth] plot ({sin(\t r)},{cos(\t r)});

	\draw (-0.95,0) -- (-1.05,0);
        \draw (0.95,0) -- (1.05,0);
        \draw[rotate=120] (-0.95,0) -- (-1.05,0);
        \draw[rotate=240] (-0.95,0) -- (-1.05,0);
        \draw[rotate=120] (0.95,0) -- (1.05,0);
        \draw[rotate=240] (0.95,0) -- (1.05,0);

    \def\u{0.366}
    \foreach \a in {-1.6,-0.5,0.8}
    {\draw[very thick, postaction={decorate},variable=\t,domain=-3:4] plot ({(1+2*(2*\u/(1+\u*\u))*tanh(-\t))/(2+(2*\u/(1+\u*\u))*tanh(-\t))},{(1.732*((1-\u*\u)/(1+\u*\u))*tanh((((1-\u*\u)/(1+\u*\u))/(2*\u/(1+\u*\u)))*(-\t)-\a))/(2+(2*\u/(1+\u*\u))*tanh(-\t))});}

    \draw[  postaction={decorate}] (0.86,-0.5) -- (-0.214,-0.97);\draw [dotted, gray]  (0.86,-0.5)--(2,0);
    \draw[  postaction={decorate}] (0.86,0.5) -- (-0.214,0.97);\draw [dotted, gray]  (0.86,0.5)--(2,0);

    \draw [dotted, thick, postaction={decorate}]  (0.866,0.5)--(0.866,-0.5);\draw [dotted, gray]  (0.866,0.5)--(0.866,1.2);
    \draw [dotted, thick, postaction={decorate}]  (-0.214,0.97)--(-0.214,-0.97);\draw [dotted, gray]  (-0.214,0.97)--(-0.214,1.2);

    \filldraw [black] (0.866,-0.5) circle (1.5pt);

    \filldraw [black] (0.866,0.5) circle (1.5pt);

    \filldraw [gray] (-0.214,-0.97) circle (1.5pt);



    \filldraw [gray] (-0.214,0.97) circle (1.5pt);

 \end{tikzpicture}
\hspace{-0.3cm}
\begin{tikzpicture}[scale=1.1]
        \draw (1,0) arc (0:360:1cm and 1cm);	 	

        \draw [gray, dashed, - ]  (1,1.2)--(1,0);
        \draw [gray, dashed, - ]  (-1,1.2)--(-1,0);
        \draw [gray,dashed, - ]  (-0.5392,-1.4660)--(0.5,-0.8660);
        \draw [gray,dashed, - ]  (-1.5392,0.2660)--(-0.5,0.8660);
	\draw[gray,dashed, -] (2,0) -- (0.5,-0.866);
	\draw[gray,dashed, -] (2,0) -- (0.5, 0.866);

       \draw [ultra thick, domain=0.52:1.570,variable=\t,smooth] plot ({sin(\t r)},{cos(\t r)});
       \draw [ultra thick, domain=-0.52:-1.570,variable=\t,smooth] plot ({sin(\t r)},{cos(\t r)});

	\draw (-0.95,0) -- (-1.05,0);
        \draw (0.95,0) -- (1.05,0);
        \draw[rotate=120] (-0.95,0) -- (-1.05,0);
        \draw[rotate=240] (-0.95,0) -- (-1.05,0);
        \draw[rotate=120] (0.95,0) -- (1.05,0);
        \draw[rotate=240] (0.95,0) -- (1.05,0);

    \def\u{0.577}
    \foreach \a in {-1.2,-0.4,0.4}
    {\draw[very thick, postaction={decorate},variable=\t,domain=-3:4.4] plot ({(1+2*(2*\u/(1+\u*\u))*tanh(-\t))/(2+(2*\u/(1+\u*\u))*tanh(-\t))},{(1.732*((1-\u*\u)/(1+\u*\u))*tanh((((1-\u*\u)/(1+\u*\u))/(2*\u/(1+\u*\u)))*(-\t)-\a))/(2+(2*\u/(1+\u*\u))*tanh(-\t))});}

    \draw[  postaction={decorate}] (0.953,0.302) -- (-0.645,0.764);\draw [dotted, gray]  (0.953,0.302)--(2,0);

    \draw[  postaction={decorate}] (0.953,-0.302) -- (-0.645,-0.764);\draw [dotted, gray]  (0.953,-0.302)--(2,0);

    \draw [dotted, thick, postaction={decorate}]  (-0.645,0.764)--(-0.645,-0.764);\draw [dotted, gray]  (-0.645,0.764)--(-0.645,1.2);

    \draw [dotted, thick, postaction={decorate}]  (0.953,0.302)--(0.953,-0.302);\draw [dotted, gray]  (0.953,0.302)--(0.953,1.2);

    \filldraw [gray] (0.953,0.302) circle (1.5pt);

    \filldraw [gray] (0.953,-0.302) circle (1.5pt);


    \filldraw [lightgray] (-0.645,-0.764) circle (1.5pt);

    \filldraw [lightgray] (-0.645,0.764) circle (1.5pt);


 \end{tikzpicture}
\hspace{-0.3cm}
\begin{tikzpicture}[scale=1.1]
        \draw (1,0) arc (0:360:1cm and 1cm);	 	

        \draw [gray, dashed, - ]  (1,1.2)--(1,0);
        \draw [gray, dashed, - ]  (-1,1.2)--(-1,0);
        \draw [gray,dashed, - ]  (-0.5392,-1.4660)--(0.5,-0.8660);
        \draw [gray,dashed, - ]  (-1.5392,0.2660)--(-0.5,0.8660);
	\draw[gray,dashed, -] (2,0) -- (0.5,-0.866);
	\draw[gray,dashed, -] (2,0) -- (0.5, 0.866);

       \draw [ultra thick, domain=0.52:1.570,variable=\t,smooth] plot ({sin(\t r)},{cos(\t r)});
       \draw [ultra thick, domain=-0.52:-1.570,variable=\t,smooth] plot ({sin(\t r)},{cos(\t r)});

	\draw (-0.95,0) -- (-1.05,0);
        \draw (0.95,0) -- (1.05,0);
        \draw[rotate=120] (-0.95,0) -- (-1.05,0);
        \draw[rotate=240] (-0.95,0) -- (-1.05,0);
        \draw[rotate=120] (0.95,0) -- (1.05,0);
        \draw[rotate=240] (0.95,0) -- (1.05,0);

    \def\u{0.732}
    \foreach \a in {-1.2,-0.4,0.4}
    {\draw[very thick, postaction={decorate},variable=\t,domain=-3:5.6] plot ({(1+2*(2*\u/(1+\u*\u))*tanh(-\t))/(2+(2*\u/(1+\u*\u))*tanh(-\t))},{(1.732*((1-\u*\u)/(1+\u*\u))*tanh((((1-\u*\u)/(1+\u*\u))/(2*\u/(1+\u*\u)))*(-\t)-\a))/(2+(2*\u/(1+\u*\u))*tanh(-\t))});}

    \draw[  postaction={decorate}] (0.97,-0.18) -- (-0.866,-0.5);\draw [dotted, gray]  (0.97,-0.18)--(2,0);

    \draw[  postaction={decorate}] (0.97,0.18) -- (-0.866,0.5);\draw [dotted, gray]  (0.97,0.18)--(2,0);

    \draw [dotted, thick, postaction={decorate}]  (-0.866,0.5)--(-0.866,-0.5);\draw [dotted, gray]  (-0.866,0.5)--(-0.866,1.2);

    \draw [dotted, thick, postaction={decorate}]  (0.984,0.178)--(0.984,-0.178);\draw [dotted, gray]  (0.984,0.178)--(0.984,1.2);

    \filldraw [lightgray] (0.984,0.178) circle (1.5pt);

    \filldraw [lightgray] (0.984,-0.178) circle (1.5pt);


    \filldraw [black] (-0.866,-0.5) circle (1.5pt);

    \filldraw [black] (-0.866,0.5) circle (1.5pt);


 \end{tikzpicture}
\hspace{-0.3cm}
\begin{tikzpicture}[scale=1.1]
        \draw (1,0) arc (0:360:1cm and 1cm);	 	

        \draw [gray, dashed, - ]  (1,1.2)--(1,0);
        \draw [gray, dashed, - ]  (-1,1.2)--(-1,0);
        \draw [gray,dashed, - ]  (-0.5392,-1.4660)--(0.5,-0.8660);
        \draw [gray,dashed, - ]  (-1.5392,0.2660)--(-0.5,0.8660);
	\draw[gray,dashed, -] (2,0) -- (0.5,-0.866);
	\draw[gray,dashed, -] (2,0) -- (0.5, 0.866);

       \draw [ultra thick, domain=0.52:1.570,variable=\t,smooth] plot ({sin(\t r)},{cos(\t r)});
       \draw [ultra thick, domain=-0.52:-1.570,variable=\t,smooth] plot ({sin(\t r)},{cos(\t r)});

	\draw (-0.95,0) -- (-1.05,0);
        \draw (0.95,0) -- (1.05,0);
        \draw[rotate=120] (-0.95,0) -- (-1.05,0);
        \draw[rotate=240] (-0.95,0) -- (-1.05,0);
        \draw[rotate=120] (0.95,0) -- (1.05,0);
        \draw[rotate=240] (0.95,0) -- (1.05,0);

    \draw [dotted, thick, postaction={decorate}]  (-0.866,0.5)--(-0.866,-0.5);\draw [dotted, gray]  (-0.866,0.5)--(-0.866,1.2);
    \draw [dotted, thick, postaction={decorate}]  (0.866,0.5)--(0.866,-0.5);\draw [dotted, gray]  (0.866,0.5)--(0.866,1.2);

    \draw [dotted, thick, postaction={decorate}]  (0,1)--(0,-1);\draw [dotted, gray]  (0,1)--(0,1.2);

    \draw [rotate=-240,dotted, thick, postaction={decorate}]  (-0.866,0.5)--(-0.866,-0.5);\draw [rotate=-240,dotted, gray]  (-0.866,0.5)--(-0.866,1.2);
    \draw [rotate=-240,dotted, thick, postaction={decorate}]  (0.866,0.5)--(0.866,-0.5);\draw [rotate=-240,dotted, gray]  (0.866,0.5)--(0.866,1.2);
    \draw [rotate=-240,dotted, thick, postaction={decorate}]  (0,1)--(0,-1);\draw [rotate=-240,dotted, gray]  (0,1)--(0,1.2);

    \draw [dotted, thick, postaction={decorate}]  (-0.738,0.674)--(-0.738,-0.674);\draw [dotted, gray]  (-0.738,0.674)--(-0.738,1.2);
    \draw [dotted, thick, postaction={decorate}]  (0.953,0.302)--(0.953,-0.302);\draw [dotted, gray]  (0.953,0.302)--(0.953,1.2);

    \draw [dotted, thick, postaction={decorate}]  (-0.214,0.97)--(-0.214,-0.97);\draw [dotted, gray]  (-0.214,0.97)--(-0.214,1.2);

    \draw [rotate=-240,dotted, thick, postaction={decorate}]  (-0.738,0.674)--(-0.738,-0.674);\draw [rotate=-240,dotted, gray]  (-0.738,0.674)--(-0.738,1.2);
    \draw [rotate=-240,dotted, thick, postaction={decorate}]  (0.953,0.302)--(0.953,-0.302);\draw [rotate=-240,dotted, gray]  (0.953,0.302)--(0.953,1.2);

    \draw [rotate=-240,dotted, thick, postaction={decorate}]  (-0.214,0.97)--(-0.214,-0.97);\draw [rotate=-240,dotted, gray]  (-0.214,0.97)--(-0.214,1.2);

    \draw [dotted, thick, postaction={decorate}]  (0.984,0.178)--(0.984,-0.178);\draw [dotted, gray]  (0.984,0.178)--(0.984,1.2);
    \draw [dotted, thick, postaction={decorate}]  (-0.645,0.764)--(-0.645,-0.764);\draw [dotted, gray]  (-0.645,0.764)--(-0.645,1.2);

    \draw [dotted, thick, postaction={decorate}]  (-0.338,0.941)--(-0.338,-0.941);\draw [dotted, gray]  (-0.338,0.941)--(-0.338,1.2);

    \draw [rotate=-240,dotted, thick, postaction={decorate}]  (0.984,0.178)--(0.984,-0.178);\draw [rotate=-240,dotted, gray]  (0.984,0.178)--(0.984,1.2);
    \draw [rotate=-240,dotted, thick, postaction={decorate}]  (-0.645,0.764)--(-0.645,-0.764);\draw [rotate=-240,dotted, gray]  (-0.645,0.764)--(-0.645,1.2);

    \draw [rotate=-240,dotted, thick, postaction={decorate}]  (-0.338,0.941)--(-0.338,-0.941);\draw [rotate=-240,dotted, gray]  (-0.338,0.941)--(-0.338,1.2);

    \draw[  postaction={decorate}] (0.86,-0.5) -- (-0.214,-0.97);\draw [dotted, gray]  (0.86,-0.5)--(2,0);
    \draw[  postaction={decorate}] (0.86,0.5) -- (-0.214,0.97);\draw [dotted, gray]  (0.86,0.5)--(2,0);

    \draw [dotted, thick, postaction={decorate}]  (0.866,0.5)--(0.866,-0.5);\draw [dotted, gray]  (0.866,0.5)--(0.866,1.2);
    \draw [dotted, thick, postaction={decorate}]  (-0.214,0.97)--(-0.214,-0.97);\draw [dotted, gray]  (-0.214,0.97)--(-0.214,1.2);

    \draw[  postaction={decorate}] (0.97,-0.18) -- (-0.866,-0.5);\draw [dotted, gray]  (0.97,-0.18)--(2,0);

    \draw[  postaction={decorate}] (0.97,0.18) -- (-0.866,0.5);\draw [dotted, gray]  (0.97,0.18)--(2,0);

    \draw [dotted, thick, postaction={decorate}]  (-0.866,0.5)--(-0.866,-0.5);\draw [dotted, gray]  (-0.866,0.5)--(-0.866,1.2);

    \draw [dotted, thick, postaction={decorate}]  (0.984,0.178)--(0.984,-0.178);\draw [dotted, gray]  (0.984,0.178)--(0.984,1.2);

    \draw[  postaction={decorate}] (0.953,0.302) -- (-0.645,0.764);\draw [dotted, gray]  (0.953,0.302)--(2,0);

    \draw[  postaction={decorate}] (0.953,-0.302) -- (-0.645,-0.764);\draw [dotted, gray]  (0.953,-0.302)--(2,0);

    \draw [dotted, thick, postaction={decorate}]  (-0.645,0.764)--(-0.645,-0.764);\draw [dotted, gray]  (-0.645,0.764)--(-0.645,1.2);

    \draw [dotted, thick, postaction={decorate}]  (0.953,0.302)--(0.953,-0.302);\draw [dotted, gray]  (0.953,0.302)--(0.953,1.2);

    \filldraw [black] (0.866,-0.5) circle (1.5pt);
    \filldraw [gray] (0.953,0.302) circle (1.5pt);
    \filldraw [lightgray] (0.984,0.178) circle (1.5pt);

    \filldraw [lightgray] (0.984,-0.178) circle (1.5pt);
    \filldraw [gray] (0.953,-0.302) circle (1.5pt);
    \filldraw [black] (0.866,0.5) circle (1.5pt);

    \filldraw [black] (0,-1) circle (1.5pt);
    \filldraw [gray] (-0.214,-0.97) circle (1.5pt);
    \filldraw [lightgray] (-0.338,-0.941) circle (1.5pt);

    \filldraw [black] (-0.866,-0.5) circle (1.5pt);
    \filldraw [lightgray] (-0.645,-0.764) circle (1.5pt);
    \filldraw [gray] (-0.738,-0.674) circle (1.5pt);

    \filldraw [black] (-0.866,0.5) circle (1.5pt);
    \filldraw [lightgray] (-0.645,0.764) circle (1.5pt);
    \filldraw [gray] (-0.738,0.674) circle (1.5pt);

    \filldraw [black] (0,1) circle (1.5pt);
    \filldraw [gray] (-0.214,0.97) circle (1.5pt);
    \filldraw [lightgray] (-0.338,0.941) circle (1.5pt);

 \end{tikzpicture}
\caption {The 18 Kasner points associated with $u_0=1+\sqrt{3}$,
with 3 points in each sector, the 6 one-parameter families of multiplie
transitions, whose boundaries yield the 24 single transitions.
 }\label{fig:period8}
\end{figure}

The dynamical importance of the heteroclinic network arising from $1+\sqrt{3}$ is
due to the eigenvalues in sector $(321)$ that are associated with the instability
of $R_1$ and $R_3$, respectively. These are given by
$\lambda_{R_1} = 3(1 - \ue_{0}^2)/f(\ue_{0})$ and
$\lambda_{R_3} = -3(1 + 2\ue_{0})/f(\ue_{0})$ according to~\eqref{linR1}
and~\eqref{linR3}, which for the value $\ue_{0} = 1 - \sqrt{3}$ leads to
$\lambda_{R_1} = \lambda_{R_3}$, while $\lambda_{R_3}>\lambda_{R_1}$
when $\ue \in (-1,1-\sqrt{3})$ and $\lambda_{R_1}>\lambda_{R_3}$
when $\ue \in (1-\sqrt{3},-\frac12)$; these eigenvalue features
presumably play a role in a possible construction of a stable manifolds,
akin to that in~\cite{Beguin10,Lieb11,Lieb13,BegDut22}.
Note that the value $\ue_{0} = 1 - \sqrt{3}$, where $\lambda_{R_1} = \lambda_{R_3}$,
corresponds to the Kasner fixed point from which the particular explicit orbit with
$R_1=R_3$ and $\Sigma_2=0$, $\Sigma_3=-\Sigma_1$, discussed previously
in connection with double frame transitions, originates.

Note that this heteroclinic network contains two special orbits, which
are in the boundary of the set $\mathcal{HO}$, see Section~\ref{sec:naa=0}; and two Kasner fixed points with $\Sigma_-=\pm 1$, denoted by $\mathrm{K}^\ocircle_\pm$,
where $\mathrm{K}^\ocircle_+$ is in sector $(312)$ with $\ue = -2 + \sqrt{3}$,
while $\mathrm{K}^\ocircle_-$ is in sector $(213)$ with $\ue = -2 - \sqrt{3}$.
Both these Kasner fixed points correspond to the Kasner parameter
$u = 1 + \sqrt{3}$, which notably characterizes the double frame transition
for which $\lambda_{R_1} = \lambda_{R_3}$.

The stable heteroclinic network, shown to the right in
Figure~\ref{FIG:graphper8}, contains five single transition heteroclinic
cycles with period 8, which form the boundaries of one-parameter families
of subnetworks containing either double-frame transitions ${\cal T}_{R_1R_3}$
or mixed curvature-frame transitions ${\cal T}_{R_1N_-}$, with a
smaller period. The five single transition heteroclinic
period 8 cycles are depicted in the $\Sigma_1+\Sigma_2+\Sigma_3=0$
plane in the projected $(\Sigma_1,\Sigma_2,\Sigma_3)$-space
representation in Figure~\ref{fig:period8network}.
\begin{figure}[H]
\centering
\begin{tikzpicture}[scale=1.1]

        \draw [gray, dashed, - ]  (1,1.2)--(1,0);
        \draw [gray, dashed, - ]  (-1,1.2)--(-1,0);
        \draw [gray,dashed, - ]  (-0.5392,-1.4660)--(0.5,-0.8660);
        \draw [gray,dashed, - ]  (-1.5392,0.2660)--(-0.5,0.8660);
	\draw[gray,dashed, -] (2,0) -- (0.5,-0.866);
	\draw[gray,dashed, -] (2,0) -- (0.5, 0.866);

        \draw (1,0) arc (0:360:1cm and 1cm);	 	

       \draw [ultra thick, domain=0.52:1.570,variable=\t,smooth] plot ({sin(\t r)},{cos(\t r)});
       \draw [ultra thick, domain=-0.52:-1.570,variable=\t,smooth] plot ({sin(\t r)},{cos(\t r)});

	\draw (-0.95,0) -- (-1.05,0);
        \draw (0.95,0) -- (1.05,0);
        \draw[rotate=120] (-0.95,0) -- (-1.05,0);
        \draw[rotate=240] (-0.95,0) -- (-1.05,0);
        \draw[rotate=120] (0.95,0) -- (1.05,0);
        \draw[rotate=240] (0.95,0) -- (1.05,0);


    \draw[  postaction={decorate}] (0.86,-0.5) -- (-0.214,-0.97);\draw [dotted, gray]  (0.86,-0.5)--(2,0);

    \draw[  postaction={decorate}] (0.953,-0.302) -- (-0.645,-0.764);\draw [dotted, gray]  (0.953,-0.302)--(2,0);

    \draw[  postaction={decorate}] (0.97,0.18) -- (-0.866,0.5);\draw [dotted, gray]  (0.97,0.18)--(2,0);

    \draw [dotted, thick, postaction={decorate}]  (0,1)--(0,-1);\draw [dotted, gray]  (0,1)--(0,1.2);

    \draw [rotate=-240,dotted, thick, postaction={decorate}]  (-0.866,0.5)--(-0.866,-0.5);\draw [rotate=-240,dotted, gray]  (-0.866,0.5)--(-0.866,1.2);
    \draw [rotate=-240,dotted, thick, postaction={decorate}]  (0.866,0.5)--(0.866,-0.5);\draw [rotate=-240,dotted, gray]  (0.866,0.5)--(0.866,1.2);

    \draw [dotted, thick, postaction={decorate}]  (-0.214,-0.97)--(0.953,-0.302);\draw [dotted, gray]  (-0.214,-0.97)--(-0.64,-1.21);

    \draw [dotted, thick, postaction={decorate}]  (-0.645,-0.764)--(0.984,0.178);\draw [dotted, gray]  (-0.645,-0.764)--(-0.85,-0.88);


    \filldraw [lightgray] (0.984,0.178) circle (1.5pt);

    \filldraw [black] (0.866,-0.5) circle (1.5pt);
    \filldraw [gray] (0.953,-0.302) circle (1.5pt);

    \filldraw [black] (0,-1) circle (1.5pt);
    \filldraw [gray] (-0.214,-0.97) circle (1.5pt);

    \filldraw [lightgray] (-0.645,-0.764) circle (1.5pt);

    \filldraw [black] (-0.866,0.5) circle (1.5pt);

    \filldraw [black] (0,1) circle (1.5pt);

 \end{tikzpicture}
\hspace{1cm}
\begin{tikzpicture}[scale=1.1]

        \draw [gray, dashed, - ]  (1,1.2)--(1,0);
        \draw [gray, dashed, - ]  (-1,1.2)--(-1,0);
        \draw [gray,dashed, - ]  (-0.5392,-1.4660)--(0.5,-0.8660);
        \draw [gray,dashed, - ]  (-1.5392,0.2660)--(-0.5,0.8660);
	\draw[gray,dashed, -] (2,0) -- (0.5,-0.866);
	\draw[gray,dashed, -] (2,0) -- (0.5, 0.866);

        \draw (1,0) arc (0:360:1cm and 1cm);	 	
       \draw [ultra thick, domain=0.52:1.570,variable=\t,smooth] plot ({sin(\t r)},{cos(\t r)});
       \draw [ultra thick, domain=-0.52:-1.570,variable=\t,smooth] plot ({sin(\t r)},{cos(\t r)});

	\draw (-0.95,0) -- (-1.05,0);
        \draw (0.95,0) -- (1.05,0);
        \draw[rotate=120] (-0.95,0) -- (-1.05,0);
        \draw[rotate=240] (-0.95,0) -- (-1.05,0);
        \draw[rotate=120] (0.95,0) -- (1.05,0);
        \draw[rotate=240] (0.95,0) -- (1.05,0);


    \draw[  postaction={decorate}] (0.86,-0.5) -- (-0.214,-0.97);\draw [dotted, gray]  (0.86,-0.5)--(2,0);

    \draw[  postaction={decorate}] (0.953,-0.302) -- (-0.645,-0.764);\draw [dotted, gray]  (0.953,-0.302)--(2,0);

    \draw[  postaction={decorate}] (0.97,0.18) -- (-0.866,0.5);\draw [dotted, gray]  (0.97,0.18)--(2,0);

    \draw [dotted, thick, postaction={decorate}]  (-0.866,0.5)--(-0.866,-0.5);\draw [dotted, gray]  (-0.866,0.5)--(-0.866,1.2);
    \draw [dotted, thick, postaction={decorate}]  (0.866,0.5)--(0.866,-0.5);\draw [dotted, gray]  (0.866,0.5)--(0.866,1.2);

    \draw [rotate=-240,dotted, thick, postaction={decorate}]  (0,1)--(0,-1);\draw [rotate=-240,dotted, gray]  (0,1)--(0,1.2);

    \draw [dotted, thick, postaction={decorate}]  (-0.214,-0.97)--(0.953,-0.302);\draw [dotted, gray]  (-0.214,-0.97)--(-0.64,-1.21);

    \draw [dotted, thick, postaction={decorate}]  (-0.645,-0.764)--(0.984,0.178);\draw [dotted, gray]  (-0.645,-0.764)--(-0.85,-0.88);


    \filldraw [black] (0.866,-0.5) circle (1.5pt);
    \filldraw [lightgray] (0.984,0.178) circle (1.5pt);

    \filldraw [gray] (0.953,-0.302) circle (1.5pt);
    \filldraw [black] (0.866,0.5) circle (1.5pt);

    \filldraw [gray] (-0.214,-0.97) circle (1.5pt);

    \filldraw [black] (-0.866,-0.5) circle (1.5pt);
    \filldraw [lightgray] (-0.645,-0.764) circle (1.5pt);

    \filldraw [black] (-0.866,0.5) circle (1.5pt);


 \end{tikzpicture}
\hspace{1cm}
\begin{tikzpicture}[scale=1.1]

        \draw [gray, dashed, - ]  (1,1.2)--(1,0);
        \draw [gray, dashed, - ]  (-1,1.2)--(-1,0);
        \draw [gray,dashed, - ]  (-0.5392,-1.4660)--(0.5,-0.8660);
        \draw [gray,dashed, - ]  (-1.5392,0.2660)--(-0.5,0.8660);
	\draw[gray,dashed, -] (2,0) -- (0.5,-0.866);
	\draw[gray,dashed, -] (2,0) -- (0.5, 0.866);

        \draw (1,0) arc (0:360:1cm and 1cm);	 	
       \draw [ultra thick, domain=0.52:1.570,variable=\t,smooth] plot ({sin(\t r)},{cos(\t r)});
       \draw [ultra thick, domain=-0.52:-1.570,variable=\t,smooth] plot ({sin(\t r)},{cos(\t r)});

	\draw (-0.95,0) -- (-1.05,0);
        \draw (0.95,0) -- (1.05,0);
        \draw[rotate=120] (-0.95,0) -- (-1.05,0);
        \draw[rotate=240] (-0.95,0) -- (-1.05,0);
        \draw[rotate=120] (0.95,0) -- (1.05,0);
        \draw[rotate=240] (0.95,0) -- (1.05,0);


    \draw[  postaction={decorate}] (0.86,0.5) -- (-0.214,0.97);\draw [dotted, gray]  (0.86,0.5)--(2,0);

    \draw[  postaction={decorate}] (0.953,-0.302) -- (-0.645,-0.764);\draw [dotted, gray]  (0.953,-0.302)--(2,0);

    \draw[  postaction={decorate}] (0.97,0.18) -- (-0.866,0.5);\draw [dotted, gray]  (0.97,0.18)--(2,0);

    \draw [dotted, thick, postaction={decorate}]  (-0.866,0.5)--(-0.866,-0.5);\draw [dotted, gray]  (-0.866,0.5)--(-0.866,1.2);

    \draw [dotted, thick, postaction={decorate}]  (-0.214,0.97)--(-0.214,-0.97);\draw [dotted, gray]  (-0.214,0.97)--(-0.214,1.2);

    \draw [rotate=-240,dotted, thick, postaction={decorate}]  (0,1)--(0,-1);\draw [rotate=-240,dotted, gray]  (0,1)--(0,1.2);

    \draw [dotted, thick, postaction={decorate}]  (-0.214,-0.97)--(0.953,-0.302);\draw [dotted, gray]  (-0.214,-0.97)--(-0.64,-1.21);

    \draw [dotted, thick, postaction={decorate}]  (-0.645,-0.764)--(0.984,0.178);\draw [dotted, gray]  (-0.645,-0.764)--(-0.85,-0.88);


    \filldraw [black] (0.866,0.5) circle (1.5pt);
    \filldraw [lightgray] (0.984,0.178) circle (1.5pt);

    \filldraw [gray] (0.953,-0.302) circle (1.5pt);

    \filldraw [gray] (-0.214,-0.97) circle (1.5pt);

    \filldraw [black] (-0.866,-0.5) circle (1.5pt);
    \filldraw [lightgray] (-0.645,-0.764) circle (1.5pt);

    \filldraw [black] (-0.866,0.5) circle (1.5pt);

    \filldraw [gray] (-0.214,0.97) circle (1.5pt);

 \end{tikzpicture}
\hspace{1cm}
\begin{tikzpicture}[scale=1.1]

        \draw [gray, dashed, - ]  (1,1.2)--(1,0);
        \draw [gray, dashed, - ]  (-1,1.2)--(-1,0);
        \draw [gray,dashed, - ]  (-0.5392,-1.4660)--(0.5,-0.8660);
        \draw [gray,dashed, - ]  (-1.5392,0.2660)--(-0.5,0.8660);
	\draw[gray,dashed, -] (2,0) -- (0.5,-0.866);
	\draw[gray,dashed, -] (2,0) -- (0.5, 0.866);

        \draw (1,0) arc (0:360:1cm and 1cm);	 	

       \draw [ultra thick, domain=0.52:1.570,variable=\t,smooth] plot ({sin(\t r)},{cos(\t r)});
       \draw [ultra thick, domain=-0.52:-1.570,variable=\t,smooth] plot ({sin(\t r)},{cos(\t r)});

	\draw (-0.95,0) -- (-1.05,0);
        \draw (0.95,0) -- (1.05,0);
        \draw[rotate=120] (-0.95,0) -- (-1.05,0);
        \draw[rotate=240] (-0.95,0) -- (-1.05,0);
        \draw[rotate=120] (0.95,0) -- (1.05,0);
        \draw[rotate=240] (0.95,0) -- (1.05,0);


    \draw[  postaction={decorate}] (0.86,-0.5) -- (-0.214,-0.97);\draw [dotted, gray]  (0.86,-0.5)--(2,0);

    \draw[  postaction={decorate}] (0.953,-0.302) -- (-0.645,-0.764);\draw [dotted, gray]  (0.953,-0.302)--(2,0);

    \draw[  postaction={decorate}] (0.97,-0.18) -- (-0.866,-0.5);\draw [dotted, gray]  (0.97,-0.18)--(2,0);

    \draw [dotted, thick, postaction={decorate}]  (0.984,0.178)--(0.984,-0.178);\draw [dotted, gray]  (0.984,0.178)--(0.984,1.2);
    \draw [dotted, thick, postaction={decorate}]  (0.866,0.5)--(0.866,-0.5);\draw [dotted, gray]  (0.866,0.5)--(0.866,1.2);

    \draw [rotate=-240,dotted, thick, postaction={decorate}]  (0,1)--(0,-1);\draw [rotate=-240,dotted, gray]  (0,1)--(0,1.2);

    \draw [dotted, thick, postaction={decorate}]  (-0.214,-0.97)--(0.953,-0.302);\draw [dotted, gray]  (-0.214,-0.97)--(-0.64,-1.21);

    \draw [dotted, thick, postaction={decorate}]  (-0.645,-0.764)--(0.984,0.178);\draw [dotted, gray]  (-0.645,-0.764)--(-0.85,-0.88);


    \filldraw [black] (0.866,-0.5) circle (1.5pt);
    \filldraw [lightgray] (0.984,0.178) circle (1.5pt);

    \filldraw [lightgray] (0.984,-0.178) circle (1.5pt);
    \filldraw [gray] (0.953,-0.302) circle (1.5pt);
    \filldraw [black] (0.866,0.5) circle (1.5pt);

    \filldraw [gray] (-0.214,-0.97) circle (1.5pt);

    \filldraw [black] (-0.866,-0.5) circle (1.5pt);
    \filldraw [lightgray] (-0.645,-0.764) circle (1.5pt);



 \end{tikzpicture}
\hspace{1cm}
\begin{tikzpicture}[scale=1.1]

        \draw [gray, dashed, - ]  (1,1.2)--(1,0);
        \draw [gray, dashed, - ]  (-1,1.2)--(-1,0);
        \draw [gray,dashed, - ]  (-0.5392,-1.4660)--(0.5,-0.8660);
        \draw [gray,dashed, - ]  (-1.5392,0.2660)--(-0.5,0.8660);
	\draw[gray,dashed, -] (2,0) -- (0.5,-0.866);
	\draw[gray,dashed, -] (2,0) -- (0.5, 0.866);

        \draw (1,0) arc (0:360:1cm and 1cm);	 	

       \draw [ultra thick, domain=0.52:1.570,variable=\t,smooth] plot ({sin(\t r)},{cos(\t r)});
       \draw [ultra thick, domain=-0.52:-1.570,variable=\t,smooth] plot ({sin(\t r)},{cos(\t r)});

	\draw (-0.95,0) -- (-1.05,0);
        \draw (0.95,0) -- (1.05,0);
        \draw[rotate=120] (-0.95,0) -- (-1.05,0);
        \draw[rotate=240] (-0.95,0) -- (-1.05,0);
        \draw[rotate=120] (0.95,0) -- (1.05,0);
        \draw[rotate=240] (0.95,0) -- (1.05,0);


    \draw[ postaction={decorate}] (0.86,0.5) -- (-0.214,0.97);\draw [dotted, gray]  (0.86,0.5)--(2,0);

    \draw[ postaction={decorate}] (0.953,-0.302) -- (-0.645,-0.764);\draw [dotted, gray]  (0.953,-0.302)--(2,0);

    \draw[ postaction={decorate}] (0.97,-0.18) -- (-0.866,-0.5);\draw [dotted, gray]  (0.97,-0.18)--(2,0);

    \draw [dotted, thick, postaction={decorate}]  (0.984,0.178)--(0.984,-0.178);\draw [dotted, gray]  (0.984,0.178)--(0.984,1.2);
    \draw [dotted, thick, postaction={decorate}]  (-0.214,0.97)--(-0.214,-0.97);\draw [dotted, gray]  (-0.214,0.97)--(-0.214,1.2);

    \draw [rotate=-240,dotted, thick, postaction={decorate}]  (0,1)--(0,-1);\draw [rotate=-240,dotted, gray]  (0,1)--(0,1.2);

    \draw [dotted, thick, postaction={decorate}]  (-0.214,-0.97)--(0.953,-0.302);\draw [dotted, gray]  (-0.214,-0.97)--(-0.64,-1.21);

    \draw [dotted, thick, postaction={decorate}]  (-0.645,-0.764)--(0.984,0.178);\draw [dotted, gray]  (-0.645,-0.764)--(-0.85,-0.88);


    \filldraw [lightgray] (0.984,0.178) circle (1.5pt);

    \filldraw [lightgray] (0.984,-0.178) circle (1.5pt);
    \filldraw [gray] (0.953,-0.302) circle (1.5pt);
    \filldraw [black] (0.866,0.5) circle (1.5pt);

    \filldraw [gray] (-0.214,-0.97) circle (1.5pt);

    \filldraw [black] (-0.866,-0.5) circle (1.5pt);
    \filldraw [lightgray] (-0.645,-0.764) circle (1.5pt);


    \filldraw [gray] (-0.214,0.97) circle (1.5pt);

 \end{tikzpicture}
\caption {The five heteroclinic single transition period 8 cycles associated
with $u=1+\sqrt{3}$. In addition, there are six cyclic subnetworks containing one-parameter
families of multiple transitions, which we refrain from plotting. 
 }\label{fig:period8network}
\end{figure}
%

\section{Comparisons between type $\mathrm{VI}_{-1/9}$ and types $\mathrm{VIII}$, $\mathrm{IX}$}\label{sec:comparison}

In order to compare the properties of Bianchi type $\mathrm{VI}_{-1/9}$
and types $\mathrm{VIII}$, $\mathrm{IX}$ let us first recall the Hubble-normalized
dynamical system for the latter. The vacuum class A models with $a_\alpha=0$, of
which Bianchi types $\mathrm{VIII}$ and $\mathrm{IX}$ are the most general ones,
see Table~\ref{BianchClass}, are naturally expressed in a spatially fixed
Fermi-frame with $\Omega^\alpha=0$ in which both the shear $\sigma_{\alpha\beta}$
and the remaining spatial commutator variables $n^{\alpha\beta}$ are diagonal.

By introducing the Hubble-normalized variables for the remaining non-zero
variables:
\begin{equation}\label{Nalphadef}
(\Sigma_1,\Sigma_2,\Sigma_3,N_1,N_2,N_3)
=
\left(
\frac{\sigma_{11}}{H},
\frac{\sigma_{22}}{H},
\frac{\sigma_{33}}{H},
\frac{n^{11}}{2\sqrt{3}\,H},
\frac{n^{22}}{2\sqrt{3}\,H},
\frac{n^{33}}{2\sqrt{3}\,H}
\right),
\end{equation}
%
%
%
which, due to~\eqref{devoleq}, lead to the dimensionless evolution equations (we use
the same dimensionless time variable as for Bianchi type $\mathrm{VI}_{-1/9}$),
\begin{subequations}\label{evolclassA}
\begin{align}
\Sigma_1^\prime &= 2(1 - \Sigma^2)- 4\left((N_2-N_3)^2 + N_1\left(N_2 + N_3 - 2N_1\right)\right),\\
\Sigma_2^\prime &= 2(1 - \Sigma^2)- 4\left((N_3-N_1)^2 + N_2\left(N_3 + N_1 - 2N_2\right)\right),\\
\Sigma_3^\prime &= 2(1 - \Sigma^2)- 4\left((N_1-N_2)^2 + N_3\left(N_1 + N_2 - 2N_3\right)\right),\\
N_1^\prime &= -2(\Sigma^2 + \Sigma_1)N_1,\\
N_2^\prime &= -2(\Sigma^2 + \Sigma_2)N_2,\\
N_3^\prime &= -2(\Sigma^2 + \Sigma_3)N_3,
\end{align}
\end{subequations}
where $\Sigma^2 := \left(\Sigma_ 1^2 + \Sigma_2^2 + \Sigma_3^2\right)/6$, and the constraints
\begin{subequations}\label{constrclassA}
\begin{align}
0 &= \Sigma_1 + \Sigma_2 + \Sigma_3,\\
1 &= \Sigma^2 + N_1^2 + N_2^2 + N_3^2 - 2N_1N_2 - 2N_2N_3 - 2N_3N_1.
\end{align}
\end{subequations}
%
%
%
The present variables $N_\alpha$ in \eqref{Nalphadef} differ from those in~\cite{waiell97} by being divided with $2\sqrt{3}$ in order to simplify the Gauss constraint.
The equations~\eqref{evolclassA} are invariant under permutations of $1, 2, 3$, and $(N_1,N_2,N_3) \mapsto -(N_1,N_2,N_3)$.
These discrete symmetries, in combination with the definitions in~\eqref{Nalphadef}, where $H>0$, and
Table~\ref{BianchClass}, lead to that the equations are characterized by the various class A Bianchi
symmetry groups according to the hierarchical stratification diagram of invariant sets
for the class A vacuum models in Figure~\ref{FIG:hierarchy2}.\footnote{Bianchi types $\mathrm{VI}_0$ and $\mathrm{VII}_0$ are denoted
by $\mathcal{T}_{N_\alpha N_\beta}$, $\alpha\neq\beta$, since the solutions on these invariant
subsets (apart from the LRS solutions in $\mathrm{VII}_0$) are given by heteroclinic transition
orbits that connect a Taub fixed point (a Taub-like line of fixed points) in type
$\mathrm{VI}_0$ ($\mathrm{VII}_0$) with a (toward the singularity) stable arc of fixed
points on the Kasner circle $\mathrm{K}^\ocircle$, see~\cite{HLU22} for details.}
\begin{figure}[H]
\centering
\begin{tikzpicture}
\node (dim) at (6, 6.6) {\underline{\footnotesize{Dimension:}}};
\node (4) at (6, 6) {\footnotesize{$4$}};
\node (3) at (6, 4.5) {\footnotesize{$3$}};
\node (2) at (6, 3) {\footnotesize{$2$}};
\node (1) at (6, 1.5) {\footnotesize{$1$}};

\node (VI19) at (-2, 6) {\boxed{\text{\footnotesize{Type $\mathrm{VIII},\mathrm{IX}$}}}};

\node (R1R3) at (-4, 4.5) {\boxed{\text{\footnotesize{$\mathcal{T}_{N_1N_2}$}}}};
\node (R1N-) at (-2, 4.5) {\boxed{\text{\footnotesize{$\mathcal{T}_{N_1N_3}$}}}};
\node (OT) at (0, 4.5) {\boxed{\text{\footnotesize{$\mathcal{T}_{N_2N_3}$}}}};

\node (R1) at (-4, 3) {\boxed{\text{\footnotesize{$\mathcal{T}_{N_1}$}}}};
\node (R3) at (-2, 3) {\boxed{\text{\footnotesize{$\mathcal{T}_{N_2}$}}}};
\node (N-) at (-0, 3) {\boxed{\text{\footnotesize{$\mathcal{T}_{N_3}$}}}};

\node (K0) at (-2, 1.5) {\boxed{\text{\footnotesize{$\mathrm{K}^{\ocircle}$}}}};


\draw[-] (VI19) -- (R1R3);\draw[-] (VI19) -- (R1N-);\draw[-] (VI19) -- (OT);


\draw[-] (R1R3) -- (R1);\draw[-] (R1R3) -- (R3);\draw[-] (R1N-) -- (N-);\draw[-] (R1N-) -- (R1);

\draw[-] (R1) -- (K0);\draw[-] (R3) -- (K0); \draw[-] (N-) -- (K0);

\draw[-] (OT) -- (N-);\draw[-] (OT) -- (R3);


\end{tikzpicture}
\captionof{figure}{The invariant set stratification diagram for the class A
vacuum models. To obtain the invariant subsets from one stratum to the next,
one of the non-zero variables $N_\alpha$, $\alpha=1,2,3$ is set to zero, where the
indices in the transition nomenclature indicate which of the variables $N_1, N_2, N_3$
that are non-zero. The top stratum is characterized by $N_1N_2N_3 \neq 0$, where one of
these variables have the opposite sign than the other in type $\mathrm{VIII}$
while they all have the same sign in type $\mathrm{IX}$; at the next strata with
$N_\alpha N_\beta \neq 0$, $\alpha\neq\beta$, the two variables have opposite
signs in Bianchi type $\mathrm{VI}_0$ while they have the same sign in type
$\mathrm{VII}_0$; in the next Bianchi type II strata there is only one non-zero
$N_\alpha$ variable; $N_1=N_2=N_3=0$ at the lowest stratum yields the Kasner
circle $\mathrm{K}^\ocircle$ of fixed points.}\label{FIG:hierarchy2}
\end{figure}

The simplicity of this stratification diagram reflects the simple hierarchical
$G_3$ Lie contraction symmetry group structure of the class A vacuum
models.\footnote{Lie contractions correspond to changing group types by setting
structure constants to zero, which here amounts to setting variables $N_\alpha$
to zero.} In contrast the $G_3$ group structure of type $\mathrm{VI}_{-1/9}$
induces an extra Codazzi constraint and breaks the permutation symmetry that
exists for class A; moreover, the subsets $\mathcal{OT}$, $\mathcal{HO}$ and
$\mathcal{D}$ are not due to Lie contractions but instead by that the $G_2$ Abelian
subgroup acts in different ways on the space-time. These features thereby lead to
the complicated type $\mathrm{VI}_{-1/9}$ stratification diagram~\ref{FIG:hierarchy},
with associated dynamical complications.

The dynamical system~\eqref{evolclassA}, \eqref{constrclassA}
admits $N_1N_2N_3$ as a monotonic function\footnote{Note that $\Sigma^2 = 0$
is not an invariant set. Hence if $\Sigma^2=0$ occurs for a solution then this only
yields an inflection point in the monotonic evolution $N_1N_2N_3 \rightarrow 0$
toward the singularity.} according to 
\begin{equation}
(N_1N_2N_3)^\prime = -6\Sigma^2(N_1N_2N_3) \qquad \Rightarrow \qquad \lim_{\tau\rightarrow\infty}(N_1N_2N_3) = 0.
\end{equation}
Thus the singularity attractor for Bianchi types $\mathrm{VIII}$ and
$\mathrm{IX}$ reside on
$\overline{\mathcal T}_{N_1N_2}\cup\overline{\mathcal T}_{N_2N_3}\cup\overline{\mathcal T}_{N_3N_1}$.
No corresponding monotonic function has been found for the vacuum
Bianchi type $\mathrm{VI}_{-1/9}$ model; this is due to the much more
complicated dynamics of these models, which is induced by their
complicated KVF structure.

If one attempts to obtain an attractor theorem of the type Ringstr{\"o}m
obtained in~\cite{rin01} for Bianchi type $\mathrm{IX}$ in the type $\mathrm{VI}_{-1/9}$
vacuum case, this requires that $\lim_{\tau\rightarrow \infty} A = 0$, since this
results in $\lim_{\tau\rightarrow \infty} R_3N_- = 0$, due to the Codazzi
constraint~\eqref{Codazzi1},
and hence that the singularity attractor, ${\cal A}$, resides
in the union of the Bianchi type I and II subsets, which would
establish Conjecture~\ref{conj1}.
That $\lim_{\tau\rightarrow \infty} A = 0$ happens is
suggested by the linear analysis of equation~\eqref{Aeq} at
$\mathrm{K}^{\ocircle}$ since this yields $A^\prime = -3(1 - p_3)A$, which implies
that $\lim_{\tau\rightarrow +\infty}A = 0$ everywhere on $\mathrm{K}^{\ocircle}$,
except at the Taub point $\mathrm{T}_3$ where $p_3=1$.

Towards this goal, one can follow section 6 in~\cite{rin01} and identify the
region in the state space where $A$ is increasing. To do so,
it is convenient to use the Misner variables in~\eqref{Misner3}, i.e., 
\begin{equation}
\Sigma_1 = \Sigma_+ + \sqrt{3}\Sigma_-,\qquad
\Sigma_2 = \Sigma_+ - \sqrt{3}\Sigma_-,\qquad
\Sigma_3 = -2\Sigma_+,
\end{equation}
which together with~\eqref{Aeq} yields
\begin{equation}\label{Aprimeconj}
A^\prime = -2\left[\left(\Sigma_+ + \frac{1}{2}\right)^2 + \Sigma_-^2 + R_1^2 + R_3^2 - \frac{1}{4}\right]A.
\end{equation}
This implies that the only region where $A$ increases is in a ball with
radius $1/2$ centered at $(\Sigma_+,\Sigma_-,R_1,R_3) = \left(-\frac{1}{2},0,0,0\right)$,
where a projection of this ball onto the plane $\Sigma_1 + \Sigma_2 + \Sigma_3 = 0$
in $(\Sigma_1,\Sigma_2,\Sigma_3)$-space  is given in Figure~\ref{fig:Ainc}. 

Alternatively, we note that $A$  is increasing if
\begin{equation}
N_-^2 + A^2 > 1 + \Sigma_+,
\end{equation}
where we have used the Gauss constraint
$1 = \Sigma_+^2 + \Sigma_-^2 + R_1^2 + R_3^2 + N_-^2 + A^2$ in
connection with~\eqref{Aprimeconj}. The challenge now is to prove that the
evolution of generic orbits are dominated by the region outside the ball where
$A$ is increasing, if true.
\begin{figure}[H]
	\centering
\begin{tikzpicture}[scale=1.1]

        \filldraw [lightgray!40,domain=0:6.28,variable=\t,smooth] plot ({(0.25)+(0.5)*sin(\t r)},{-(0.433)+(0.5)*cos(\t r)});

        \draw [domain=0:6.28,variable=\t,smooth] plot ({(0.25)+(0.5)*sin(\t r)},{-(0.433)+(0.5)*cos(\t r)});

	\draw [very thin] (1,0) arc (0:360:1cm and 1cm);	 	
	
   	\draw (-0.95,0) -- (-1.05,0) node[anchor= east] {\scriptsize{$\mathrm{T}_1$}};
        \draw (0.95,0) -- (1.05,0);

        \draw[rotate=120] (-0.95,0) -- (-1.05,0) node[anchor= north] {\scriptsize{$\mathrm{T}_3$}};
        \draw[rotate=240] (-0.95,0) -- (-1.05,0) node[anchor= south] {\scriptsize{$\mathrm{T}_2$}};

        \draw[rotate=120] (0.95,0) -- (1.05,0);
        \draw[rotate=240] (0.95,0) -- (1.05,0);

    \draw[color=gray,->]
    (0,0) -- (-0.343,0)  node[anchor= east] {\scriptsize{$\Sigma_{1}$}};
    \draw[rotate=240,color=gray,->]
    (0,0) -- (-0.343,0)  node[anchor= south] {\scriptsize{$\Sigma_{2}$}};
    \draw[rotate=-240,color=gray,->]
    (0,0) -- (-0.343,0)  node[anchor= north] {\scriptsize{$\Sigma_{3}$}};



%
\end{tikzpicture}
\caption{The shaded region is the projection onto the plane $\Sigma_1 + \Sigma_2 + \Sigma_3 = 0$
in $(\Sigma_1,\Sigma_2,\Sigma_3)$-space of the region where $A$ increases.
Note that this ball only intersects the Kasner circle at the
Taub point $\mathrm{T}_3$, where they are tangential. } \label{fig:Ainc}
\end{figure}
\vspace{-0.3cm}
However, any seeming similarity with the road to the proof for Bianchi type $\mathrm{IX}$ is a mirage. 
There is no known monotonic function that pushes the dynamics from the general type $\mathrm{VI}_{-1/9}$ vacuum model to the next strata and each of the subsets of this strata is different. 
In Bianchi type $\mathrm{VI}_{-1/9}$ there is no permutation symmetry; there exists an additional Codazzi constraint~\eqref{Codazzi1}. 
All these facts further complicate a possible singularity proof.

The new types of proofs in~\cite{Beguin10,Lieb11,Lieb13,ReitererTrubowitz,bre16,BegDut22}
about the generic singularity for the Bianchi types $\mathrm{VIII}$ and $\mathrm{IX}$
vacuum models use a different strategy than that in~\cite{rin01,heiugg09} for the
Bianchi type $\mathrm{IX}$ attractor. These works use the solutions
on the Bianchi type I and II subsets in the class A stratification hierarchy,
which in the Hubble-normalized dynamical systems approach implies demonstrating
the stability of various heteroclinic chains that are obtained by concatenation
of heteroclinic Bianchi type II orbits on these subsets. The effect of the
permutation symmetry is illustrated by the heteroclinic cycles induced by the
golden (silver) ratio $u_0= (1 + \sqrt{5})/2$ ($u_0=1 + \sqrt{2}$), which gives
rise to two (three) heteroclinic cycles with period 3 (4), all related by axis
permutations, see Figure~\ref{fig:period3IX}.
\begin{figure}[H]
\centering
\begin{tikzpicture}[scale=1.1]
        \draw (1,0) arc (0:360:1cm and 1cm);	 	

        \draw[gray, thin,dashed] (2,0)--(-1,1.732) ;
        \draw[gray, thin,dashed] (-1,1.732) --(-1,-1.732) ;
        \draw[gray, thin,dashed] (-1,-1.732) --(2,0);

	\draw (-0.95,0) -- (-1.05,0) node[anchor=east] {\scriptsize $\mathrm{T}_1$};
        \draw (0.95,0) -- (1.05,0);
        \draw[rotate=120] (-0.95,0) -- (-1.05,0) node[anchor=north] {\scriptsize $\mathrm{T}_3$};;
        \draw[rotate=240] (-0.95,0) -- (-1.05,0)  node[anchor=south] {\scriptsize $\mathrm{T}_2$};
        \draw[rotate=120] (0.95,0) -- (1.05,0);
        \draw[rotate=240] (0.95,0) -- (1.05,0);

    \draw[ lightgray, postaction={decorate}] (0.96,-0.28) -- (-0.71,-0.71);\draw [dotted, gray]  (0.96,-0.28)--(2,0);
    \draw[ lightgray, rotate=120,  postaction={decorate}] (0.96,-0.28) -- (-0.71,-0.71);\draw [rotate=120,dotted, gray]  (0.96,-0.28)--(2,0);
    \draw[ lightgray, rotate=240, postaction={decorate}] (0.96,-0.28) -- (-0.71,-0.71);\draw [rotate=240,dotted, gray]  (0.96,-0.28)--(2,0);

    \draw[  postaction={decorate}] (0.96,0.28) -- (-0.71,0.71);\draw [dotted, gray]  (0.96,0.28)--(2,0);
    \draw[ rotate=120, postaction={decorate}] (0.96,0.28) -- (-0.71,0.71);\draw [rotate=120,dotted, gray]  (0.96,0.28)--(2,0);
    \draw[ rotate=240, postaction={decorate}] (0.96,0.28) -- (-0.71,0.71);\draw [rotate=240,dotted, gray]  (0.96,0.28)--(2,0);

    \filldraw [black] (0.96,-0.28) circle (1.5pt);
    \filldraw [black] (0.96,0.28) circle (1.5pt);
    \filldraw [rotate=240,black] (-0.71,-0.71) circle (1.5pt);
    \filldraw [black] (-0.71,0.71) circle (1.5pt);
    \filldraw [black] (-0.71,-0.71) circle (1.5pt);
    \filldraw [rotate=120,black] (-0.71,0.71) circle (1.5pt);

 \end{tikzpicture}
\hspace{2.5cm}
\begin{tikzpicture}[scale=1.1]
        \draw (1,0) arc (0:360:1cm and 1cm);	 	

        \draw[gray, thin,dashed] (2,0)--(-1,1.732) ;
        \draw[gray, thin,dashed] (-1,1.732) --(-1,-1.732) ;
        \draw[gray, thin,dashed] (-1,-1.732) --(2,0);

	\draw (-0.95,0) -- (-1.05,0) node[anchor=east] {\scriptsize $\mathrm{T}_1$};
        \draw (0.95,0) -- (1.05,0);
        \draw[rotate=120] (-0.95,0) -- (-1.05,0) node[anchor=north] {\scriptsize $\mathrm{T}_3$};;
        \draw[rotate=240] (-0.95,0) -- (-1.05,0)  node[anchor=south] {\scriptsize $\mathrm{T}_2$};
        \draw[rotate=120] (0.95,0) -- (1.05,0);
        \draw[rotate=240] (0.95,0) -- (1.05,0);

    \draw[ lightgray, postaction={decorate}] (0.98,-0.198) -- (-0.837,-0.547);\draw [dotted, gray]  (0.98,-0.198)--(2,0); 

    \draw[ lightgray, postaction={decorate}] (0.98,0.198) -- (-0.837,0.547);\draw [dotted, gray]  (0.98,0.198)--(2,0);

    \draw[ lightgray, rotate=120, postaction={decorate}] (0.891,-0.454) -- (-0.32,-0.947);\draw [rotate=120,dotted, gray]  (0.891,-0.454)--(2,0);

    \draw[ lightgray, rotate=120, postaction={decorate}] (0.98,0.198) -- (-0.837,0.547);\draw [rotate=120,dotted, gray]  (0.98,0.198)--(2,0); 

    \draw[ lightgray, rotate=120, postaction={decorate}] (0.891,0.454) -- (-0.32,0.947);\draw [rotate=120,dotted, gray]  (0.891,0.454)--(2,0);

    \draw[ lightgray, rotate=240, postaction={decorate}] (0.891,-0.454) -- (-0.32,-0.947);\draw [rotate=240,dotted, gray]  (0.891,-0.454)--(2,0); 

    \draw[ lightgray, rotate=240, postaction={decorate}] (0.98,-0.198) -- (-0.837,-0.547);\draw [rotate=240,dotted, gray]  (0.98,-0.198)--(2,0);

    \draw[ lightgray, rotate=240, postaction={decorate}] (0.891,0.454) -- (-0.32,0.947);\draw [rotate=240,dotted, gray]  (0.891,0.454)--(2,0); 


    \draw[ postaction={decorate}] (0.891,-0.454) -- (-0.32,-0.947);\draw [dotted, gray]  (0.891,-0.454)--(2,0);

    \draw[ postaction={decorate}] (0.891,0.454) -- (-0.32,0.947);\draw [dotted, gray]  (0.891,0.454)--(2,0);

    \draw[ rotate=120, postaction={decorate}] (0.98,-0.198) -- (-0.837,-0.547);\draw [rotate=120,dotted, gray]  (0.98,-0.198)--(2,0);

    \draw[ rotate=240, postaction={decorate}] (0.98,0.198) -- (-0.837,0.547);\draw [rotate=240,dotted, gray]  (0.98,0.198)--(2,0);

    \filldraw [black] (0.891,0.454) circle (1.5pt);
    \filldraw [lightgray] (0.98,0.198) circle (1.5pt);

    \filldraw [lightgray] (0.98,-0.198) circle (1.5pt);
    \filldraw [black] (0.891,-0.454) circle (1.5pt);

    \filldraw [lightgray] (-0.32,-0.947) circle (1.5pt);
    \filldraw [black] (-0.054,-0.998) circle (1.5pt);

    \filldraw [lightgray] (-0.660,-0.751) circle (1.5pt);
    \filldraw [black] (-0.837,-0.547) circle (1.5pt);

    \filldraw [black] (-0.837,0.547) circle (1.5pt);
    \filldraw [lightgray] (-0.660,0.751) circle (1.5pt);

    \filldraw [lightgray] (-0.32,0.947) circle (1.5pt);
    \filldraw [black] (-0.054,0.998) circle (1.5pt);
 \end{tikzpicture}
\caption{Projection onto the $\Sigma_1 + \Sigma_2 + \Sigma_3 = 0$ plane in
$(\Sigma_1,\Sigma_2,\Sigma_3)$-space of the heteroclinic cycles induced
by the golden ($u_0=\sfrac{1 + \sqrt{5}}{2}$)  and silver
($u_0=1 + \sqrt{2}$) ratios for the Kasner parameter $u$
on the Bianchi type $\mathrm{I}$ and $\mathrm{II}$ boundary subsets of
Bianchi types $\mathrm{VIII}$ and $\mathrm{IX}$.
\textbf{Left:} For the gold ratio, there are two heteroclinic cycles related
by axis permutations (one in black and one in gray).
\textbf{Right:} For the silver ratio there are three heteroclinic cycles
related by axis permutations (one in black and two in gray).
These two examples illustrate that by quoting out the permutation symmetery
there is a one-to-one correspondence between spatially frame-invariant
Kasner sequences and heteroclinic chains, and hence, in contrast to
type $\mathrm{VI}_{-1/9}$, no isolated heteroclinic chains (cf. the
above Figure for the silver ratio and those for type $\mathrm{VI}_{-1/9}$
in section~\ref{sec:silverrat}).
}\label{fig:period3IX}
\end{figure}
\vspace{-0.3cm}
These simple heteroclinic structures are in stark contrast to the complicated networks of heteroclinic cycles induced by the gold and silver ratios
in type $\mathrm{VI}_{-1/9}$, illustrated in Figures~\ref{FIG:graphper3},
\ref{fig:period3network} (right), \ref{fig:period3}
and Figures~\ref{FIG:graphper5}, \ref{fig:period5network} (lower right),
\ref{fig:period5}, respectively, where the latter silver ratio
illustrates the new phenomenon of stable heteroclinic subnetworks, due to the
lack of permutation symmetries and the existence of multiple transitions.

As the silver ratio is the simplest cyclic heteroclinic network that admits a removal process in Bianchi type $\mathrm{VI}_{-1/9}$, we discuss its consequences and compare with Bianchi types $\mathrm{VIII}$ and $\mathrm{IX}$. 
For Bianchi types $\mathrm{VIII}$ and $\mathrm{IX}$, there exist three period-4 heteroclinic cyles (see Figure~\ref{fig:period3IX}, right), each of which admits a codimension-one stable manifold as shown by Liebscher et al., see~\cite{Lieb11}.
Note that these three disjoint period-4 cycles exhaust the full heteroclinic network of the silver ratio for these models.
In contrast, for Bianchi type $\mathrm{VI}_{-1/9}$, while one may expect that a suitable modification of the result of Liebscher et al. would still provide stable manifolds for certain heteroclinic chains, our result in Theorem~\ref{thm:cyclic} identifies an isolated heteroclinic structure within the silver ratio network that cannot be asymptotically approximated in such a way as described by Liebscher et al., see Figures~\ref{FIG:graphper5}, \ref{fig:period5network}, \ref{fig:period5}.

These examples also illustrate that the methods used in the new generation
of singularity proofs for the Bianchi types $\mathrm{VIII}$ and $\mathrm{IX}$
vacuum models do not work for Bianchi type $\mathrm{VI}_{-1/9}$. These
proofs use the permutation symmetries on the type I and II subsets, which when
quotient out yields a one-to-one correspondence between heteroclinic chains
and frame-invariant Kasner sequences $\{u_0,u_1,u_2\dots\}$, making it possible to
use $u$ as a state space coordinate. Due to the lack of permutation
symmetries, this is no longer possible for type
$\mathrm{VI}_{-1/9}$. Furthermore, the one-to-one correspondence between Kasner
sequences $\{u_0,u_1,u_2\dots\}$ and heteroclinic chains in
Bianchi types $\mathrm{VIII}$ and $\mathrm{IX}$ results in that
there are no isolated parts of the heteroclinic network of these
vacuum models.

In addition, the proofs in Bianchi types $\mathrm{VIII}$ and $\mathrm{IX}$
rely on a certain ordering and relative sizes of the eigenvalues of
the linearization at $\mathrm{K}^\ocircle$. These structures
do not exist for Bianchi type $\mathrm{VI}_{-1/9}$,\footnote{Note that not
all eigenvalues in Table~\ref{tab:per3EV} pertain to the physical state space
due to the constraints; in particular, there is one extra eigenvalue associated
with the Codazzi constraint which does not exist for Bianchi type $\mathrm{VIII}$
and $\mathrm{IX}$. Note further, that the tangentialities to $\mathrm{K}^\ocircle$
of the single frame transitions are different than that for the single curvature
transitions, see Figures~\ref{KMaps} and~\ref{fig:CurvatureTrans}, which,
e.g., implies that the structure near the Taub point $\mathrm{T}_3$
is different and more problematic for type $\mathrm{VI}_{-1/9}$ than it is
for type $\mathrm{VIII}$ and $\mathrm{IX}$.} where the corresponding linearization
even give rise to \emph{two} unstable variables in two sectors of the Kasner
circle $\mathrm{K}^\ocircle$, and associated multiple transitions. Furthermore,
one also has to deal with the new phenomenon of isolated heteroclinic chains in
Bianchi type $\mathrm{VI}_{-1/9}$, which, e.g., leads to that cyclic networks
corresponding to Kasner sequences $\{u_0,u_1,\dots\}$ with a period two or more result
in stable heteroclinic subnetworks.\footnote{The existence of isolated heteroclinic
networks in the Bianchi type $\mathrm{VI}_{-1/9}$ vacuum model suggests that
the Conjectures~\ref{conj1} or~\ref{conj2} may not fully capture the generic
asymptotic dynamics toward the type $\mathrm{VI}_{-1/9}$ singularity.}

Due to the lack of permutation symmetries in Bianchi type $\mathrm{VI}_{-1/9}$,
era changes, and thereby also the critical value $u=2$, play a different
role for these models than in types $\mathrm{VIII}$ and $\mathrm{IX}$. In
contrast to the latter models, the stability properties of each sector of
$\mathrm{K}^\ocircle$ in type $\mathrm{VI}_{-1/9}$ are different and thus
the division of each sector with the six values of $\ue$ corresponding to
$u=2$, given in Table~\ref{ueandu}, have different, complicated,
dynamical consequences.

Finally, note that since there is no removal procedure and no
multiple transitions for the Bianchi types $\mathrm{VIII}$ and $\mathrm{IX}$
models, there is no use for the hexagonal representation, which
has been crucial for identifying the heteroclinic Bianchi
type I and II structures for Bianchi type $\mathrm{VI}_{-1/9}$.

\begin{appendix}

\section{A toy model\label{app:toy}}

In contrast to Bianchi types $\mathrm{VIII}$ and $\mathrm{IX}$, the Bianchi type
$\mathrm{VI}_{-1/9}$ vacuum model does not admit a Hamiltonian formulation, see
e.g.~\cite{jan01} and references therein. However, in~\cite{heietal09} a `dominant'
Hamiltonian was presented yielding a toy model that leads to a Hubble-normalized
dynamical system with an invariant subset stratification that includes the same
Bianchi type $\mathrm{I}$ and $\mathrm{II}$ subsets as exhibited by the Bianchi type
$\mathrm{VI}_{-1/9}$ vacuum model (we will use these relationships in order to obtain
monotonic functions for these subsets in the next appendix);
moreover, heuristic considerations in~\cite{heietal09} suggest that the toy and
type $\mathrm{VI}_{-1/9}$ models share the past generic asymptotic features.
The toy Hamiltonian can be written as follows:\footnote{Obtained by using the
Misner parametrization $-b_1 = \beta^0 + \beta^+ - \sqrt{3}\beta^-$,
$-b_2 = \beta^0 - 2\beta^+$,
$-b_3 = \beta^0 + \beta^+ + \sqrt{3}\beta^-$ in~\cite{heietal09},
and hence $\Sigma_1 = \Sigma_+ - \sqrt{3}\Sigma_-$,
$\Sigma_2 = -2\Sigma_+$, $\Sigma_3 = \Sigma_+ + \sqrt{3}\Sigma_-$.}
\begin{equation}\label{HamDom}
\begin{split}
H_\mathrm{Dom} &= \frac12\left(-p_0^2 + p_+^2 + p_-^2\right) \\ &\quad + \frac12e^{2\sqrt{3}\beta^-}
\left(A_1^2 e^{6\beta^+} + A_3^2 e^{-6\beta^+}\right) + \frac12A_-^2e^{4(\beta^0 + \beta^+ - \sqrt{3}\beta^-)} = 0,
\end{split}
\end{equation}
where 
the future directed time variable $t_T$ is called the Taub time. Introducing the variables
\begin{subequations}
\begin{alignat}{4}
\Sigma_+ &= \frac{p_+}{-p_0}, &\qquad \Sigma_- &= \frac{p_-}{-p_0}, &\qquad & \\
R_1 &= \frac{A_1e^{2\sqrt{3}\beta^- + 3\beta^+}}{-p_0},&\qquad R_3 &= \frac{A_3e^{3\beta^+ - \sqrt{3}\beta^-}}{-p_0}, &\qquad
N_- &= \frac{A_-e^{2(\beta^0 + \beta^+ - \sqrt{3}\beta^-}}{-p_0},
\end{alignat}
\end{subequations}
and the past directed time variable $\tau$, defined by
\begin{equation}
\frac{d\tau}{dt_T} = -\frac{d\beta^0}{dt_T} = p_0,
\end{equation}
leads to the following dominant system of evolution equations (after calculating
and using the Hamiltonian equations from~\eqref{HamDom})
for the state vector $(\Sigma_+,\Sigma_-,R_1,R_3,N_-)$:
\begin{subequations}\label{Domdynsys}
\begin{align}
\Sigma_+^\prime &= 2(1 - \Sigma^2)\Sigma_+ - 3R_3^2 + 3R_1^2 + 2N_-^2,\\
\Sigma_-^\prime &= 2(1 - \Sigma^2)\Sigma_- + \sqrt{3}\left(R_1^2 + R_3^2\right) - 2\sqrt{3}N_-^2,\\
R_1^\prime &= [2(1 - \Sigma^2) - 3\Sigma_+ - \sqrt{3}\Sigma_-]R_1,\\
R_3^\prime &= [2(1 - \Sigma^2) + 3\Sigma_+ - \sqrt{3}\Sigma_-]R_3,\label{R3toy}\\
N_-^\prime &= -2(\Sigma^2 + \Sigma_+ - \sqrt{3}\Sigma_-)N_-,\label{N-toy}
\end{align}
\end{subequations}
and the constraint
\begin{equation}\label{app:const}
1 - \Sigma^2 - N_-^2 = 0,
\end{equation}
where $\Sigma^2 := \Sigma_+^2 + \Sigma_-^2 + R_1^2 + R_3^2$.
%
%
Replacing $\Sigma_\pm$ with $\Sigma_1, \Sigma_2, \Sigma_3$ according to
\begin{equation}
\Sigma_1 = \Sigma_+ - \sqrt{3}\Sigma_-,\qquad
\Sigma_2 = -2\Sigma_+,\qquad \Sigma_3 = \Sigma_+ + \sqrt{3}\Sigma_-,
\end{equation}
results in the evolution equations
\begin{subequations}\label{BilliardeqsolevolHnormVItoy}
\begin{align}
\Sigma_1^\prime &= 2(1 - \Sigma^2)\Sigma_1 - 6R_3^2 + 8N_-^2,\label{Sig1eqtoy}\\
\Sigma_2^\prime &= 2(1 - \Sigma^2)\Sigma_2 + 6R_3^2 - 6R_1^2 - 4N_-^2,\label{Sig2eqtoy}\\
\Sigma_3^\prime &= 2(1 - \Sigma^2)\Sigma_3 + 6R_1^2 - 4N_-^2,\label{Sig3eqtoy}\\
R_1^\prime &= [2(1 - \Sigma^2) + \Sigma_2 - \Sigma_3]R_1,\label{R1eqtoy}\\
R_3^\prime &= [2(1 - \Sigma^2) + \Sigma_1 - \Sigma_2]R_3,\\
N_-^\prime &= -2(\Sigma^2 + \Sigma_1)N_-,
\end{align}
\end{subequations}
where $(\Sigma_1, \Sigma_2, \Sigma_3, R_1, R_3, N_-)\in\mathbb{R}^6$
is subject to the constraint equations:
\begin{subequations}\label{billiardeqsolconsHnormvacVItoy}
\begin{align}
1 - \Sigma^2 - N_-^2 &= 0,\label{Gausstoy}\\
\Sigma_1 + \Sigma_2 + \Sigma_3 &= 0.\label{Sigconstrtoy}
\end{align}
\end{subequations}
These equations are the same as the corresponding ones
in~\eqref{BilliardeqsolevolHnormVI} and~\eqref{billiardeqsolconsHnormvacVI}
for Bianchi type $\mathrm{VI}_{-1/9}$ when one sets $A=0$, but note the lack
of a constraint equation corresponding to the Codazzi
constraint~\eqref{Codazzi1} in Bianchi type $\mathrm{VI}_{-1/9}$.
In spite of this, due to that $A=0$, the state space is
4D, just as that for the Bianchi type
$\mathrm{VI}_{-1/9}$ vacuum model.

The dynamical system for the toy model share the same Bianchi type $\mathrm{I}$ and
$\mathrm{II}$ subsets as the type $\mathrm{VI}_{-1/9}$ vacuum model,
obtained by setting $N_-=0$ and $R_3 =0$, respectively, and intersections thereof.
However, it also admits a different invariant 3D subset obtained by setting $R_1=0$, which
we refer to as the $\mathcal{T}_{R_3N_-}$ subset.

The toy dynamical system is invariant under the transformations
\begin{equation}
R_1\mapsto - R_1,\qquad R_3\mapsto - R_3,\qquad N_-\mapsto - N_-.
\end{equation}
%
These discrete symmetries lead to an invariant set stratification that resembles the simple
hierarchical set stratification of the class A models rather than the more
complicated one for the Bianchi type $\mathrm{VI}_{-1/9}$ vacuum model. 
This is due to that taking the intersection of the three
invariant 3D subsets, $\mathcal{T}_{R_1R_3}$ with $N_-=0$,
$\mathcal{T}_{R_1N_-}$ with $R_3=0$, $\mathcal{T}_{R_3N_-}$ with $R_1=0$,
yield the 2D subsets $\mathcal{T}_{R_1}$, $\mathcal{T}_{R_3}$,
$\mathcal{T}_{N_-}$, where intersections of any of these 2D
subsets yield the 1D Kasner circle $\mathrm{K}^\ocircle$ with
$N_-=R_1=R_3=0$; see Figure~\ref{FIG:hierarchyToy}.
\begin{figure}[H]
\centering
\begin{tikzpicture}
\node (dim) at (6, 6.6) {\underline{\footnotesize{Dimension:}}};
\node (4) at (6, 6) {\footnotesize{$4$}};
\node (3) at (6, 4.5) {\footnotesize{$3$}};
\node (2) at (6, 3) {\footnotesize{$2$}};
\node (1) at (6, 1.5) {\footnotesize{$1$}};

\node (VI19) at (-2, 6) {\boxed{\text{\footnotesize{Toy Model}}}};

\node (R1R3) at (-4, 4.5) {\boxed{\text{\footnotesize{$\mathcal{T}_{R_1R_3}$}}}};
\node (R1N-) at (-2, 4.5) {\boxed{\text{\footnotesize{$\mathcal{T}_{R_1N_-}$}}}};
\node (OT) at (0, 4.5) {\boxed{\text{\footnotesize{$\mathcal{T}_{R_3N_-}$}}}};

\node (R1) at (-4, 3) {\boxed{\text{\footnotesize{$\mathcal{T}_{R_1}$}}}};
\node (R3) at (-2, 3) {\boxed{\text{\footnotesize{$\mathcal{T}_{R_3}$}}}};
\node (N-) at (-0, 3) {\boxed{\text{\footnotesize{$\mathcal{T}_{N_-}$}}}};

\node (K0) at (-2, 1.5) {\boxed{\text{\footnotesize{$\mathrm{K}^{\ocircle}$}}}};


\draw[-] (VI19) -- (R1R3);\draw[-] (VI19) -- (R1N-);\draw[-] (VI19) -- (OT);


\draw[-] (R1R3) -- (R1);\draw[-] (R1R3) -- (R3);\draw[-] (R1N-) -- (N-);\draw[-] (R1N-) -- (R1);

\draw[-] (R1) -- (K0);\draw[-] (R3) -- (K0); \draw[-] (N-) -- (K0);

\draw[-] (OT) -- (N-);\draw[-] (OT) -- (R3);


\end{tikzpicture}
\captionof{figure}{The invariant set stratification diagram for the toy model,
which is more similar to the stratification diagram for the class A vacuum models
in Figure~\ref{FIG:hierarchy2} than that of type $\mathrm{VI}_{-1/9}$ in Figure~\ref{FIG:hierarchy}.
The difference with the type $\mathrm{VI}_{-1/9}$ stratification diagram
is that the set $\mathcal{T}_{R_3N_-}$ replaces the subsets $\mathcal{OT}$, $\mathcal{HO}$ and
$\mathcal{D}$. To obtain the invariant subsets from one strata to the next,
one of the non-zero variables $R_1,R_3,N_-$ is set to zero, where the indices in the
transition nomenclature indicate which of these variables that are non-zero.}\label{FIG:hierarchyToy}
\end{figure}

Linearization of the system~\eqref{BilliardeqsolevolHnormVItoy} at an arbitrary point
$(p_1,p_2,p_3)\in\mathrm{K}^{\ocircle}$ yields the same results as for type $\mathrm{VI}_{-1/9}$
for $R_1$, $R_3$ and $N_-$, given by equations~\eqref{linR1}, \eqref{linR3} and \eqref{linN-},
since $A\equiv 0$. 
%
%
The analysis of the stability of the Kasner circle $\mathrm{K}^{\ocircle}$
for these models is thereby also summarized by Figure~\ref{fig:sectors}.

As in the Bianchi type $\mathrm{VI}_{-1/9}$ vacuum case, there is a
fixed point that is a local source, which we refer to as $\mathrm{RT}_*$,
in analogy with the nomenclature for the fixed point $\mathrm{RT}$ in type
$\mathrm{VI}_{-1/9}$;\footnote{However, note that $R_3 \neq 0$ for
$\mathrm{RT}_*$, which is in contrast to $\mathrm{RT}$ for which $R_3=0$; as
opposed to the type $\mathrm{VI}_{-1/9}$ vacuum case where $\mathrm{RT}$
is a source for the $\mathcal{HO}$ and $\mathcal{D}$ subsets, $R_1R_3N_- \neq 0$
implies that $\mathrm{RT}_*$ is not a local source for any of the invariant
toy model subsets since it does not belong to these.} it is given by
\begin{equation}\label{Toyfixedpoint}
\mathrm{RT}_* := \left\{ (\Sigma_+, \Sigma_-, R_1, R_3, N_-)
= \left(0,\frac{2}{3\sqrt{3}},\frac{2}{3\sqrt{3}},\frac{\sqrt{10}}{3\sqrt{3}},\frac{1}{\sqrt{3}}\right) \right\}.
\end{equation}

Next we derive a monotonic function using the methods in~\cite[ch. 10]{waiell97} and
in~\cite{heiugg10}. 
As the first step we therefore make a boost transformation in the $\beta^-$ direction in the
projected diagonal minisuperspace characterized by the metric $\eta_{AB} = \mathrm{diag}[-1,1,1]$
of the kinetic part of the Hamiltonian (whose form thereby is preserved), i.e.,
\begin{equation}
\bar{\beta}^0 = \gamma(\beta^0 - v\beta^-)\qquad \text{ and } \qquad
\bar{\beta}^- = \gamma(-v\beta^0 + \beta^-).
\end{equation}
We then set
\begin{equation}
v = 2/(3\sqrt{3}) \qquad \text{and} \qquad \gamma = \frac{1}{\sqrt{1-v^2}} = \sqrt{\sfrac{27}{23}},
\end{equation}
in order for the potential to explicitly exhibit a conformal exponential factor with a
timelike variable with respect to the minisuperspace metric $\eta_{AB}$. The Hamiltonian
now takes the form:
\begin{equation}\label{HamDomconf}
\begin{split}
H_\mathrm{Dom} &= \frac12\left(-\bar{p}_0^2 + \bar{p}_+^2 + \bar{p}_-^2\right) \\
&\quad +
\frac12 e^{\sfrac{4\sqrt{3}}{\sqrt{23}}\bar{\beta}^0}
\left(e^{\sfrac{18}{\sqrt{23}}\bar{\beta}^-}\left(A_1^2 e^{6\bar{\beta^+}} + A_3^2 e^{-6\beta^+}\right)
+ A_-^2 e^{4\bar{\beta}^+ - \sfrac{28}{\sqrt{23}}\bar{\beta}^-}\right) = 0,
\end{split}
\end{equation}
Finally, we exploit the results in~\cite[ch. 10]{waiell97} and in~\cite{heiugg10}
to use the above form of the potential to derive the following monotonic function
\begin{equation}\label{Delta}
\Delta := \frac{(1 -v\Sigma_-)^2}{(R_1^4R_3^{10}N_-^{9})^{\frac{2}{23}}},
\end{equation}
which evolves according to
\begin{equation}
\Delta^\prime = \frac{4}{69}\left(\frac{27(\Sigma_- - v)^2 + 23\Sigma_+^2}{1 - v\Sigma_-}\right) \Delta ,
\end{equation}
where we recall that $v = 2/(3\sqrt{3})$. Hence $\Delta$ is monotonically growing,
except at the fixed point $\mathrm{RT}_*$ given by~\eqref{Toyfixedpoint},
where $\Delta$ takes its minimum value, which happens asymptotically
when $\tau\rightarrow -\infty$. Thus, using the monotonicity principle
in~\cite{waiell97} we find that the fixed point $\mathrm{RT}_*$ is not
just a local source but a global one --- all interior toy model solutions
originate from this fixed point.

Moreover, $\Delta\rightarrow\infty$ when $\tau\rightarrow +\infty$
toward the initial singularity, which implies that
\begin{equation}\label{PROD}
\lim_{\tau\rightarrow +\infty}R_1R_3N_- = 0.
\end{equation}
As a consequence this monotonic function plays a similar role as the monotonic
function $N_1N_2N_3$ does for the Bianchi types $\mathrm{VIII}$ and $\mathrm{IX}$
models, which shows that the attractor toward the singularity in these models
must reside on the union of the type $\mathrm{VII}_0$ subsets in type $\mathrm{IX}$
and the union of the single type $\mathrm{VII}_0$ subset and the two type $\mathrm{VI}_0$
subsets in type $\mathrm{VIII}$. Here equation~\eqref{PROD} leads to the following Proposition:
\begin{proposition}
The global attractor $\mathcal{A}$ toward the singularity of the present toy model resides in
$\overline{\mathcal{T}}_{R_1R_3}\cup\overline{\mathcal{T}}_{R_1N_-}\cup\overline{\mathcal{T}}_{R_3N_-}$,
where $N_-=0$, $R_3=0$, or $R_1=0$ in respective subset.
\end{proposition}
To obtain an attractor that resides in the subsets that coincide with
the conjectured Bianchi type $\mathrm{VI}_{-1/9}$ attractor, i.e.,
$\overline{\mathcal{T}}_{R_1R_3}\cup\overline{\mathcal{T}}_{R_1N_-}$,
by using similar methods as in~\cite{rin01,heiugg09} for Bianchi
type $\mathrm{IX}$, requires ruling out that $\mathcal{T}_{R_3N_-}$ is part
of the attractor.

Next, we derive more information about the subset $\mathcal{T}_{R_3N_-}$, where $R_1=0$.
In this subset, we use the constraint~\eqref{Gausstoy} to solve for $N_-^2$ in equation~\eqref{Sig3eqtoy}, which
implies that 
\begin{equation}
(2 - \Sigma_3)^\prime = 2(1 - \Sigma^2)(2 - \Sigma_3),\label{Sig3eqtoyR10}
\end{equation}
where $\Sigma^2 = \frac16(\Sigma_1^2 + \Sigma_2^2 + \Sigma_3^2) + R_3^2 = 1 - N_-^2<1$
on $\mathcal{T}_{R_3N_-}$. Since $2-\Sigma_3 > 0$ it follows that 
$2-\Sigma_3$ is strictly monotonically increasing on this subset, 
but $2 - \Sigma_3$ is constant on the subset $\mathcal{T}_{R_3}$ where
$\Sigma^2 = \frac16(\Sigma_1^2 + \Sigma_2^2 + \Sigma_3^2) + R_3^2 = 1$ and, of course,
also on $\mathrm{K}^\ocircle$ where $R_3 = N_- = 0$ and
$\Sigma^2 = \frac16(\Sigma_1^2 + \Sigma_2^2 + \Sigma_3^2) = 1$, which includes the Taub point
$\mathrm{T}_3$ for which $\Sigma_3 = 2$. It also follows that $2-\Sigma_3$ is strictly
monotonically increasing on the $\mathcal{T}_{N_-}$ subset where
$\Sigma^2 = \frac16(\Sigma_1^2 + \Sigma_2^2 + \Sigma_3^2)= 1 - N_-^2<1$.
Additional information comes from linearization of the equations at $\mathrm{K}^\ocircle$,
which shows that $\mathrm{K}^{\ocircle}$ has a single unstable variable, $R_3$ or $N_-$, 
everywhere except at $\mathrm{T}_3$ and on the sector $(312)$, which is stable for the subset
$\mathcal{T}_{R_3N_-}$, and its boundary points $\mathrm{Q}_3$ and $\mathrm{T}_2$, 
which is similar to the $\mathcal{OT}$ vacuum case, see Figure~\ref{OTasymp}
(although the toy model has no arc of $\mathrm{PW}^\pm$ fixed points).
Hence, applying the monotonicity principle in~\cite{waiell97} to the monotonically
increasing function $2 - \Sigma_3$, in combinations with using the properties of
the $\mathcal{T}_{R_3}$ and $\mathcal{T}_{N_-}$ subsets and the local stability
analysis of $\mathrm{K}^\ocircle$, leads to the following lemma:
\begin{lemma}
The $\alpha$-limit set (in $\tau$) for all orbits on $\mathcal{T}_{R_3N_-}$ is the
Taub fixed point $\mathrm{T}_3$. The $\omega$-limit set for all orbits on
$\mathcal{T}_{R_3N_-}$ is the stable sector $(312)$.
\end{lemma}
Note that $\mathrm{T}_3$ here replaces $\overline{PW}^\pm$ in
the $\mathcal{OT}$ Bianchi type $\mathrm{VI}_{-1/9}$
vacuum subset as the $\alpha$-limit set.
Unfortunately, this result does not suffice to establish a proof that the
toy model's equations~\eqref{BilliardeqsolevolHnormVItoy}, 
\eqref{billiardeqsolconsHnormvacVItoy} obey one of the following conjectures (similar to
Conjectures~\ref{conj1} and~\ref{conj2} for the Bianchi type $\mathrm{VI}_{-1/9}$
vacuum model) for the global attractor ${\cal A}$:
\begin{conjecture}\label{conj1toy}
%
${\cal A} \subseteq 
\mathrm{K}^\ocircle\cup{\cal T}_{R_1}\cup{\cal T}_{R_3}\cup{\cal T}_{R_1R_3}\cup{\cal T}_{N_-}\cup{\cal T}_{R_1N_-}$,
%
\end{conjecture}
or the stronger conjecture:
\begin{conjecture} \label{conj2toy}
%
${\cal A} \subseteq 
\mathrm{K}^\ocircle\cup{\cal T}_{R_1}\cup{\cal T}_{R_3}\cup{\cal T}_{N_-}$.
%
\end{conjecture}
Similar heuristic dynamical arguments as in~\cite{heietal09} leads to the billiard formulation of~\cite{dametal03}, which is the configuration space version of Conjecture~\ref{conj2toy}.
This is further supported by the analysis in~\cite{HLU22}.
Hence, this suggests that Conjecture~\ref{conj2toy} is the correct description of the singularity attractor of this toy model.
Since we expect that the toy model faithfully reproduces the generic asymptotic dynamics of the Bianchi $\mathrm{VI}_{-1/9}$ model, this results in further support for Conjecture~\ref{conj2}.

To prove that the attractor at least resides 
on ${\cal A}$ in Conjecture~\ref{conj1toy} requires that one establishes 
that $\lim_{\tau\rightarrow\infty}(R_3N_-)=0$.
This entails investigating the evolution of $R_3N_-$
in the full state space. Using the variables $\Sigma_\pm$ and
eq.~\eqref{R3toy} and~\eqref{N-toy}, yields
\begin{equation}
|R_3N_-|^\prime = -4\left[(\Sigma_+ -1/8)^2 + (\Sigma_- - \sqrt{3}/8)^2 + R_1^2 + R_3^2 - \frac{9}{16}\right]|R_3N_-|,\label{crossR3N-}
\end{equation}
which shows that $|R_3N_-|$ only increases inside a ball of radius $3/4$ centered at
$(\Sigma_+,\Sigma_-,R_1,R_3) = (1/8,\sqrt{3}/8,0,0)$, depicted in
Figure~\ref{fig:unstableCROSSinc}.
\begin{figure}[H]
	\centering
\begin{tikzpicture}[scale=1.1]

        \filldraw [gray!25,domain=0:6.28,variable=\t,smooth] plot ({(0.125)+(0.75)*sin(\t r)},{-(0.216)+(0.75)*cos(\t r)});

        \draw [black,domain=0:6.28,variable=\t,smooth] plot ({(0.125)+(0.75)*sin(\t r)},{-(0.216)+(0.75)*cos(\t r)});

	\draw [very thin] (1,0) arc (0:360:1cm and 1cm);	 	
	
   	\draw (-0.95,0) -- (-1.05,0) node[anchor= east] {\scriptsize{$\mathrm{T}_1$}};
        \draw (0.95,0) -- (1.05,0);

        \draw[rotate=120] (-0.95,0) -- (-1.05,0) node[anchor= north] {\scriptsize{$\mathrm{T}_3$}};
        \draw[rotate=240] (-0.95,0) -- (-1.05,0) node[anchor= south] {\scriptsize{$\mathrm{T}_2$}};

        \draw[rotate=120] (0.95,0) -- (1.05,0);
        \draw[rotate=240] (0.95,0) -- (1.05,0);

    \draw[color=gray,->]
    (0,0) -- (-0.343,0)  node[anchor= east] {\scriptsize{$\Sigma_{1}$}};
    \draw[rotate=240,color=gray,->]
    (0,0) -- (-0.343,0)  node[anchor= south] {\scriptsize{$\Sigma_{2}$}};
    \draw[rotate=-240,color=gray,->]
    (0,0) -- (-0.343,0)  node[anchor= north] {\scriptsize{$\Sigma_{3}$}};

\end{tikzpicture}

\caption{
The shaded region is a projection onto the plane
$\Sigma_1 + \Sigma_2 + \Sigma_3 = 0$ in $(\Sigma_1,\Sigma_2,\Sigma_3)$-space of
the region where the cross-term $|R_3N_-|$ increases on the $\mathcal{T}_{R_3N_-}$
subset. Note that the corresponding ball in the state space only intersects the Kasner circle at the Taub point $\mathrm{T}_3$, where they are tangential.}
\label{fig:unstableCROSSinc}
\end{figure}
%
The challenge now is to establish that the evolution of generic solutions 
is dominated by the decaying state space region for $|R_3N_-|$, if true. 
In this context, note that it follows from~\eqref{crossR3N-}, 
visualized in Figure~\ref{fig:unstableCROSSinc}, that $|R_3N_-|$ is 
decreasing everywhere on $\mathrm{K}^\ocircle$ except at the Taub fixed 
point $\mathrm{T}_3$, where all eigenvalues are zero except for the 
one associated with $R_1$ since linearization then yields 
$R_1^\prime = 3R_1$, i.e., $\mathrm{T}_3$ is a centre saddle.

The above is analogous to the first steps in the proof of the Bianchi type
$\mathrm{IX}$ attractor theorem in~\cite{rin01}, although in that case
there were three identical Bianchi type $\mathrm{VII}_0$ subsets where only
one of them therefore needed to be perturbed. Note also that for type
$\mathrm{VII}_0$ the Taub points are replaced with Taub-like lines of fixed
points, see~\cite{heiugg09} for a detailed local analysis of these lines of
fixed points and an abbreviated type $\mathrm{IX}$ attractor proof. However,
the present situation more resembles the relationship between the Bianchi 
type $\mathrm{VI}_0$ subsets in Bianchi type $\mathrm{VIII}$ than  
that for Bianchi type $\mathrm{IX}$ and its $\mathrm{VII}_0$ subsets.  
Since there are no similar proofs for an attractor theorem for 
Bianchi type $\mathrm{VIII}$ as those in~\cite{rin01,heiugg09} for 
type $\mathrm{IX}$, and since the multiple transitions for the 
$\mathcal{T}_{R_1R_3}$ and $\mathcal{T}_{R_1N_-}$ subsets cause 
difficulties for proofs of the types given 
in~\cite{Beguin10,Lieb11,Lieb13,ReitererTrubowitz,bre16,BegDut22}, 
the present toy model poses considerable difficulties when it 
comes to singularity theorems, even though it is a much simpler 
model than the Bianchi type $\mathrm{VI}_{-1/9}$ vacuum model.

\section{Dynamics in invariant subsets\label{app:inv}}

\subsection{The Kasner subset $\mathcal{T}_{R_1R_3}$}\label{app:Kasnersubset}

This invariant subset occurs when $N_-=0$ in~\eqref{Domdynsys} and 
it coincides with the subset when $A=N_-=0$ in the original 
system~\eqref{BilliardeqsolevolHnormVI}. We will here give an explicit 
description of the solutions of this subset using their Hamiltonian structure.
Setting $A_-=0$ (and hence $N_-=0$) in the dominant Hamiltonian~\eqref{HamDom} leads
to
\begin{equation}\label{KHam}
H_{{\cal K}} = \sfrac12\left(-p_0^2 + p_+^2 + p_-^2\right) + \sfrac12e^{2\sqrt{3}\beta^-}
\left(A_1^2 e^{6\beta^+} + A_3^2 e^{-6\beta^+}\right) = 0,
\end{equation}
which is a Hamiltonian that correctly describes the Kasner subset ${\cal K}$. By
setting $B^2 = |A_1A_3|$, $C = \ln(|A_1/A_3|)$, this Hamiltonian can be written as
\begin{equation}\label{KHam2}
H_{{\cal K}} = \sfrac12\left(-p_0^2 + p_+^2 + p_-^2\right) + B^2e^{2\sqrt{3}\beta^-}
\cosh(6\beta^+ + C) = 0,
\end{equation}
Since $\beta^0$ is a cyclic variable, it follows that the conjugate momentum
$p_0$ is conserved, which leads to the reduced Hamiltonian
\begin{equation}\label{KHam3}
H_\mathrm{red} = \sfrac12\left(p_+^2 + p_-^2\right) + B^2e^{2\sqrt{3}\beta^-}
\cosh(6\beta^+ + C) = \sfrac12p_0^2 = E = \mathrm{constant}.
\end{equation}
We note that since $\cosh(\cdot )$ is a symmetric function it follows that there exists
a discrete symmetry, which in the dynamical system~\eqref{BilliardeqsolevolHnormVI},
with $A=0=N_-$, i.e., the Kasner subset ${\cal K}$, corresponds to
invariance under interchanging $1$ and $3$ and making the transformation
$(\Sigma_1,\Sigma_2,\Sigma_3) = - (\Sigma_1,\Sigma_2,\Sigma_3)$
(or letting $\tau-\rightarrow - \tau$). Moreover,
there exists a special solution corresponding to the minimum of
$\cosh(6\beta^+ + C)$, i.e., at $6\beta^+ = -C = -\ln(|A_1/A_3|)$ with $p_+=0$.
This leads to that $\Sigma_+=0$ with the present Misner parametrization, which
corresponds to that $\Sigma_2=0$ and hence $\Sigma_1= -\Sigma_3$,
while $R_1 = R_3 >0$, as also follows from the equations obtained by
specializing~\eqref{BilliardeqsolevolHnormVI} to the Kasner subset
${\cal K}$.

The above Hamiltonian problem is not obviously solvable. However, as pointed out in~\cite{lim15},
the solutions, which correspond to double frame transitions, can be obtained from the
original Kasner solution in a diagonal and Fermi-propagated frame by means of a
coordinate transformation. We here present the solution for $\Sigma_\alpha$, $R_1$ and $R_3$
in the time variable $\tau$ used in the main text in a simple and transparent form:\footnote{The
time variable $\tau=\tau_\mathrm{Lim}$ in~\cite{lim15} is related to the present time
$\tau$ according to $\tau = \frac{1}{12}(w^2 + 3)\tau_\mathrm{Lim}$, where the parameter
$w$ is related to $\ue$ according to $w = (1-\ue)/(1+\ue)$, $\ue \in (-1,-\frac12)$.}
\begin{subequations}\label{AppB:solTR1R3}
\begin{align}
\Sigma_1 &= -1 + 3\left[\frac{p_1n_{20}^2 e^{-6p_1\tau} + p_2n_{30}^2 e^{-6p_2\tau} + p_3 e^{-6p_3\tau}}
{n_{20}^2 e^{-6p_1\tau} + n_{30}^2 e^{-6p_2\tau} + e^{-6p_3\tau}}\right],\\
\Sigma_3 &= -1 + 3\left[\frac{p_1 e^{6p_1\tau} + p_2n_{10}^2 e^{6p_2\tau} + p_3(n_{10}n_{30} - n_{20})^2e^{6p_3\tau}}
{e^{6p_1\tau} + n_{10}^2 e^{6p_2\tau} + (n_{10}n_{30} - n_{20})^2e^{6p_3\tau}}\right],\\
R_1 &= \frac{1}{\sqrt{3}}e^{-3\tau}\left[\frac{(3p_2-1-\Sigma_3)n_{10} e^{6p_2\tau} + (3p_3-1-\Sigma_3)n_{30}(n_{10}n_{30} - n_{20})e^{6p_3\tau} }
{\sqrt{n_{20}^2e^{-6p_1\tau} + n_{30}^2 e^{-6p_2\tau} + e^{-6p_3\tau}}}\right],\\
R_3 &= \frac{1}{\sqrt{3}}e^{3\tau}\left[\frac{(3p_1+1+\Sigma_1)n_{10}n_{20}e^{-6p_1\tau} + (3p_3+1+\Sigma_3)n_{30}e^{6p_3\tau}}
{\sqrt{e^{6p_1\tau} + n_{10}^2 e^{6p_2\tau} + (n_{10}n_{30} - n_{20})^2e^{6p_3\tau}}}\right],
\end{align}
\end{subequations}
while $\Sigma_2 = - \Sigma_1 - \Sigma_3$, and where $p_1, p_2, p_3$ are
Kasner parameters that belong to sector $(321)$ on
$\mathrm{K}^\ocircle$, i.e., $p_3<p_2<p_1$. These parameters are conveniently
parametrized with $\ue$ according to~\eqref{ueeq}, where we recall that
$\ue \in (-1,-\frac12)$ in sector $(321)$. The parameter $n_{20}$ is related to the
parameters $n_{10}$ and $n_{30}$ according to
\begin{equation}
n_{20} = - \left[\frac{1-\ue^2}{(2 + \ue)\ue}\right]n_{10}n_{30}.
\end{equation}
By performing a translation in $\tau$, it is possible to scale the remaining parameters
$n_{10}$ and $n_{30}$ and obtain expressions that only involves $\ue$ and one independent
parameter in the above expressions, i.e., there originates a one-parameter set of double
frame transition orbits from each Kasner point in sector $(321)$.

\subsection{The Bianchi type II subset $\mathcal{T}_{R_1N_-}$}\label{app:BII}

This invariant subset corresponds to $R_3=0$ in~\eqref{Domdynsys} 
and it coincides with the subset when $A=R_3=0$ in the original 
system~\eqref{BilliardeqsolevolHnormVI}. We will here use the Hamiltonian 
formulation in order to obtain a conserved quantity for these solutions.
By setting $A_3=0$ in the Hamiltonian~\eqref{HamDom}
(and hence $R_3=0$) we obtain a Hamiltonian that describes the Bianchi type II
subset correctly. In this case it is beneficial to adapt the metric
variables to the first direction rather than the second\footnote{We use the Misner parametrization
$-b_1 = \beta^0 - 2\beta^+$, $-b_2 = \beta^0 + \beta^+ + \sqrt{3}\beta^-$,
$-b_3 = \beta^0 + \beta^+ - \sqrt{3}\beta^-$ in~\cite{heietal09}
and hence $\Sigma_1 = -2\Sigma_+$, $\Sigma_2 = \Sigma_+ + \sqrt{3}\Sigma_-$,
$\Sigma_3 = \Sigma_+ - \sqrt{3}\Sigma_-$.}
which leads to
\begin{equation}\label{HII}
H_\mathrm{II} = \sfrac12\left(-p_0^2 + p_+^2 + p_-^2\right) + \sfrac12A_1^2 e^{-4\sqrt{3}\beta^-}
+ \sfrac12A_-^2e^{4(\beta^0 - 2\beta^+)} = 0,
\end{equation}
where, without loss of generality, $A_1>0$, and the time variable is
the Taub time, $t_T$.

Next, following~\cite{uggetal95}, we make a boost with $v = 1/2$ in the $\beta^+$-direction
in the projected diagonal minisuperspace characterized by the metric
$\eta_{AB} = \mathrm{diag}[-1,1,1]$ of the kinetic part of the Hamiltonian, i.e.,
\begin{equation}
(\bar{\beta}^0,\bar{\beta}^+,\bar{\beta}^-) =
\left(\sfrac{2}{\sqrt{3}}\left(\beta^0 - \sfrac12\beta^+\right),\sfrac{2}{\sqrt{3}}\left(-\sfrac12\beta^0 + \beta^+\right),\beta^-\right).
\end{equation}
This leads to the Hamiltonian
\begin{equation}\label{IIHam2}
H_\mathrm{II} = -\underbrace{\sfrac12\bar{p}_0^2}_{E} + \underbrace{\sfrac12\bar{p}_+^2 +
\sfrac12A_-^2e^{-4\sqrt{3}\bar{\beta}^+}}_{E_+} +
\underbrace{\sfrac12\bar{p}_-^2 + \sfrac12A_1^2 e^{-4\sqrt{3}\bar{\beta}^-}}_{E_-} = 0,
\end{equation}
where $E$, $E_\pm $ are constants due to the fact that $\bar{\beta}^0$ is a 
cyclic coordinate and because the Hamiltonian is separable in $\bar{\beta}^+$ 
and $\bar{\beta}^-$. Since the $E_\pm$ parts are formally the same, they 
result in an equivalent first order (non-linear) ODE which can be solved 
by making the variable transformation
$x_\pm = e^{2\sqrt{3}\bar{\beta}^\pm}$. The solution is given by
$\bar{\beta}^0 = -\bar{p}_0t_T$, 
while
\begin{subequations}\label{BIIintegrals1}
\begin{align}
e^{2\sqrt{3}\bar{\beta}^+} &= \sfrac{|A_-|}{\sqrt{2E_+}}\cosh(\sqrt{24E_+}t_T - \alpha_+),\\
\bar{p}_+ &= \sqrt{2E_+}\tanh(\sqrt{24E_+}t_T - \alpha_+),\\
e^{2\sqrt{3}\bar{\beta}^-} &= \sfrac{A_1}{\sqrt{2E_-}}\cosh(\sqrt{24E_-}t_T - \alpha_-),\\
\bar{p}_- &= \sqrt{2E_-}\tanh(\sqrt{24E_-}t_T - \alpha_-).
\end{align}
\end{subequations}

To express these results in the dynamical systems setting, we first
introduce the variables
\begin{subequations}
\begin{alignat}{4}
\Sigma_+ &= \frac{p_+}{-p_0}, &\qquad \Sigma_- &= \frac{p_-}{-p_0}, &\qquad & \\
R_1 &= \frac{A_1e^{-2\sqrt{3}\beta^-}}{-p_0}, &\qquad
N_- &= \frac{A_-e^{2(\beta^0 - 2\beta^+)}}{-p_0},
\end{alignat}
\end{subequations}
and the time variable $\tau$, as before defined by
$d\tau/dt_T = -d\beta^0/dt_T = p_0$.
%
%
%
%
%
%
Using the Misner parametrization 
$\Sigma_1 = -2\Sigma_+$, $\Sigma_2 = \Sigma_+ + \sqrt{3}\Sigma_-$,
$\Sigma_3 = \Sigma_+ - \sqrt{3}\Sigma_-$ solves the constraint
$\Sigma_1 + \Sigma_2 + \Sigma_3 = 0$, while using $\Sigma_1,\Sigma_2,\Sigma_3$ then
yields the equations in~\eqref{BilliardeqsolevolHnormVI} and~\eqref{Gauss1} for the
Bianchi type II case, obtained by setting $A=N_-=0$.

However, before translating the results in~\eqref{BIIintegrals1} into our original
state space variables, it is advantageous to first introduce the following variables
\begin{subequations}\label{barvarII}
\begin{alignat}{2}
\bar{\Sigma}_+ &= \frac{\bar{p}_+}{-\bar{p}_0},&\qquad \bar{\Sigma}_- &= \frac{\bar{p}_-}{-\bar{p}_0},\\
\bar{R}_1 &=  \frac{A_1e^{-2\sqrt{3}\bar{\beta}^-}}{-\bar{p}_0},&\qquad
\bar{N}_- &= \frac{A_-e^{-2\sqrt{3}\bar{\beta}^+}}{-\bar{p}_0},
\end{alignat}
\end{subequations}
and the constants
\begin{equation}
\epsilon_\pm = \sqrt{\frac{E_\pm}{E}} = \frac{\sqrt{2E_\pm}}{-\bar{p}_0},\qquad \bar{A}_1 = \frac{A_1}{-\bar{p}_0},\qquad
\bar{A}_- = \frac{A_-}{-\bar{p}_0}.
\end{equation}
This leads to
\begin{equation}
\epsilon_+ = \frac{2\ue}{1 + \ue^2},\qquad \epsilon_- = \frac{1 - \ue^2}{1 + \ue^2},
\end{equation}
and consequently $\epsilon_+^2 + \epsilon_-^2 =1$, where $\ue = \ue_- \in (0,1)$
describes the past (in $\tau$) initial state for the orbits in the original
state space in sector $(132)$, where both $R_1$ and $N_-$ are
unstable with respect to the past time direction $\tau$. Hence,
\begin{subequations}\label{solII}
\begin{align}
e^{2\sqrt{3}\bar{\beta}^+} &= \sfrac{|\bar{A}_-|}{\epsilon_+}\cosh(T),\\
\bar{\Sigma}_+ &= \epsilon_+\tanh(T),\\
e^{2\sqrt{3}\bar{\beta}^-} &= \sfrac{\bar{A}_1}{\epsilon_-}\cosh\left(\left(\sfrac{\epsilon_-}{\epsilon_+}\right)T - \alpha\right),\\
\bar{\Sigma}_- &= \epsilon_-\tanh\left(\left(\sfrac{\epsilon_-}{\epsilon_+}\right)T - \alpha\right),
\end{align}
\end{subequations}
where
\begin{equation}
\frac{\epsilon_-}{\epsilon_+} = \frac{1 - \ue^2}{2\ue},
\end{equation}
while the time parameter $T\in(-\infty,\infty)$ is defined
by\footnote{Note that $d\bar{\tau}/dt_T = -d\bar{\beta}^0/dt_T = \bar{p}_0$,
and thus $\bar{\tau}$ is a time variable with the same direction as $\tau$.}
$T := -2\sqrt{3}\epsilon_+\bar{\tau}$,
while $\alpha = \alpha_- - \alpha_+\epsilon_-/\epsilon_+$.
Here $\alpha\in (-\infty,\infty)$ yields a one-parameter set of orbits coming
from each Kasner point characterized by $\ue = \ue_- \in (0,1)$.

We then note that $\epsilon_+ = \epsilon_-$ implies that $\ue = \sqrt{2} - 1$ which
corresponds to the silver ratio $u = 1 + \sqrt{2}$, see Section~\ref{sec:silverrat}.
Moreover, at $\ue = \sqrt{2} - 1$ the two unstable eigenvalues $\lambda_{R_1}$
and $\lambda_{N_-}$ are equal in sector $(132)$. At this value of $\ue$, as for all values
of $\ue$ in sector $(132)$ there originates a one-parameter set of curvature-frame transitions.
There is, however, a special orbit in this set, namely the one with $\alpha=0$. In this case
$\bar{\Sigma}_+ = \bar{\Sigma}_-$, which leads to an orbit with
$\sqrt{3}\Sigma_- = 2\Sigma_+ - 1$ and hence $\Sigma_1 = -2(1 - \Sigma_3)$.

Apart from the constraint (which follows from the Hamiltonian~\eqref{IIHam2}
and the definitions~\eqref{barvarII}),
\begin{equation}
1 = \bar{\Sigma}_+^2 + \bar{\Sigma}_-^2 + \bar{R}_1^2 + \bar{N}_-^2,
\end{equation}
we obtain the following integral by using the hyperbolic identity
\begin{equation}\label{int1}
\bar{\Sigma}_+^2 + \bar{N}_-^2 = \epsilon_+^2.
\end{equation}
The above formulas can be transformed into the previous
variables $\Sigma_\pm$, $R_1$ and $N_-$ by using the relations
\begin{equation}
\Sigma_+ = \frac{1 + 2\bar{\Sigma}_+}{2 + \bar{\Sigma}_+},\qquad \Sigma_- = \frac{\sqrt{3}\bar{\Sigma}_-}{2 + \bar{\Sigma}_+},\qquad
R_1 = \frac{\sqrt{3}\bar{R}_1}{2 + \bar{\Sigma}_+},\qquad N_- = \frac{\sqrt{3}\bar{N}_-}{2 + \bar{\Sigma}_+}.
\end{equation}
In particular, this leads to that~\eqref{int1} can be written as
\begin{equation}\label{int2}
\frac{(1 - 2\Sigma_+)^2 + 3N_-^2}{(2 - \Sigma_+)^2} = 4\left(\frac{(1 + \Sigma_1)^2 + 3N_-^2}{(4 + \Sigma_1)^2}\right) = \epsilon_+^2.
\end{equation}
Note that $\ue = \ue_-\in (0,1)$ yields the initial
(in $\tau$, i.e., $T\rightarrow + \infty$)
point for a one-parameter set of heteroclinic orbits
in sector $(132)$ for a mixed curvature-frame transition,
described by the parameter $\alpha$ in~\eqref{solII}. However, the
integral~\eqref{int2} also holds for sector $(123)$
where $\ue \in (1,\infty)$ and for the point $\mathrm{Q}_1$
where $\ue = \ue_- = 1$, where, in this case, the conserved quantity~\eqref{int2}
determines the one-parameter set of orbits.


\subsection{The ${\cal HO}$ subset}\label{app:D}

We will now derive a monotonic function for the $\mathcal{HO}$ subset. Fortunately,
these models admit a Hamiltonian formulation, which we will now introduce. First
we use a metric parametrization adapted to the first direction, just as done in the 
previous Bianchi type II case. Then we take the Hamiltonian in~\cite{uggetal95} 
for the Fermi-propagated diagonal case $n^\alpha\!_\alpha=0$, i.e., ${\cal D}$, 
and add the potential term $\sfrac12A_1^2 e^{-4\sqrt{3}\beta^-}$, which 
corresponds to spatial frame rotations generated by $R_1$, see~\eqref{HII}. 
This results in the following Hamiltonian:
\begin{equation}\label{HamHO}
H_\mathcal{HO} = \frac12\left(-p_0^2 + p_-^2\right) + \frac12A_1^2e^{-4\sqrt{3}\beta^-}
 + \frac12A_a^2e^{4\beta^0 +  2\sqrt{3}\beta^-} = 0.
\end{equation}
Introducing the variables
\begin{equation}
\Sigma_- = \frac{p_-}{-p_0},\qquad R_1^2 = A_1^2\left(\frac{e^{-4\sqrt{3}\beta^-}}{p_0^2}\right),
\end{equation}
and the usual time variable $\tau$, leads, via the Hamiltonian equations, to the 2D
dynamical system given in~\eqref{ODE:D}.
%
%
%

Before establishing global results for ${\cal HO}$, we first recall that the
system~\eqref{ODE:D} has four fixed points
\begin{subequations}
\begin{align}
\mathrm{RT} &:= \left\{ (\Sigma_-, R_1) = \sfrac{1}{3\sqrt{3}}\left(-2,\sqrt{5}\right) \right\},\\
\mathrm{PW}^0 &:= \left\{ (\Sigma_-, R_1) = \left(-\sfrac{\sqrt{3}}{2},0\right)\right\},\\
\mathrm{K}^{\ocircle}_\pm &:= \left\{ (\Sigma_-,R_1) = (\pm 1,0)\right\}.
\end{align}
\end{subequations}
Linear fixed point analysis reveals the following: The fixed point $\mathrm{RT}$
is a local source; $\mathrm{PW}^0$ is a saddle with one orbits entering the interior
state space (its stable manifold is the invariant ${\cal OT}$ subset $R_1=0$);
$\mathrm{K}^{\ocircle}_+$ with $(\Sigma_-,R_1) = (1,0)$ is a saddle
while $\mathrm{K}^{\ocircle}_-$ with $(\Sigma_-,R_1) = (-1,0)$ is a sink.

Next we proceed as in the previous subsections in order to obtain a monotonic
function and therefore perform a boost in the $\beta^-$ direction with
$v = - 2/(3\sqrt{3}) = \Sigma_-|_\mathrm{RT}$
%
%
%
%
which leads to that the Hamiltonian in~\eqref{HamHO} takes the form
\begin{equation}\label{HamD}
H_\mathcal{HO} = \frac12\left(-\bar{p}_0^2 + \bar{p}_-^2\right)  +
\frac12e^{\frac{8\sqrt{3}}{\sqrt{23}}\bar{\beta}^0}\left(A_1^2e^{-\frac{36}{\sqrt{23}}\bar{\beta}^-}
+ A_a^2e^{\frac{10}{\sqrt{23}}\bar{\beta}^-}\right) = 0.
\end{equation}
We then use that the potential has a conformal exponential factor
$e^{\frac{8\sqrt{3}}{\sqrt{23}}\bar{\beta}^0}$ with a timelike
variable $\bar{\beta}^0$ with respect to the minisuperspace metric $\eta_{AB}$
to derive the monotonic function
\begin{equation}
M = \frac{(1 - v\Sigma_-)^2}{[R_1^5(1 - \Sigma^2)^{9}]^{\sfrac{2}{23}}},
\end{equation}
which evolves according to
\begin{equation}
M^\prime = \frac{27 \left(\Sigma_- - v\right)^2}{23\left(1 - v\Sigma_-\right)}M.
\end{equation}
Thus, $M$ is monotonically increasing, which enables a global dynamical description
of the invariant ${\cal HO}$ subset, as discussed in Section~\ref{sec:naa=0}.

\end{appendix}

\textbf{Acknowledgments.}
PL was supported by Marie Skłodowska-Curie Actions, H2020 Cofund, UNA4CAREER, 847635, 
with the project DynCosmos. 

\textbf{Competing interest and data availability.}
The authors have no conflict of interest to declare. 
Moreover, the authors confirm that the data supporting 
the findings of this study are available within the article.



\bibliographystyle{plain}
\bibliography{G2_papers}

\begin{thebibliography}{10}

\bibitem{andren01}
L.~Andersson and A.~D. Rendall.
\newblock Quiescent cosmological singularities.
\newblock {\em Commun. Math. Phys.}, \textbf{218}:479, 2001.

\bibitem{Beguin10}
F.~B\'eguin.
\newblock Aperiodic oscillatory asymptotic behavior for some {B}ianchi
  spacetimes.
\newblock {\em Class. Quantum Grav.}, \textbf{27}:185005, 2010.

\bibitem{BegDut22}
F.~B\'eguin and T.~Dutilleul.
\newblock Chaotic dynamics of spatially homogeneous spacetimes.
\newblock {\em Commun. Math. Phys.}, \textbf{399}:737--927, 2023.

\bibitem{BelHenneaux17}
V.~Belinski and M.~Henneaux.
\newblock {\em The Cosmological Singularity}.
\newblock Cambridge University Press, 2017.

\bibitem{bkl82}
V.~A. Belinski, I.~M. Khalatnikov, and E.~M. Lifshitz.
\newblock A general solution of the {E}instein equations with a time
  singularity.
\newblock {\em Adv. Phys.}, \textbf{31}:639, 1982.

\bibitem{bkl70}
V.~A. Belinski, I.M. Khalatnikov, and E.~M. Lifshitz.
\newblock Oscillatory approach to a singular point in the relativistic
  cosmology.
\newblock {\em Adv. Phys.}, \textbf{19}:525, 1970.

\bibitem{BN73}
O.~I. Bogoyavlenskii and S.~P. Novikov.
\newblock Singularities of the cosmological model of the {B}ianchi {IX} type
  according to the qualitative theory of differential equations.
\newblock {\em JETP}, \textbf{37}(5):747--755, 1973.

\bibitem{BradySmith}
P.~R. Brady and J.~D. Smith.
\newblock Black hole singularities: A numerical approach.
\newblock {\em Phys. Rev. Lett.}, \textbf{75}:1256, 1995.

\bibitem{bre16}
B.~Brehm.
\newblock {B}ianchi {VIII} and {IX} vacuum cosmologies: Almost every solution
  forms particle horizons and converges to the {M}ixmaster attractor.
\newblock {\em \href{https://arxiv.org/abs/1606.08058}{arXiv:1606.08058}},
  2016.

\bibitem{buc13}
J.~Buchner.
\newblock The 18-cycle in {B}ianchi {VI}$^*_{-1/9}$: Combined linear local
  passage and numerical simulation.
\newblock {\em \href{https://arxiv.org/abs/2503.02684}{arXiv:2503.02684}},
  2025.

\bibitem{Chesleretal}
P.~M. Chesler, R.~Narayan, and E.~Curiel.
\newblock Singularities in {R}eissner--{N}ordström black holes.
\newblock {\em Class. Quantum Grav.}, \textbf{37}(2):025009, 2020.

\bibitem{Church}
K.~E.~M. Church, O.~Hénot, P.~Lappicy, J.-P. Lessard, and H.~Sprink.
\newblock Periodic orbits in {H}ořava--{L}ifshitz cosmologies.
\newblock {\em Gen. Relativ. Gravit.}, \textbf{55}:2, 2023.

\bibitem{collins71}
C.~B. Collins.
\newblock More qualitative cosmology.
\newblock {\em Commun. Math. Phys.}, \textbf{23}(2):137--158, 1971.

\bibitem{Dafermos}
M.~Dafermos.
\newblock Black holes without spacelike singularities.
\newblock {\em Commun. Math. Phys.}, \textbf{332}:729--757, 2014.

\bibitem{dametal03}
T.~Damour, M.~Henneaux, and H.~Nicolai.
\newblock Cosmological billiards.
\newblock {\em Class. Quantum Grav.}, \textbf{20}:R145, 2003.

\bibitem{ellmac69}
G.~F.~R Ellis and M.~A.~H. MacCallum.
\newblock A class of homogeneous cosmological models.
\newblock {\em Commun. Math. Phys.}, \textbf{12}:108, 1969.

\bibitem{FourLuk}
G.~Fournodavlos and J.~Luk.
\newblock Asymptotically {K}asner-like singularities.
\newblock {\em Am. J. Math.}, \textbf{145}(4):1183--1272, 2023.

\bibitem{FourRodSpeck}
G.~Fournodavlos, I.~Rodnianski, and J.~Speck.
\newblock Stable big bang formation for {E}instein's equations: The complete
  sub-critical regime.
\newblock {\em J. Amer. Math. Soc.}, \textbf{36}:827--916, 2023.

\bibitem{Geroch66}
R.~P. Geroch.
\newblock Singularities in closed universes.
\newblock {\em Phys. Rev. Lett.}, \textbf{17}:445, 1966.

\bibitem{Groeniger}
H.~O. Groeniger.
\newblock Quiescence for the exceptional {B}ianchi cosmologies.
\newblock {\em \href{https://arxiv.org/abs/2311.05522}{arXiv:2311.05522}},
  2023.

\bibitem{RingstromQuiet}
H.~O. Groeniger, O.~Petersen, and H.~Ringström.
\newblock Formation of quiescent big bang singularities.
\newblock {\em \href{https://arxiv.org/abs/2309.11370}{arXiv:2309.11370}},
  2023.

\bibitem{HawkingEllis73}
S.~W. Hawking and G.~F.~R. Ellis.
\newblock {\em The Large Scale Structure of Space-Time}.
\newblock Cambridge University Press, 1973.

\bibitem{HawkingPenrose70}
S.~W. Hawking and R.~Penrose.
\newblock The singularities of gravitational collapse and cosmology.
\newblock {\em Proc. R. Soc. Lond. Ser. A}, \textbf{314}:529--548, 1970.

\bibitem{heietal09}
J.~M. Heinzle, N.~R{\"o}hr, and C.~Uggla.
\newblock The cosmological billiard attractor.
\newblock {\em Adv. Theor. Math. Phys.}, \textbf{13}(2):293--407, 2009.

\bibitem{heiugg09a}
J.~M. Heinzle and C.~Uggla.
\newblock Mixmaster: Fact and belief.
\newblock {\em Class. Quantum Grav.}, \textbf{26}:075016, 2009.

\bibitem{heiugg09}
J.~M. Heinzle and C.~Uggla.
\newblock A new proof of the {B}ianchi type {IX} attractor theorem.
\newblock {\em Class. Quantum Grav.}, \textbf{26}:075015, 2009.

\bibitem{heiugg10}
J.~M. Heinzle and C.~Uggla.
\newblock Monotonic functions in {B}ianchi models: why they exist and how to
  find them.
\newblock {\em Class. Quantum Grav.}, \textbf{27}:015009, 2010.

\bibitem{heietal12}
J.~M. Heinzle, C.~Uggla, and W.~C. Lim.
\newblock Spike oscillations.
\newblock {\em Phys. Rev. D}, \textbf{86}(10):104049, 2012.

\bibitem{HLU22}
J.~Hell, P.~Lappicy, and C.~Uggla.
\newblock Bifurcations and chaos in {H}o{ř}ava-{L}ifshitz cosmology.
\newblock {\em Adv. Theor. Math. Phys.}, \textbf{26}(7):2095, 2022.

\bibitem{heretal07}
S.~Hervik, R.~J. van~den Hoogen, W.~C. Lim, and A.~A. Coley.
\newblock Late-time behaviour of the tilted {B}ianchi type {VI}$_{-1/9}$
  models.
\newblock {\em Class. Quantum Grav.}, \textbf{25}:015002, 2008.

\bibitem{hew91}
C.~G. Hewitt.
\newblock An investigation of the dynamical evolution of a class of {B}ianchi
  {VI}$_{-1/9}$ cosmological models.
\newblock {\em Gen. Relativ. Gravit.}, \textbf{23}:691--712, 1991.

\bibitem{hewetal03}
C.~G. Hewitt, J.~T. Horwood, and J.~Wainwright.
\newblock Asymptotic dynamics of the exceptional {B}ianchi cosmologies.
\newblock {\em Class. Quantum Grav.}, \textbf{20}:1743, 2003.

\bibitem{hewetal93}
C.~G. Hewitt and J.~Wainwright.
\newblock A dynamical systems approach to {B}ianchi cosmologies: orthogonal
  models of class {B}.
\newblock {\em Class. Quantum Grav.}, \textbf{10}:99, 1993.

\bibitem{HodPiran}
S.~Hod and T.~Piran.
\newblock Mass inflation in dynamical gravitational collapse of a charged
  scalar field.
\newblock {\em Phys. Rev. Lett.}, \textbf{81}:1554, 1998.

\bibitem{jan01}
R.~T. Jantzen.
\newblock {\em Spatially Homogeneous Dynamics: A Unified Picture}.
\newblock Proc.\ Int.\ Sch.\ Phys.\ ``E. Fermi" Course LXXXVI on ``Gamov
  Cosmology", R. Ruffini, F. Melchiorri, Eds. (North Holland, Amsterdam, 1987)
  and in Cosmology of the Early Universe, R. Ruffini, L.Z. Fang, Eds. World
  Scientific, Singapore, 1984, 1984.

\bibitem{janugg99}
R.~T. Jantzen and C.~Uggla.
\newblock The kinematical role of automorphisms in the orthonormal frame
  approach to {B}ianchi cosmology.
\newblock {\em J. Math. Phys.}, \textbf{40}:353, 1999.

\bibitem{khaetal85}
I.~M. Khalatnikov, E.~M. Lifshitz, K.~M. Khanin, L.~N. Shur, and Y.~G. Sinai.
\newblock On the stochasticity in relativistic cosmology.
\newblock {\em J. Stat. Phys.}, \textbf{38}:97, 1985.

\bibitem{kingell73}
A.~R. King and G.~F.~R. Ellis.
\newblock Tilted homogeneous cosmological models.
\newblock {\em Commun. Math. Phys.}, \textbf{31}:209, 1973.

\bibitem{Kraetal03}
A.~Krasiński, C.~G. Behr, E.~Schücking, F.~B. Estabrook, H.~D. Wahlquist,
  G.~F.~R. Ellis, R.~Jantzen, and W.~Kundt.
\newblock The {B}ianchi classification in the {S}chücking-{B}ehr approach.
\newblock {\em Gen. Relativ. Gravit.}, \textbf{35}(3):475, 2003.

\bibitem{lapp22}
P.~Lappicy and V.~H. Daniel.
\newblock Chaos in spatially homogeneous {H}o{ř}ava--{L}ifshitz subcritical
  cosmologies.
\newblock {\em Class. Quantum Grav.}, \textbf{39}(13):135017, 2022.

\bibitem{Li}
W.~Li.
\newblock {BKL} bounces outside homogeneity: {G}owdy symmetric spacetimes.
\newblock {\em \href{https://arxiv.org/abs/2408.12427}{arXiv:2408.12427}},
  2024.

\bibitem{Lieb11}
S.~Liebscher, J.~Härterich, K.~Webster, and M.~Georgi.
\newblock Ancient dynamics in {B}ianchi models: Approach to periodic cycles.
\newblock {\em Commun. Math. Phys.}, \textbf{305}:59--83, 2011.

\bibitem{Lieb13}
S.~Liebscher, A.~Rendall, and S.~B. Tchapnda.
\newblock Oscillatory singularities in {B}ianchi models with magnetic fields.
\newblock {\em Ann. Henri Poincar\'e}, \textbf{14}:1043--1075, 2013.

\bibitem{lk63}
E.~M. Lifshitz and I.M. Khalatnikov.
\newblock Investigations in relativistic cosmology.
\newblock {\em Adv. Phys.}, \textbf{12}:185, 1963.

\bibitem{lim04}
W.~C. Lim.
\newblock {\em The Dynamics of inhomogeneous cosmologies}.
\newblock PhD Thesis, University of Waterloo, 2004.

\bibitem{lim15}
W.~C. Lim.
\newblock Non-orthogonally transitive {G2} spike solution.
\newblock {\em Class. Quantum Grav.}, \textbf{32}:162001, 2015.

\bibitem{lim22}
W.~C. Lim.
\newblock Numerical confirmations of joint spike transitions in {G2}
  cosmologies.
\newblock {\em Class. Quantum Grav.}, \textbf{39}(6):065001, 2022.

\bibitem{limmou22}
W.~C. Lim and M.~Z.~A. Moughal.
\newblock Transition analysis of the non-{OT} {G2} stiff fluid spike solution.
\newblock {\em Class. Quantum Grav.}, \textbf{39}(2):025010, 2022.

\bibitem{limetal06}
W.~C. Lim, C.~Uggla, and J.~Wainwright.
\newblock Asymptotic silence-breaking singularities.
\newblock {\em Class. Quantum Grav.}, \textbf{23}:2607, 2006.

\bibitem{Luk}
J.~Luk.
\newblock Weak null singularities in general relativity.
\newblock {\em J. Amer. Math. Soc.}, \textbf{31}(1):1--63, 2018.

\bibitem{Luk2}
J.~Luk.
\newblock {\em Singularities in general relativity}.
\newblock Proc. International Congress of Mathematicians (D.Beliaev and S.
  Smirnov, eds.), ICM 2022, Volume 5, Sections 9--11. Berlin: European
  Mathematical Society (EMS), 2023.

\bibitem{mac73}
M.~A.~H. MacCallum.
\newblock Cosmological models from a geometric point of view.
\newblock {\em Carg\`{e}se lectures, ed. E. Schatzman.}, \textbf{6}:61, 1973.

\bibitem{mis69a}
C.~W. Misner.
\newblock Mixmaster universe.
\newblock {\em Phys. Rev. Lett.}, \textbf{22}:1071, 1969.

\bibitem{Nutzi}
A.~Nützi, M.~Reiterer, and E.~Trubowitz.
\newblock Semiglobal non-oscillatory big bang singular spacetimes for the
  {E}instein-scalar field system.
\newblock {\em \href{https://arxiv.org/abs/2005.03395}{arXiv:2005.03395}},
  2020.

\bibitem{OriFlanagan}
A.~Ori and É.~É. Flanagan.
\newblock How generic are null spacetime singularities?
\newblock {\em Phys. Rev. D}, \textbf{53}:R1754(R), 1996.

\bibitem{Penrose65}
R.~Penrose.
\newblock Gravitational collapse and space-time singularities.
\newblock {\em Phys. Rev. Lett.}, \textbf{14}:57--59, 1965.

\bibitem{Radermacher}
K.~Radermacher.
\newblock Strong cosmic censorship in orthogonal {B}ianchi class {B} perfect
  fluids and vacuum models.
\newblock {\em Ann. Henri Poincar\'e}, \textbf{20}:689--796, 2019.

\bibitem{ReitererTrubowitz}
M.~Reiterer and E.~Trubowitz.
\newblock The {BKL} conjectures for spatially homogeneous spacetimes.
\newblock {\em \href{https://arxiv.org/abs/1005.4908}{arXiv:1005.4908}}, 2010.

\bibitem{RendallSurvey}
A.~Rendall.
\newblock {\em The nature of spacetime singularities}.
\newblock In Ashtekar, A.: 100 years of relativity. Spacetime structure:
  {E}instein and beyond. World Scientific, Singapore., 2005.

\bibitem{rin00}
H.~Ringstr\"om.
\newblock Curvature blow up in {B}ianchi {VIII} and {IX} vacuum spacetimes.
\newblock {\em Class. Quantum Grav.}, \textbf{17}:713, 2000.

\bibitem{rin01}
H.~Ringstr\"om.
\newblock The {B}ianchi {IX} attractor.
\newblock {\em Ann. Henri Poincar\'e}, \textbf{2}:405, 2001.

\bibitem{rin25}
H.~Ringstr\"om.
\newblock Cosmology, the big bang and the {BKL} conjecture.
\newblock {\em Comptes Rendus. Mécanique}, \textbf{353}:405, 2025.

\bibitem{RodSpeck}
I.~Rodnianski and J.~Speck.
\newblock A regime of linear stability for the {E}instein-scalar field system
  with applications to nonlinear big bang formation.
\newblock {\em Ann. of Math.}, \textbf{187}(1):65--156, 2018.

\bibitem{rohugg05}
N.~R\"ohr and C.~Uggla.
\newblock Conformal regularization of {E}instein's field equations.
\newblock {\em Class. Quantum Grav.}, \textbf{22}:3775, 2005.

\bibitem{Schmidt69}
W.~Schmidt.
\newblock Badly approximable systems of linear forms.
\newblock {\em J. Number Theory}, \textbf{1}:139--154, 1969.

\bibitem{Schmidt}
W.~Schmidt.
\newblock {\em Diophantine Approximation}.
\newblock Lecture Notes in Mathematics, 785. Springer, Berlin, 1980.

\bibitem{Senovilla15}
J.~M.~M. Senovilla and D.~Garfinkle.
\newblock The 1965 {P}enrose singularity theorem.
\newblock {\em Class. Quantum Grav.}, \textbf{32}(12):124008, 2015.

\bibitem{Fishman}
D.~Simmons, L.~Fishman, and M.~Urbański.
\newblock Diophantine properties of measures invariant with respect to the
  {G}auss map.
\newblock {\em J. d'Analyse Mathématique}, \textbf{122}:289--315, 2014.

\bibitem{steetal03}
H.~Stephani, D.~Kramer, M.~MacCallum, C.~Hoenselaers, and E.~Herlt.
\newblock {\em Exact Solutions of {E}instein's Field Equations}.
\newblock Cambridge University Press, 2 edition, 2003.

\bibitem{ugg13}
C.~Uggla.
\newblock Recent developments concerning generic spacelike singularities.
\newblock {\em Gen. Relativ. Gravit.}, \textbf{45}:1669--1710, 2013.

\bibitem{ugg13c}
C.~Uggla.
\newblock Spacetime singularities: Recent developments.
\newblock {\em Int. J. Mod. Phys. D}, \textbf{22}:1330002, 2013.

\bibitem{uggetal95}
C.~Uggla, R.~T. Jantzen, and K.~Rosquist.
\newblock Exact hypersurface-homogeneous solutions in cosmology and
  astrophysics.
\newblock {\em Phys. Rev. D}, \textbf{51}:5522, 1995.

\bibitem{VandeMoortel}
M.~{Van de Moortel}.
\newblock The breakdown of weak null singularities inside black holes.
\newblock {\em Duke Math. J.}, \textbf{172}(15):2957--3012, 2023.

\bibitem{waiell97}
J.~Wainwright and G.~F.~R. Ellis.
\newblock {\em Dynamical systems in cosmology}.
\newblock Cambridge University Press, 1997.

\bibitem{waihsu89}
J.~Wainwright and L.~Hsu.
\newblock A dynamical systems approach to {B}ianchi cosmologies: orthogonal
  models of class {A}.
\newblock {\em Class. Quantum Grav.}, \textbf{6}:1409, 1989.

\end{thebibliography}

\end{document}